\documentclass[aps,reprint,superscriptaddress,nofootinbib,twocolumn]{revtex4-2}
\usepackage[utf8]{inputenc}\usepackage[T1]{fontenc}
\usepackage{amsmath}
\usepackage{amssymb}
\usepackage{graphicx}
\usepackage{slashed}
\usepackage{cases}
\usepackage[toc,page]{appendix}
\usepackage{empheq}
\usepackage[normalem]{ulem} 
\usepackage{xcolor} 
\usepackage{afterpage}
\usepackage{float}
\usepackage{bbold}
\usepackage{hypbmsec}
\usepackage[colorlinks = true,
            linkcolor = blue,
            urlcolor  = blue,
            citecolor = blue,
            anchorcolor = blue]{hyperref}
\allowdisplaybreaks
\usepackage[export]{adjustbox}

\makeatletter
\newcommand\fake@math{}
\def\fake@math#1\){[math]}
\makeatother

\begin{document}

\title{Interplay between Resonant Leptogenesis, Neutrinoless Double Beta Decay and Collider Signals in  a Model with Flavor and CP Symmetries}

\author{Garv Chauhan}
    \email{gchauhan@vt.edu}
    \affiliation{Center for Neutrino Physics, Department of Physics, Virginia Tech, Blacksburg, VA 24061, USA}
    \affiliation{Centre for Cosmology, Particle Physics and Phenomenology (CP3), 
    Universit\'{e} Catholique de Louvain, Chemin du Cyclotron 2, 
    B-1348 Louvain-la-Neuve, Belgium}
    \affiliation{Department of Physics and McDonnell Center for the Space Sciences,  Washington University, 
    St. Louis, MO 63130, USA}
\author{P. S. Bhupal Dev}
    \email{bdev@wustl.edu}
    \affiliation{Department of Physics and McDonnell Center for the Space Sciences,  Washington University, 
    St. Louis, MO 63130, USA}

\begin{abstract}
We present a low-scale type-I seesaw scenario with discrete flavor and CP symmetries. This scenario not only explains the measured values of the lepton mixing angles, but also makes predictions for leptonic CP violation, and connects the low-energy CP phases relevant for neutrino oscillation and neutrinoless double beta decay experiments with the high-energy CP phases relevant for leptogenesis. We show that the three right-handed Majorana neutrinos in this scenario have (almost) degenerate masses and their decays can explain the observed baryon asymmetry of the Universe via resonant leptogenesis. We study the correlation of the predicted baryon asymmetry with  lepton-number-violating signals at high-energy colliders, including both prompt and displaced vertex/long-lived signatures, as well as in low-energy neutrinoless double beta decay experiments. We find that the normal ordering of light neutrino masses leads to an enhanced collider signal, whereas the neutrinoless double beta decay provides a promising probe in the inverted  ordering case.
\end{abstract}


\maketitle

\section{Introduction}
\label{sec:intro}
The origin of the observed baryon asymmetry of the Universe (BAU)~\cite{Aghanim:2018eyx} remains an open question, as it calls for some beyond the Standard Model (BSM) physics. A dynamical generation of the BAU requires fulfilling the three necessary (but not sufficient) Sakharov conditions~\cite{Sakharov:1967dj}, namely, baryon number (B)  violation, charge conjugation (C) and charge conjugation-parity (CP) violation, and departure from thermal equilibrium. There exist several viable baryogenesis mechanisms satisfying these basic Sakharov conditions; see Ref.~\cite{Bodeker:2020ghk} for a recent review. A particularly attractive mechanism is leptogenesis~\cite{Fukugita:1986hr}, which can potentially link the BAU to another outstanding puzzle that also necessarily requires BSM physics, namely, the origin of neutrino mass; see Refs.~\cite{Buchmuller:2005eh, Davidson:2008bu, Fong:2012buy} for reviews on various aspects of leptogenesis. 

The central idea of thermal leptogenesis is the production of a net leptonic asymmetry in the early Universe, via the CP-violating out-of-equilibrium decays of heavy right-handed neutrinos (RHNs), which is then converted to a net baryon asymmetry through non-perturbative electroweak sphaleron processes~\cite{Kuzmin:1985mm}. The same complex Yukawa interactions ($Y_D$) of the RHNs with the SM lepton ($L$) and Higgs ($H$) doublets, in conjunction  with the Majorana masses ($M_R$) of the RHNs, are also responsible for the neutrino mass generation via the type-I seesaw mechanism~\cite{Minkowski:1977sc, Mohapatra:1979ia, Yanagida:1979as, GellMann:1980vs, Glashow:1979nm, Schechter:1980gr}:
\begin{align}
    M_\nu \ \simeq \ -M_D M_R^{-1} M_D^T \, ,
    \label{eq:seesaw}
\end{align}
where $M_D=vY_D$ is the Dirac neutrino mass matrix and $v\simeq 174$ GeV is the electroweak vacuum expectation value (VEV). 
In the so-called vanilla leptogenesis scenario with hierarchical RHNs, the maximum CP violation from the complex Yukawa interactions can be related to the lightest RHN mass through the seesaw formula, and the requirement of successful leptogenesis imposes a lower limit on $M_R\gtrsim 10^9$ GeV, known as the Davidson-Ibarra bound~\cite{Davidson:2002qv}. Although this limit can be somewhat relaxed by including flavor effects and with some cancellation between the tree and one-loop level contributions to the light neutrino mass matrix~\cite{Moffat:2018wke}, it still remains beyond the reach of any foreseeable laboratory experiment. An appealing alternative is the resonant leptogenesis mechanism~\cite{Pilaftsis:2003gt} (see Ref.~\cite{Dev:2017wwc} for a review) in which the CP asymmetry can be resonantly enhanced~\cite{Flanz:1994yx,Covi:1996wh, Flanz:1996fb}, even up to order one~\cite{Pilaftsis:1997dr, Pilaftsis:1997jf}, if at least two RHNs are quasi-degenerate, thereby allowing  significantly lower values of $M_R$ all the way down to the electroweak scale~\cite{Pilaftsis:2005rv, Deppisch:2010fr},\footnote{The RHNs can be even lighter, as low as GeV-scale, in the parametric regime of ARS leptogenesis~\cite{Drewes:2017zyw}.}  while maintaining agreement with neutrino oscillation data.
 Moreover, embedding such a resonant leptogenesis mechanism into a gauge theory, in which the RHNs exist naturally for anomaly cancellation, makes it testable at the colliders via searches for RHNs and new gauge/scalar bosons~\cite{Frere:2008ct, Blanchet:2009bu, Blanchet:2010kw, Iso:2010mv, Okada:2012fs, Fong:2013gaa, Dev:2014hro, Dev:2015khe,Dhuria:2015cfa, Heeck:2016oda,Caputo:2017pit, Geng:2017foe, Dev:2017xry,Gu:2017gra, Dev:2019ljp,Borah:2021mri, Liu:2021akf}; see Ref.~\cite{Chun:2017spz} for a review on the testable signatures of leptogenesis.

Although the seesaw connection to leptogenesis is qualitatively attractive, it can make quantitative predictions consistent with the low-energy neutrino data only after the flavor structures of $Y_D$ and $M_R$ are specified~\cite{Xing:2020ijf}. Without loss of generality, we can always choose the basis in which the charged-lepton mass matrix $M_\ell$ and the RHN mass matrix $M_R$ are diagonal. In this case the Dirac Yukawa coupling matrix $Y_D$ can be conveniently written in the so-called Casas-Ibarra (CI) parametrization~\cite{Casas:2001sr}:
\begin{align}
    Y_D \, = \, \frac{i}{v}U\sqrt{\hat{D}_\nu}O\sqrt{\hat{D}_R} \, ,
    \label{eq:CI}
\end{align}
where $U$ is the Pontecorvo-Maki-Nakagawa-Sakata (PMNS) neutrino mixing matrix used to diagonalize $M_\nu$ [cf.~Eq.~\eqref{eq:seesaw}] in the chosen basis, i.e.~$U^\dag M_\nu U^\star=\hat{D}_\nu={\rm diag}(m_1,m_2,m_3)$; $\hat{D}_R={\rm diag}(M_1,M_2,M_3)$ is the diagonal RHN mass matrix; and $O$ is an arbitrary complex orthogonal matrix. It is the complex phases hidden in $Y_D$ that govern the CP asymmetries $\varepsilon_{i\alpha}$ between the lepton-number-violating (LNV) decays of the RHNs $N_i\to L_\alpha+H$ and $N_i\to \bar L_\alpha+H^c$ (where $H^c\equiv \epsilon H^\star$, $\epsilon$ being the $SU(2)$ antisymmetric tensor). In particular, the flavored CP asymmetries $\varepsilon_{i\alpha}$ depend on both $(Y_D^\star)_{\alpha i}(Y_D)_{\alpha j}$ and $(Y_D^\dag Y_D)_{ij}$ (for $j\neq i$), whereas the unflavored asymmetries $\varepsilon_i\equiv \sum_{\alpha} \varepsilon_{i\alpha}$ only depend on the combination $(Y_D^\dag Y_D)_{ij}$~\cite{Dev:2017trv}. Given the CI parametrization of $Y_D$ [cf.~Eq.~\eqref{eq:CI}], one can immediately see that the CP asymmetries $\varepsilon_i$ have nothing to do with the PMNS mixing matrix $U$. In other words, due to the arbitrariness of the $O$ matrix in Eq.~\eqref{eq:CI}, the high-energy CP phases present in the Yukawa couplings $Y_D$ that are responsible for leptogenesis are in general unrelated to the low-energy CP phases in $U$~\cite{Branco:2001pq,Rebelo:2002wj, Pascoli:2003uh, Xing:2009vb, Rodejohann:2009cq, Antusch:2009gn}, i.e.~the Dirac phase $\delta$ that is measurable in neutrino oscillation experiments~\cite{Feldman:2012jdx} and a combination of the two Majorana phases $\alpha_1$ and $\alpha_2$ that is potentially measurable in neutrinoless double beta decay $(0\nu\beta\beta)$ experiments~\cite{Dolinski:2019nrj}. It is only when we assume $O$ in Eq.~\eqref{eq:CI} to be real, the flavored CP asymmetries $\varepsilon_{i\alpha}$ will depend directly on the PMNS phases~\cite{Pascoli:2006ie,Branco:2006ce, Pascoli:2006ci, Anisimov:2007mw,  Molinaro:2009lud,Molinaro:2008cw, Moffat:2018smo}; see Ref.~\cite{Hagedorn:2017wjy} for a review on the implications of low-energy leptonic CP violation for leptogenesis.

Since the experiments are only sensitive to the low-energy CP phases, it is desirable to have a theoretical setup where the high- and low-energy CP phases can be related, which can provide a highly predictive leptogenesis scenario. Examples are models with residual flavor and CP symmetries~\cite{Chen:2016ptr, Hagedorn:2016lva, Li:2017zmk, Samanta:2018efa, Fong:2021tqj}. In particular, a non-abelian discrete flavor symmetry $G_f$ combined with a CP symmetry, both acting non-trivially on flavor space, turns out to be highly constraining for the lepton mixing angles and both low- and high-energy CP phases,  because in this case the PMNS mixing matrix depends on a single free parameter~\cite{Feruglio:2012cw, Holthausen:2012dk, Chen:2014tpa}. In this paper we will adopt this setup with $G_f$ being a member of the series of groups $\Delta(3\, n^2)$~\cite{Luhn:2007uq} or $\Delta(6\, n^2)$~\cite{Escobar:2008vc} which are known to give several interesting neutrino mixing patterns~\cite{King:2014rwa, Hagedorn:2014wha, Ding:2014ora, Ding:2015rwa}. In particular, we adopt the model framework presented in Refs.~\cite{Hagedorn:2014wha, Hagedorn:2016lva} and explore the possibility of low-scale leptogenesis and its correlation with collider and $0\nu\beta\beta$ signals. Specifically, the predicted BAU is shown to be correlated with the low-energy $0\nu\beta\beta$ rate that can in principle be tested in future tonne-scale $0\nu\beta\beta$ 
experiments. In addition, we identify points of {\it enhanced residual symmetry} (ERS), where one of the three RHNs is long-lived and can be detected in dedicated long-lived particle (LLP) searches~\cite{Alimena:2019zri, Alimena:2021mdu}, while at the same time the remaining two RHNs can be searched for via either prompt or displaced vertex signals at the LHC~\cite{Deppisch:2015qwa, Cai:2017mow}, provided the RHNs are charged under a new gauge group, so that their production cross section is not necessarily suppressed by the small light-heavy neutrino mixing. We take this new gauge group to be $U(1)_{B-L}$ with the associated $Z'$ boson in the multi-TeV range and scan the parameter space in the $(M_{Z'},M_R)$ plane to find regions that lead to successful leptogenesis. The detection prospects of these regions using LNV signals at the high-luminosity LHC~\cite{ZurbanoFernandez:2020cco}, as well as at a futuristic 100 TeV hadron collider~\cite{FCC:2018vvp}, are also discussed.

Th rest of the paper is organized as follows: In section~\ref{sec:frame}, we discuss the embedding of the charged-lepton sector and light neutrinos (with masses arising from the type-I seesaw mechanism) with three RHNs in the flavor group $G_f=\Delta(6\, n^2)$ and a CP symmetry group, with the Yukawa sector presented in section~\ref{sec:2.1}, RHN mass spectrum in section~\ref{sec:2.2}, and light neutrino masses and mixing in section~\ref{sec:2.3}. In section~\ref{sec:cases}, we discuss the residual symmetries and the form of the corresponding representation matrices for the different cases, along with  additional constraints imposed from light neutrino masses. In section~\ref{sec:CPasymmetries}, we discuss the CP asymmetries produced in our scenario through out-of-equilibrium decays of the quasi-degenerate RHNs via resonant leptogenesis. In section~\ref{sec:colliderLep}, we study the collider phenomenology of our scenario, starting with the production of RHNs in section~\ref{subsec:productionRHnu}, their  decay lengths in  section~\ref{subsec:RHnuwidths} and branching ratios (BRs) in section~\ref{subsec:RHnuBRs}, same-sign dilepton signals in section~\ref{subsec:SSleptonsLHCRHnu}, and finally the correlation with leptogenesis in section~\ref{subsec:colliderleptogenesis}. The correlation of BAU with $0\nu\beta\beta$ is studied in section~\ref{sec:leptogenesis0nubb}. Our conclusions are given in  section~\ref{sec:concl}. Appendix~\ref{appA} reviews the group theory of $\Delta(6 \, n^2)$ and representation matrices. Appendix~\ref{appC} discusses the CP symmetries and the form of CP transformations. Appendix~\ref{appB} gives the form of the representation matrices for residual symmetries. Appendix~\ref{appD} gives the explicit form of the Yukawa couplings in terms of the other model parameters.

\section{Framework}
\label{sec:frame}

We focus on the lepton sector and assume that light neutrino masses arise from the type-I seesaw mechanism  with three RHNs [cf.~Eq.~\eqref{eq:seesaw}]. 
 We consider a scenario with a flavor symmetry $G_f$ and a CP symmetry which are broken into residual groups $G_\ell$ and $G_\nu$ that determine the forms of the charged lepton and neutrino mass matrices, respectively.
 For a review on lepton flavor symmetries, see e.g.~Ref.~\cite{Feruglio:2019ybq}.

For flavor symmetry $G_f$, following the approach to lepton mixing presented in Ref.~\cite{Hagedorn:2014wha}, we use a discrete, finite, non-abelian group of the form $\Delta (6 \, n^2)$~\cite{Escobar:2008vc} with $n$ even and not divisible by three.\footnote{We could also consider a group of the form $\Delta (3 \, n^2)$~\cite{Luhn:2007uq}, but since this is a subgroup of $\Delta (6 \, n^2)$, it is sufficient to focus on the latter group only.} Note that $\Delta(6n^2)\sim (Z_n\times Z_n)\rtimes S_3$ is one of the simplest non-abelian finite subgroups of $SU(3)$. The group theory and relevant representation matrices of $\Delta (6 \, n^2)$ are given in Appendix~\ref{appA}. The groups $\Delta (6 \, n^2)$ for $n \geq 2$ are particularly interesting, as they possess at least one irreducible, faithful, complex three-dimensional 
representation ${\bf 3}$. 
In the following, we assign the three generations of left-handed (LH) lepton doublets $L_\alpha$ (with $\alpha=e , \, \mu, \, \tau$) to this ${\bf 3}$. The three generations of RHNs $N_i$ (with $i=1, 2, 3$) are unified in an irreducible, in general unfaithful, real representation ${\bf 3^\prime}$
of $G_f$ which requires the index $n$ of the group $\Delta (6 \, n^2)$ to be even; see Appendix~\ref{appA} for details.\footnote{Only for $n=2$ this representation is faithful. This, however, does not affect our discussion.} Assigning LH leptons and RHNs to these in general different
three-dimensional representations of $G_f$ is crucial, as the assignment $L_\alpha \sim {\bf 3}$ allows to fully explore the predictive power of $G_f$ (and not only of one of its subgroups),
while $N_i \sim {\bf 3^\prime}$ permits the RHNs to have a flavor-universal mass term without breaking $G_f$ and the CP symmetry.  The right-handed charged leptons $\alpha_R$ are assigned to the trivial one-dimensional representation
${\bf 1}$ of $G_f$. In order to distinguish them, we assume the existence of a $Z_3$
symmetry, called $Z_3^{(\mathrm{aux})}$, under which $e_R$, $\mu_R$ and $\tau_R$ are assigned charges $1$, $\omega$ and $\omega^2$ respectively (with $\omega=e^{2 \pi {i}/3}$ being the third root of unity), whereas LH leptons and RHNs are assumed to be invariant under $Z_3^{(\mathrm{aux})}$. The relevant particle content and the flavor symmetries are summarized in Table~\ref{tab:sym}.

\begin{table}[]
    \label{table:charges}
    \centering
    \begin{tabular}{|c | c | c | c | c |} 
    \hline
    &  $SU(2)_L$ & $U(1)_{Y}$ & $G_f=\Delta (6 \, n^2) $ & $Z_3^{(\mathrm{aux})}$  \\ \hline
    {$L_\alpha\equiv \begin{pmatrix}\nu_{L\alpha} \\ \alpha_{L}\end{pmatrix}$} & $\mathbf{2}$ & $-1$ & $\mathbf{3}$ & -  \\ \hline
    {$e_R$} & $\mathbf{1}$ & $-2$ & $\mathbf{1}$ & 1  \\ 
    {$\mu_R$} & $\mathbf{1}$ & $-2$ & $\mathbf{1}$ & $\omega$  \\ 
    {$\tau_R$} & $\mathbf{1}$ & $-2$ & $\mathbf{1}$ & $\omega^2$  \\ \hline
    $N_i$  & $\mathbf{1}$ & 0 & $\mathbf{3^\prime}$ & -   \\ \hline
    \end{tabular}
    \caption{The relevant particle content and their transformations under the given groups used in this work.}
    \label{tab:sym}
\end{table}

The CP symmetry
imposed on the theory corresponds to an automorphism of $G_f$~\cite{Holthausen:2012dk,Chen:2014tpa}. They are represented by the CP transformation $X (\mathrm{\mathbf{r}})$
in the different (irreducible) representations ${\mathrm{\mathbf{r}}}$ of $G_f$ and depend on the parameters determining the automorphism. 
 For completeness, we show the form of the automorphisms and of $X (\mathrm{\mathbf{r}})$
for the relevant representations in Appendix~\ref{appC}.

The residual symmetry in the charged lepton sector is chosen as $G_\ell=Z_3^{(\mathrm{D})}$ which is the diagonal abelian subgroup of the $Z_3$ group contained in $G_f$ and the additional auxiliary group $Z_3^{(\mathrm{aux})}$ (see Appendix~\ref{appA}).  In the neutrino sector, we use the residual symmetry $G_\nu=Z_2 \times {\rm CP}$, where the $Z_2$ symmetry is a subgroup of $G_f$ and the CP symmetry the one of the underlying theory (see Appendix~\ref{appC}). 
In the following, the generator $Z$ of the residual $Z_2$ symmetry in the different representations $\mathrm{\mathbf{r}}$ is denoted as $Z (\mathrm{\mathbf{r}})$.
The $Z_2$ symmetry and CP commute, i.e. they fulfill
\begin{equation}
X (\mathrm{\mathbf{r}}) \, Z (\mathrm{\mathbf{r}}) - Z (\mathrm{\mathbf{r}})^\star \, X (\mathrm{\mathbf{r}}) \, = \, 0
\end{equation}
for all representations $\mathrm{\mathbf{r}}$ of $G_f$. The mismatch of the residual symmetries $G_\ell$ and $G_\nu$ determines the form of lepton mixing~\cite{Hagedorn:2014wha, Hagedorn:2016lva}. Specifically, it has been found that lepton mixing patterns can be classified into four distinct types, called Case 1, Case 2, Case 3a and Case 3b.1 in Ref.~\cite{Hagedorn:2016lva}, depending on the choices of $Z (\mathrm{\mathbf{r}})$ and $X (\mathrm{\mathbf{r}})$. The forms of the lepton mixing matrices for 
these four different types are shown in section~\ref{sec:cases}; see also Appendix~\ref{appB}. The aim of this paper is to study resonant leptogenesis and its correlation with $0\nu\beta\beta$ and collider signals for these four representative cases, which have different mixing patterns.  

\subsection{Yukawa Sector}
\label{sec:2.1}
The forms of the charged lepton mass matrix $M_\ell$, the neutrino Yukawa coupling matrix $Y_D$ and the RHN Majorana mass matrix $M_R$ are determined by $G_\ell$ and $G_\nu$. The form of the relevant Lagrangian is 
\begin{align}
   - {\cal L} \ \supset \ (Y_\ell)_{\alpha\beta} \bar{L}_\alpha H \beta_R + (Y_D)_{\alpha i}\bar{L}H^c N_i \nonumber \\ + \frac{1}{2}(M_R)_{ij}\overline{N_i^c}N_j+{\rm H.c.} \, ,
   \label{eq:lag}
\end{align}
where the superscript $c$ denotes charge conjugation. 
In our chosen basis (see Appendix~\ref{appA}), the charged lepton mass matrix $M_\ell=vY_\ell$ is diagonal and contains three independent parameters corresponding to the three charged lepton
masses, i.e.~$M_\ell={\rm diag}(m_e, m_\mu,  m_\tau)$. As $M_\ell$ is diagonal, there is no contribution to lepton mixing from the charged lepton sector. As for the neutrino sector, we take the neutrino Yukawa coupling matrix $Y_D$
to be invariant under $G_\nu$, whereas the matrix $M_R$ does neither break $G_f$ nor CP.  Being invariant under $Z_2 \times {\rm CP}$, the matrix $Y_D$, in the basis in which LH fields are on the left and RH
ones on the right [cf.~Eq.~\eqref{eq:lag}], fulfills the following relations 
\begin{equation}
\label{YDZX}
Z ({\bf 3})^\dagger \, Y_D \, Z ({\bf 3^\prime}) \, = \,  Y_D \;\;\; \mbox{and} \;\;\; X ({\bf 3})^\star \, Y_D \, X ({\bf 3^\prime}) \, = \, Y_D^\star \; .
\end{equation}
We can rewrite the conditions in Eq.~(\ref{YDZX}) using the unitary matrices
$\Omega ({\bf 3})$ and $ \Omega ({\bf 3^\prime})$, which are determined by the form of the CP transformations $X ({\bf 3})$ and $X ({\bf 3^\prime})$ in the representations of LH leptons and RHNs, i.e. they fulfil
\begin{equation}
X ({\bf 3}) \, = \, \Omega ({\bf 3}) \, \Omega ({\bf 3})^T \;\; \mbox{and} \;\; X ({\bf 3^\prime}) \, = \, \Omega ({\bf 3^\prime}) \, \Omega ({\bf 3^\prime})^T \, .
\end{equation}
We find that $\Omega ({\bf 3})^\dagger \, Y_D \,  \Omega ({\bf 3^\prime})$ 
is real and can be diagonalized by two rotation matrices from the left and right, respectively: 
\begin{equation}
 \Omega ({\bf 3})^\dagger \, Y_D \,  \Omega ({\bf 3^\prime}) \, = \, R_{ij} (\theta_L) \, \left(
\begin{array}{ccc}
 y_1 & 0 & 0\\
 0 & y_2 & 0\\
 0 & 0 & y_3
 \end{array}
 \right) \, R_{kl} (-\theta_R) \, ,
 \label{eq:2.4}
\end{equation}
and the form of $Y_D$ is thus given by 
\begin{equation}
\label{eq:YDgen}
Y_D \, = \, \Omega ({\bf 3}) \, R_{ij} (\theta_L) \, \left(
\begin{array}{ccc}
 y_1 & 0 & 0\\
 0 & y_2 & 0\\
 0 & 0 & y_3
 \end{array}
 \right) \, R_{kl} (-\theta_R) \, \Omega ({\bf 3^\prime})^\dagger \; .
\end{equation}
The matrices $R_{ij} (\theta_L)$ and $R_{kl} (\theta_R)$ denote
rotations in the $(ij)$ and $(kl)$ planes (where $i,j,k,l=1,2,3$ with $i< j$ and $k< l$) through the angles $\theta_L$ and $\theta_R$, respectively. For instance,  
\begin{align}
R_{12} (\theta) & = \left(
\begin{array}{ccc}
 \cos\theta & \sin\theta & 0\\
 -\sin\theta & \cos\theta & 0\\
 0 & 0 & 1
\end{array}
\right) \, , 
\end{align}
The angles $\theta_L$ and $\theta_R$ in Eq.~\eqref{eq:YDgen} are free parameters,
i.e. not fixed by the residual symmetry $G_\nu$, and can take values in the range $[0,\pi)$. The planes, in which the rotations $R_{ij} (\theta_L)$ and $R_{kl} (\theta_R)$ act, are determined 
by the $(ij)$- and $(kl)$-subspaces of degenerate eigenvalues of the generator $Z$ in the representation ${\bf 3}$ and ${\bf 3^\prime}$, when transformed with the matrix $\Omega ({\bf 3})$
and $\Omega ({\bf 3^\prime})$, respectively.\footnote{Examples can be found in the discussion of the different cases (Case 1 through Case 3b.1) in section~\ref{sec:cases}.} 
In addition to these two angles, $Y_D$ contains  three more real parameters, namely, the Yukawa
couplings $y_f$ (with $f=1,2,3$). As the choice of CP symmetry and thus the corresponding CP transformations $X ({\bf 3})$ is in general indicated by natural numbers [see e.g. the parameter $s$ in Eq.~\eqref{Xs3}], also
the matrices  $\Omega ({\bf 3})$ and $\Omega ({\bf 3^\prime})$ (potentially) depend on these parameters; see section~\ref{sec:cases} for more details. 

\subsection{RHN Mass Spectrum}
\label{sec:2.2}
As the RHN Majorana mass matrix $M_R$ leaves $G_f$ and CP invariant, its form is simply 
\begin{equation}
M_R \, = \, M_N \, \left(
\begin{array}{ccc}
1 & 0 & 0\\
0 & 0 & 1\\
0 & 1 & 0
\end{array}
\right) \, ,
\label{eq:MR}
\end{equation}
with $M_N > 0$, setting the overall mass scale of the RHNs. From Eq.~\eqref{eq:MR}, it is clear that the three RHNs are exactly degenerate in the flavor symmetry limit. However, in order to successfully generate the BAU via resonant leptogenesis, the masses of at least two of the RHNs have to be slightly different, such that when the mass difference is of the order of their decay width, we can get the resonant enhancement in the self-energy contribution to the CP asymmetry~\cite{Dev:2017wwc}.
This can be achieved by small corrections $\delta M_R$ to the RHN Majorana mass matrix in Eq.~\eqref{eq:MR}. These corrections are expected to arise by (higher order) residual symmetry breaking effects which are generically present in concrete model realizations~\cite{Ishimori:2010au, King:2013eh, Feruglio:2019ybq}. In the following, without specifying any particular breaking 
mechanism or flavon dynamics, we consider general corrections to $M_R$ which are invariant under the residual symmetry $G_\ell$. The generator of $G_\ell$ is represented in the representation
of the RHNs $N_i$ as
\begin{equation}
 a ({\bf 3^\prime}) \, = \, \left( \begin{array}{ccc}
1 & 0 & 0\\
0 & \omega & 0\\
0 & 0 & \omega^2
\end{array}
\right) \; ,
\end{equation}
since $N_i$'s are not charged under the auxiliary symmetry $Z_3^{(\mathrm{aux})}$. The correction $\delta M_R$ must thus fulfil
\begin{equation}
a({\bf 3^\prime})^T \, \delta M_R \, a ({\bf 3^\prime}) \, = \, \delta M_R \; ,
\end{equation}
meaning that it is of the form
\begin{equation}
\label{dMRtilde}
\delta M_R \, = \, \kappa \, M_N \, \left( \begin{array}{ccc}
2 & 0 & 0\\
0 & 0 & -1\\
0 & -1 & 0 
\end{array}
\right) \, ,
\end{equation}
with $\kappa\ll 1$ being a small dimensionless symmetry breaking parameter. The RHN masses $M_i$ (with $i=1,2,3$) then acquire a small correction, as follows:
\begin{equation}
M_1 \, = \, M_N \, (1+ 2 \, \kappa) \;\; \mbox{and} \;\; M_2 \, = \, M_3 \, = \, M_N \, (1-\kappa) \, .
\label{eq:massesNi}
\end{equation}
It is the mass splitting between $N_1$ and $N_{2,3}$ that turns out to be relevant for resonant leptogenesis in our model. In the following numerical analysis, we will treat $\kappa$ as a free parameter and choose its value suitably in order to maximize the CP asymmetry for each case.  

We emphasize here that the higher-dimensional operators connecting different sectors of the theory
are responsible for the eventual breaking of the residual symmetries $G_\nu$ and $G_\ell$ and thus
 affect the given forms of $M_D$, $M_\ell$ and $M_R$. In particular, they are the source of
corrections leading to the small splitting in the RHN masses which  is crucial for resonant leptogenesis. As we will see in section~\ref{sec:cases}, the same higher-dimensional operators also source corrections to the tribimaximal (TB) form of the PMNS mixing matrix which is needed to explain the nonzero reactor mixing angle. Therefore, this  provides a natural setup for motivating the quasi-degeneracy of the RHNs for resonant leptogenesis. This is reminiscent of radiative resonant leptogenesis models~\cite{GonzalezFelipe:2003fi, Turzynski:2004xy, Branco:2005ye, Ahn:2006rn, Babu:2008kp, Dev:2015wpa, Zhao:2020bzx} where the RHN mass matrix is $O(N)$-symmetric at high scale, and small $O(N)$-breaking effects are induced naturally at low scale due to renormalization group evolution effects. 

\subsection{Light Neutrino Masses and Mixing}
\label{sec:2.3}

The light neutrino mass matrix $M_\nu$ follows from the type-I seesaw formula, cf.~Eq.~\eqref{eq:seesaw}. 
As the charged lepton mass matrix $M_\ell$ is diagonal, lepton mixing arises from the diagonalization of $M_\nu$ only. In general, the resulting lepton mixing angles involve
a combination of all parameters appearing in $Y_D$. However, if
\begin{equation}
[ \Omega ({\bf 3^\prime})^T \, M_R^{-1} \, \Omega ({\bf 3^\prime})^\star, R_{kl} (\theta_R) ] \, = \, 0 \, 
\end{equation}
(see section~\ref{sec:cases} for such cases), the lepton mixing angles only depend on the free parameter $\theta_L$ and the parameters describing the flavor and CP symmetry
as well as the residual symmetry $G_\nu$. In these cases, we find that the PMNS mixing matrix is given by 
\begin{equation} 
U \, = \, \Omega ({\bf 3}) \, R_{ij} (\theta_L) \, K_\nu \, ,
\label{eq:upmnsform}
\end{equation}
where $K_\nu$ is a diagonal matrix with entries equal to $\pm 1$ and $\pm {i}$, and is necessary to make neutrino masses positive. This matrix is generally parametrized in the following form: 
\begin{equation}
\label{eq:Knu}
 K_\nu \, = \, \left( \begin{array}{ccc}
1 & 0 & 0\\
0 & {i}^{k_1} & 0\\
0 & 0 & {i}^{k_2}
\end{array}
\right) \;  \;\; \mbox{with} \;\;  k_{1,2}=0,1,2,3 \, .
\end{equation}
We can easily verify that $U$ fulfills 
\begin{equation}
U^\dagger \, M_\nu \, U^\star \, = \, \mathrm{diag} \, (m_1, m_2, m_3) \, ,
\end{equation}
with the mass spectrum of the light neutrinos being determined by the Yukawa couplings $y_f$ in Eq.~\eqref{eq:YDgen},~i.e.
\begin{equation}
m_f \, = \, \frac{y_f^2 \, v^2}{M_N} \;\; \mbox{for} \;\;  f=1,2,3 \, .
\label{eq:2.17}
\end{equation}
As the Yukawa couplings are not constrained other than being real, the light neutrino mass spectrum is not predicted in our scenario. It can thus accommodate both normal ordering (NO) and inverted ordering (IO), depending in general on the parameters
encoded in $Y_D$. In turn, we can constrain these parameters by the experimental
information available on neutrino masses,~i.e.~the measurement of the two mass squared differences~\cite{ParticleDataGroup:2020ssz} and the upper bound on the sum of the 
neutrino masses~\cite{Aghanim:2018eyx}. 

For NO, the three light neutrino masses $m_i$ are parametrized as
\begin{eqnarray}
\label{massesNO}
& m_1 \, = \, m_0 \;\; , \;\;\; m_2 \, = \, \sqrt{m_0^2 + \Delta m_{\mathrm{sol}}^2} \;\; , \;\;\; \nonumber \\ & m_3 \, = \,  \sqrt{m_0^2 + \Delta m_{\mathrm{atm}}^2} \, ,
\end{eqnarray}
 with $m_0$ denoting the lightest neutrino mass. 
For IO, the masses $m_i$ are written as 
\begin{eqnarray}
\label{massesIO}
m_1 \, = \, \sqrt{m_0^2 + |\Delta m_{\mathrm{atm}}^2| - \Delta m_{\mathrm{sol}}^2 } \, , \;\; \nonumber \\ m_2 \, = \, \sqrt{m_0^2 + |\Delta m_{\mathrm{atm}}^2| } \, , 
\;\; m_3 \, = \, m_0 \, ,
\end{eqnarray}
where $m_0$ is the lightest neutrino mass. In each case, we use the $3\sigma$ allowed range of the solar and atmospheric mass-squared differences $\Delta m_{\mathrm{sol}}^2$ and $\Delta m_{\mathrm{atm}}^2$ respectively from a recent global fit result~\cite{Esteban:2020cvm, NUFIT}. 
Note that the current oscillation data has a slight preference for NO over IO at about $2\sigma$ level~\cite{ParticleDataGroup:2020ssz}. 

Similarly, the 95\% CL Planck (TT,TE,EE,lowE+lensing+BAO) upper limit on the sum of neutrino masses $\sum_i m_i<0.12$ eV~\cite{Aghanim:2018eyx} translates into an upper bound on the lightest neutrino mass
\begin{equation}
 \label{m0bound}
m_0 \ \lesssim \ 0.030\; (0.016) \;\text{eV}\;
 \end{equation}
for NO (IO). In our numerical analysis, we will fix $m_0$ which then determines $m_f$ from Eqs.~\eqref{massesNO} and \eqref{massesIO} for NO and IO respectively. Furthermore, for a given value of $M_N$, the Yukawa coupling parameters $y_f$ are calculated using Eq.~\eqref{eq:2.17}, which are used to get the form of $Y_D$ in Eq.~\eqref{eq:YDgen} that goes as an input into the leptogenesis calculation.

Note that the resulting PMNS mixing matrix in Eq.~\eqref{eq:upmnsform} is fixed by the symmetries $G_f$, CP, $G_\ell$ and $G_\nu$ up to the free real parameter $\theta_L$. Consequently, all mixing angles and CP phases are strongly correlated, because they all only depend on $\theta_L$. Since all lepton mixing angles $\theta_{ij}$ (with $i,j=1,2,3$ and $i<j$) have been measured with some accuracy~\cite{ParticleDataGroup:2020ssz}, the admitted values of $\theta_L$ are usually constrained to a rather narrow range~\cite{Hagedorn:2016lva}.

As for the parametrization of the PMNS mixing matrix in terms of the low-energy lepton mixing angles, we take~\cite{Hagedorn:2014wha}  
\begin{equation}
\label{UPMNSdef}
U \, = \, \widetilde{U} \, {\rm diag}\left(1, e^{i \alpha_1/2}, e^{i (\alpha_2/2 + \delta)}\right) \, ,
\end{equation}
with $\widetilde{U}$ being of the form of the Cabibbo-Kobayashi-Maskawa (CKM) matrix $V_{\mathrm{CKM}}$~\cite{ParticleDataGroup:2020ssz}
\begin{widetext}
\begin{equation}
\widetilde{U} \, = \, 
\begin{pmatrix}
c_{12} c_{13} & s_{12} c_{13} & s_{13} e^{- i \delta} \\
-s_{12} c_{23} - c_{12} s_{23} s_{13} e^{i \delta} & c_{12} c_{23} - s_{12} s_{23} s_{13} e^{i \delta} & s_{23} c_{13} \\
s_{12} s_{23} - c_{12} c_{23} s_{13} e^{i \delta} & -c_{12} s_{23} - s_{12} c_{23} s_{13} e^{i \delta} & c_{23} c_{13}
\end{pmatrix} \, ,
\end{equation}
\end{widetext}
and $s_{ij}\equiv \sin\theta_{ij}$ and $c_{ij}\equiv \cos\theta_{ij}$.  The mixing angles $\theta_{ij}$ range from $0$ to $\pi/2$, while the Majorana phases 
$\alpha_1, \alpha_2$ as well as the Dirac phase $\delta$ take values between $0$ and $2 \pi$. 
Note that one of the Majorana phases becomes unphysical, if the lightest neutrino mass $m_0$ vanishes.

\section{Different Cases}
\label{sec:cases}

In this section, we discuss the residual symmetries and the form of the corresponding representation matrices for four different cases, namely, Case 1, Case 2, Case 3a and Case 3b.1, as singled out in Ref.~\cite{Hagedorn:2014wha}.
We discuss additional constraints imposed from light neutrino masses and the constraints on the neutrino mass spectrum arising from imposing the condition in Eq.~(\ref{YDZX}).
Furthermore, we briefly review the results for lepton mixing, as found in~\cite{Hagedorn:2016lva}. We also comment on the special points corresponding
to specific choices of the parameters $\theta_L$ and $\theta_R$, that lead to ERS of the Dirac neutrino Yukawa couplings.

\subsection{Case 1}
\label{subsec:case1}

\subsubsection{Residual Symmetries}
\label{subsec:case1ressymm}
In this case, the residual $Z_2$ symmetry in the neutrino sector is generated by 
\begin{equation}
Z \, = \, c^{n/2} \, ,
\label{eq:Z1}
\end{equation}
where $c$ is one of the group generators (cf.~Appendix~\ref{appA}) and $n$ is the index of the flavor group $\Delta (6 \, n^2)$. Eq.~\eqref{eq:Z1} 
requires $n$ to be even. The explicit form of $Z$ in the irreducible, faithful, complex three-dimensional representation ${\bf 3}$
and in the irreducible, unfaithful, (in general) real three-dimensional representation ${\bf 3^\prime}$ can be found in Appendix~\ref{appA}.
As we will see in section~\ref{subsec:case1numass}, due to the form of the generator $Z$ in ${\bf 3^\prime}$ for $n$ divisible by four the Dirac neutrino Yukawa coupling matrix $Y_D$ becomes singular
and the light neutrino mass is not viable. For this reason, we will only focus on $4\nmid n$ (in addition to $3\nmid n$) for Case 1. 

The CP symmetry corresponds to the automorphism, given in Eq.~(\ref{XP23auto}),
conjugated with the inner automorphism associated with the group transformation $a \, b \, c^s \, d^{2 s}$ with $s=0,1,...,n-1$. The corresponding CP transformations $X (s)$ read
\begin{subequations}
\label{Xs3}
\begin{alignat}{2}
X(s) ({\bf 3}) \ & = \ a ({\bf 3}) \, b({\bf 3}) \, c ({\bf 3})^s \, d({\bf 3})^{2 s} \, X_0 ({\bf 3}) \, , \\ 
X(s) ({\bf 3^\prime}) \ & = \ a ({\bf 3^\prime}) \, b({\bf 3^\prime}) \, c ({\bf 3^\prime})^s \, d({\bf 3^\prime})^{2 s} \, X_0 ({\bf 3^\prime}) \, ,
\end{alignat}
\end{subequations}
in ${\bf 3}$ and ${\bf 3^\prime}$ respectively. 
The explicit forms of $X(s) ({\bf 3})$ and $X(s) ({\bf 3^\prime})$ can be obtained using the group generators $a,b,c,d$ given in Appendix~\ref{appA}.

The unitary matrix $\Omega(s) ({\bf 3})$, derived from $X ({\bf 3}) (s)$, can be chosen as 
\begin{equation}
\label{case1Omegain3}
\Omega(s) ({\bf 3}) \, = \, e^{i \, \phi_s} \, U_{\mathrm{TB}} \,
\left( \begin{array}{ccc}
 1 & 0 & 0 \\
 0 & e^{-3 \, i \, \phi_s} & 0\\
 0 & 0 & -1
\end{array}
\right) \, ,
\end{equation}
with $\phi_s=\pi s/n$ and the TB form for~\cite{Harrison:2002er} 
\begin{equation}
U_{\mathrm{TB}} \, = \, 
 \left( \begin{array}{ccc}
  \sqrt{2/3} & \sqrt{1/3} & 0\\
 -\sqrt{1/6} & \sqrt{1/3} & \sqrt{1/2} \\
 -\sqrt{1/6} & \sqrt{1/3} & -\sqrt{1/2} 
 \end{array}
 \right) \, .
\end{equation}

Based only on theoretical requirements, the form of the matrix $\Omega(s) ({\bf 3^\prime})$ depends on whether $s$ is even or odd, i.~e.
\begin{align}
 \Omega(s \, \mbox{even}) ({\bf 3^\prime}) \, = \, U_{\mathrm{TB}} \; , \:\:    
 \Omega(s \, \mbox{odd}) ({\bf 3^\prime}) \, = \,  U_{\mathrm{TB}} \, \left(
 \begin{array}{ccc}
 i & 0 & 0\\
 0 & 1 & 0\\
 0 & 0 & i
\end{array} 
 \right)
 \; .
 \label{Omega3p}
\end{align}
\label{Omega3p_seven}
Comparing these forms to the form of $\Omega(s) ({\bf 3})$ in Eq.~\eqref{case1Omegain3}, we observe that they have the same
structure, but the crucial difference lies in the phase matrix multiplied from the right (overall phases are clearly irrelevant).

In order to determine the plane in which the rotation $R_{ij} (\theta_L)$ acts, we look at
\begin{equation}
\label{Omegaeqs}
 \Omega(s) ({\bf 3})^\dagger \, Z ({\bf 3}) \,  \Omega(s) ({\bf 3}) \, = \, \left( \begin{array}{ccc}
 -1 & 0 & 0\\
 0 & 1 & 0\\
 0 & 0 & -1
 \end{array}
 \right) \, ,
\end{equation}
implying that the rotation through $\theta_L$ will be in the $(13)$-plane~\cite{Hagedorn:2014wha}. Therefore, the PMNS mixing matrix is given by 
\begin{align}
    U \, = \, \Omega(s) ({\bf 3}) \, R_{13}(\theta)\, K_\nu \, ,
    \label{eq:PMNS1}
\end{align}
where the rotation angle $\theta$ is a free real parameter (related to $\theta_L$) which is to be adjusted to its best-fit value $\theta_{\rm bf}$ in order to reproduce the best-fit with the measured lepton mixing angles.

Similarly, we can find the plane in which the rotation $R_{kl} (\theta_R)$ acts. 
The representation matrix $Z ({\bf 3^\prime})$ for $4\nmid n$ reads, after the transformation with $\Omega (s) ({\bf 3^\prime})$
for both $s$ even and $s$ odd,
\begin{equation}
 \Omega(s) ({\bf 3^\prime})^\dagger \, Z ({\bf 3^\prime})  \,  \Omega(s) ({\bf 3^\prime}) \, = \, \left( \begin{array}{ccc}
 -1 & 0 & 0\\
 0 & 1 & 0\\
 0 & 0 & -1
 \end{array}
 \right)
\; ,
\end{equation}
meaning that $R_{kl} (\theta_R)$ also acts in the $(13)$-plane.

\subsubsection{Constraints from and on Light Neutrino Sector}
\label{subsec:case1numass}

First, we discuss constraints on the possible choices of the residual symmetry $G_\nu$ arising from the light neutrino mass spectrum.
In order to find these we consider the form of the Dirac neutrino Yukawa coupling matrix $Y_D$ fulfilling the conditions in Eq.~(\ref{YDZX}).
 For $n$ divisible by four, $Z ({\bf 3^\prime})$ is nothing but the identity matrix [cf.~Eq.~(\ref{Zcn2_n2even})] and we find that the form of $Y_D$ needs to be
\begin{equation}
\label{YDnotworking}
Y_D = \left(
\begin{array}{ccc}
 y_{11} & y_{12} & y_{13}\\
 y_{11} & y_{12} & y_{13}\\
 y_{11} & y_{12} & y_{13}
\end{array}
\right) \, ,
\end{equation}
with $y_{1i}$ complex, $i=1,2,3$. It is clear from the form of $Y_D$ in Eq.~\eqref{YDnotworking} that the determinant vanishes and that $Y_D$
 has two zero eigenvalues. As a consequence, the light neutrino mass matrix arising from the type-I seesaw mechanism [cf.~Eq.~\eqref{eq:2.17}] also 
has two zero eigenvalues. Furthermore, we can check that the non-zero eigenvalue has to correspond to the second 
light neutrino mass, since it is always associated with the eigenvector proportional to $\left( 1,1,1 \right)^T$ which can only be identified
with the second column of the PMNS mixing matrix. It is, however, experimentally highly disfavored that such a form can be the dominant contribution to 
light neutrino masses. Indeed, we can show that, if $Z({\bf 3^\prime})$ is the identity matrix and $Z ({\bf 3})$ is any generator of a $Z_2$
symmetry, i.e.~it can be represented by a matrix $Z ({\bf 3})$ that fulfills 
\begin{align}
    V^\dagger \, Z({\bf 3}) \, V \, = \, \mbox{diag} \, (1,-1,-1) \, ,
\end{align}
with $V$ being a unitary matrix, then
\begin{equation}
Z ({\bf 3})^\dagger \, Y_D \, = \,  V \, \mbox{diag} \, (1,-1,-1) \, V^\dagger \, Y_D \, = \, Y_D \, ,
\end{equation}
or we can rewrite this condition as 
\begin{equation}
\mbox{diag} \, (1,-1,-1) \, \left[ V^\dagger \, Y_D \right] = \left[ V^\dagger \, Y_D \right] \; .
\end{equation}
Consequently, the combination $V^\dagger \, Y_D$ must have two vanishing rows, namely the second and the third ones.
In particular, the determinant of $V^\dagger \, Y_D$ vanishes. From the latter, we can conclude for $Y_D$ itself that its determinant must vanish,
since the determinant of $V$ cannot be zero. In addition, we can also know that $Y_D$ must have two vanishing eigenvalues. So, in general knowing 
that $Z ({\bf 3^\prime})$ is given by the identity matrix is sufficient in order to discard this case as realistic without corrections which can induce, at least, one further non-vanishing neutrino mass. We thus do not discuss further the case where $n$ is divisible by four.  

For $n$ not divisible by four, the form of the matrix $Z ({\bf 3^\prime})$ is shown in Eq.~(\ref{Zcn2_n2odd}). Again, we can compute the constraints on $Y_D$,
arising from imposing the conditions in Eq.~(\ref{YDZX}). In particular, we see that the first condition in Eq.~(\ref{YDZX}) reduces the number of free
(complex) parameters in $Y_D$ from nine to five, meaning the other four can be expressed in terms of these, e.g.
\begin{eqnarray}
\begin{array}{ll}
y_{23} \, = \, y_{11} + y_{12} + y_{13} - y_{21} - y_{22} \, , \\  y_{31} \, = \, y_{12} + y_{13} - y_{21} \, ,\\
y_{32} \, = \, y_{11} + y_{13} - y_{22} \, , \\    y_{33} \, = \, -y_{13} + y_{21} + y_{22} \, . 
\end{array}
\end{eqnarray}
The five free complex parameters in $Y_D$ are further constrained by requiring that also the second condition in Eq.~(\ref{YDZX}) is fulfilled. 
As a consequence, these parameters have to be real. This is consistent with the findings in the general case where $Y_D$ contains three real Yukawa couplings $y_f$, $f=1,2,3$ 
and two angles $\theta_L$ and $\theta_R$. In general, such a matrix $Y_D$ has a non-vanishing determinant and three different eigenvalues, namely (proportional to) $y_f$.

We know from the type-I seesaw formula [cf.~Eq.~\eqref{eq:seesaw}] that for eventually relating the parameters of $Y_D$ to the light neutrino masses, we have to look at the following expression
\begin{equation}
\label{type1seesawcomb}
 \Omega(s) ({\bf 3^\prime})^\dagger \, M_R^{-1} \,  \Omega(s) ({\bf 3^\prime})^\star \, ,
\end{equation}
with $M_R$ as in Eq.~(\ref{eq:MR}). For $ \Omega(s) ({\bf 3^\prime})$ in (\ref{Omega3p}) we find
\begin{subequations}
\label{type1seesawcomb_comp}
\begin{alignat}{2}
 \Omega(s\, \mbox{even}) ({\bf 3^\prime})^\dagger \, M_R^{-1} \,  \Omega(s\, \mbox{even}) ({\bf 3^\prime})^\star \ & = \
 \frac{1}{M_N}  \left(
 \begin{array}{ccc}
  1 & 0 & 0\\
  0 & 1 & 0\\
  0 & 0 & -1
 \end{array}
 \right), \\
 \Omega(s\, \mbox{odd}) ({\bf 3^\prime})^\dagger \, M_R^{-1} \,  \Omega(s\, \mbox{odd}) ({\bf 3^\prime})^\star \ & = \ 
\frac{1}{M_N} \, \left(
 \begin{array}{ccc}
  -1 & 0 & 0\\
  0 & 1 & 0\\
  0 & 0 & 1
 \end{array}
 \right).
\label{type1seesawcomb_comp2}
\end{alignat}
\end{subequations}
Note that in both cases the resulting structure is simple but does not commute with the arbitrary rotations $R_{13} (\theta_{L,R})$. Hence, this
has to be taken into account when computing the light neutrino masses from the type-I seesaw formula [cf.~Eq.~\eqref{eq:seesaw}]. Indeed only the light neutrino mass $m_2$
is related to $y_2$ and $M_N$ in a trivial way [cf.~Eq.~\eqref{eq:y2case1}], 
while for the full matrix part, we calculate
\begin{align}
M_\nu \ & = \ v^2 \left(
\begin{array}{ccc}
 y_1 & 0 & 0\\
 0 & y_2 & 0\\
 0 & 0 & y_3
 \end{array}
 \right) \, R_{13} (-\theta_R) \, \Omega(s) ({\bf 3^\prime})^\dagger \, 
 M_R^{-1} \, \nonumber \\ & \qquad \times \Omega(s) ({\bf 3^\prime})^\star \, R_{13} (\theta_R) \,
\left(
\begin{array}{ccc}
 y_1 & 0 & 0\\
 0 & y_2 & 0\\
 0 & 0 & y_3
 \end{array}
 \right) 
\end{align}
which yields
\begin{subequations}
\label{MRfullpart}
\begin{alignat}{2}
M_\nu\, (s~{\rm even}) \, = \, & \frac{v^2}{M_N} 
 \left( \begin{array}{ccc}
 y_1^2 \, \cos 2\theta_R & 0 & y_1 y_3 \sin 2\theta_R\\
0& y_2^2 & 0\\
y_1 y_3 \sin 2 \theta_R & 0 & - y_3^2 \cos 2\theta_R
\end{array}
\right) \, , \\
M_\nu\, (s~{\rm odd}) \, = \, & \frac{v^2}{M_N} 
\left( \begin{array}{ccc}
- y_1^2 \, \cos 2\theta_R & 0 & -y_1 y_3 \sin 2\theta_R\\
0& y_2^2 & 0\\
-y_1 y_3 \sin 2 \theta_R & 0 &  y_3^2 \cos 2\theta_R
\end{array}
\right) \, .
\end{alignat}
\end{subequations}
The difference is just the overall sign so we can nicely treat both cases of $s$ even and $s$ odd at once, as shown in Appendix~\ref{appD1}. 

We note a few things regarding the matrices in Eq.~(\ref{MRfullpart}): if we set $y_1=0$, then $m_1$ vanishes and  we obtain NO with the matrix being automatically diagonal and 
does not need a further rotation; on the other hand, if we set $y_3=0$, $m_3=0$ follows, and we obtain IO and again the matrix
is automatically diagonal with no further rotation required. We can also set $\sin 2\theta_R=0$ leading to no further
rotation needed as well, but in this case there are also no constraints on the neutrino masses. Some values of $\theta_R$
are not admitted, e.g.~$\cos 2\theta_R=0$ and consequently, $\sin2 \theta_R= \pm 1$ (meaning $\theta_R=\pi/4$, $3\pi/4$, etc),
since then two of the neutrino masses are degenerate.\footnote{For the matrices in Eq.~(\ref{MRfullpart}), these two are the first and
the third neutrino masses, and thus, the spectrum becomes completely unrealistic.} Similar statements hold in the other cases that
have matrices like in Eq.~(\ref{MRfullpart}) as part of the light neutrino mass matrix, because the combination in (\ref{type1seesawcomb}) is not trivial in flavor space. 

The general solution for both cases goes as follows: For $s$ even, the PMNS lepton mixing matrix [cf.~Eq.~\eqref{eq:PMNS1}] is given by 
\begin{equation}
	U \, = \, \Omega(s)(\mathbf{3})\,\,R_{13}\left(\theta_L-\psi\right)\,\text{diag}\left(1,\,1,\,\pm i\right) \,,
\end{equation}
with
\begin{equation}
\label{psiE}
	\tan^2\psi \; \equiv \; \frac{m_1\,+\,m_3\,-\,\sqrt{m_1^2\,+\,m_3^2\,+\,2\,m_1\,m_3\,\cos\left(4\,\theta_R\right)}}
	{m_1\,+\,m_3\,+\,\sqrt{m_1^2\,+\,m_3^2\,+\,2\,m_1\,m_3\,\cos\left(4\,\theta_R\right)}}\,.
\end{equation}
The Yukawa matrix $Y_D$ is constructed from Eq.~(\ref{eq:YDgen}), using the expressions of $\Omega(s)(\mathbf{3})$ and $\Omega(s)(\mathbf{3^\prime})$ corresponding to $s$-even, as given in Eqs.~\eqref{case1Omegain3} and \eqref{Omega3p} respectively. 
The Yukawa parameters $y_f$ in this case are explicitly given in Appendix~\ref{appD1}. For $s$ odd, the PMNS lepton mixing matrix is 
\begin{equation}
	U \, = \, \Omega(s)(\mathbf{3})\,\,R_{13}\left(\theta_L-\psi\right)\,\text{diag}\left(\pm i,\,1,\,1\right) \,,
\end{equation}
with $\psi$ introduced in Eq.~(\ref{psiE}). 
The Yukawa matrix $Y_D$ is constructed using the expressions of $\Omega(s)(\mathbf{3})$ and $\Omega(s)(\mathbf{3^\prime})$ corresponding to $s$-odd. The parameters $y_f$
are same as in the $s$ even case, cf. Appendix~\ref{appD1}. Notice that  $y_f$ are real quantities, provided $-\pi/4<\theta_R<\pi/4$.

For our numerical study of Case 1 in this paper, we will choose $n=26$ as an example which fulfills all the constraints on $n$, i.e.~$n$ even and not divisible by three or four.\footnote{As we will see later, it is the quantity $s/n$ (which varies between 0 and 1) that is relevant to the phenomenology, and not the actual value of $n$.} 
The corresponding form of $Y_D$ can be easily computed from Eq.~(\ref{eq:YDgen}), but we can also explicitly check by applying the conditions in Eq.~(\ref{YDZX}) to a general complex $3\times 3$ matrix
$Y_D$ that this is the correct form of the Dirac mass matrix of the neutrinos. The actual expressions are quite lengthy and not very illuminating, thus we do not display them explicitly here,
but they can be easily derived with the information given above. 

We notice that only five real parameters $y_f$ (with $f=1,2,3$), $\theta_L$ and $\theta_R$
appear in $Y_D$ and that lepton mixing depends effectively only on one free parameter $\theta=\theta_L-\psi$, which has to be adjusted to $\theta_{\mbox{\footnotesize bf}}$ in order to obtain the best-fit with the measured mixing angles~\cite{Esteban:2020cvm, NUFIT}. If the expression in Eq.~(\ref{type1seesawcomb}) is proportional to the identity matrix, $\theta$ is simply given by $\theta_L$ (since $\psi=0$ in this case)
and $y_f$ can be directly matched to the light neutrino masses $m_f$. If this is not true and we find a situation like in Eq.~(\ref{type1seesawcomb_comp}),
there is only one coupling $y_{f^\prime}$ directly proportional to one light neutrino mass $m_{f^\prime}$, whereas the other two together with $\theta_R$ determine the 
remaining two light neutrino masses, as shown in Appendix~\ref{appD1}. In addition, these three parameters determine another mixing angle $\psi$ given by Eq.~\eqref{psiE}, that together with
$\theta_L$ gives $\theta_{\mbox{\footnotesize bf}}\equiv \theta_L-\psi$. Hence, in both cases there are four experimentally constrained quantities (three neutrino masses and
$\theta_{\mbox{\footnotesize bf}}$) which determine five free parameters, namely, $y_f$, $\theta_L$ and $\theta_R$. Thus, only one of them (usually $\theta_R$)
can be chosen freely. 

We would like to emphasize here that there are only six free parameters in this model, namely, $m_0$, $M_N$, $\theta_R$, $\kappa$ and $s/n$. This is in contrast with the ordinary type-I seesaw-based models with three RHNs where without any flavor symmetry, there are 18 free parameters which get reduced to 13 after imposing the neutrino oscillation constraints. Therefore, the flavor model being considered here is more predictive than the ordinary type-I seesaw model.

We give here a numerical example from Ref.~\cite{Hagedorn:2014wha} (recently updated in Ref.~\cite{Drewes:2022kap} using the current global fit data~\cite{Esteban:2020cvm, NUFIT}) which leads to the mixing pattern of Case 1. The characteristics of this mixing
pattern are the following: the mixing angles can always be fitted well, independent of the choice of the group $\Delta (6 \, n^2)$ as well as the 
CP symmetry $X (s)$, if we choose the free parameter $\theta_{\rm bf}$ correctly. In the limit of residual symmetries $G_\nu$ and $G_\ell$, we obtain from a simple $\chi^2$-analysis that the lepton mixing angles can be accommodated at the $3\sigma$ level or better with $\theta_L \approx 0.183 \, (0.184)$ for NO (IO). This yields 
$\sin^2 \theta_{13}\approx 0.0220\, (0.0222)$,  $\sin^2 \theta_{12}\approx 0.341$ and $\sin^2 \theta_{23} \approx  0.605 \, (0.605)$, corresponding to $\Delta \chi^2\approx 11.9\,(11.2)$.\footnote{When computing the $\Delta \chi^2$ for IO, we subtract the overall $\Delta \chi^2=\chi^2_{\rm IO}-\chi^2_{\rm NO}=2.6$ for IO with respect to NO, that is favored by the current global fit data~\cite{Esteban:2020cvm, NUFIT}.} The results of the CP phases are simple: the Dirac phase $\delta$ as well as 
one of the Majorana phases $\alpha_2$ are trivial,~i.e.~$\sin\delta=0$\footnote{Although there is a mild preference for a nonzero $\delta$ in the global fit of neutrino oscillation data~\cite{Esteban:2020cvm, NUFIT}, $\delta=0$ is still allowed at 3\,$\sigma$ confidence level~\cite{ParticleDataGroup:2020ssz}; therefore our Case 1 is not excluded yet. } and $\sin\alpha_2=0$, 
while the other Majorana phase $\alpha_1$ depends on the chosen CP symmetry $X (s)$. For strong NO (IO), we get 
\begin{align}
\label{case1sina}
\sin\alpha_1 \, &= \, (-1)^{k+r+s} \, \sin (6 \, \phi_s) \, ,  \;\; \nonumber \\ \;\; \cos\alpha_1 \, &= \, (-1)^{k+r+s+1} \, \cos (6 \, \phi_s) \;, 
\end{align}
where $k=0$ ($k=1$) for $\cos 2 \, \theta_R > 0$ ($ \cos 2 \, \theta_R < 0$)
and $r=0$ ($r=1$). 

\subsubsection{Enhanced Residual Symmetries}
\label{subsec:case1specialpoints}

For particular values of $\theta_L$ and $\theta_R$, the residual symmetry $G_\nu=Z_2 \times {\rm CP}$ can be enhanced.
 If $\theta_L=0$ or $\pi$, the combination $M_D M_D^\dagger$ becomes invariant under a further $Z_2$ subgroup of $G_f$.
Similarly, for the choices $\theta_R=0, \pi/2, \pi$ and $3 \pi/2$, the combination $M_D^\dagger M_D$ preserves a symmetry larger than $G_\nu$.
 This symmetry is also larger than the one of $M_D M_D^\dagger$ for $\theta_L=0, \pi$ since RHNs transform as the real representation ${\bf 3^\prime}$ of $G_f$
that is unfaithful for $n > 2$.

These points of ERS are of particular relevance for phenomenology, since $\theta_L$ deviating
from $\theta_{L,0} =$ $0$ or $\pi$ leads to a non-zero value of the reactor mixing angle $\theta_{13}$, as confirmed experimentally~\cite{ParticleDataGroup:2020ssz}. Similarly, $\theta_R$ close to $\theta_{R,0} =$ $0$, $\pi/2$, $\pi$ or $3\pi/2$ makes it possible for
the RHN $N_3$ to be long-lived enough for being detected with the LLP searches (see section~\ref{subsec:RHnuwidths}), while simultaneously maximizing the CP asymmetries $\varepsilon_{i \alpha}$ relevant for leptogenesis (see section~\ref{sec:CPasymmetries}). One can argue that the larger the ERS is, the smaller the deviation
from points of ERS  will be, i.e.~$\theta_R$ is expected to deviate from $\theta_{R,0}$ by $\delta\theta_R=|\theta_R-\theta_{R,0}|\lesssim 0.01$, while $\theta_L$
can deviate from $\theta_{L,0}$ up to $\delta\theta_L=|\theta_L-\theta_{L,0}|\sim 0.2$.

In one type of explicit models~\cite{King:2013eh}, the flavor and CP symmetry are spontaneously broken to the residual symmetries $G_\nu$ and $G_\ell$
with the help of flavor symmetry breaking fields (flavons) and a peculiar alignment of their VEVs, achieved with a particular form of the potential. Depending on the fields and the form of the potential, an ERS larger than $G_\nu$ and $G_\ell$
 can be preserved at leading order. Higher-dimensional operators then induce small
deviations from these points of ERS, thus explaining the particular sizes of $\theta_L$ and $\theta_R$.
 An example can be found in Ref.~\cite{Feruglio:2013hia}, where the correct size of $\theta_L$ and thus the observed reactor mixing angle $\theta_{13}$ are generated in this way.

\subsection{Case 2}
\label{subsec:case2}

\subsubsection{Residual Symmetries}
\label{subsec:case2ressymm}
The residual $Z_2$ symmetry in the neutrino sector is generated by the same element $Z=c^{n/2}$ as in Case 1 [cf.~Eq.~\eqref{eq:Z1}]; therefore, $n$ must be even. Thus, all comments made in the context of Case 1, and in particular, the forms of $Z ({\bf 3})$ and $Z ({\bf 3^\prime})$ in Eqs.~(\ref{Zcn2in3}) and (\ref{Zcn2_n2even}), (\ref{Zcn2_n2odd}), apply respectively. 

The CP symmetry is given by the automorphism in Eq.~(\ref{XP23auto}) and the inner
automorphism $h= c^s d^t$ with $0 \leq s, t \leq n-1$ and thus depends on two parameters: $X (s,t)$.
In the three-dimensional representations ${\bf 3}$ and ${\bf 3^\prime}$, $X (s,t)$ is respectively given by
\begin{align}
X (s,t) ({\bf 3}) \, &= \, c ({\bf 3})^s \, d ({\bf 3})^t \, X_0 ({\bf 3}) \;\; \nonumber \\ \;\; X (s,t) ({\bf 3^\prime}) \, &= \, c ({\bf 3^\prime})^s \, d ({\bf 3^\prime})^t \, X_0 ({\bf 3^\prime}) \, .
\end{align}
The explicit forms can be found in Appendix~\ref{appC}.

In the analysis
of lepton mixing patterns for Case 2 in Ref.~\cite{Hagedorn:2014wha}, it turned out to be more convenient to use the parameters $u$ and $v$ that are linearly
related to $s$ and $t$ as follows: 
\begin{equation}
\label{defuv}
u \, = \,  2 \, s -t \;\; \mbox{and} \;\; v \, = \, 3 \, t \; .
\end{equation}
Since $0 \leq s, t \leq n-1$, the admitted intervals for $u$ and $v$ are $-(n-1)\leq u\leq 2(n-1)$ and $0\leq v\leq 3(n-1)$. 
Here we use $(s,t)$ and $(u,v)$ interchangeably as needed. A suitable choice of the matrix $\Omega (s,t) ({\bf 3})$ is given by
\begin{equation}
\Omega (s,t) ({\bf 3}) \, = \, e^{i \phi_v/6} \, U_{\mathrm{TB}} \, R_{13} \left( -\frac{\phi_u}{2} \right) \, \left( \begin{array}{ccc}
 1 & 0 & 0\\
 0 & e^{- i \phi_v/2} & 0\\
 0 & 0 & -i
\end{array}
\right) \, ,
\end{equation}
with $\phi_u=\pi \, u/n$ and $\phi_v=\pi \, v/n$. 

The form of the matrix  $\Omega (s,t) ({\bf 3^\prime})$, derived from $X ({\bf 3}) (s,t)$, depends like the latter on whether $s$ and $t$ are even or odd.
The explicit form of $\Omega (s,t) ({\bf 3^\prime})$, however, does neither contain $s$ nor $t$ as parameters, and is given by 
\begin{widetext}
\begin{subequations}
\begin{alignat}{2}
 \Omega (s \, \mbox{even},t \, \mbox{even}) ({\bf 3^\prime}) \, = \, & U_{\mathrm{TB}} \, \left( \begin{array}{ccc}
 1 & 0 & 0\\
0 & 1 & 0\\
0 & 0 & i
 \end{array}
 \right) \; , \\
 \Omega (s \, \mbox{odd},t \, \mbox{even}) ({\bf 3^\prime}) \, = \, & U_{\mathrm{TB}} \, \left( \begin{array}{ccc}
i & 0 & 0\\
0 & 1 & 0\\
0 & 0 & 1
 \end{array}
 \right) \; , \\
  \Omega (s \, \mbox{even},t \, \mbox{odd}) ({\bf 3^\prime}) \, = \, & e^{-i \pi/4} \, U_{\mathrm{TB}} \, R_{13} \left( \frac{\pi}{4} \right) \, \left(
 \begin{array}{ccc}
  e^{- i \pi/2} & 0 & 0\\
  0 & e^{- i \pi/4} & 0\\
  0 & 0 & 1
 \end{array}
 \right) \; , \\
 \Omega (s \, \mbox{odd},t \, \mbox{odd}) ({\bf 3^\prime}) \, = \, & e^{- 3 \, i \, \pi/4} \, U_{\mathrm{TB}} \, R_{13} \left( \frac{\pi}{4} \right) \, \left(
 \begin{array}{ccc}
  e^{- i \, \pi/2} & 0 & 0\\
  0 & e^{i \, \pi/4} & 0\\
  0 & 0 & 1
 \end{array}
 \right) \; .
 \end{alignat}
 \end{subequations}
Similar as in Case 1, the rotation associated with the representation ${\bf 3}$ and thus with LH leptons is always $R_{13} (\theta_L)$. Therefore, the PMNS mixing matrix is given by a form similar to Eq.~\eqref{eq:PMNS1}, i.e.
\begin{align}
    U \, = \, \Omega (s,t)({\bf 3})\, R_{13} (\theta)\, K_\nu \, .
    \label{eq:PMNScase2}
\end{align}
\end{widetext}
For all choices of $(s,t)$ above, $\Omega (s,t)({\bf 3^\prime})$ fulfills the two equations, i.e.~$Z ({\bf 3^\prime})$ always like in Eq.~(\ref{Zcn2_n2odd}) and 
\begin{equation}
 \Omega (s,t) ({\bf 3^\prime})^\dagger \, Z ({\bf 3^\prime}) \, \Omega (s,t) ({\bf 3^\prime}) \, = \, \left( \begin{array}{ccc}
-1 & 0 & 0\\
0 & 1 & 0\\
0 & 0 & -1
\end{array}
\right) \, 
\end{equation}
and hence, also for the representation ${\bf 3^\prime}$ for RHNs, the relevant rotation is in the (13)-plane, namely $R_{13} (\theta_R)$.
We observe that for none of the above combinations of $X$ and $Z$ in ${\bf 3}$ and ${\bf 3^\prime}$, we find zero eigenvalues for $Y_D$ as long as we only consider cases in which $n$ is not divisible by four so that
$Z ({\bf 3^\prime})$ is not the identity matrix; see discussion in section~\ref{subsec:case1ressymm}. 

\subsubsection{Constraints from and on Light Neutrino Sector}
\label{subsec:case2numass}

As a further step, we present the form of the relevant matrix
combination appearing in the type-I seesaw formula, involving $\Omega (s,t) ({\bf 3^\prime})$ and $M_R$ [cf.~(\ref{type1seesawcomb})]. We find that  
\begin{widetext}
\begin{subequations}
\begin{alignat}{2}
   \Omega (s~{\rm even},t~{\rm even}) ({\bf 3^\prime})^\dagger \, M_R^{-1} \,  \Omega(s~{\rm even},t~{\rm even}) ({\bf 3^\prime})^\star \, = \, &  \frac{1}{M_N}
 \begin{pmatrix}
   1 & 0 & 0\\
  0 & 1 & 0\\
  0 & 0 & 1
   \end{pmatrix}
 \\
 \Omega (s~{\rm odd},t~{\rm even}) ({\bf 3^\prime})^\dagger \, M_R^{-1} \,  \Omega(s~{\rm odd},t~{\rm even}) ({\bf 3^\prime})^\star \, = \, &  \frac{1}{M_N}
 \begin{pmatrix}
 -1 & 0 & 0\\
  0 & 1 & 0\\
  0 & 0 & -1
 \end{pmatrix}
 \\
 \Omega (s~{\rm even},t~{\rm odd}) ({\bf 3^\prime})^\dagger \, M_R^{-1} \,  \Omega(s~{\rm even},t~{\rm odd}) ({\bf 3^\prime})^\star \, = \, &  \frac{1}{M_N}
 \begin{pmatrix}
  0 & 0 & -1\\
  0 & -1 & 0\\
  -1 & 0 & 0
 \end{pmatrix}
 \\
 \Omega (s~{\rm odd},t~{\rm odd}) ({\bf 3^\prime})^\dagger \, M_R^{-1} \,  \Omega(s~{\rm odd},t~{\rm odd}) ({\bf 3^\prime})^\star \, = \, &  \frac{1}{M_N}
  \begin{pmatrix}
   0 & 0 & 1\\
  0 & -1 & 0\\
  1 & 0 & 0
  \end{pmatrix}
   \; .
\end{alignat}
\end{subequations}
\end{widetext}
Using this, we thus obtain the form of the light neutrino mass matrix for the different choices of $s$ and $t$:
\begin{widetext}
\begin{subequations}
\begin{alignat}{2}
    M_\nu \, (s~{\rm even},~t~{\rm even}) \, = \, & \frac{v^2}{M_N}
    \begin{pmatrix}
    y_1^2 & 0 & 0\\
 0 & y_2^2 & 0\\
 0 & 0 & y_3^2
    \end{pmatrix} 
 \\
   M_\nu \, (s~{\rm odd},~t~{\rm even}) \, = \, & \frac{v^2}{M_N}
 \begin{pmatrix}
      -y_1^2 & 0 & 0\\
 0 & y_2^2 & 0\\
 0 & 0 & -y_3^2
 \end{pmatrix}
 \\
  M_\nu \, (s~{\rm even},~t~{\rm odd}) \, = \, & \frac{v^2}{M_N}
 \begin{pmatrix}
     y_1^2 \, \sin 2 \theta_R & 0 & - y_1\, y_3 \, \cos 2 \theta_R\\
0 & - y_2^2 & 0\\
- y_1 \, y_3 \, \cos 2 \theta_R & 0 & - y_3^2 \, \sin 2 \theta_R
 \end{pmatrix}
 \\
  M_\nu \, (s~{\rm odd},~t~{\rm odd}) \, = \, & \frac{v^2}{M_N}
 \begin{pmatrix}
     -y_1^2 \, \sin 2 \theta_R & 0 &  y_1\, y_3 \, \cos 2 \theta_R\\
0 & - y_2^2 & 0\\
 y_1 \, y_3 \, \cos 2 \theta_R & 0 & y_3^2 \, \sin 2 \theta_R
 \end{pmatrix} \, .
\end{alignat}
\end{subequations}
\end{widetext}
These forms are very similar to those encountered before in Case 1 [cf.~Eq.~\eqref{MRfullpart}] and thus can be treated in the same way to obtain PMNS lepton mixing matrix. It turns out that the expressions for the $y_f$'s only depend on whether $t$ is even or odd, irrespective of the choice of $s$.  For $t$ even, the $y_f$'s are trivially related to the $m_f$'s; see Eq.~\eqref{eq:ycase2teven}. For $t$-odd, $s$-even case, the PMNS lepton mixing matrix~\eqref{eq:PMNScase2} is 
\begin{equation}
	U \, = \, \Omega(s,t)(\mathbf{3})\,R_{13}\left(\theta_L-\eta\right)\,\text{diag}\left(\pm i,\,\pm i,\,1\right) \,,
\end{equation}
with 
\begin{equation}
	\tan^2\eta \; \equiv \; \frac{m_1\,+\,m_3\,+\,\sqrt{m_1^2\,+\,m_3^2\,-\,2\,m_1\,m_3\,\cos\left(4\,\theta_R\right)}}
	{m_1\,+\,m_3\,-\,\sqrt{m_1^2\,+\,m_3^2\,-\,2\,m_1\,m_3\,\cos\left(4\,\theta_R\right)}}\label{eq:eta}\,.
\end{equation}
The Yukawa matrix $Y_D$ is constructed from Eq.~(\ref{eq:YDgen}), using the expressions of $\Omega(s,t)(\mathbf{3})$ and $\Omega(s,t)(\mathbf{3^\prime})$ corresponding to $s$-even and $t$-odd. 
The parameters $y_f$ in this case are given in Eq.~\eqref{eq:ycase2todd}. We get the same $y_f$ expressions for the $t$-odd, $s$-odd case. Note that these Yukawa parameters are real for $0<\theta_R<\pi/2$.

The results for lepton mixing in Case 2 are much richer
than in Case 1, and indeed, in general all CP phases are non-trivial. We can observe the following approximate dependence of the different CP phases
on the parameters $u$ and $v$ (for $k_{1,2}=0$ in Eq.~\eqref{eq:Knu} and no shift in $u$):
\begin{eqnarray}
\label{eq:CPcase2}
&\sin\delta \ \approx \ \pm 1 \mp 3.3 \, \phi_u^2, \quad 
\sin\alpha_1 \ \approx \ -\sin \phi_v, \nonumber \\ 
&\sin\alpha_2 \ \approx \  \mp 5.6 \, \phi_u \pm 23 \, \phi_u^3.
\end{eqnarray}
Detailed numerical results, including tables with examples of $n$, $u$, $v$ and $\theta_{\rm bf}$ that permit agreement of the three lepton mixing angles with experimental observations at the $3 \, \sigma$ level or better can be
found in Ref.~\cite{Hagedorn:2014wha} (and recently updated in Ref.~\cite{Drewes:2022kap}). In our numerical analysis, we will use $n=14$ and $u=0$ as the representative example for Case 2. In this case, we get $\theta_L\approx 0.184$, and  $\sin^2\theta_{12}\approx 0.341$, $\sin^2\theta_{13}\approx 0.0222\,(0.0224)$, $\sin^2\theta_{23}\approx 0.5$ for NO (IO) with $\Delta\chi^2=10.8\,(12.5)$. As for the CP phases, we find $\sin\delta=-1$ and $\sin\alpha_2=0$, whereas $\alpha_1$ is non-trivial and depends on the choice of $v$ [cf.~Eq.~\eqref{eq:CPcase2}]. 

As for the ERS points, we will see in section~\ref{subsec:RHnuwidths} that the ERS points only occur for $t$ even (irrespective of whether $s$ is even or odd) and at the same $\theta_R$ values as in Case 1.

\subsection{Case 3}
\label{subsec:case3}

\subsubsection{Residual Symmetries}
\label{subsec:case3ressymm}

In this case the $Z_2$ symmetry in 
the neutrino sector is generated by 
\begin{equation}
Z \, = \, b c^m d^m \;\; \mbox{with} \;\; 0\leq m\leq n-1 \, .
\end{equation}
Since $Z$ involves the generator $b$, this case can only be realized for the flavor group $\Delta (6 \, n^2)$. 
We have in general $n$ different choices for the generator $Z$. However, as  discussed in Ref.~\cite{Hagedorn:2014wha}, preferred values of $m$ are either around 
$m\approx 0$ and $m\approx n$ for Case 3a, or $m\approx n/2$ for Case 3b.1, as long as the charged lepton masses are ordered canonically.
The form of $Z$ in the representations ${\bf 3}$ and ${\bf 3^\prime}$ can be found in Appendix~\ref{appA}. 

The CP symmetry is induced by the automorphism, shown in Eq.~(\ref{XP23auto}), conjugated with the
inner one, represented by the group transformation $h= b \, c^s \, d^{n-s}$ with $0\leq s\leq n-1$. The corresponding CP transformations $X (s)$ in ${\bf 3}$ and ${\bf 3^\prime}$
are respectively given by
\begin{align}
X (s) ({\bf 3}) \, & = \, b ({\bf 3})^s \, c ({\bf 3})^s \, d ({\bf 3})^{n-s} \, X_0 ({\bf 3}) \, , \nonumber \\
X (s) ({\bf 3^\prime}) \, & = \, b ({\bf 3^\prime})^s \, c ({\bf 3^\prime})^s \, d ({\bf 3^\prime})^{n-s} \, X_0 ({\bf 3^\prime}) \, .
\end{align}
The explicit forms of $X (s) ({\bf 3})$ and $X (s) ({\bf 3^\prime})$ can be found in Appendix~\ref{appC}.

The form of the matrix $\Omega (s, m) ({\bf 3})$, derived from $X  (s, m) ({\bf 3})$ in Eq.~(\ref{eq:Xs3Case3}), is given by 
\begin{align}
\Omega (s, m) ({\bf 3}) \, = \, & e^{i \, \phi_s} \,  \left( \begin{array}{ccc}
 1 & 0 & 0\\
 0 & \omega & 0 \\
 0 & 0 & \omega^2
\end{array}
\right) \, \nonumber \\ \times & \: U_{\mathrm{TB}} \,
 \left( \begin{array}{ccc}
 1 & 0 & 0\\
 0 & e^{-3 \, i \, \phi_s} & 0 \\
 0 & 0 & -1
\end{array}
\right) \, R_{13} \left( \phi_m \right) \, ,
\end{align}
with $\phi_s=\pi s/n$ and $\phi_m=\pi  m/s$.
The form of the matrix $\Omega (s) ({\bf 3^\prime})$ only depends on whether $s$ is even or odd and is  independent of the choice of the parameter $m$. 
We use
\begin{subequations}
\begin{alignat}{2}
\Omega (s \, \mbox{even}) ({\bf 3^\prime}) \, = \, & \left(
\begin{array}{ccc}
 1 & 0 & 0\\
 0 & \omega & 0\\
 0 & 0 & \omega^2
\end{array}
\right) \, U_{\mathrm{TB}} \, \left(
\begin{array}{ccc}
 1 & 0 & 0\\
 0 & 1 & 0\\
 0 & 0 & -1
\end{array}
\right) \; , \\
\Omega (s \, \mbox{odd}) ({\bf 3^\prime}) \, = \, & \left(
\begin{array}{ccc}
 1 & 0 & 0\\
 0 & \omega & 0\\
 0 & 0 & \omega^2
\end{array}
\right) \, U_{\mathrm{TB}} \, \left(
\begin{array}{ccc}
 i & 0 & 0\\
 0  & -1 & 0\\
 0 & 0 & -i
\end{array}
\right) \; .
\end{alignat}
\end{subequations}
We note that the form of $\Omega (s \, \mbox{even}) ({\bf 3^\prime})$ coincides with 
$\Omega (s, m) ({\bf 3})$ for the special choices $s=0$ and $m=0$. Similarly, 
$\Omega (s \, \mbox{odd}) ({\bf 3^\prime})$ coincides with $\Omega (s, m) ({\bf 3})$ for $s=n/2$ and $m=0$.

We have to compute the form of the matrix $Z (m) ({\bf 3})$ in the basis rotated via $\Omega (s, m) ({\bf 3})$ for the representation ${\bf 3}$
which means
\begin{equation}
\Omega (s, m) ({\bf 3})^\dagger \, Z (m) ({\bf 3}) \, \Omega (s, m) ({\bf 3}) \, = \, 
\left( \begin{array}{ccc}
  1 & 0 & 0\\
  0 & 1 & 0\\
  0 & 0 & -1
\end{array}
\right) \; .
\end{equation}
Note that this holds for all choices of $s$, $m$ and $n$.
 So, we know that LH leptons, being in the representation ${\bf 3}$, are always
accompanied with a rotation $R_{12} (\theta_L)$ with the rotation angle $\theta_L$ related to the fitting of the lepton mixing angles. Therefore, the PMNS mixing matrix in this case is 
\begin{align}
    U \, = \, \Omega(s,m)({\bf 3})\, R_{12}(\theta)\, K_\nu \, .
    \label{eq:PMNScase3a}
\end{align}

In the next step we consider the form of $Z ({\bf 3^\prime})$ [cf.~Eq.~(\ref{Z3pmevenodd})] in the basis
rotated by $\Omega  (s \, \mbox{even}) ({\bf 3^\prime})$ and $\Omega  (s \, \mbox{odd}) ({\bf 3^\prime})$,
respectively. The matrix $Z (m \, \mbox{even}) ({\bf 3^\prime})$ reads as follows in the basis rotated by 
 $\Omega  (s \, \mbox{even}) ({\bf 3^\prime})$:
 \begin{equation}
\Omega  (s \, \mbox{even}) ({\bf 3^\prime})^\dagger \, Z (m \, \mbox{even}) ({\bf 3^\prime}) \, \Omega  (s \, \mbox{even}) ({\bf 3^\prime}) \, = \, 
\left(
\begin{array}{ccc}
 1 & 0 & 0\\
 0 & 1 & 0\\
 0 & 0 & -1
\end{array}
\right)
\end{equation}
and in the basis rotated by  $\Omega  (s \, \mbox{odd}) ({\bf 3^\prime})$ it reads the same:
\begin{equation}
\Omega  (s \, \mbox{odd}) ({\bf 3^\prime})^\dagger \, Z (m \, \mbox{even}) ({\bf 3^\prime}) \, \Omega  (s \, \mbox{odd}) ({\bf 3^\prime}) \, = \, 
\left(
\begin{array}{ccc}
 1 & 0 & 0\\
 0 & 1 & 0\\
 0 & 0 & -1
\end{array}
\right) \; .
\end{equation}
Hence, in both cases we need a rotation $R_{12} (\theta_R)$
for the RHN fields in the representation ${\bf 3^\prime}$.
Doing the same for the matrix  $Z (m \, \mbox{odd}) ({\bf 3^\prime})$
in the basis rotated with $\Omega  (s \, \mbox{even}) ({\bf 3^\prime})$, we get 
 \begin{equation}
\Omega  (s \, \mbox{even}) ({\bf 3^\prime})^\dagger \, Z (m \, \mbox{odd}) ({\bf 3^\prime}) \, \Omega  (s \, \mbox{even}) ({\bf 3^\prime}) \, = \, 
\left(
\begin{array}{ccc}
 -1 & 0 & 0\\
 0 & 1 & 0\\
 0 & 0 & 1
\end{array}
\right) \, , 
\end{equation}
and in the basis rotated by $\Omega  (s \, \mbox{odd}) ({\bf 3^\prime})$, we find as well 
\begin{equation}
\Omega  (s \, \mbox{odd}) ({\bf 3^\prime})^\dagger \, Z (m \, \mbox{odd}) ({\bf 3^\prime}) \, \Omega  (s \, \mbox{odd}) ({\bf 3^\prime}) \, = \, 
\left(
\begin{array}{ccc}
 -1 & 0 & 0\\
 0 & 1 & 0\\
 0 & 0 & 1
\end{array}
\right) \; .
\end{equation}
Thus, in both bases the free rotation due to $ Z (m \, \mbox{odd}) ({\bf 3^\prime})$ is given by $R_{23} (\theta_R)$
among the RHNs.

\subsubsection{Constraints from and on Light Neutrino Sector}
\label{subsec:case3numass}
A further step is to check the relevant combination in (\ref{type1seesawcomb}) 
for which we find
\begin{subequations}
\begin{alignat}{2}
 \Omega (s \, \mbox{even}) ({\bf 3^\prime})^\dagger \, M_R^{-1} \,  \Omega (s \, \mbox{even}) ({\bf 3^\prime})^\star \, = \, & 
 \frac{1}{M_N} \, \left(
 \begin{array}{ccc}
  1 & 0 & 0\\
  0 & 1 & 0\\
  0 & 0 & -1
 \end{array}
 \right) \, , \\
 \Omega (s \, \mbox{odd}) ({\bf 3^\prime})^\dagger \, M_R^{-1} \,  \Omega (s \, \mbox{odd}) ({\bf 3^\prime})^\star \, = \, &  
 \frac{1}{M_N} \, \left(
 \begin{array}{ccc}
  -1 & 0 & 0\\
  0 & 1 & 0\\
  0 & 0 & 1
 \end{array}
 \right) \; .
\end{alignat}
\end{subequations}
Using these results, we see that for $m$ even, $s$ even, and for $m$ odd, $s$ odd, the structure of the light neutrino mass matrix is trivial, i.e.~diagonal, and we obtain a direct relation between the Yukawa couplings and
the light neutrino masses, cf.~Eq.~\eqref{eq:ycase3aeven}. The same is true in Case 3b.1, except for an additional permutation of the rows of the PMNS mixing matrix, cf.~Eq.~\eqref{eq:ycase3beven}.  
For the other two combinations of $m$ and $s$, the structure is analogous to the ones shown in Eq.~(\ref{MRfullpart}), although we have to change the rotation plane for $R_{ij} (\theta_R)$). 
\begin{widetext}
\begin{eqnarray}
M_\nu (m \, \mbox{even},s \, \mbox{odd}) \ & = & \ v^2\left(
\begin{array}{ccc}
 y_1 & 0 & 0\\
 0 & y_2 & 0\\
 0 & 0 & y_3
 \end{array}
 \right) \, R_{12} (-\theta_R) \, \Omega (s \, \mbox{odd}) ({\bf 3^\prime})^\dagger \, M_R^{-1} \,  \Omega (s \, \mbox{odd}) ({\bf 3^\prime})^\star \, R_{12} (\theta_R) \,
\left(
\begin{array}{ccc}
 y_1 & 0 & 0\\
 0 & y_2 & 0\\
 0 & 0 & y_3
 \end{array}
 \right) \nonumber 
 \\ 
  \ & = & \  \frac{v^2}{M_N} \, \left(
 \begin{array}{ccc}
 -y_1^2 \, \cos 2\theta_R & - y_1 \, y_2 \, \sin 2 \theta_R & 0\\
 - y_1 \, y_2 \, \sin 2 \theta_R & y_2^2 \, \cos 2 \theta_R & 0\\
 0 & 0 & y_3^2
 \end{array}
 \right) \, ,
\end{eqnarray}
\begin{eqnarray}
M_\nu (m \, \mbox{odd},s \, \mbox{even}) \ & = & \  v^2\left(
\begin{array}{ccc}
 y_1 & 0 & 0\\
 0 & y_2 & 0\\
 0 & 0 & y_3
 \end{array}
 \right) \, R_{23} (-\theta_R) \, \Omega (s \, \mbox{even}) ({\bf 3^\prime})^\dagger \, M_R^{-1} \,  \Omega (s \, \mbox{even}) ({\bf 3^\prime})^\star \, R_{23} (\theta_R) \,
\left(
\begin{array}{ccc}
 y_1 & 0 & 0\\
 0 & y_2 & 0\\
 0 & 0 & y_3
 \end{array}
 \right) \nonumber \\
\ & = & \ \frac{v^2}{M_N} \, \left(
 \begin{array}{ccc}
 y_1^2 & 0 & 0\\
 0 & y_2^2 \, \cos 2 \theta_R & y_2 \, y_3 \, \sin 2\theta_R\\
 0 & y_2 \, y_3 \, \sin 2\theta_R & - y_3^2 \, \cos 2\theta_R
 \end{array}
 \right) \, .
\end{eqnarray}
\end{widetext}
In these latter two cases, the relation~\eqref{eq:2.17}
holds for only one of the three neutrino generations, whereas the other two belong to a sub-sector that
requires further diagonalization, in a way discussed already for Case 1. As an example, we present the general result for one of the cases ($m$ even and $s$ odd) in which 
the lepton mixing matrix is 
\begin{equation}
	U \, = \, \Omega(m,s)(\mathbf{3})\,R_{12}\left(\theta_L-\zeta\right)\,\text{diag}(\pm i,1,1)\, ,
\end{equation}
with 
\begin{equation}
	\tan^2\zeta \; \equiv \; \frac{m_1\,+\,m_2\,-\,\sqrt{m_1^2\,+\,m_2^2\,+\,2\,m_1\,m_2\,\cos\left(4\,\theta_R\right)}}
	{m_1\,+\,m_3\,+\,\sqrt{m_1^2\,+\,m_3^2\,+\,2\,m_1\,m_2\,\cos\left(4\,\theta_R\right)}}\label{eq:zeta}\,.
\end{equation}
The Yukawa matrix $Y_D$ is constructed from (\ref{eq:YDgen}), using the expressions of $\Omega(m,s)(\mathbf{3})$ and $\Omega(m,s)(\mathbf{3^\prime})$ corresponding to $m$-even and $s$-odd.
The parameters $y_f$ in this case are given in Eq.~\eqref{eq:ycase3aevenodd} which are real for $-\pi/4<\theta_R<\pi/4$. The remaining case with $m$-odd and $s$-even, as well as those in Case 3b.1 can be similarly analyzed and the results are given in Eqs.~\eqref{eq:ycase3aoddeven}, \eqref{eq:ycase3bevenodd} and \eqref{eq:ycase3boddeven} respectively.

As for the lepton mixing results, note that this time there is no additional constraint on the choice of the index $n$, i.e.~both $n$ even and odd are allowed. As for the choice of $m$, in Case 3a effectively a small ratio $m/n$ (or close to one) is needed for achieving small
$\theta_{13}$. Regarding the choice of the CP transformation $X (s)$ there are, indeed, in total 16 choices~\cite{Hagedorn:2014wha} and all of them lead
to a reasonable agreement of the lepton mixing angles with experimental data. Some of them like $s=0$ and $s=8$ lead to CP conservation (either due to symmetry or rather accidentally) and 
values $s > n/2$ usually reproduce results like the corresponding value $s^\prime= n-s < n/2$. For our numerical analysis of Case 3a, we will use the example of $n=16$. In this case, the ERS points can only be achieved for NO and for $m$ even, $s$ even. As we will see later, leptogenesis in this scenario cannot be done at the ERS points, because $N_1$ becomes long-lived. Therefore, we will use an $m$-odd, $s$-odd example for leptogenesis with $m=s=1$. In this case, we have $\theta_L\approx 2.00$, which gives $\sin^2\theta_{12}\approx 0.305$, $\sin^2\theta_{13}\approx 0.0254$ and $\sin^2\theta_{23}\approx 0.613$ for NO. We note that the solar mixing angle fits very well with the corresponding $\Delta \chi_{12}^2<10^{-3}$. However, the results for reactor and atmospheric mixing angles are outside the $3\sigma$ ranges of the global fit~\cite{Esteban:2020cvm, NUFIT}, i.e. $\Delta \chi^2_{13}\approx 21.8$ and $\Delta \chi^2_{23}\approx 6.25$. However, they can be brought into the $3\sigma$ allowed range by taking into account other corrections in an explicit model (as in e.g.~Ref.~\cite{Lin:2009bw} with $A_4$ flavor symmetry, where $\theta_{13}$ is purely generated from corrections only).  As for the CP  phases, we get $\sin\delta\approx 0.458$, $\sin\alpha_1\approx 0.939$ and $\sin\alpha_2\approx 0.662$. 

For the mixing pattern of Case 3b.1, with the particular choice $m/n=1/2$, the sines of the two Majorana phases
$\alpha_1$ and $\alpha_2$ have the same magnitude~\cite{Hagedorn:2014wha}. The particular choice $s=n/2$ gives trivial Majorana phases and
maximal Dirac phase, as well as maximal atmospheric mixing.
Again, some values of $s$ like $s=0$ lead to no CP violation at all
and other values of $s$ like $s^\prime=n -s > n/2$ only produce results equivalent to those of $s < n/2$. For our numerical analysis of Case 3b.1, we will use the example of $n=10$.  In this case, the ERS points can only be achieved for IO and for $m$ even, $s$ even. Like in Case 3a, leptogenesis cannot be done at the ERS points in this case, because $N_1$ is long-lived. Therefore, we will choose an example from $m$-odd, $s$-odd with $m=s=5$. In this case, we have $\theta_L\approx 1.31$, which gives $\sin^2\theta_{23}\approx 0.5$, $\sin^2\theta_{12}\approx 0.318$, $\sin^2\theta_{13}\approx 0.0220$ with $\Delta \chi^2=4.12$. As for the CP phases, we get $\sin\delta=-1$, whereas the Majorana phases are trivial, i.e.~$\sin\alpha_1=\sin\alpha_2=0$.

\section{CP Asymmetries for Resonant Leptogenesis}
\label{sec:CPasymmetries}
In the minimal framework of resonant leptogenesis, the lepton asymmetry is generated from the CP-violating on-shell decays of the RHNs $N_i\to L_\alpha H$ and $N_i\to \bar{L}_\alpha H^c$ via the Yukawa couplings $(Y_D)_{\alpha i}$. The lepton asymmetry is obtained from the interference of tree- and self-energy diagrams for $N_i$ decay, which is resonantly enhanced if the intermediate state $N_j$ ($j\neq i$) in the self-energy diagram is quasi-degenerate with $N_i$~\cite{Pilaftsis:1997dr}. In the semi-analytic Boltzmann approach, the flavored lepton asymmetry (or lepton-to-photon ratio) can be approximated as~\cite{Buchmuller:2004nz, Deppisch:2010fr, Dev:2014laa}  
\begin{align}
    \eta_{L_\alpha} \ \simeq \ \frac{3}{2z_cK_\alpha^{\rm eff}}\sum_i \varepsilon_{i\alpha}d_i \, ,
    \label{eq:etaL}
\end{align}
where $z_c=M_N/T_c$ ($T_c$ being the critical temperature below which the electroweak sphalerons freeze-out and $M_N$ being the average mass of $N_i$ and $N_j$), $K_\alpha^{\rm eff}$ are the effective washout factors in presence of Yukawa and any additional interactions present in the model, and $d_i$ are the corresponding dilution factors given in terms of ratios of thermally-averaged rates for decays and scatterings involving $N_i$ (see Ref.~\cite{Dev:2014laa} for details). This final lepton asymmetry at temperature $T_c$ is then converted to a baryon asymmetry via (B+L)-violating electroweak sphaleron processes~\cite{Kuzmin:1985mm}. The conversion of the total lepton-to-photon ratio $\sum_\alpha \eta_{ L_\alpha}$ to the current baryon-to-photon ratio $\eta_{B}$ is given by the relation 
\begin{align}
    \eta_B \ \simeq \ -0.01\sum _\alpha \eta_{ L_\alpha} \, ,
    \label{eq:etaB}
\end{align}
where the prefactor contains the product of the sphaleron conversion rate of 28/79~\cite{Harvey:1990qw} and the entropy dilution factor of 1/27.3~\cite{Dev:2014laa}. The theoretical prediction for $\eta_B$ in Eq.~\eqref{eq:etaB} is to be compared (in both magnitude and sign) with the observed baryon-to-photon ratio~\cite{Aghanim:2018eyx} 
\begin{align}
    \eta_{B}^{\rm obs} \, = \, (6.12\pm 0.08)\times 10^{-10} \, .
    \label{eq:etaBobs}
\end{align}

Using the analytic forms of the $Y_D$ matrix given in section~\ref{sec:cases}, we can calculate the flavored CP asymmetries analytically. The general formula for the flavored CP asymmetry reads
\begin{align}
    \varepsilon_{i\alpha} \, = \, \frac{\Gamma(N_i\to L_\alpha H)-\Gamma(N_i\to \bar{L}_\alpha H^c)}{\Gamma(N_i\to L_\alpha H)+\Gamma(N_i\to \bar{L}_\alpha H^c)} \, ,
    \label{eq:eps1}
\end{align}
where $\Gamma$ stands for the RHN decay rate. In the resonant regime, Eq.~\eqref{eq:eps1} can be written in a compact form~\cite{Dev:2017wwc} 
\begin{align}
    \varepsilon_{i\alpha} \ \simeq & \ \frac{1}{8\pi \left(Y_D^\dag Y_D\right)_{ii}} \sum_{j\neq i} {\rm Im}\left[\left(Y_D^\star\right)_{\alpha i} \left(Y_D\right)_{\alpha j}\right] \nonumber \\ \times & \, {\rm Re}\left[\left(Y_D^\dag Y_D\right)_{ij}\right]\frac{M_i M_j (M_i^2-M_j^2)}{(M_i^2-M_j^2)^2+A_{ij}^2}  \, , 
    \label{eq:eps2}
\end{align}
where the $Y_D$ matrices are evaluated in the RHN mass basis,\footnote{Normally this is denoted by $Y_D$, but we drop the hat for brevity.} and $A_{ij}$ is a regulator that controls the behavior of the CP asymmetry in the limit $\Delta M\equiv |M_1-M_2|\to 0$. Moreover, as pointed out in Refs.~\cite{Dev:2014laa, Dev:2014oar}, in the resonant regime there are two distinct contributions to the lepton asymmetry from RHN mixing and oscillation effects, which can be effectively captured by the same form of $\varepsilon_{i\alpha}$ as in Eq.~\eqref{eq:eps2} but with different regulators:
\begin{align}
    A_{ij}^{\rm mix} \, &= \, M_i \Gamma_j \, , \quad \nonumber \\ \quad A_{ij}^{\rm osc} \, &= \, (M_1\Gamma_1+M_2\Gamma_2)\left[\frac{{\rm det}\left({\rm Re}\left(Y_D^\dag Y_D\right)\right)}{\left(Y_D^\dag Y_D\right)_{ii}\left(Y_D^\dag Y_D\right)_{jj}} \right]^{1/2} \, .
    \label{eq:regulator}
\end{align}
The net CP asymmetry that goes into Eq.~\eqref{eq:etaL} is then the sum of the mixing and oscillation contributions. This analytic approximation tends to agree well with the full quantum kinetic treatment~\cite{Dev:2014oar, Kartavtsev:2015vto, Klaric:2021cpi} in the strong washout regime.  

From Eq.~\eqref{eq:eps2}, we observe the following important features:
\begin{itemize}
    \item If ${\rm Re}\left[\left(Y_D^\dag Y_D\right)_{ij} \right]=0$ for some $i$ and all $j\neq i$, then $\varepsilon_{i\alpha}=0$, i.e.~that particular RHN flavor does not contribute to the CP asymmetry. 
    
    \item If ${\rm Re}\left[\left(Y_D^\dag Y_D\right)_{ij} \right]=0$ for some $i$ and one $j$, then $\varepsilon_{i\alpha}$ has only one contribution. 
    
    \item Since the regulator part is independent of the lepton flavor $\alpha$, we can sum over $\alpha$ to obtain the total CP asymmetry for a given RHN $N_i$:
    \begin{align}
        \varepsilon_i \ \equiv \ \sum_\alpha \varepsilon_{i\alpha} \, = \, & \frac{1}{8\pi \left(Y_D^\dag Y_D\right)_{ii}} \sum_{j\neq i} {\rm Im}\left[\left(Y_D^\dag Y_D\right)_{ij}\right] \nonumber \\ \times & \, {\rm Re}\left[\left(Y_D^\dag Y_D\right)_{ij}\right]{\cal F}_{ij}  \, , 
    \label{eq:eps3}
    \end{align}
    where we have defined the dimensionless quantity
    \begin{align}
        {\cal F}_{ij} \, = \, \frac{M_i M_j (M_i^2-M_j^2)}{(M_i^2-M_j^2)^2+A_{ij}^2} \, .
    \end{align}
     Eq.~\eqref{eq:eps3} implies that if ${\rm Im}\left[\left(Y_D^\dag Y_D\right)_{ij}\right]=0$ for some $i$ and all $j\neq i$, then $\varepsilon_i=0$. This is reflected in the vanishing weak-basis CP-odd invariants~\cite{Branco:1986gr, Branco:2005jr, Jenkins:2009dy}.  
\end{itemize}
We can exemplify the usefulness of these results by applying them to Case 1 with $s$ even. We find that ${\rm Re}\left[\left(Y_D^\dag Y_D\right)_{ij}\right]=0$ for either $i$ or $j$ being 3, but not both. This implies that $\varepsilon_{3\alpha}=0$, i.e.~the RHN mass eigenstate $N_3$ does not contribute to the CP asymmetry. We also have ${\rm Im}\left[\left(Y_D^\dag Y_D\right)_{ij}\right]=0$ if either $i=j$ or $i,j$ are both $1,2$. This implies that $\varepsilon_{1\alpha}$ and $\varepsilon_{2\alpha}$ both depend on one term only, e.g.~
    \begin{align}
    \varepsilon_{1\alpha} \ \simeq \ &\frac{1}{8\pi \left(Y_D^\dag Y_D\right)_{11}}  {\rm Im}\left[\left(Y_D^\star\right)_{\alpha 1} \left(Y_D\right)_{\alpha 2}\right] \nonumber \\ \times & \, {\rm Re}\left[\left(Y_D^\dag Y_D\right)_{12}\right] {\cal F}_{12} \, , 
    \label{eq:eps4}
\end{align}
with ${\rm Im}\left[\left(Y_D^\star\right)_{\alpha 1} \left(Y_D\right)_{\alpha 2}\right]\propto \sin(3 \, \phi_s)$ for all $\alpha$. By evaluating Eq.~\eqref{eq:eps4} in the strong NO and IO limits, we obtain the following compact analytical expressions for $\varepsilon_{1\alpha}$: for strong NO, we get
\begin{align}
\label{eps1NO}
\varepsilon_{1 \alpha} \ \approx \ &\frac{y_2 \, y_3}{9} \, \left[-2 \, y_2^2+y_3^2 \, (1- \cos 2 \, \theta_R)\right] \, \sin 3 \, \phi_s \nonumber \\ \times & \, \sin\theta_R \, \sin\theta_{L, \alpha} \, \mathcal{F}_{12} \, ,
\end{align}
and for strong IO, we get
\begin{align}
\label{eps1IO}
\varepsilon_{1 \alpha} \ \approx \ &\frac{y_1 \, y_2}{9} \, \left[-2 \, y_2^2+y_1^2 \, (1+ \cos 2 \, \theta_R)\right] \, \sin 3 \, \phi_s \nonumber \\ \times & \, \cos\theta_R \, \cos\theta_{L, \alpha} \, \mathcal{F}_{12} \, ,
\end{align}
with $\theta_{L, \alpha} = \theta_L + \rho_\alpha \, 4 \pi/3$ and $\rho_e=0$, $\rho_\mu=1$, $\rho_\tau=-1$.
For strong NO (IO) $\varepsilon_{i \alpha}$ becomes very small, if $\theta_R \approx 0 , \, \pi$ ($\theta_R\approx\pi/2,3\pi/2$). In addition, $\mathcal{F}_{ij}$
vanishes for $\cos 2 \, \theta_R=0$.
The CP asymmetries $\varepsilon_{2 \alpha}$ are the negatives of $\varepsilon_{1 \alpha}$ with $\mathcal{F}_{12}$ being replaced by $\mathcal{F}_{21}$.
For $s$ odd, similar expressions are obtained with $\sin(3\, \phi_s)$ being replaced by $-\cos(3\, \phi_s)$. All these analytic results have been verified numerically. 

We note that different values of $s$ can lead to the same value of $\varepsilon_{i \alpha}$. In particular, 
\begin{align}
\label{eps_srel}
\varepsilon_{i\alpha} (s) \,&  = \, (-1)^s \, \varepsilon_{i\alpha} (n-s) \, = \, \varepsilon_{i\alpha} (n/2-s) \nonumber \\ &  =\, (-1)^{s+1}\, \varepsilon_{i\alpha} (n/2+s) \;\; \mbox{for} \;\; s \ \leq \ n/2 \, .
\end{align}
Eqs.~\eqref{case1sina}, \eqref{eps1NO} and \eqref{eps1IO} show the close correlation between CP violation at low and high energies due to the flavor and CP symmetries chosen here. 

The analytic expressions for $\varepsilon_{i\alpha}$ in Cases 2 and 3 are more involved and not very illuminating. We can simplify them somewhat by taking ${\cal F}_{23}={\cal F}_{32}=0$ because $M_2=M_3$ in our scenario [cf.~Eq.~\eqref{eq:massesNi}]. As a consequence, $\varepsilon_{i\alpha}$ for $i=2$ and $i=3$ (and all $\alpha$) only have one contribution. However, the expressions for $\varepsilon_{i\alpha}$ turn out to be different for different charged lepton flavors. Just as an example, we present below the Case 2 result for $s$ even, $t$ even, and assuming that both ${\cal F}_{12}$ and ${\cal F}_{13}$ are of similar size, which we commonly write as ${\cal F}$: \begin{widetext}
\begin{subequations}
\begin{alignat}{2}
    \varepsilon_{1e} \  =  \ & \frac{4{\cal F}}{9}\Big(-y_1y_3(y_1^2-y_3^2)\sin\phi_u \cos\theta_R\sin\theta_R \nonumber \\
    & +\cos\left(\frac{\phi_v}{2}\right)\sin\left(\frac{\phi_u}{2}\right)\left[-y_1y_2(y_1^2-y_2^2)\cos\theta_R\sin\theta_L-y_2y_3(y_2^2-y_3^2)\cos\theta_L\sin\theta_R\right] \nonumber \\
    & + \cos\left(\frac{\phi_u}{2}\right)\sin\left(\frac{\phi_v}{2}\right)\left[y_1y_2(y_1^2-y_2^2)\cos\theta_R\cos\theta_L-y_2y_3(y_2^2-y_3^2)\sin\theta_L\sin\theta_R \right]\Big) \, , \\
    \varepsilon_{1\mu} \  =  \ & \frac{4{\cal F}}{9}\Big(y_1y_3(y_1^2-y_3^2)\sin\phi_{u,-} \cos\theta_R\sin\theta_R \nonumber \\
    & +\cos\left(\frac{\phi_v}{2}\right)\cos\left(\frac{\phi_{u,-}}{2}\right)\left[y_1y_2(y_1^2-y_2^2)\cos\theta_R\sin\theta_L+y_2y_3(y_2^2-y_3^2)\cos\theta_L\sin\theta_R\right] \nonumber \\
    & + \sin\left(\frac{\phi_{u,-}}{2}\right)\sin\left(\frac{\phi_v}{2}\right)\left[y_1y_2(y_1^2-y_2^2)\cos\theta_R\cos\theta_L-y_2y_3(y_2^2-y_3^2)\sin\theta_L\sin\theta_R \right]\Big) \, , \\
    \varepsilon_{1\tau} \  =  \ & \frac{4{\cal F}}{9}\Big(y_1y_3(y_1^2-y_3^2)\sin\phi_{u,+} \cos\theta_R\sin\theta_R \nonumber \\
    & +\cos\left(\frac{\phi_v}{2}\right)\cos\left(\frac{\phi_{u,+}}{2}\right)\left[-y_1y_2(y_1^2-y_2^2)\cos\theta_R\sin\theta_L-y_2y_3(y_2^2-y_3^2)\cos\theta_L\sin\theta_R\right] \nonumber \\
    & + \sin\left(\frac{\phi_{u,+}}{2}\right)\sin\left(\frac{\phi_v}{2}\right)\left[-y_1y_2(y_1^2-y_2^2)\cos\theta_R\cos\theta_L+y_2y_3(y_2^2-y_3^2)\sin\theta_L\sin\theta_R \right]\Big) \, , 
\end{alignat}
\end{subequations}
\end{widetext}
where $\phi_{u,\pm}=\phi_u\pm \pi/3$. Similar expressions for $\varepsilon_{i\alpha}$ (with $\alpha=2,3$), as well as for other choices of $s$ and $t$ can be obtained, but we do not show them here. 

Similarly for Case 3, we only show one subcase with $m$ and $s$ even just for illustration, again assuming that both ${\cal F}_{12}$ and ${\cal F}_{13}$ are of similar size:
\begin{widetext}
\begin{subequations}
\begin{alignat}{2}
    \varepsilon_{1e} \, = \, & \frac{2{\cal F}}{9}y_1y_2(y_1^2-y_2^2)\cos\phi_m \sin 3\phi_s \left(4\cos2\theta_R+\sqrt 2 \sin 2\theta_R\right) \, , \\ 
    \varepsilon_{1\mu} \, = \, & -\frac{2{\cal F}}{9}y_1y_2(y_1^2-y_2^2)\cos\phi_{m,+} \sin 3\phi_s \left(4\cos2\theta_R+\sqrt 2 \sin 2\theta_R\right) \, , \\ 
    \varepsilon_{1\tau} \, = \, & -\frac{2{\cal F}}{9}y_1y_2(y_1^2-y_2^2)\cos\phi_{m,-} \sin 3\phi_s \left(4\cos2\theta_R+\sqrt 2 \sin 2\theta_R\right) \, , 
\end{alignat}
\end{subequations}
\end{widetext}
where $\phi_{m,\pm}=\phi_m\pm \pi/3$. 

Plugging in these CP asymmetries into Eq.~\eqref{eq:etaL} which feeds into Eq.~\eqref{eq:etaB}, we will calculate the BAU predictions in all the four cases considered here, and compare those with the observed value in Eq.~\eqref{eq:etaBobs} to identify the parameter space for successful leptogenesis (see section~\ref{subsec:colliderleptogenesis}) and to correlate the high- and low-energy CP phases (see section~\ref{sec:leptogenesis0nubb}). We will choose a value for the mass splitting between the RHNs $N_1$ and $N_{2,3}$, i.e.~$\Delta\, M_N=3\kappa M_N$ [cf.~Eq.~\eqref{eq:massesNi}], which maximizes the CP asymmetry in Eq.~\eqref{eq:eps3}. We find it to happen at  
\begin{equation}
   \frac{\Delta\, M_N}{M_N} \, = \, 3\, \kappa_{\rm max} \sim 1.23\, \frac{\Gamma_N}{M_N} \, ,
   \label{eq:kappamax}
\end{equation}
where $\Gamma_N$ stands for the average decay width of the $N_i$-pair participating in resonant leptogenesis, and the factor 1.23 is obtained numerically by maximizing the sum of the regulator parts for the mixing and oscillation contributions. 
Since $\Gamma_N$ scales as $M_N^2$, $\kappa_{\rm max}$ increases linearly with $M_N$. We will use this fact to choose the appropriate $\kappa_{\rm max}$ for our leptogenesis scans presented in section~\ref{subsec:colliderleptogenesis}. 

\section{Collider Signatures}
\label{sec:colliderLep}
In this section, we discuss the collider signatures of the RHNs in our scenario and show how they can be used to test resonant leptogenesis in our model. Firstly, we will discuss in section~\ref{subsec:productionRHnu} the production of heavy RHNs at hadron colliders and our use of a simple gauge extension of the SM for enhancing the production cross section, without affecting the low-energy neutrino oscillation phenomena. Next, in section~\ref{subsec:RHnuwidths} we discuss the RHN decay lengths and identify the ERS points where one of the RHNs becomes long-lived. In section~\ref{subsec:RHnuBRs}, we analyze how the RHN decay BRs into charged leptons of different flavors vary in different cases. Using these BRs, we study the smoking-gun same-sign dilepton signal of RHNs in section~\ref{subsec:SSleptonsLHCRHnu}. Finally, in section~\ref{subsec:colliderleptogenesis} we show the correlation between the collider signal and leptogenesis, and study the detection prospects of the parameter space for successful leptogenesis at the HL-LHC, as well as at a future $100$-TeV collider. 

\subsection{Production of RHNs}
\label{subsec:productionRHnu}
As already mentioned in section~\ref{sec:intro}, a nice feature of the resonant leptogenesis mechanism is that it allows the RHNs to be as light as the electroweak scale~\cite{Pilaftsis:2005rv, Deppisch:2010fr, Dev:2014laa}, thus making it {\it testable} in laboratory experiments~\cite{Chun:2017spz}.\footnote{In contrast, the high-scale leptogenesis mechanism can only be {\it falsified}~\cite{Deppisch:2013jxa, Deppisch:2015yqa, Deppisch:2017ecm}.} However, in the minimal type-I seesaw within the SM gauge group, the constraints from light neutrino masses and mixing usually restrict the Yukawa couplings $y_f\lesssim 10^{-7}$ for TeV-scale RHNs, as in our case.\footnote{Exceptions can be made by choosing specific textures of $M_D$ and/or $M_R$, thereby allowing some of the Yukawa entries to be large~\cite{Pilaftsis:1991ug, Tommasini:1995ii, Gluza:2002vs, Kersten:2007vk, Xing:2009in, Gavela:2009cd, He:2009ua, Adhikari:2010yt, Ibarra:2010xw, Ibarra:2011xn, Mitra:2011qr, Lee:2013htl, CarcamoHernandez:2019kjy}, but at the expense of making the relevant RHNs quasi-Dirac and thus suppressing the corresponding LNV signal~\cite{Deppisch:2015qwa,Lopez-Pavon:2015cga,Fernandez-Martinez:2015hxa, Fernandez-Martinez:2016lgt, Das:2017nvm, Bolton:2019pcu}.} This makes the RHN production cross section for the smoking-gun collider signature of same-sign dilepton plus two jets without missing transverse energy~\cite{Keung:1983uu, Datta:1993nm, Han:2006ip, delAguila:2007qnc,Atre:2009rg, Dev:2013wba, Alva:2014gxa, Das:2015toa, Das:2016hof,Das:2017gke} too small to be accessible at the LHC~\cite{CMS:2018jxx, ATLAS:2019kpx}. For the same reason, the charged lepton flavor violating processes such as $\mu\to e\gamma$ are also suppressed in this scenario; see e.g.~Refs.~\cite{Alonso:2012ji,Deppisch:2013cya, Abada:2021zcm}.   

\begin{figure}[t!]
\centering
\includegraphics[width=0.7\linewidth]{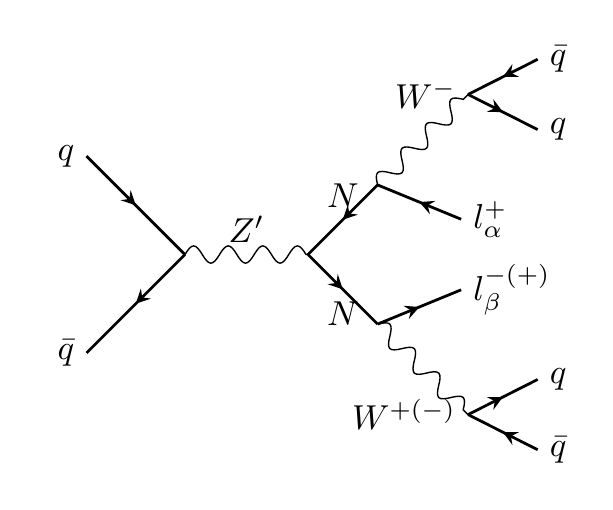}
\caption{Smoking-gun (LNV) signature of RHNs at hadron colliders in the $U(1)_{B-L}$ model. The final-state lepton flavors are dictated by the same Yukawa coupling structure $Y_D$ that governs leptogenesis.}
\label{fig:feyn} 
\end{figure}

Therefore, for the collider tests of the RHNs to be feasible, one needs to find a more efficient production mechanism that is not suppressed by the neutrino Yukawa couplings. One way is to extend the SM gauge group and make the RHNs as well as the SM quarks (and leptons) charged under this new group, so that the RHN production can occur via mediation of the new gauge bosons~\cite{Deppisch:2015qwa}. Here we will consider one such simple $U(1)_{B-L}$ extension of the SM~\cite{Davidson:1978pm, Marshak:1979fm}, which also provides a simple  ultraviolet-completion for the RHNs, which are in turn now required to cancel the gauge anomalies. In this model, the RHNs can be pair-produced via the $U(1)_{ B-L}$ gauge-boson mediation: $pp\to Z'\to N_i N_i$~\cite{Buchmuller:1991ce,Basso:2008iv, FileviezPerez:2009hdc, Kang:2015uoc, Cox:2017eme, Han:2021pun} (see Figure~\ref{fig:feyn}).
This production channel is only kinematically suppressed by the mass of the new gauge
boson, $M_{Z'}$. If $M_i<M_{Z'}/2$, the two RHNs are produced on-shell. Very similar production cross sections are expected for all $N_i$ in our scenario, since their masses are (almost) degenerate,
 see Eq.~\eqref{eq:massesNi}.

There exist stringent limits on the $Z'$ mass and the corresponding gauge coupling $g_{B-L}$ from existing collider data. The contact interaction bound from $e^+e^-\to f\bar{f}$ data at LEP-II requires $M_{Z'}/g_{B-L}\gtrsim 7.0$ TeV~\cite{ALEPH:2013dgf}. For $Z'$ masses kinematically accessible at the LHC, more stringent bounds are obtained from high-mass dilepton resonance searches~\cite{ATLAS:2019erb, CMS:2019tbu}. The LHC limits are usually derived in the so-called sequential SM where the $Z'$ couplings are the same as those of the SM $Z$ boson. In the $U(1)_{B-L}$ model with RHNs, these limits are slightly modified~\cite{Das:2021esm}. In fact, in a general $U(1)_X$ gauge group which is a linear combination of the $U(1)_{\rm Y}$ and $U(1)_{B-L}$~\cite{Appelquist:2002mw}, the $Z'$ limits depend on the choice of two scalars charges $x_H$ and $x_\Phi$. In fact, the RHN collider signal can be enhanced for special values of $(x_H,x_\Phi)$ due to an enhancement in the ${\rm BR}(Z'\to N_iN_i)$~\cite{Das:2017flq}. There also exist other $U(1)_X$ models where the $Z'$ is leptophobic~\cite{Faraggi:1996kk, GomezDumm:1997br, Malinsky:2005bi, Buckley:2011mm, Deppisch:2013cya}, thus avoiding the LHC dilepton bounds, and can be lighter, as long as it satisfies the weaker dijet bounds~\cite{Sirunyan:2018xlo, ATLAS:2019bov}. In this work, we only consider the simplest case with  $(x_H,x_\Phi)=(0,2)$, which corresponds to the minimal $B-L$ model with flavor-diagonal and flavor-universal $Z'$ couplings to leptons. Furthermore, we will use a benchmark value of $g_{B-L}=0.1$ for which the current LHC limit is $M_{Z'}\gtrsim 4.1$ TeV. 
\begin{figure}[t!]
\centering
\includegraphics[width=0.47\textwidth]{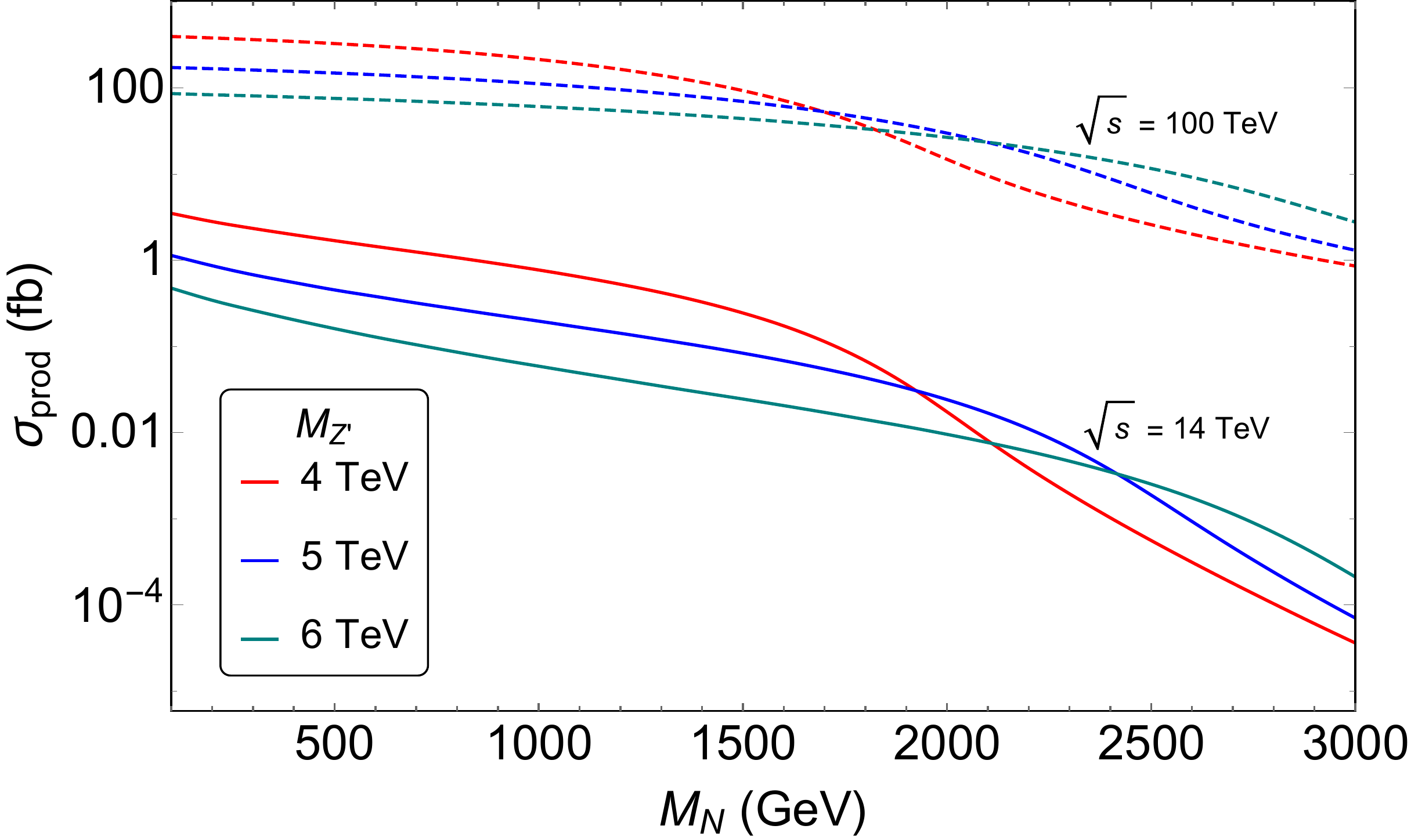}
\caption{RHN pair production ($pp\to Z' \to N_iN_i$) cross section as a function of the RHN mass $M_N$ at $\sqrt s=14$ TeV LHC (solid lines) and $\sqrt s=100$ TeV future collider (dotted lines) for $M_{Z'}= 4$ TeV (red), 5 TeV (blue) and 6 TeV (green). Here we have normalized the cross-section for $g_{B-L}=1$.}
\label{fig:collider}
\end{figure}

To calculate the hadron collider production cross sections 
\begin{align}
    \sigma_{\mathrm{prod}} \, \equiv \, \sigma(pp\to Z' \to N_iN_i) \, ,
    \label{eq:prod}
\end{align}
we implement the model Lagrangian into {\tt FeynRules}~\cite{Alloul:2013bka} and generate the Universal FeynRules Output (UFO)  file for the $B-L$ embedding~\cite{Basso:2008iv,Amrith:2018yfb}. This is then imported to the {\tt MadGraph\_aMC@NLO-v2.8.3} event generator~\cite{Alwall:2014hca} with the default PDF set to calculate $\sigma_{\mathrm{prod}}$ at parton level. Our results for the gauge coupling $g_{B-L}=1$ are shown in Figure~\ref{fig:collider} for three different values of $M_{Z'}=$ 4,5,6 TeV (red, blue and green, respectively). The solid lines are for the LHC center-of-energy $\sqrt s=14$ TeV, whereas the dotted lines are for a future 100 TeV collider. The change in the slope of the curves occurs near the kinematic threshold for on-shell pair-production,~i.e.~$M_N=M_{Z'}/2$. For $M_N<M_{Z'}/2$, the cross-section scales as $g_{B-L}^2$, whereas above the $M_{Z'}/2$ threshold, the cross-section scales as $g_{B-L}^4$. From Figure~\ref{fig:collider}, we find that the cross section is below fb-level at the LHC for realistic values of the $Z'$ mass and coupling consistent with the dilepton bound mentioned above, which makes it challenging to find sizable number of events even at the HL-LHC, as we will see explicitly in section~\ref{subsec:colliderleptogenesis}. Therefore, we have included the 100 TeV option which has far better sensitivity to the leptogenesis parameter space discussed in section~\ref{subsec:colliderleptogenesis}.

\subsection{Decay Lengths}
\label{subsec:RHnuwidths}
After being produced, the RHNs decay into SM final states through their Yukawa couplings $Y_D$. The total decay width $ \Gamma_i$ of the RHN $N_i$ at tree-level is given by
\begin{equation}
    \Gamma_i \, = \, \frac{(Y_D^\dagger \, Y_D)_{ii}}{8 \, \pi } M_i \, ,
\end{equation}
where the form of $Y_D$ in our model is determined by the choice for generator $Z$ of the $Z_2$ symmetry and the choice of the CP transformation $X$, as discussed in section~\ref{sec:cases}. Despite this dependence on the generators of $Z_2$ symmetry and CP transformation, we will see that $\Gamma_i$ is independent of the value of $n$ and depends only on odd/even behavior of the parameters $s,t,m$ and on the rotation angle $\theta_R$.  Even in some cases, $\Gamma_i$ is completely independent of all these parameters, as shown below.

\subsubsection{Case 1}
\begin{figure*}[t!]
\centering
\includegraphics[width=0.49\textwidth]{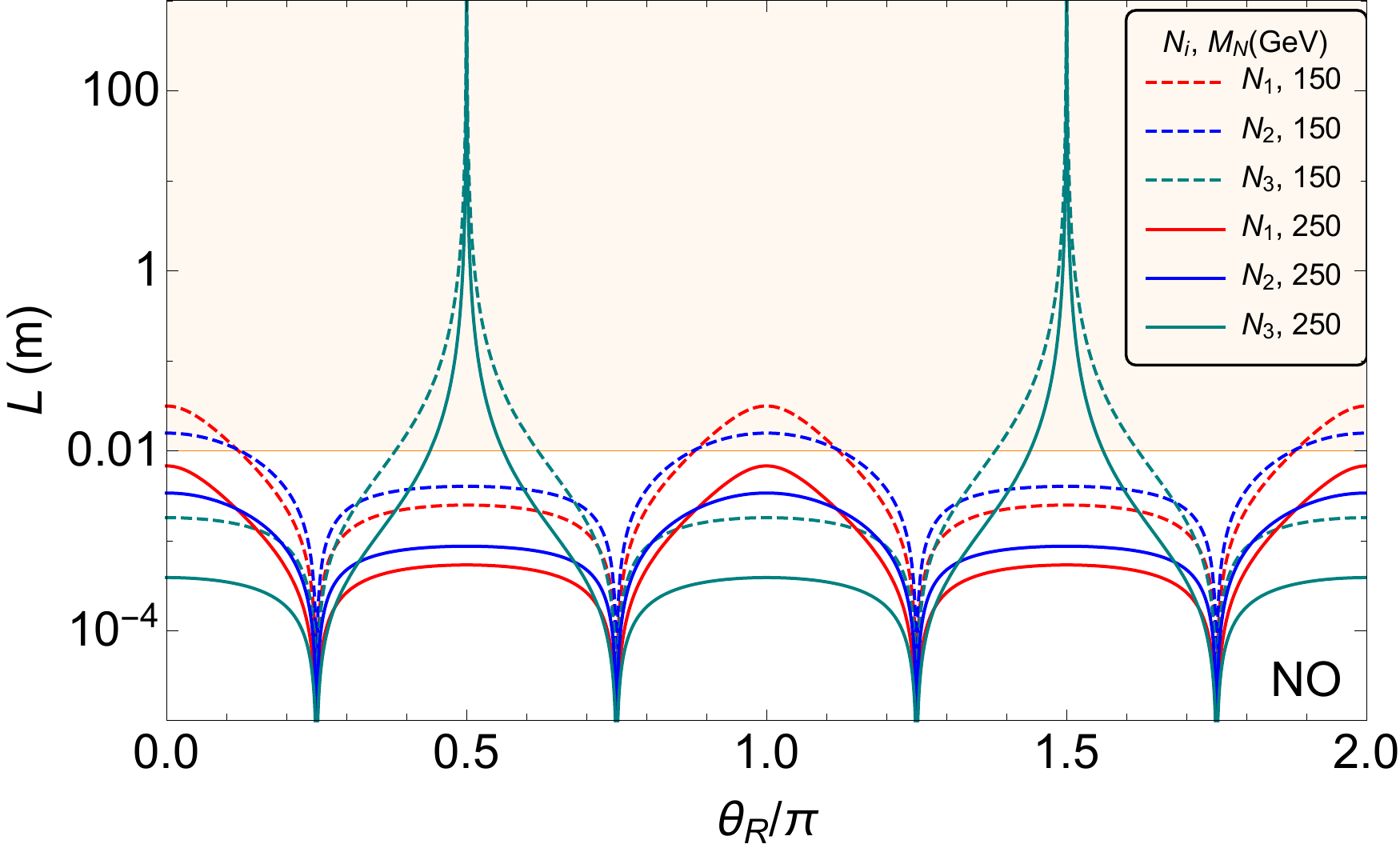}
\includegraphics[width=0.49\textwidth]{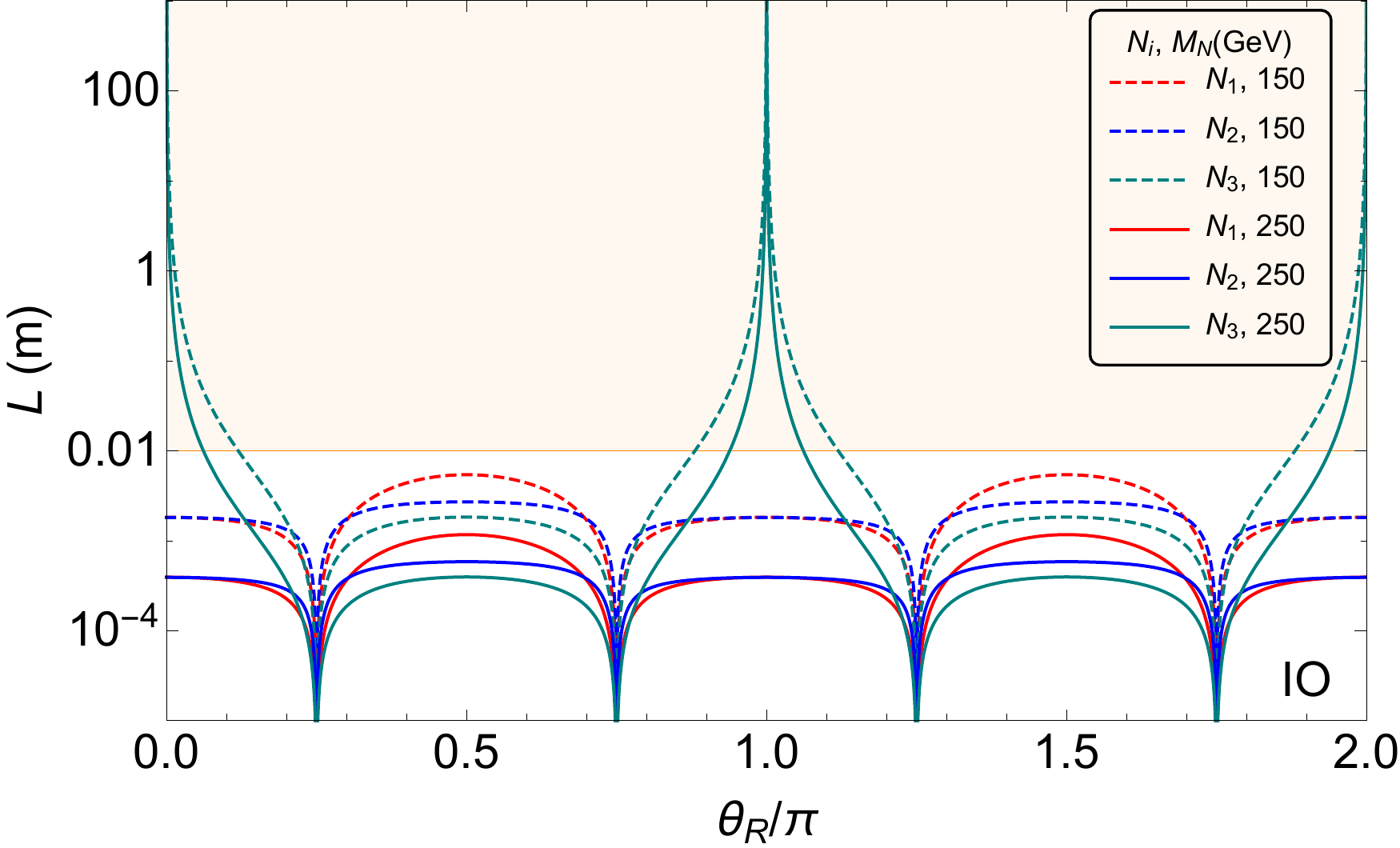}\\
\includegraphics[width=0.49\textwidth]{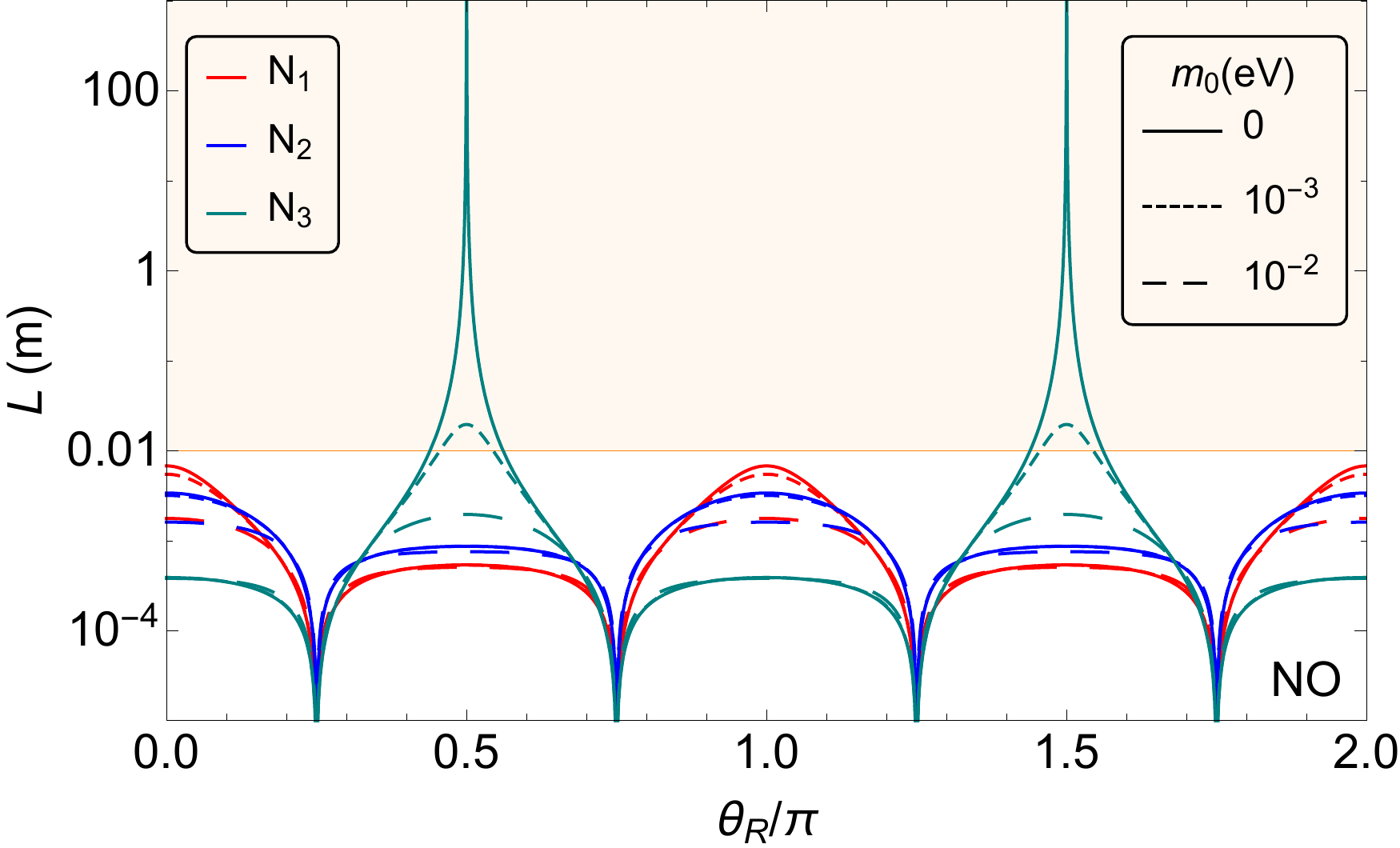}
\includegraphics[width=0.49\textwidth]{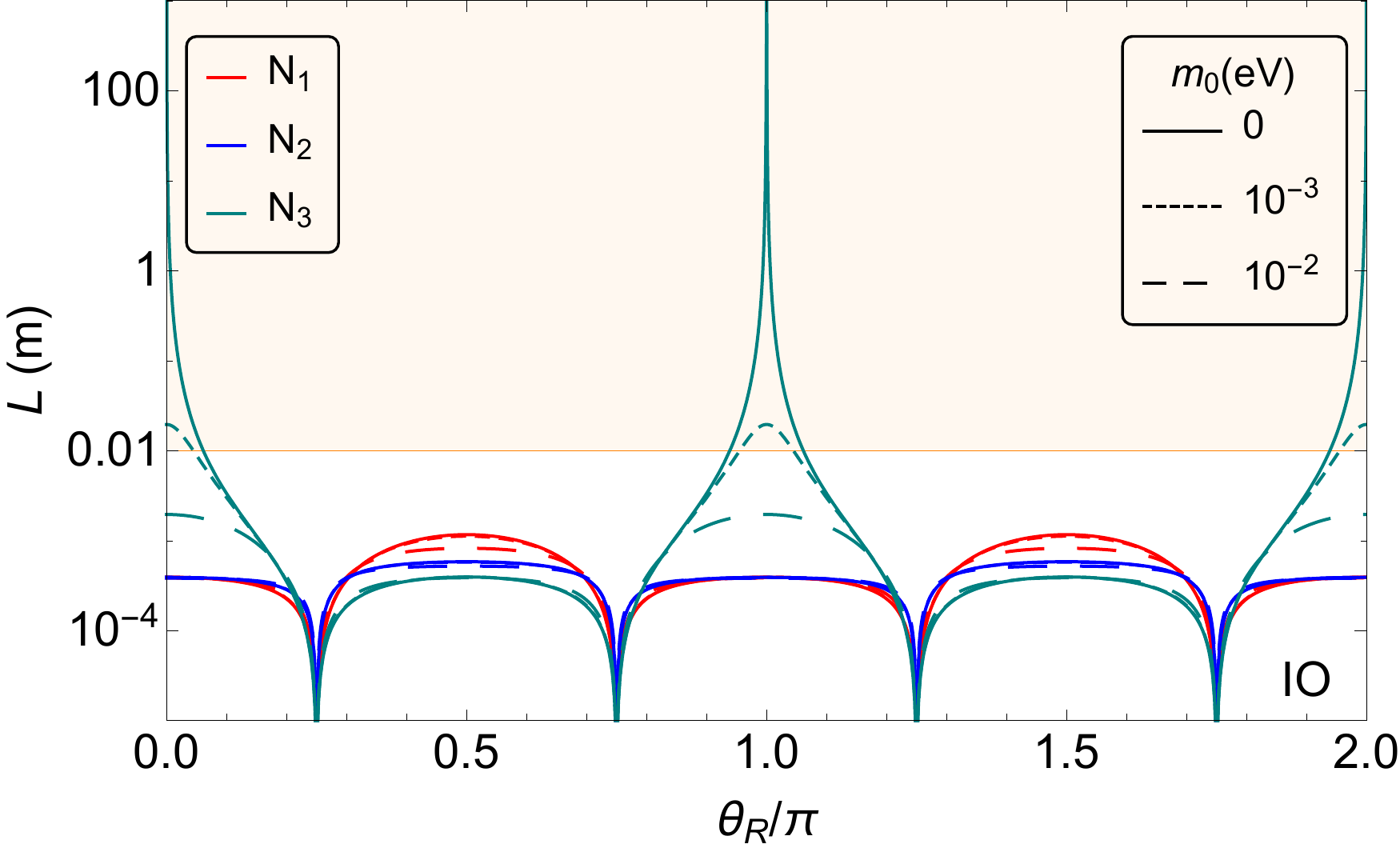}
\caption{{\it Case 1}. $N_{1,2,3}$ decay lengths are plotted against $\theta_R$ for different values of the RHN mass scale $M_N$ (upper panels, with $m_0=0$) and light neutrino mass $m_0$ (lower panels, with $M_N=250$ GeV). The left (right) panels are for NO (IO). The unshaded (shaded) region indicates the prompt (displaced/long-lived) signal regime.}
\label{fig:dln1}
\end{figure*}

The expressions for the decay widths of the three  heavy RHNs in this case do not depend on the values of $s$, and only depend on the Yukawa couplings $y_f$ and the angle $\theta_R$: 
\begin{subequations}\label{gammaC1}
\begin{alignat}{2}
\Gamma_1 \, & =  \, \frac{M_N}{24 \, \pi} \, \left( 2\, y_1^2 \, \cos^2 \theta_R + y_2^2 + 2\, y_3^2 \, \sin^2 \theta_R \right)  \, , \\
\Gamma_2 \, & =  \,  \frac{M_N}{24 \, \pi} \, \left( y_1^2 \, \cos^2\theta_R + 2 \, y_2^2 + y_3^2\, \sin^2 \theta_R  \right) \, ,  \\
\Gamma_3 \ & =  \, \frac{M_N}{8 \, \pi} \, \left( y_1^2 \, \sin^2 \theta_R + y_3^2 \, \cos^2 \theta_R \right) \, .
\end{alignat}
\end{subequations}
We calculate the corresponding decay lengths in the laboratory frame and plot them as a function of $\theta_R$ in Figure~\ref{fig:dln1} for two different values of $M_N$ (upper panels, with $m_0=0$) and three different values of the lightest neutrino mass $m_0$ (lower panels, with $M_N=250$ GeV) for both NO (left panels) and IO (right panels). In doing so, we have assumed that $N_i$ are produced via $Z^\prime$ with  mass $M_{Z^\prime}=4$ TeV, meaning 
 the Lorentz boost factor is given by $\gamma=M_{Z^\prime}/2M_N=8~(13.3)$ for $M_N=250~(150)$ GeV. 

The Yukawa couplings are fixed by the light neutrino mass spectrum [cf.~Eq.~\eqref{eq:ycase1}]. Therefore, we show the decay lengths for NO and IO in the left and right panels of Figure~\ref{fig:dln1}, respectively. 
In Case 1, strong NO and strong IO correspond to $y_1=0$ and $y_3=0$ respectively, when the lightest neutrino becomes massless, i.e.~$m_0=0$. For the ERS points $\theta_R \to \pi/2, \, 3 \pi/2$ (NO) or $\theta_R \to 0, \, \pi$ (IO), we see from Eq.~\eqref{gammaC1} that $\Gamma_3\to 0$, or $N_3$ becomes long-lived, as shown by the shaded region corresponding to $L>1$ cm.  Thus, for sufficiently large production cross section, $N_3$ can be searched for either with displaced vertex searches at the LHC or with dedicated LLP detectors like FASER~\cite{FASER:2018eoc} and MATHUSLA~\cite{Curtin:2018mvb}, depending on the amount of deviation from the ERS point which can be parametrized by $\delta \theta_R=|\theta-\theta_{\rm ERS}|$. For $M_N$ in the few hundred GeV range and $10^{-4} \lesssim \delta\theta_R \lesssim 10^{-2}$, $N_3$ can have decay lengths of a few hundred m which is in the range of the LLP detectors like MATHUSLA.
If $10^{-3} \lesssim \delta \theta_R \lesssim 10^{-1}$, $N_3$ can be detected either with the LLP searches or can be probed at the LHC via displaced vertex signatures, similar to the scenarios studied in Refs.~\cite{Das:2019fee, Chiang:2019ajm}. There are also accompanying signals from $N_{1,2}$ decays, which are mostly prompt (but can also be displaced, depending on the choice of $\theta_R$), as shown in the upper panels of Figure~\ref{fig:dln1}. For $m_0\neq 0$, the ERS becomes less pronounced, as shown by the flattening of the $N_3$ peaks in the lower panels of Figure~\ref{fig:dln1}. For $m_0$ close to its maximum allowed value [cf.~Eq.~\eqref{m0bound}], the three RHN decay widths become almost indistinguishable.

\subsubsection{Case 2}
\begin{figure*}[t!]
\centering
\includegraphics[width=0.49\textwidth]{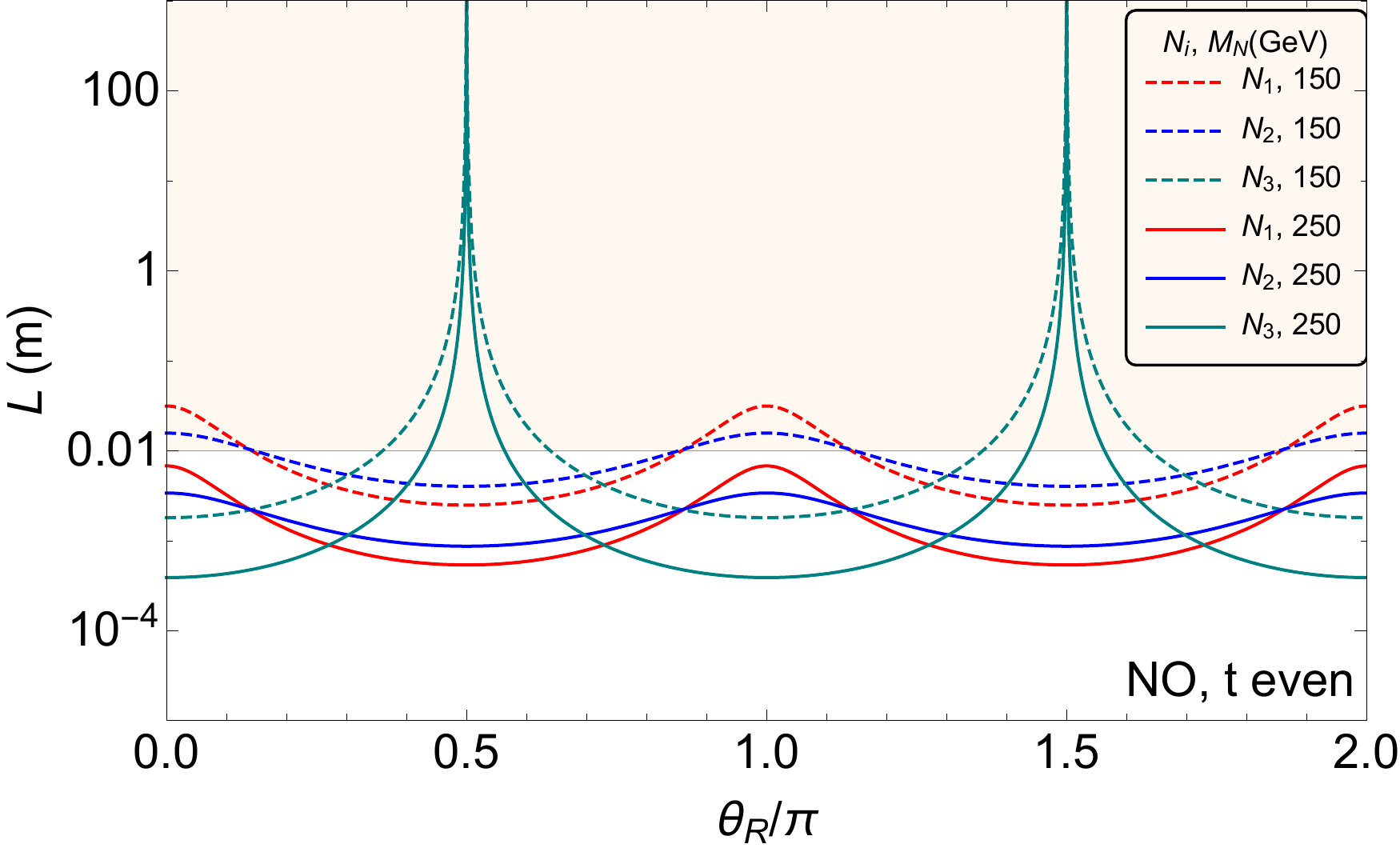}
\includegraphics[width=0.49\textwidth]{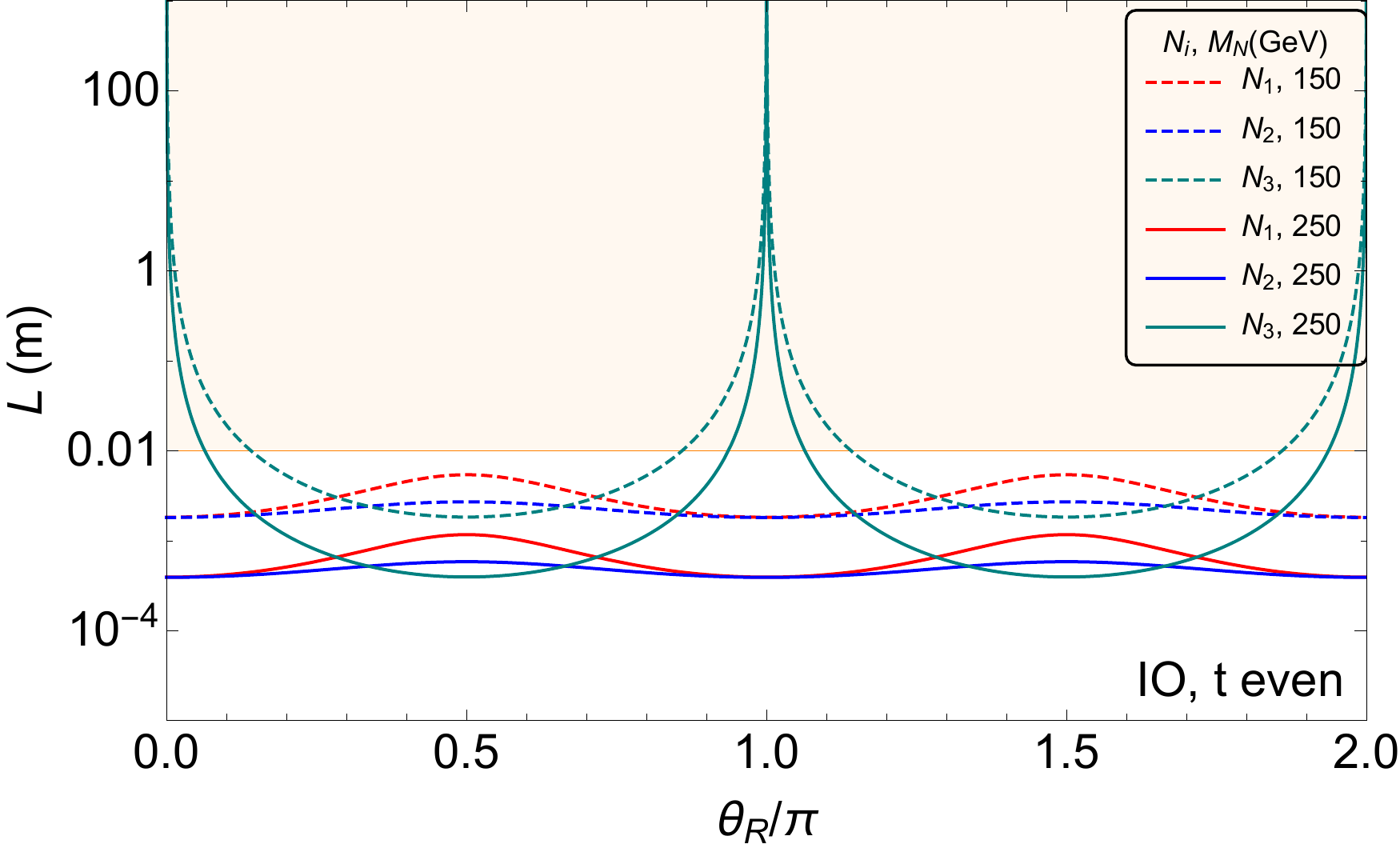}
\caption{{\it Case 2} with $t$ even. $N_{1,2,3}$ decay lengths are plotted against $\theta_R$ for  different values of the RHN mass scale $M_N$ with $m_0=0$. The left (right) panels are for NO (IO). The unshaded (shaded) region indicates the prompt (displaced/long-lived) signal regime.}
\label{fig:dlc2}
\end{figure*}

The decay widths of the RHNs in this case just depend on  whether $t$ is even/odd and are independent of $s$. \\
For $t$-even,
\begin{subequations}\label{eq:case2even}
\begin{alignat}{2}
\Gamma_1 \, &  =  \,  \frac{M_N}{24 \, \pi} \, \left( 2\, y_1^2 \, \cos^2 \theta_R + y_2^2 + 2\, y_3^2 \, \sin^2 \theta_R \right)  \, , \\
\Gamma_2 \, &  =  \,  \frac{M_N}{24 \, \pi} \, \left( y_1^2 \, \cos^2\theta_R + 2 \, y_2^2 + y_3^2\, \sin^2 \theta_R  \right) \, ,  \\
\Gamma_3  \, &  =  \, \frac{M_N}{8 \, \pi} \, \left( y_1^2 \, \sin^2 \theta_R + y_3^2 \, \cos^2 \theta_R \right) \, . 
\end{alignat}
\end{subequations}
For $t$-odd, 
\begin{subequations}\label{eq:case2odd}
\begin{alignat}{2}
\Gamma_1 \, & =  \, \frac{M_N}{24 \, \pi} \, \left( y_1^2  + y_2^2 + y_3^2 \right)  \, , \\
 \Gamma_2 \,  & =  \,  \frac{M_N}{24 \, \pi} \, \left( y_1^2  + 4\, y_2^2 + y_3^2 \right)  \, , \\ 
\Gamma_3  \, &  =  \, \frac{M_N}{24 \, \pi} \, \left( y_1^2  + y_3^2 \right)  \, .
\end{alignat}
\end{subequations}
As can be seen above,  the decay lengths are independent of $\theta_R$ for odd values of $t$ and non-zero in all cases including strong NO and strong IO. Thus in Case 2, the ERS points are present only for even values of $t$ and lie at the same values of $\theta_R$ as in Case 1 for both mass orderings, irrespective of the $s$ values as shown in Figure~\ref{fig:dlc2}. Note that similar to Case 1, strong NO and strong IO in this case correspond to $y_1=0$ and $y_3=0$ respectively. We do  not show the variation with respect to $m_0$ in this case because it follows the same trend as in Case 1. In what follows, we will set $m_0=0$ for concreteness, unless otherwise specified. 

\subsubsection{Case 3}
\label{subsec:decay3}
The decay widths of the RHNs in Case 3a and 3b.1 depend on the combination of $(m,s)$ being even/odd. There is an important distinction to be noted that unlike other cases where $N_3$ becomes long-lived, in Case 3a and 3b.1, it is $N_1$ for which the decay length is the longest near  the ERS points. The decay widths are explicitly given as follows: 
\begin{widetext} 
For $m$-even, $s$-even, 
\begin{subequations}
\begin{alignat}{2}
\Gamma_1 \, = \, & \frac{M_N}{48 \, \pi} \, \left( 3(y_1^2  + y_2^2) + (y_1^2  - y_2^2)(\cos 2\theta_R-2 \sqrt{2}\,\sin 2\theta_R) \right)  \, ,  \\
\Gamma_2 \, = \, & \frac{M_N}{192 \, \pi} \, \left( 3(y_1^2  + y_2^2 + 6\, y_3^2) - (y_1^2  - y_2^2)(\cos 2\theta_R-2 \sqrt{2}\,\sin 2\theta_R) \right) \, ,  \\
\Gamma_3  \, = \,  & \frac{M_N}{64 \, \pi} \, \left( 3(y_1^2  + y_2^2 )+ 2\, y_3^2 - (y_1^2  - y_2^2)(\cos 2\theta_R-2 \sqrt{2}\,\sin 2\theta_R) \right) \, .  
\end{alignat}
\end{subequations}
For $m$-even, $s$-odd, 
\begin{subequations}
\begin{alignat}{2}
\Gamma_1 \, = \, & \frac{M_N}{48 \, \pi} \, \left( 3(y_1^2  + y_2^2) + (y_1^2  - y_2^2)\cos 2\theta_R \right)  \, ,  \\
\Gamma_2 \, = \, & \frac{M_N}{192 \, \pi} \, \left( 3(y_1^2  + y_2^2 + 6\, y_3^2) - (y_1^2  - y_2^2)\cos 2\theta_R \right) \, ,  \\
\Gamma_3  \, = \,  & \frac{M_N}{64 \, \pi} \, \left( 3(y_1^2  + y_2^2 )+ 2\, y_3^2 - (y_1^2  - y_2^2)\cos 2\theta_R \right) \, .  
\end{alignat}
\end{subequations}
For $m$-odd, $s$-even,
\begin{subequations}
\begin{alignat}{2}
\Gamma_1 \, = \, & \frac{M_N}{48 \, \pi} \, \left( 4y_1^2  + y_2^2 + y_3^2 + (y_2^2  - y_3^2)\cos 2\theta_R \right)  \, ,  \\
\Gamma_2 \, = \,  & \frac{M_N}{192 \, \pi} \, \left(2 \, y_1^2 + 11(y_2^2  + y_3^2 ) - 7(y_2^2  - y_3^2)\cos 2\theta_R \right) \, ,  \\
\Gamma_3 \, = \, & \frac{M_N}{64 \, \pi} \, \left(2 \, y_1^2 + 3(y_2^2  + y_3^2 ) +(y_2^2  - y_3^2)\cos 2\theta_R \right) \, .  
\end{alignat}
\end{subequations}
For $m$-odd, $s$-odd, 
\begin{subequations}
\begin{alignat}{2}
\Gamma_1 \, = \, & \frac{M_N}{48 \, \pi} \, \left( 4y_1^2  + y_2^2 + y_3^2 + (y_2^2  - y_3^2)\cos 2\theta_R \right)  \, ,  \\
\Gamma_2 \, = \, & \frac{M_N}{192 \, \pi} \, \left(2 \, y_1^2 + 11(y_2^2  + y_3^2 ) - (y_2^2  - y_3^2)(7\,\cos 2\theta_R+6 \sqrt{2}\,\sin 2\theta_R) \right) \, ,  \\
\Gamma_3  \, = \,  & \frac{M_N}{64 \, \pi} \, \left(2 \, y_1^2 + 3(y_2^2  + y_3^2 ) +(y_2^2  - y_3^2)(\cos 2\theta_R+2 \sqrt{2}\,\sin 2\theta_R) \right) \, . 
\end{alignat}
\end{subequations}
\end{widetext}
\begin{figure*}[t!]
\centering
\includegraphics[width=0.49\textwidth]{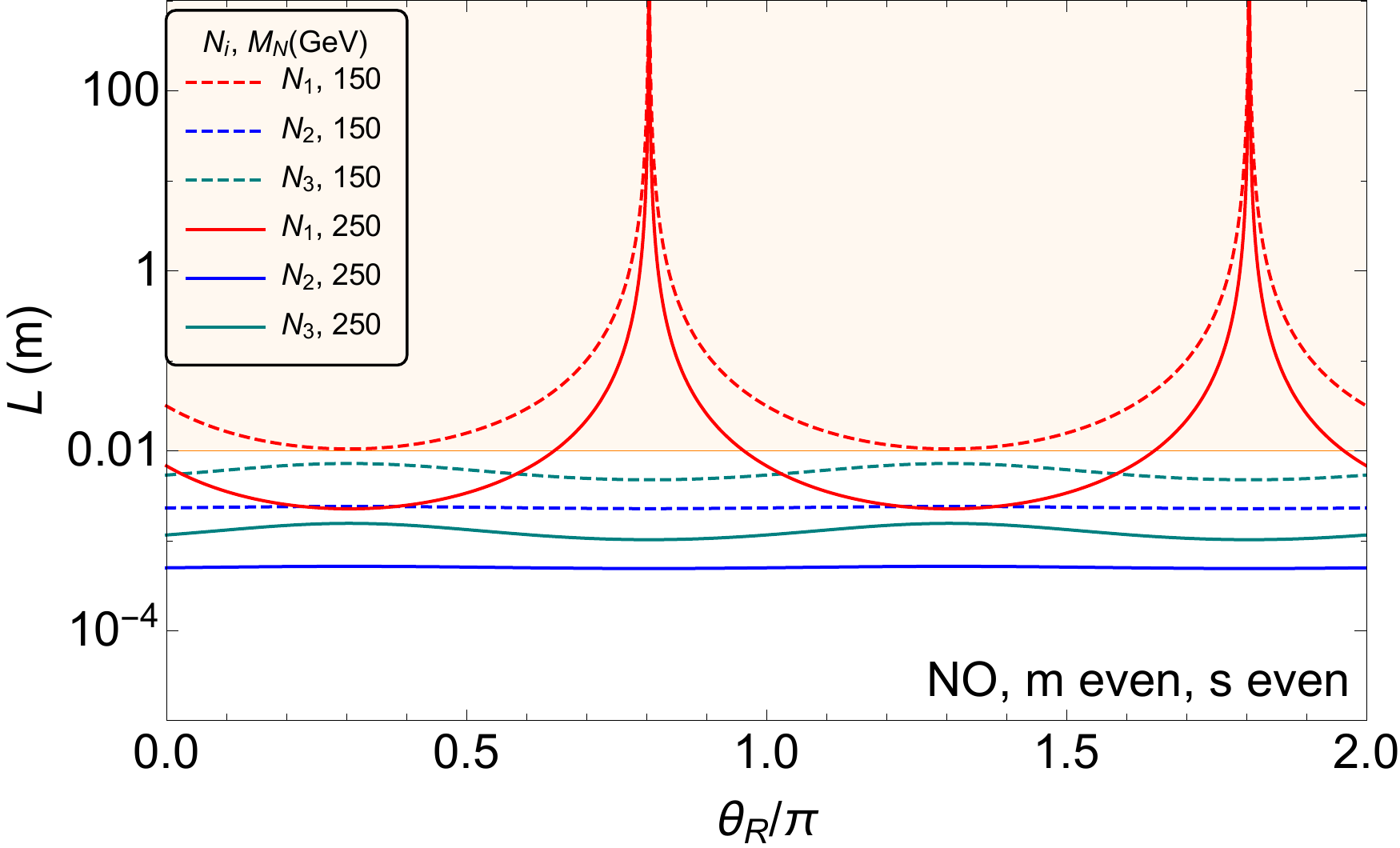}
\includegraphics[width=0.49\textwidth]{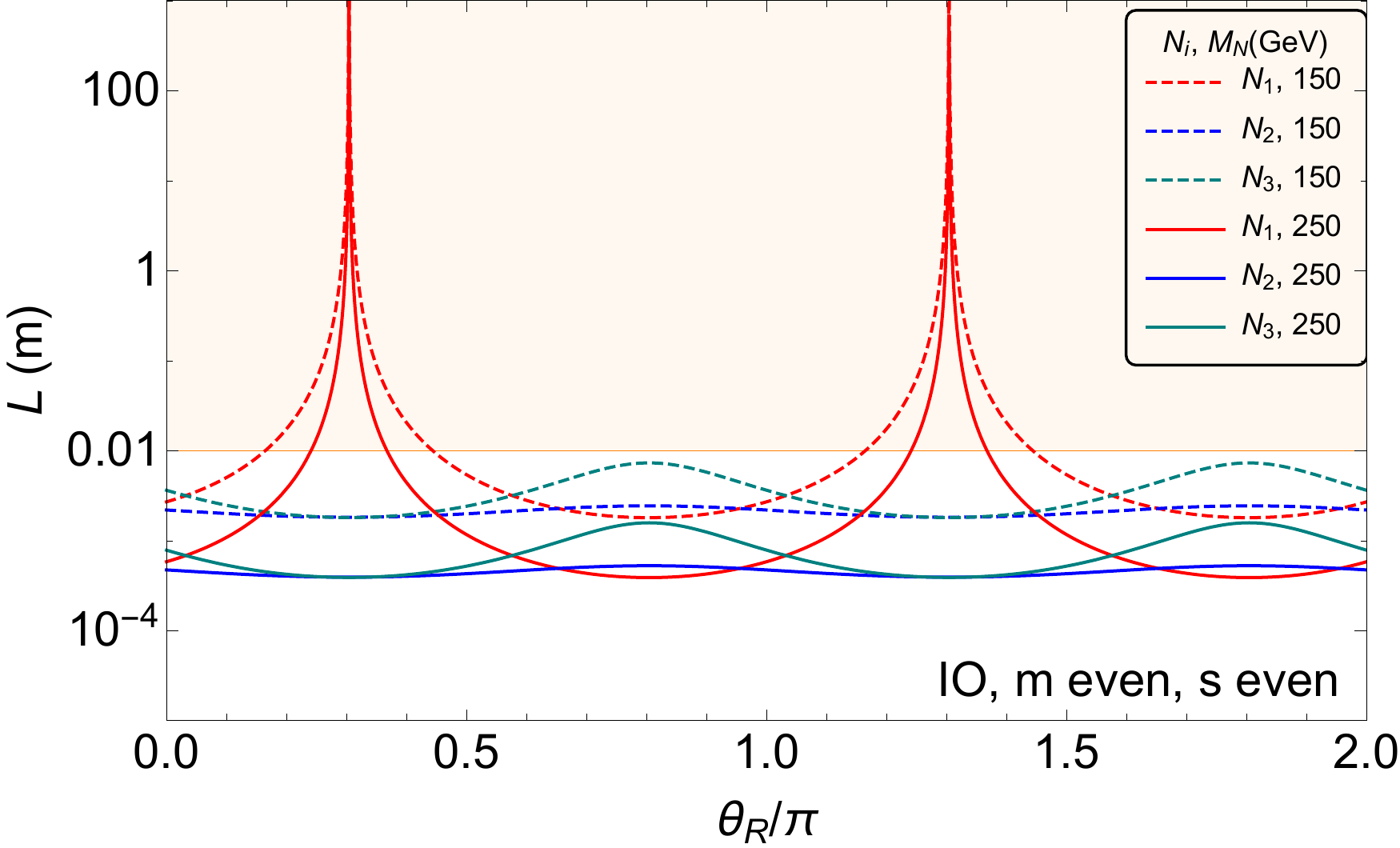}
\caption{{\it Case 3a} with NO (left panel) and {\it Case 3b.1} with IO (right panel), both with $m$ even, $s$ even. $N_{1,2,3}$ decay lengths are plotted against $\theta_R$ for two different values of the RHN mass scale $M_N$, assuming $m_0=0$. The unshaded (shaded) region indicates the prompt (displaced/long-lived) signal regime. }
\label{fig:dlc3}
\end{figure*}

In Case 3a, strong NO and strong IO correspond to $y_1=0$ and $y_3=0$ respectively. As can be verified, ERS points are exhibited only for even values of $(m,s)$  with NO mass ordering, as shown in the left panel of Figure~\ref{fig:dlc3}. These points correspond to values of $\theta_R \approx 0.8\pi, \, 1.8\pi$. 

In Case 3b.1, strong NO and strong IO corresponds to $y_3=0$ and $y_2=0$ respectively. In this case, ERS points are also exhibited only for even values of $(m,s)$ but with IO mass ordering, as shown in the right panel of Figure~\ref{fig:dlc3}. These points correspond to values of $\theta_R \approx 0.3\pi, \, 1.3\pi$. 

\subsection{Branching Ratios}\label{subsec:RHnuBRs}
\begin{figure*}
\centering
\includegraphics[width=0.325\textwidth]{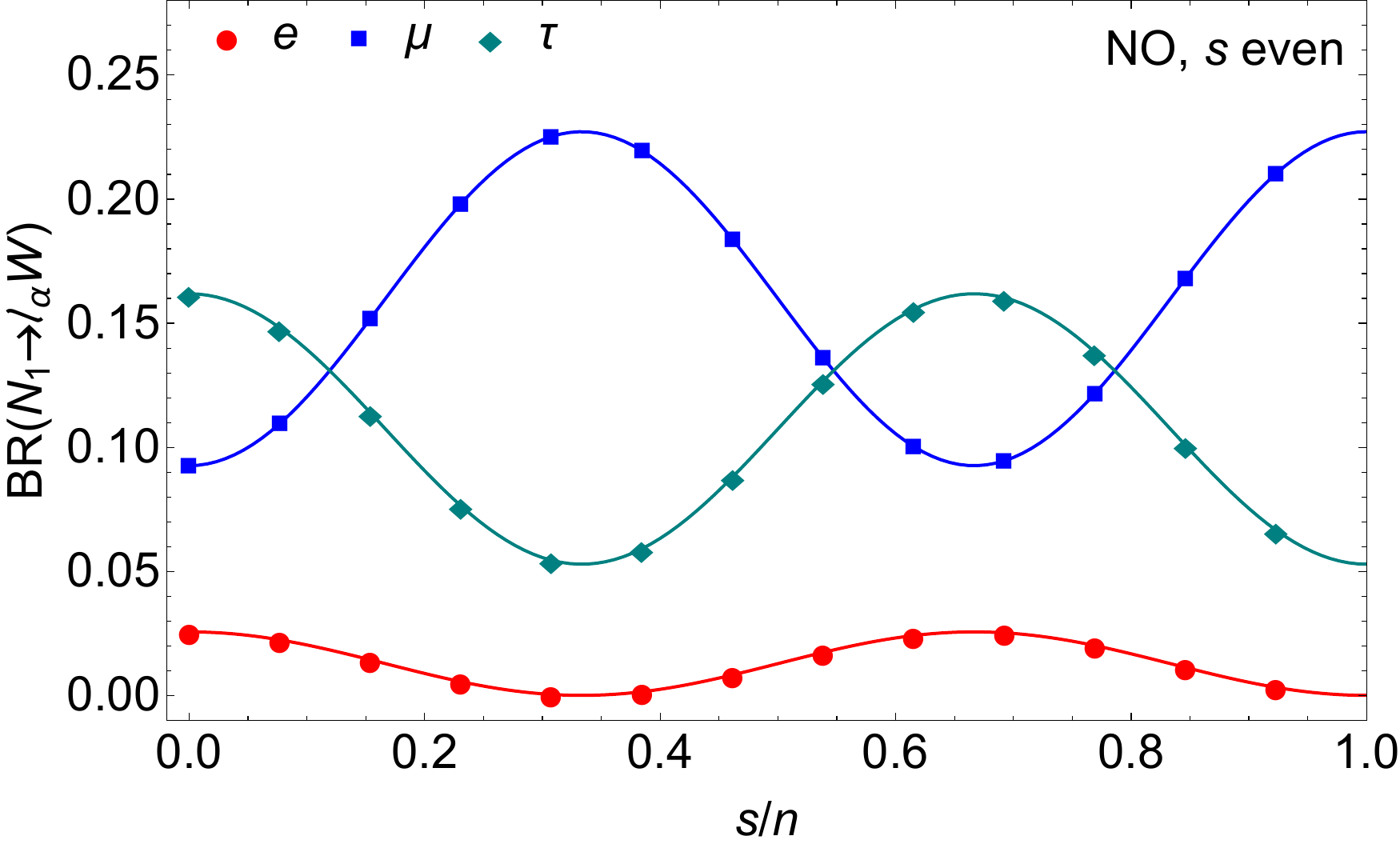}
\includegraphics[width=0.325\textwidth]{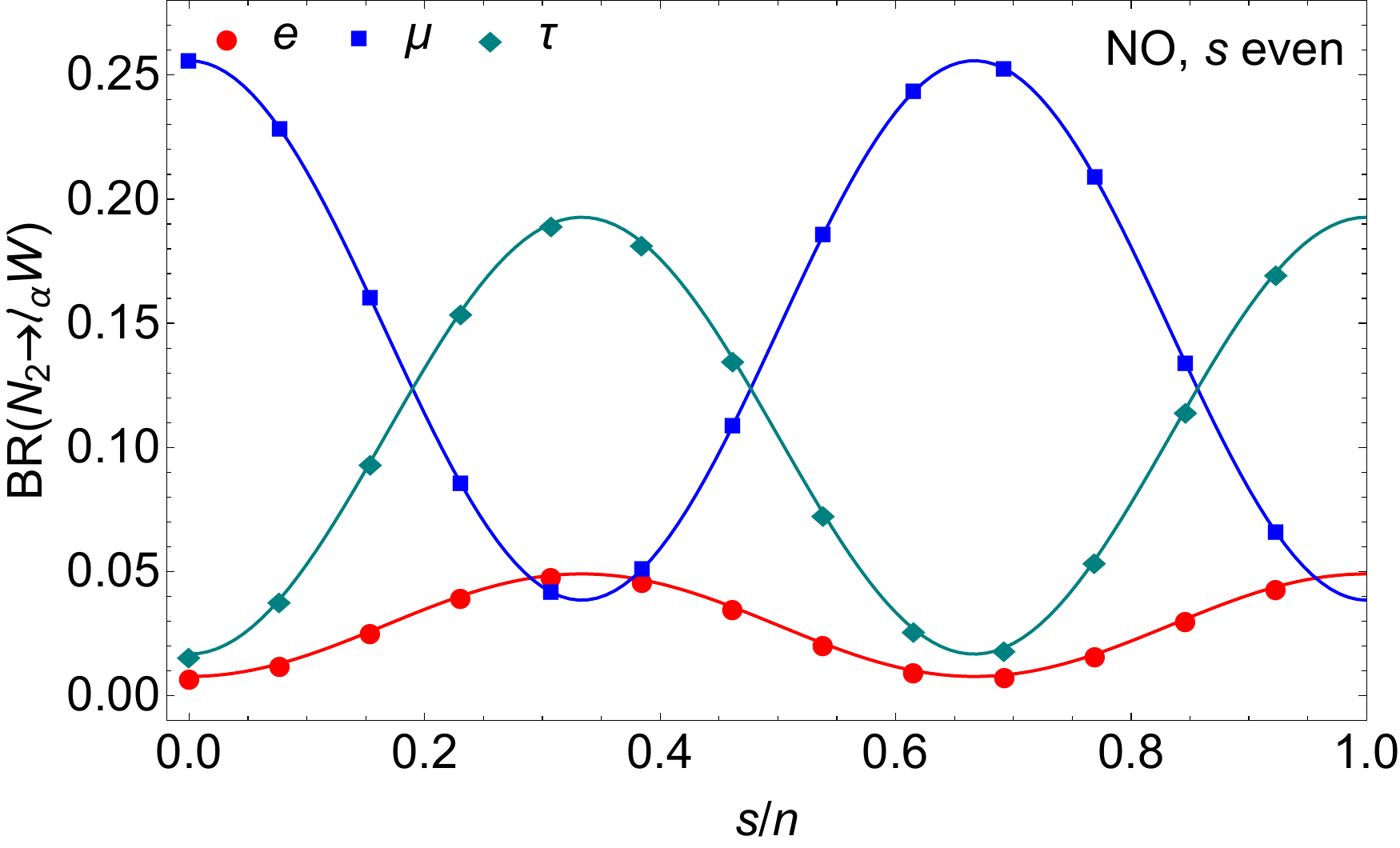}
\includegraphics[width=0.325\textwidth]{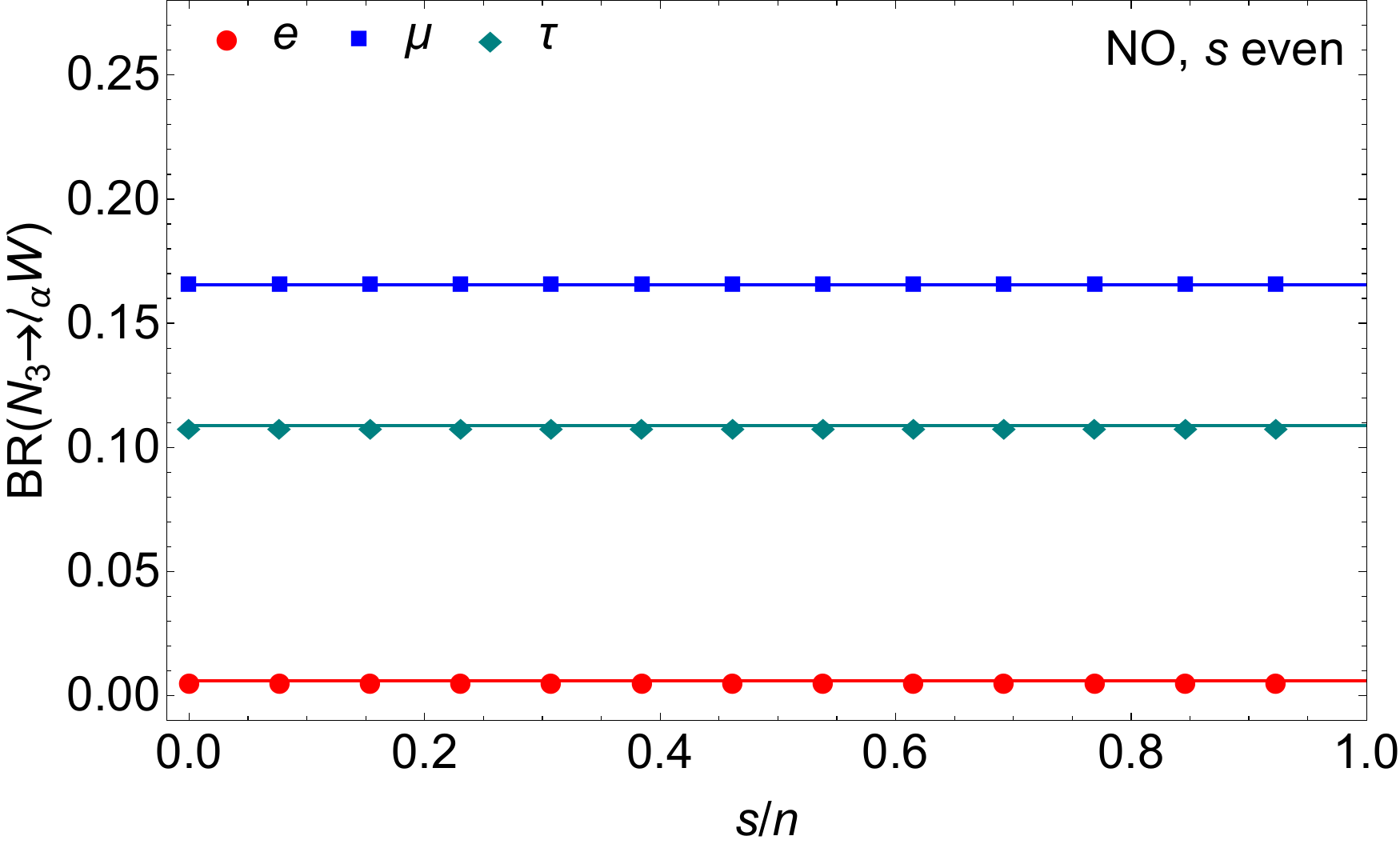} \\
\includegraphics[width=0.325\textwidth]{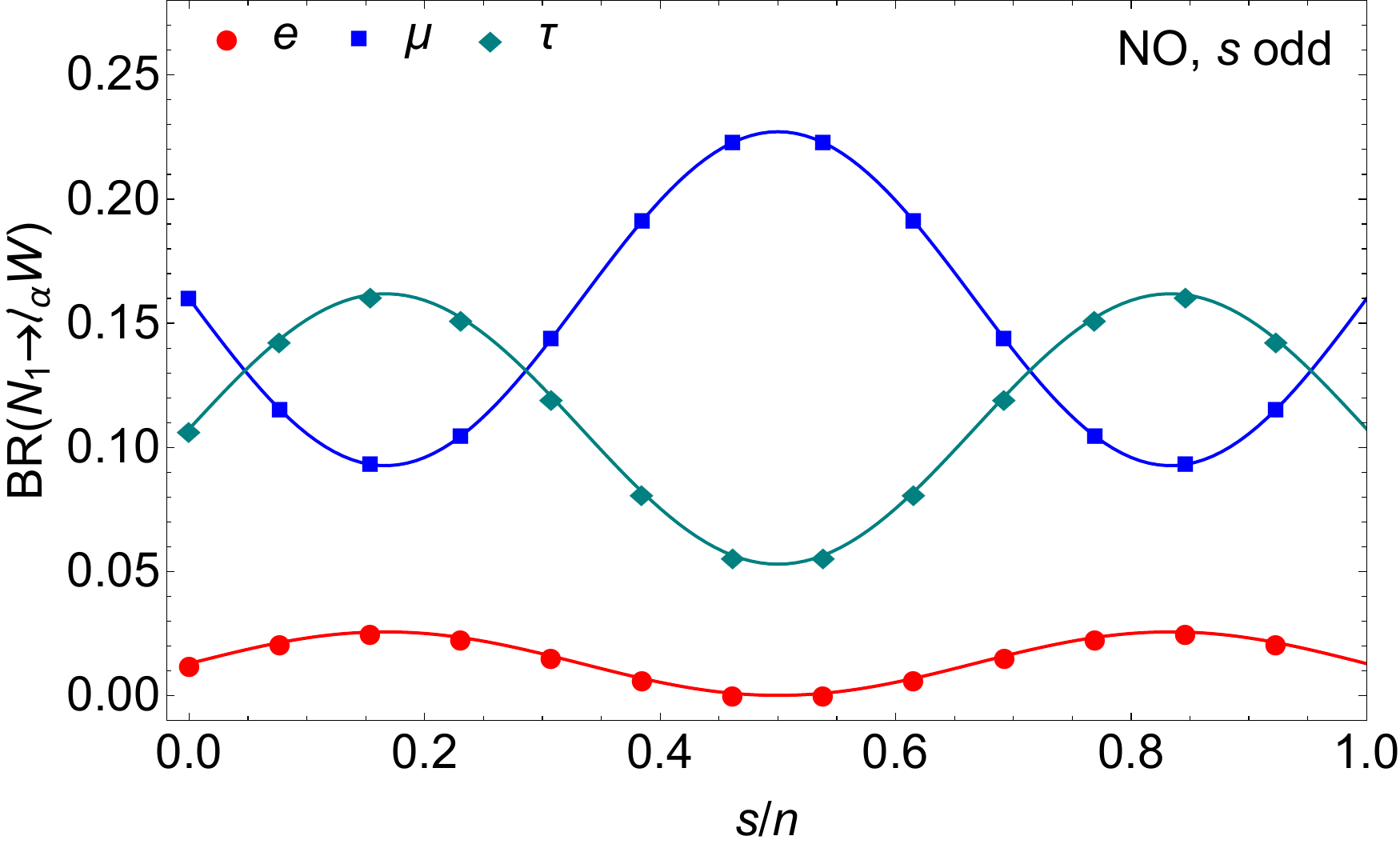}
\includegraphics[width=0.325\textwidth]{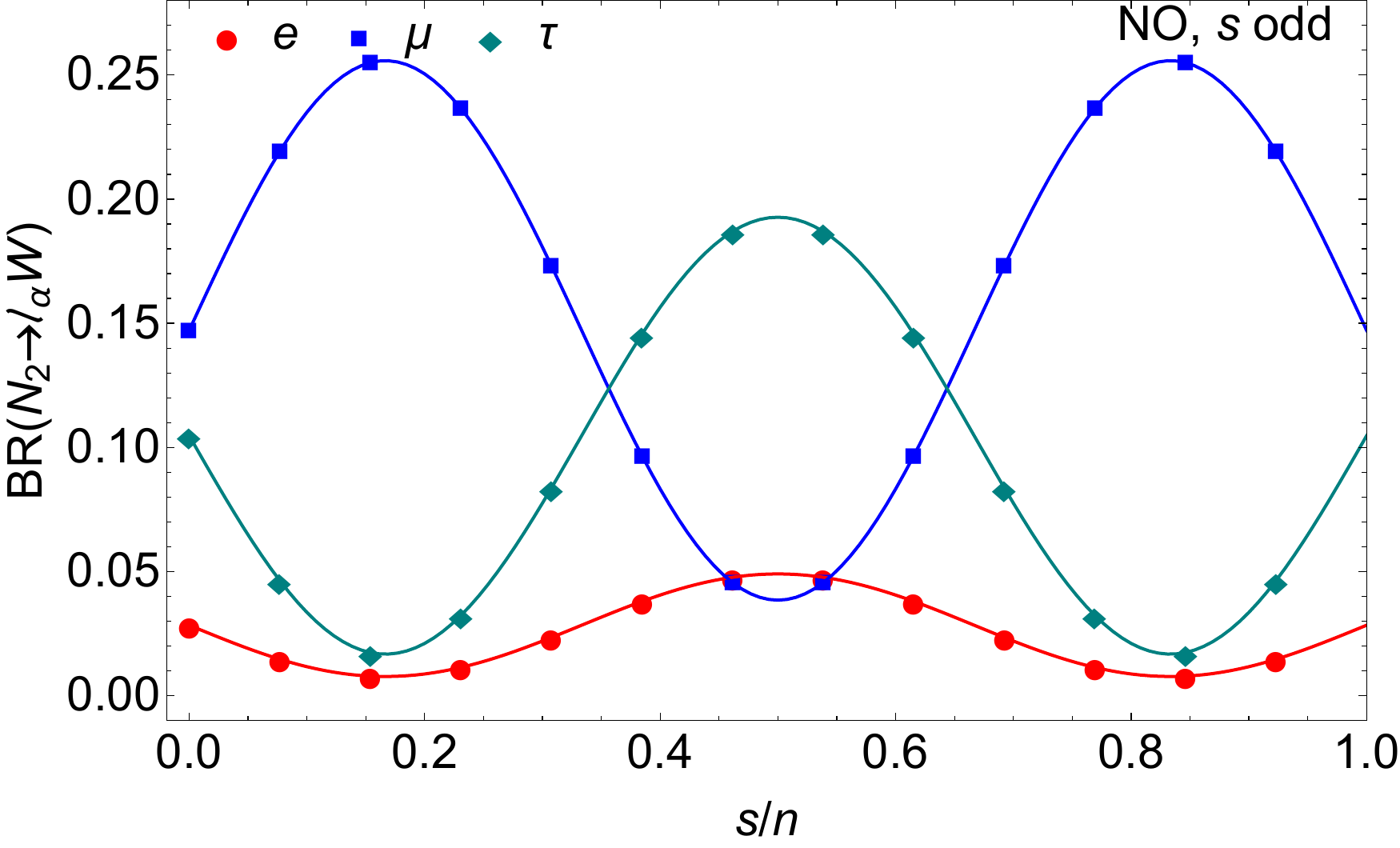}
\includegraphics[width=0.325\textwidth]{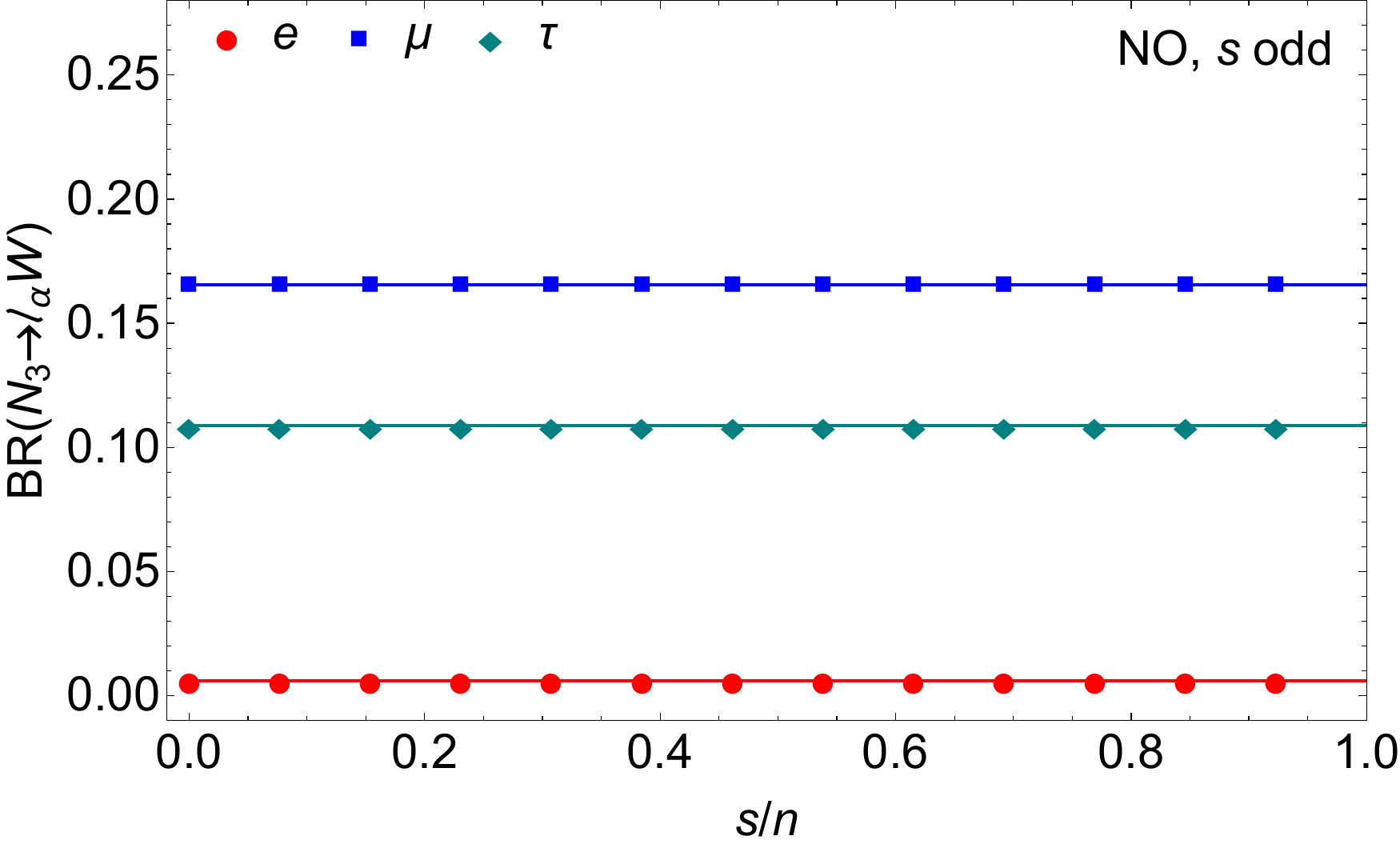} \\
\includegraphics[width=0.325\textwidth]{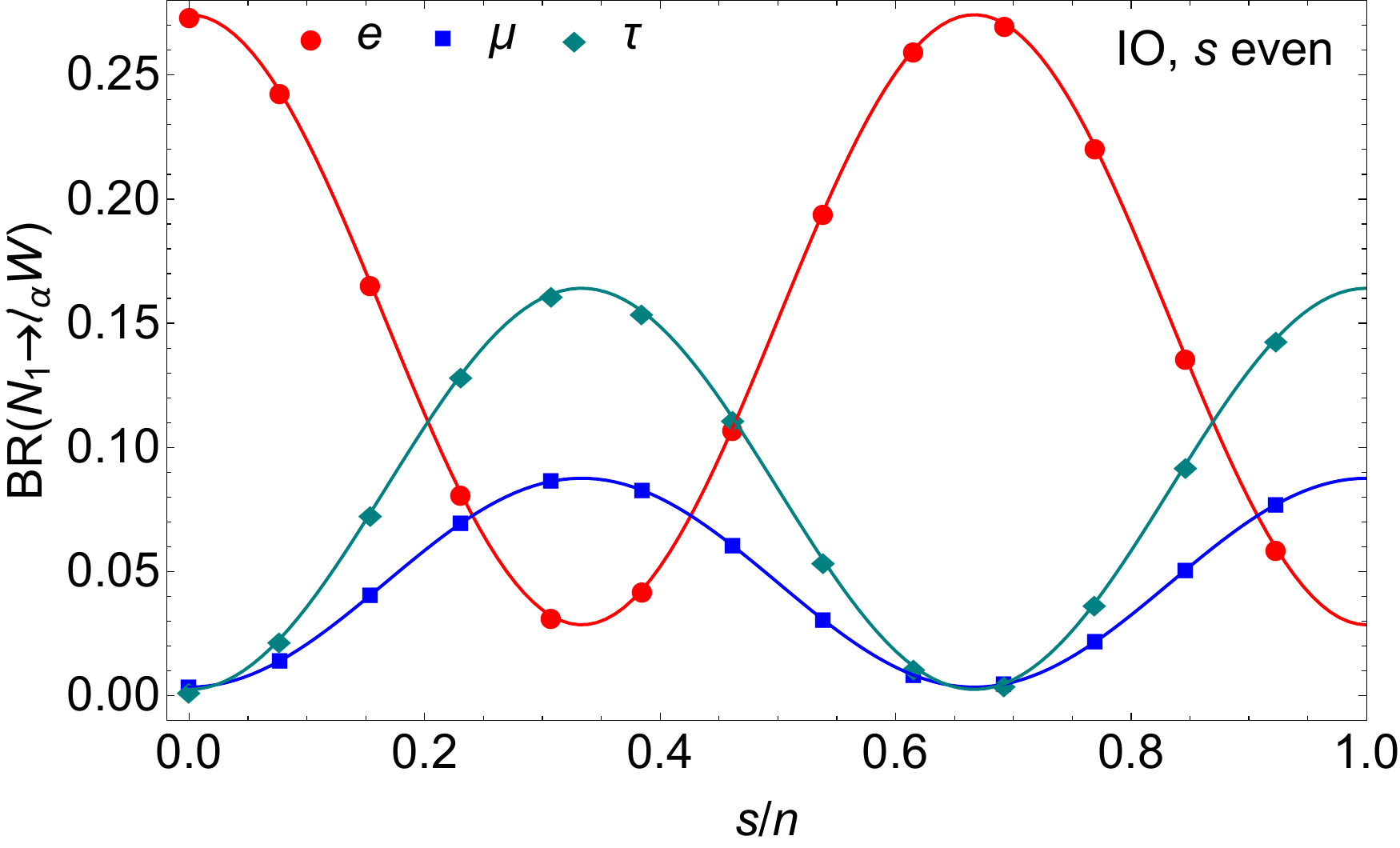}
\includegraphics[width=0.325\textwidth]{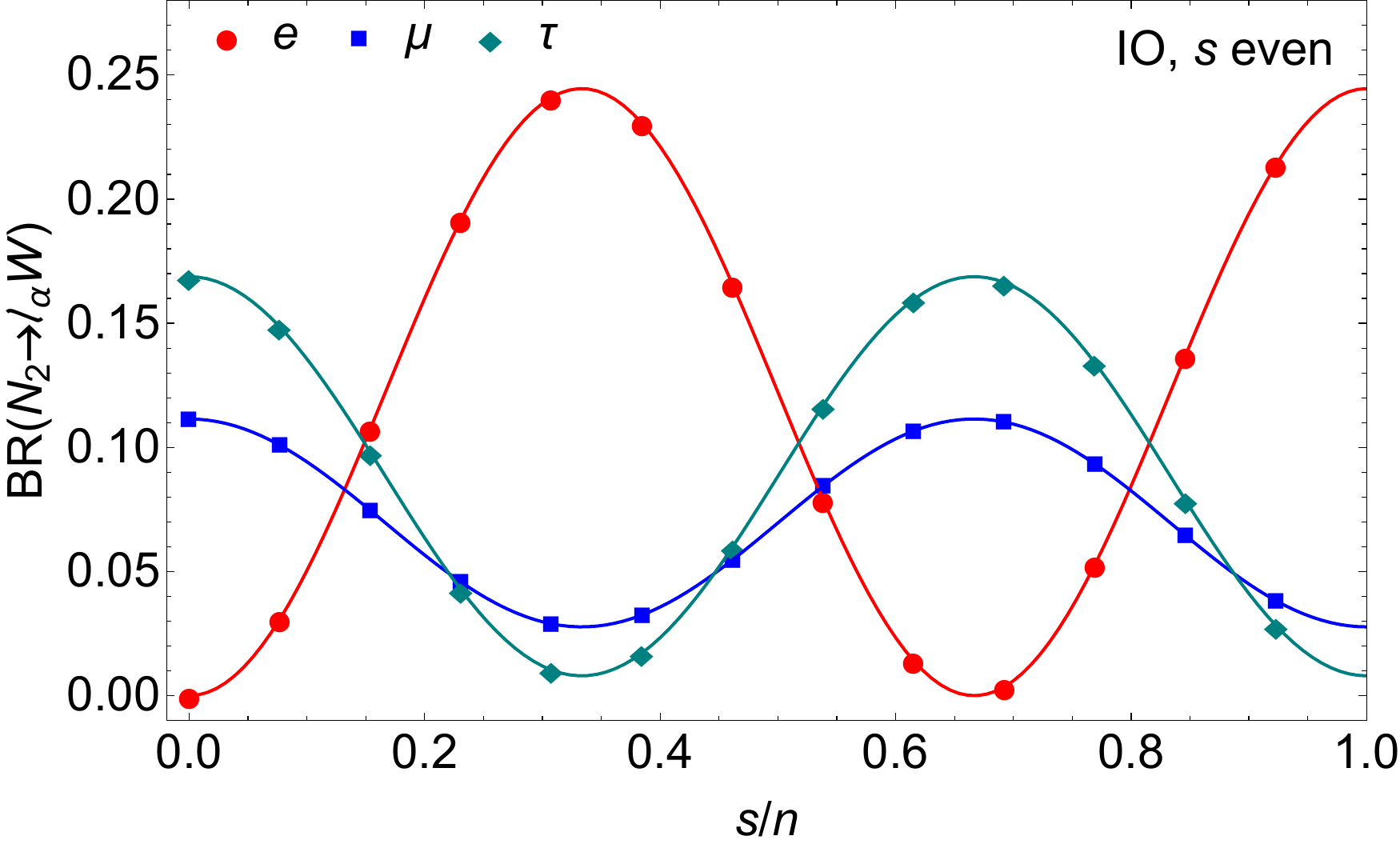}
\includegraphics[width=0.325\textwidth]{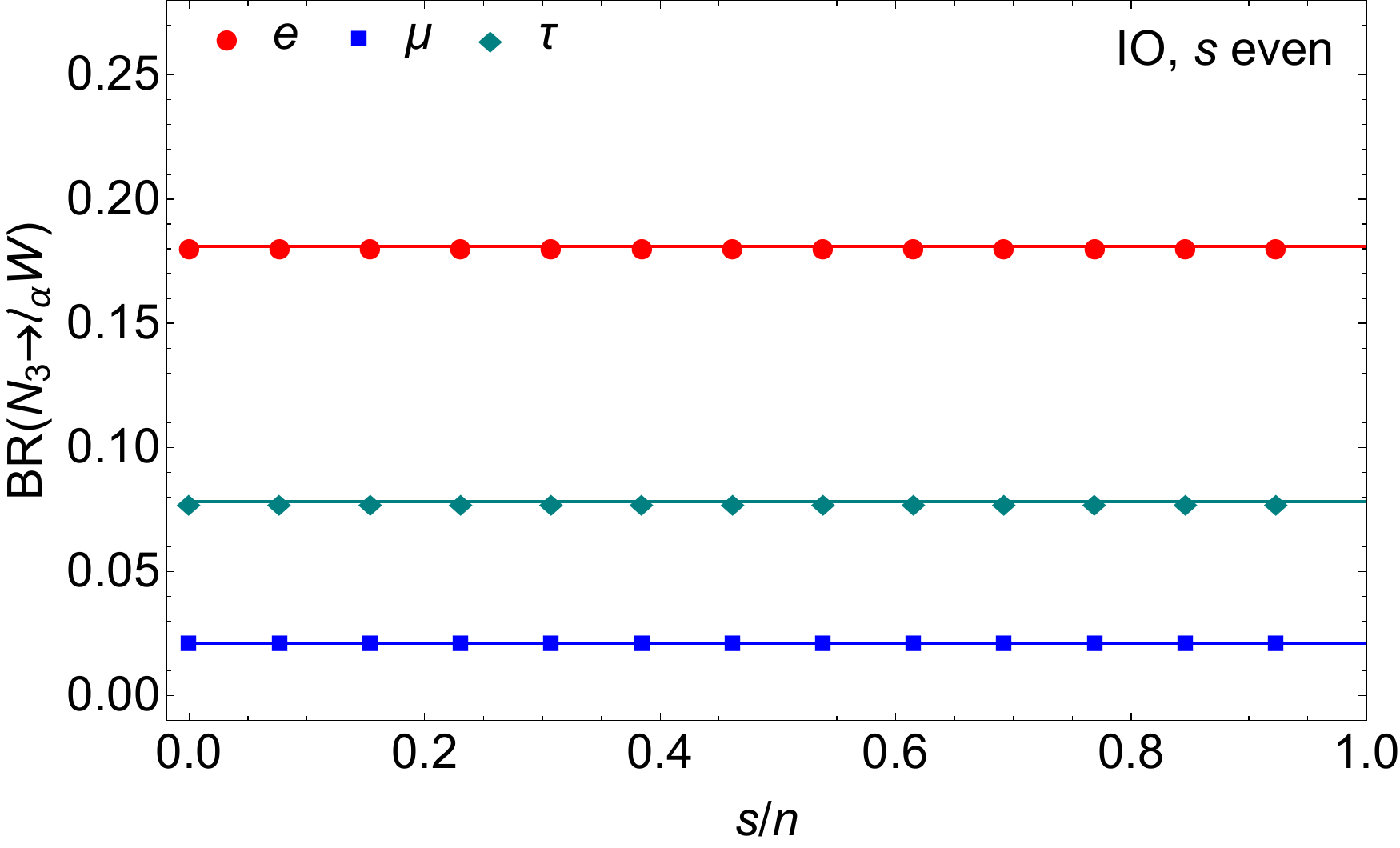}\\
\includegraphics[width=0.325\textwidth]{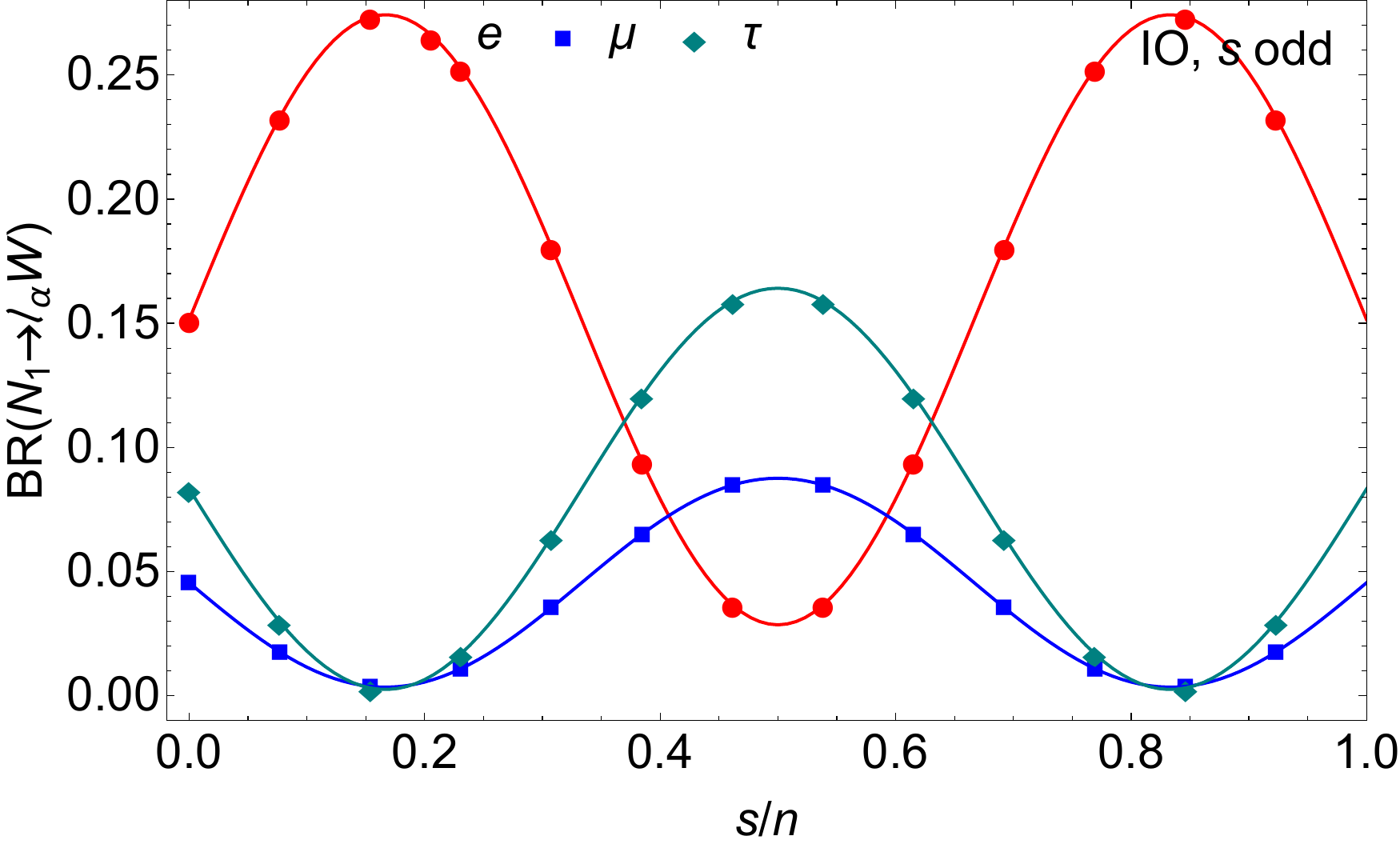}
\includegraphics[width=0.325\textwidth]{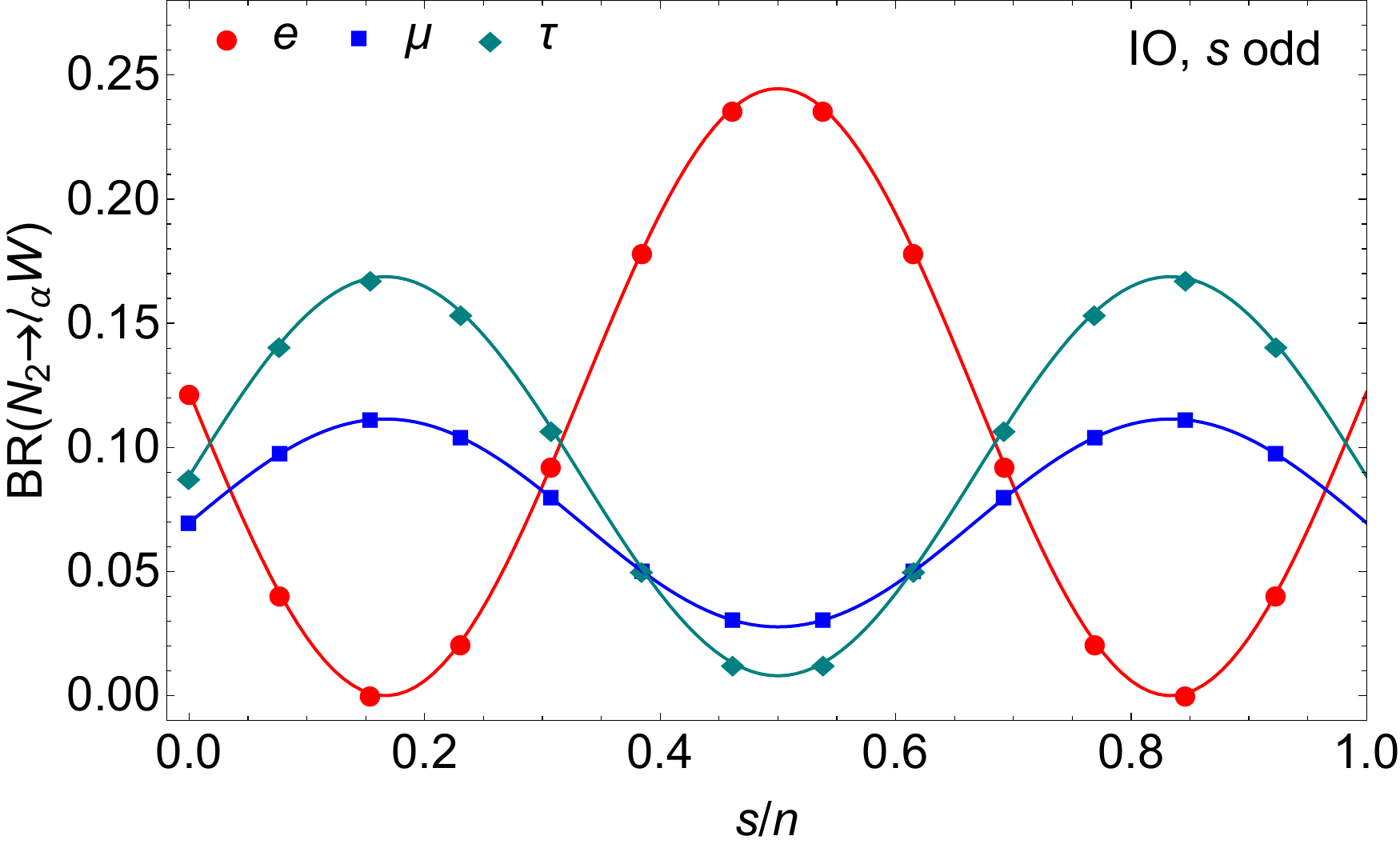}
\includegraphics[width=0.325\textwidth]{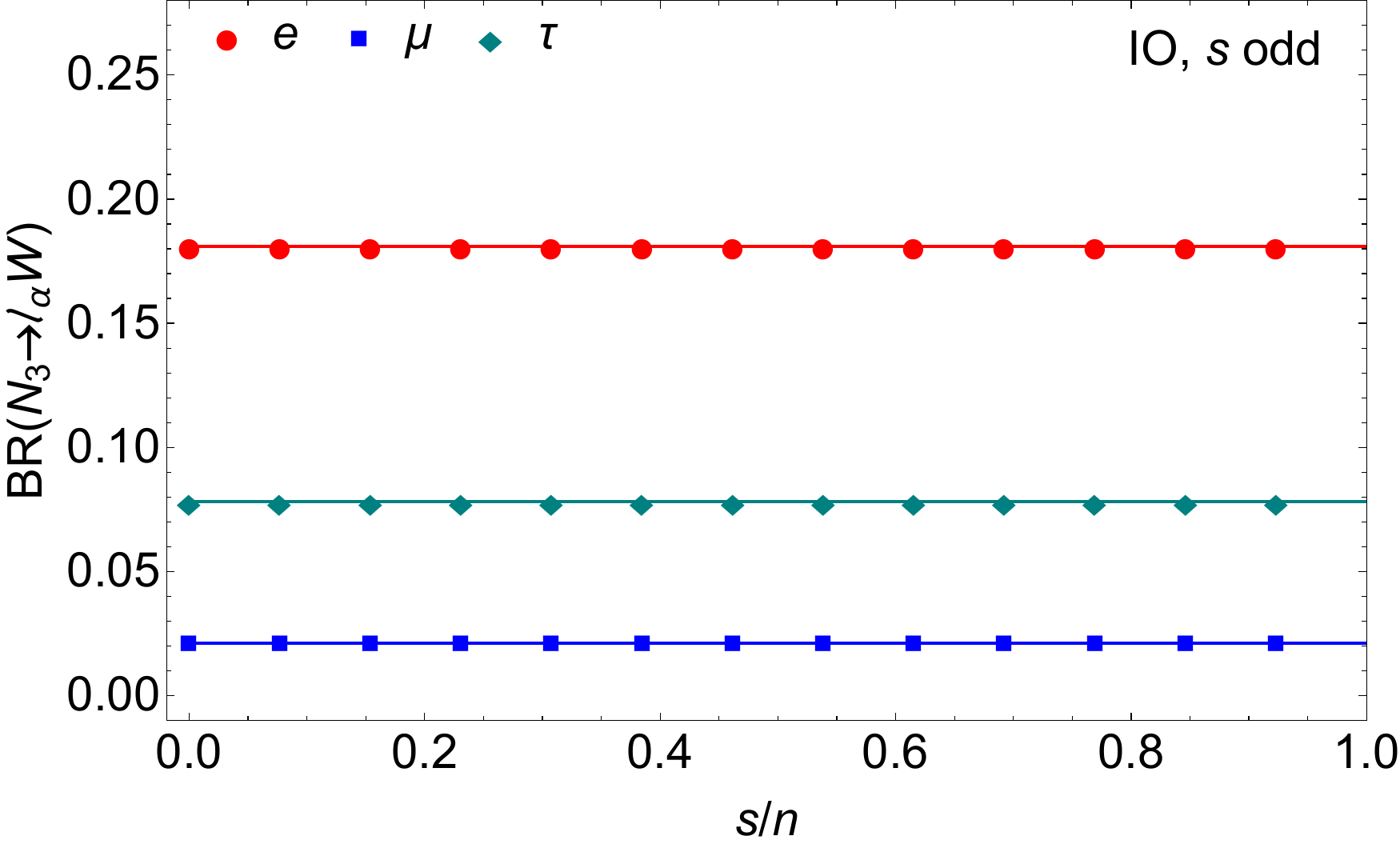}
\caption{For {\it Case 1}. ${\rm BR}(N_i \rightarrow \ell_\alpha\, W)$ for $\alpha=e,\mu,\tau$ are shown as a function of $s/n$. The left, middle and right columns are for $N_1$, $N_2$ and $N_3$ respectively. The top (bottom) two rows are for strong NO (IO) with $s$ even and $s$ odd respectively. These results do not depend on the specific choice of $n$, but for  illustration, we show the results for $n=26$ by the discrete points (13 in total, corresponding to $0\leq s\leq n-1$, either even or odd). We have fixed  $M_N = 250$ GeV and $\theta_R$ at its ERS value in each case.}
\label{fig:brc1}
\end{figure*}
\begin{figure*}[t!]
\centering
\includegraphics[width=0.325\textwidth]{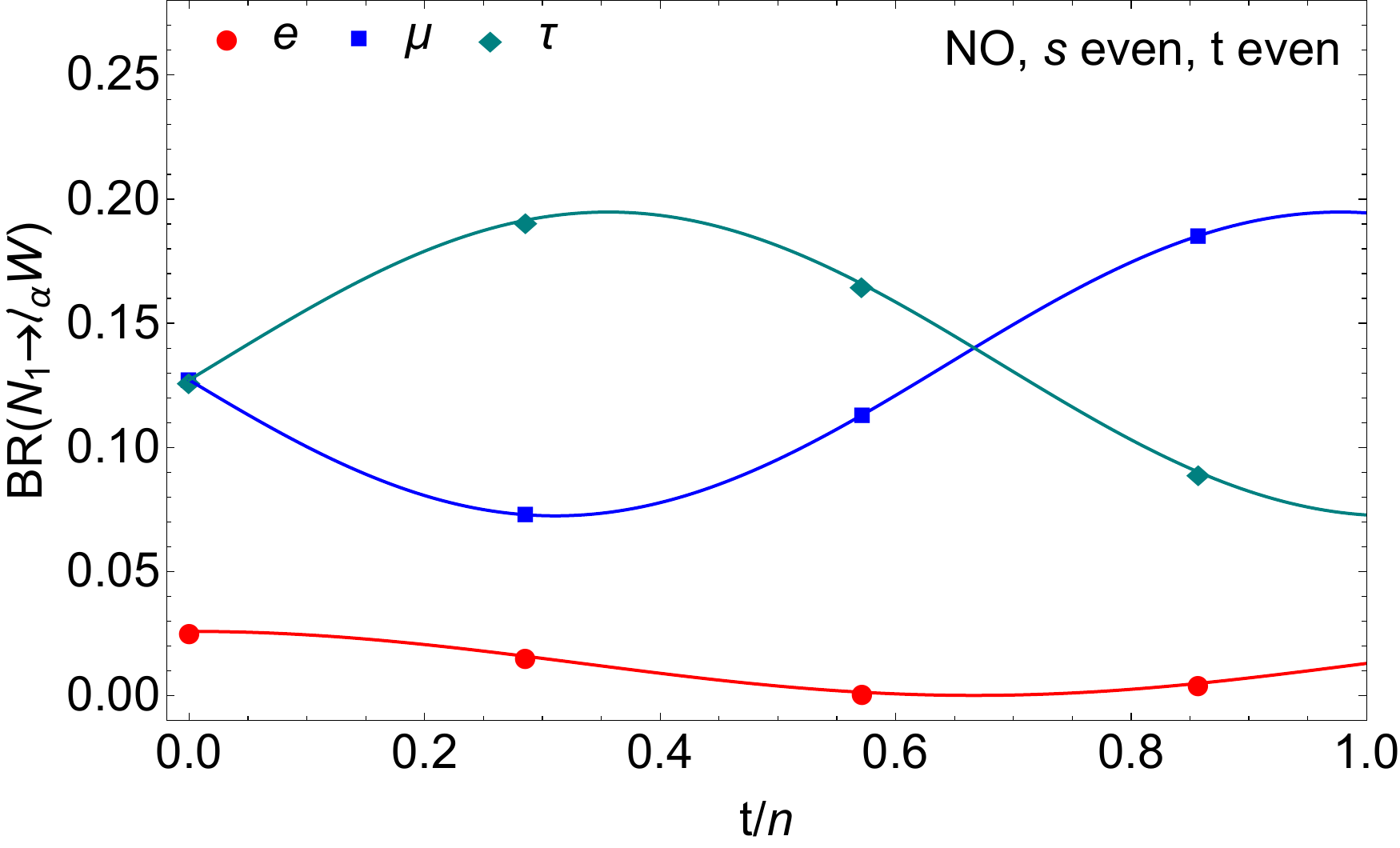}
\includegraphics[width=0.325\textwidth]{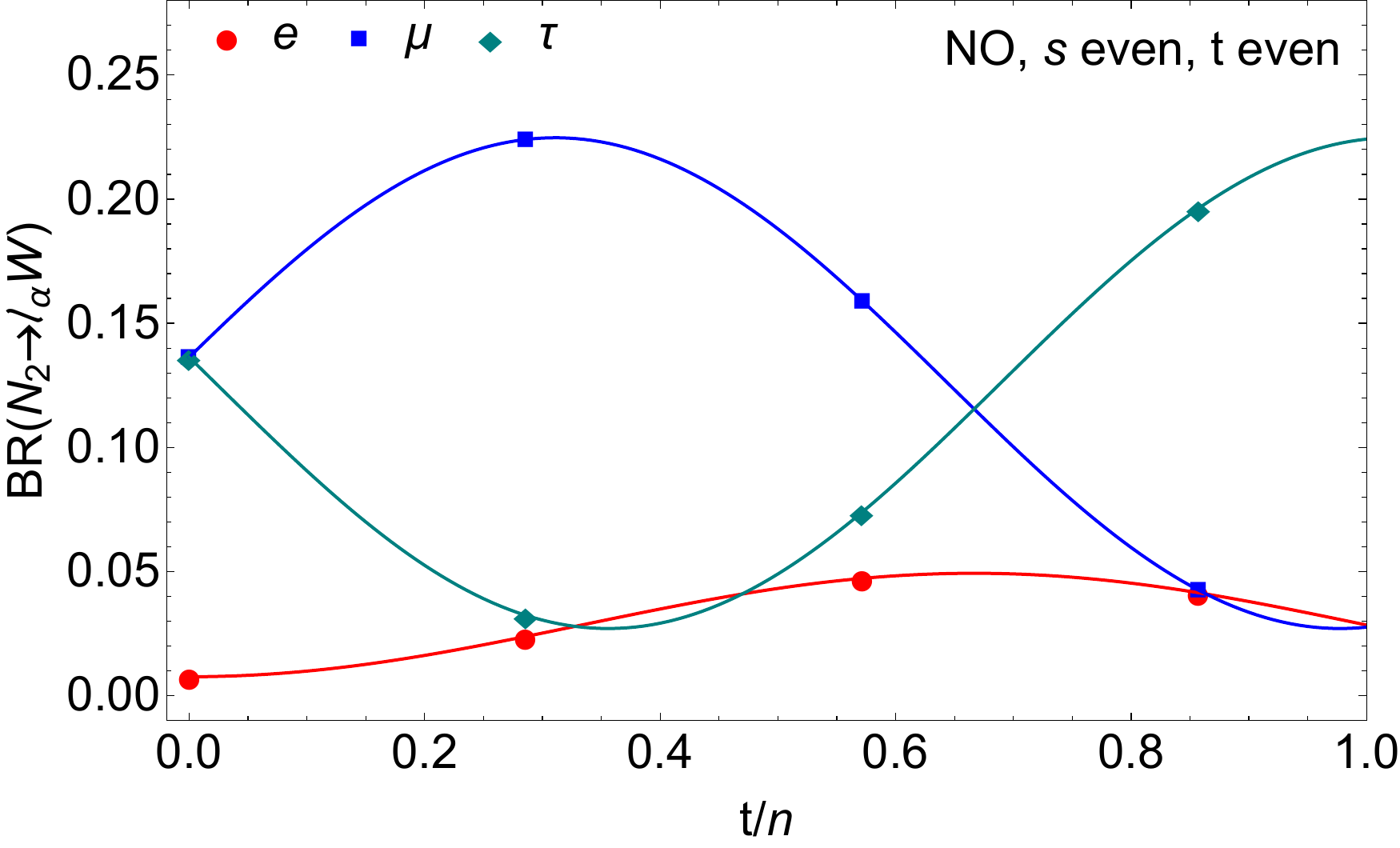}
\includegraphics[width=0.325\textwidth]{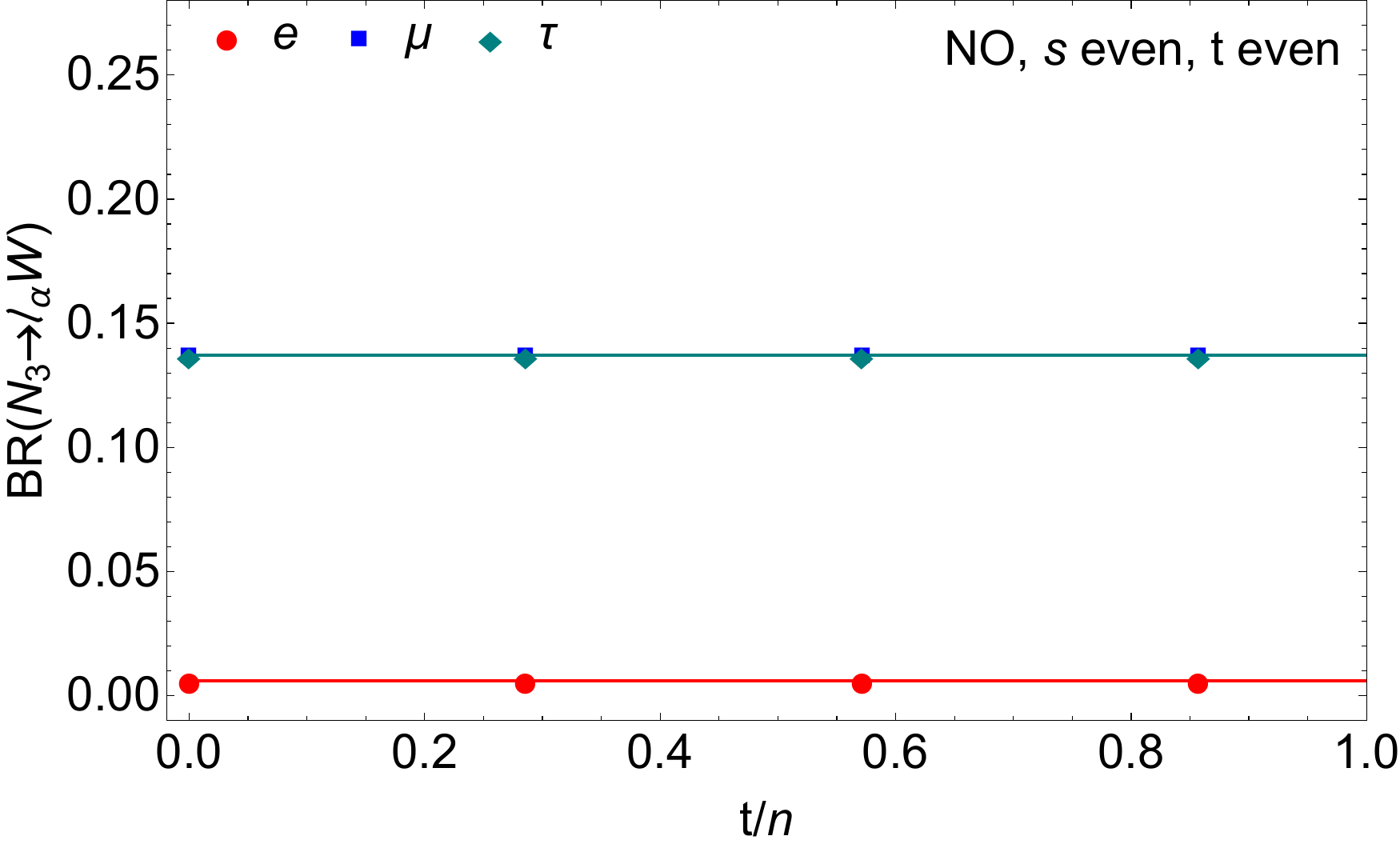} \\
\includegraphics[width=0.325\textwidth]{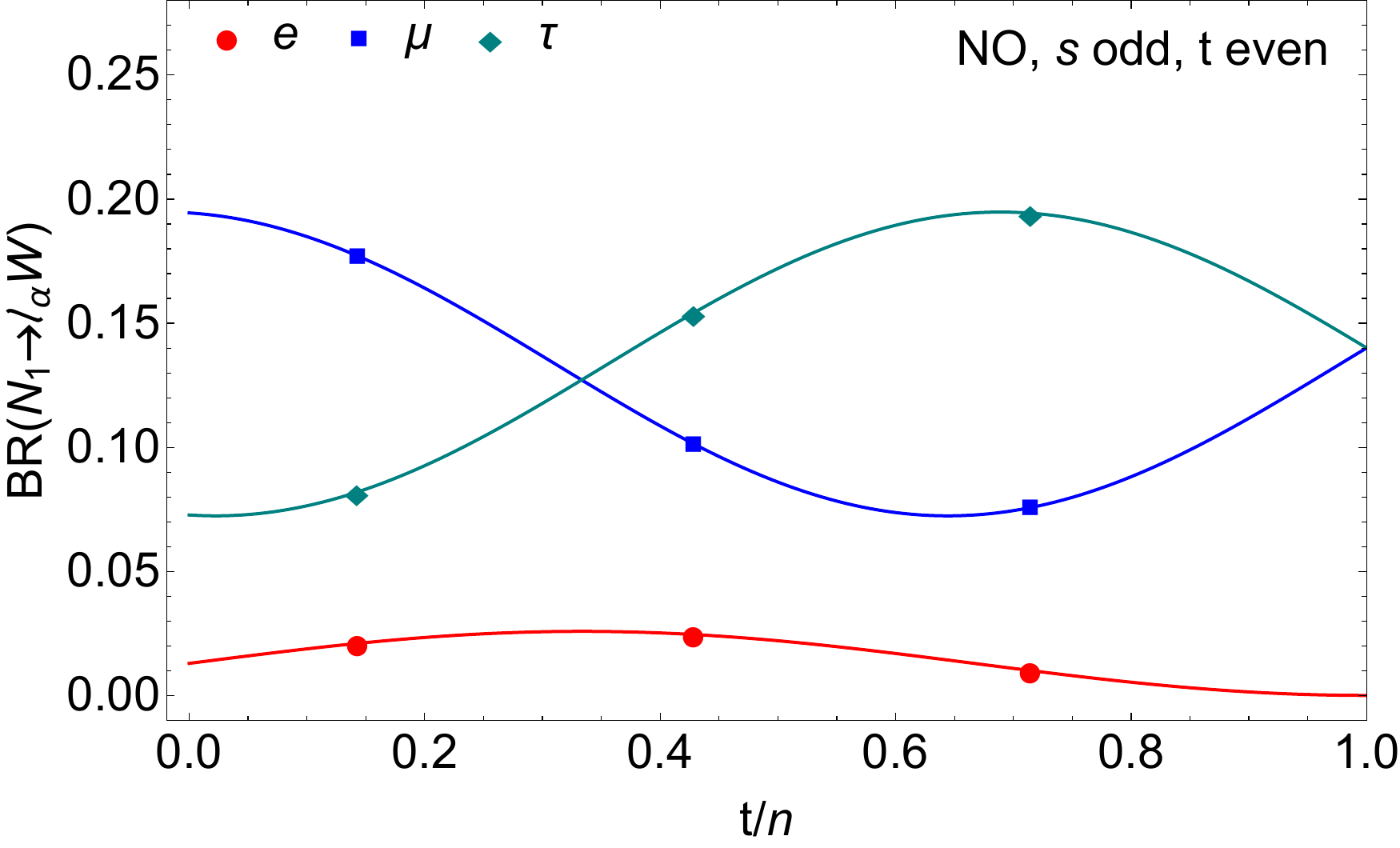}
\includegraphics[width=0.325\textwidth]{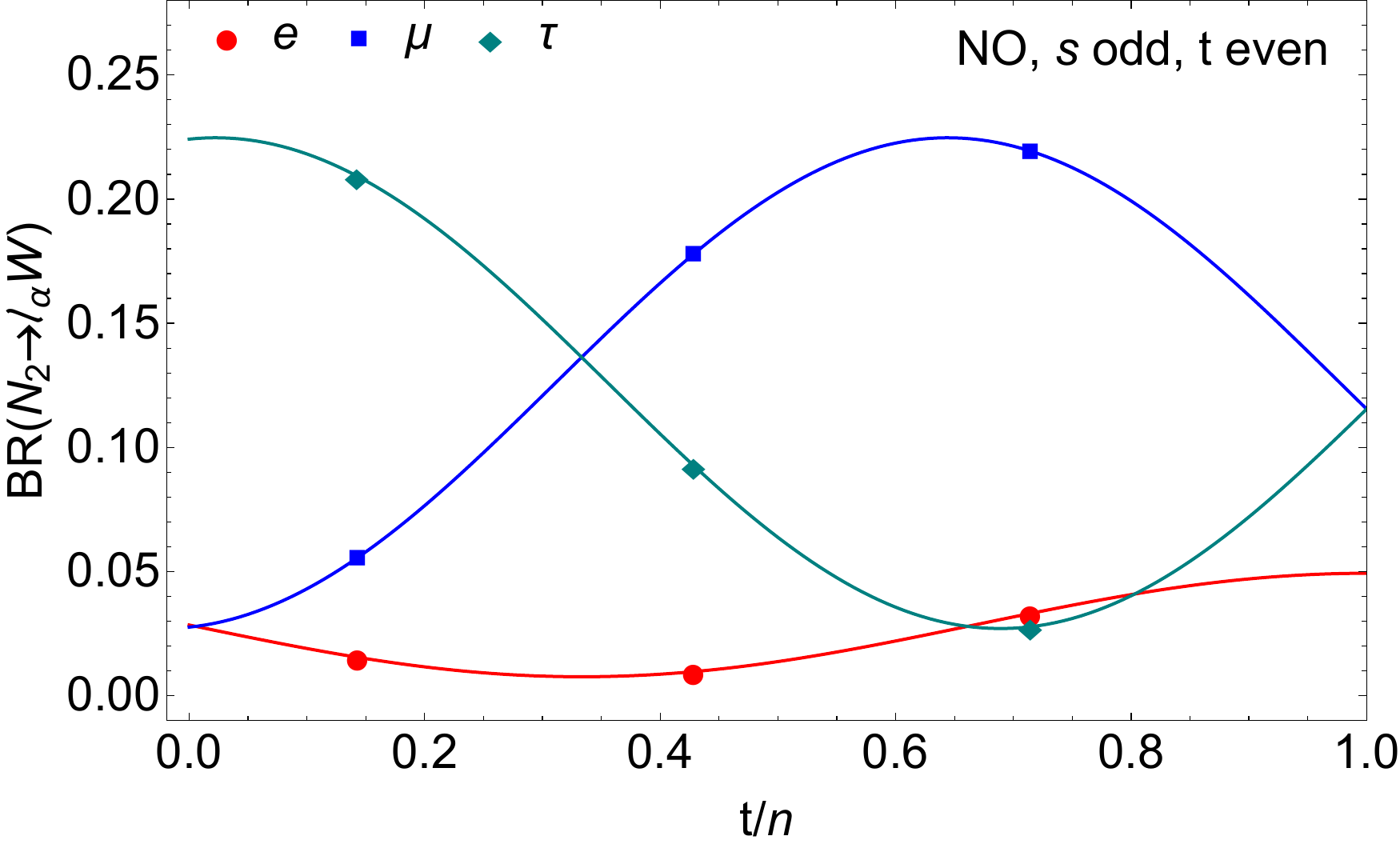}
\includegraphics[width=0.325\textwidth]{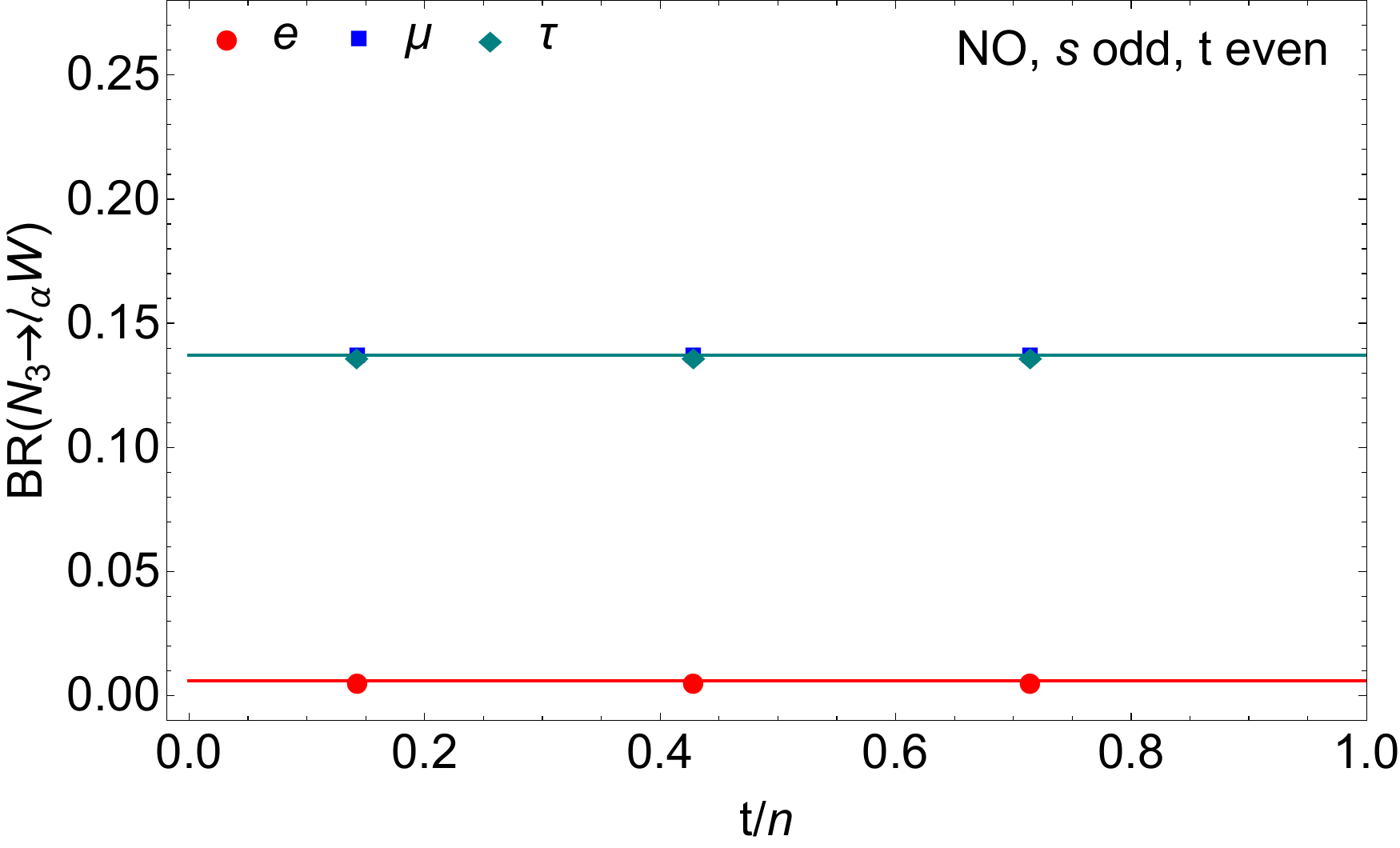}
\\
\includegraphics[width=0.325\textwidth]{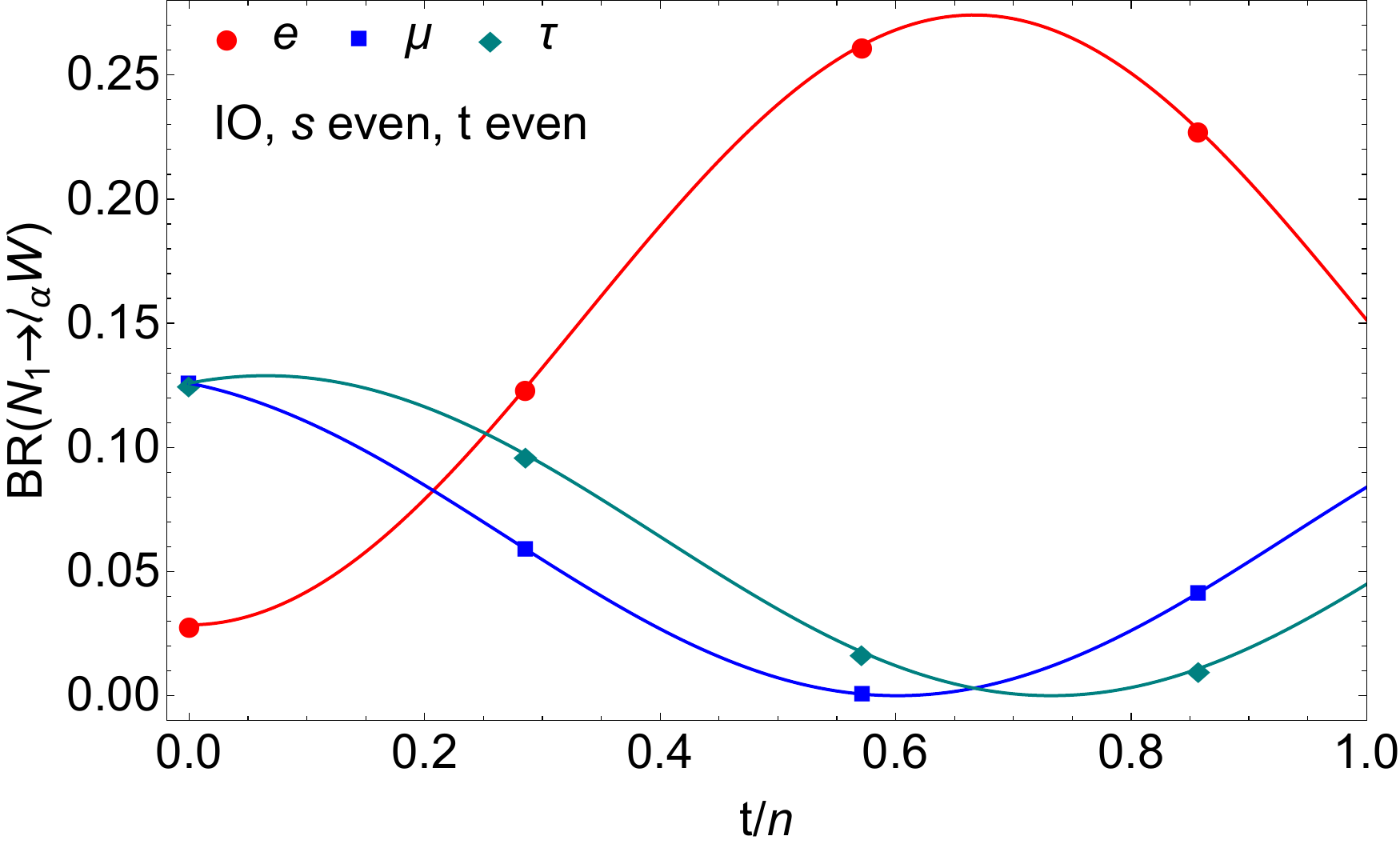}
\includegraphics[width=0.325\textwidth]{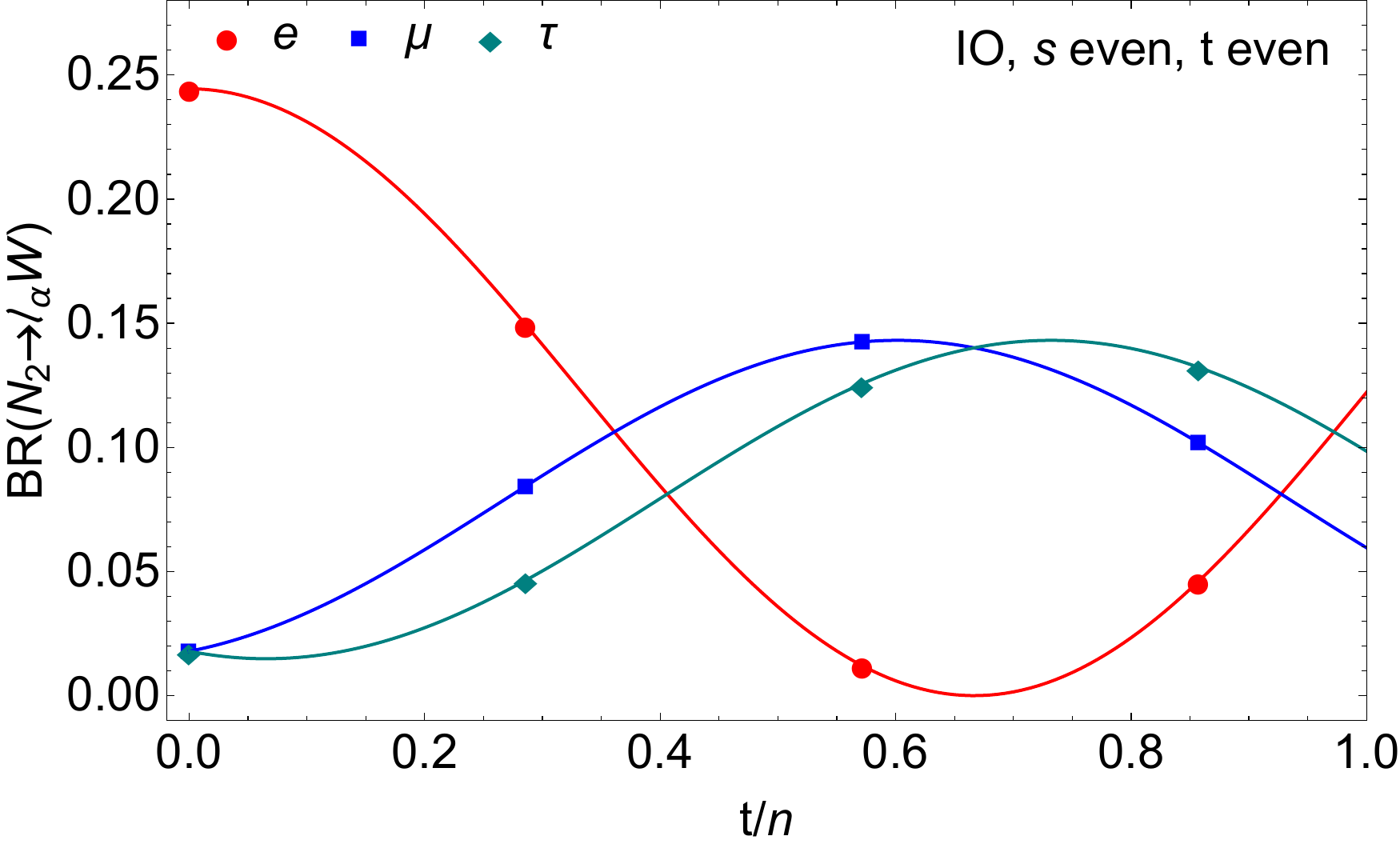}
\includegraphics[width=0.325\textwidth]{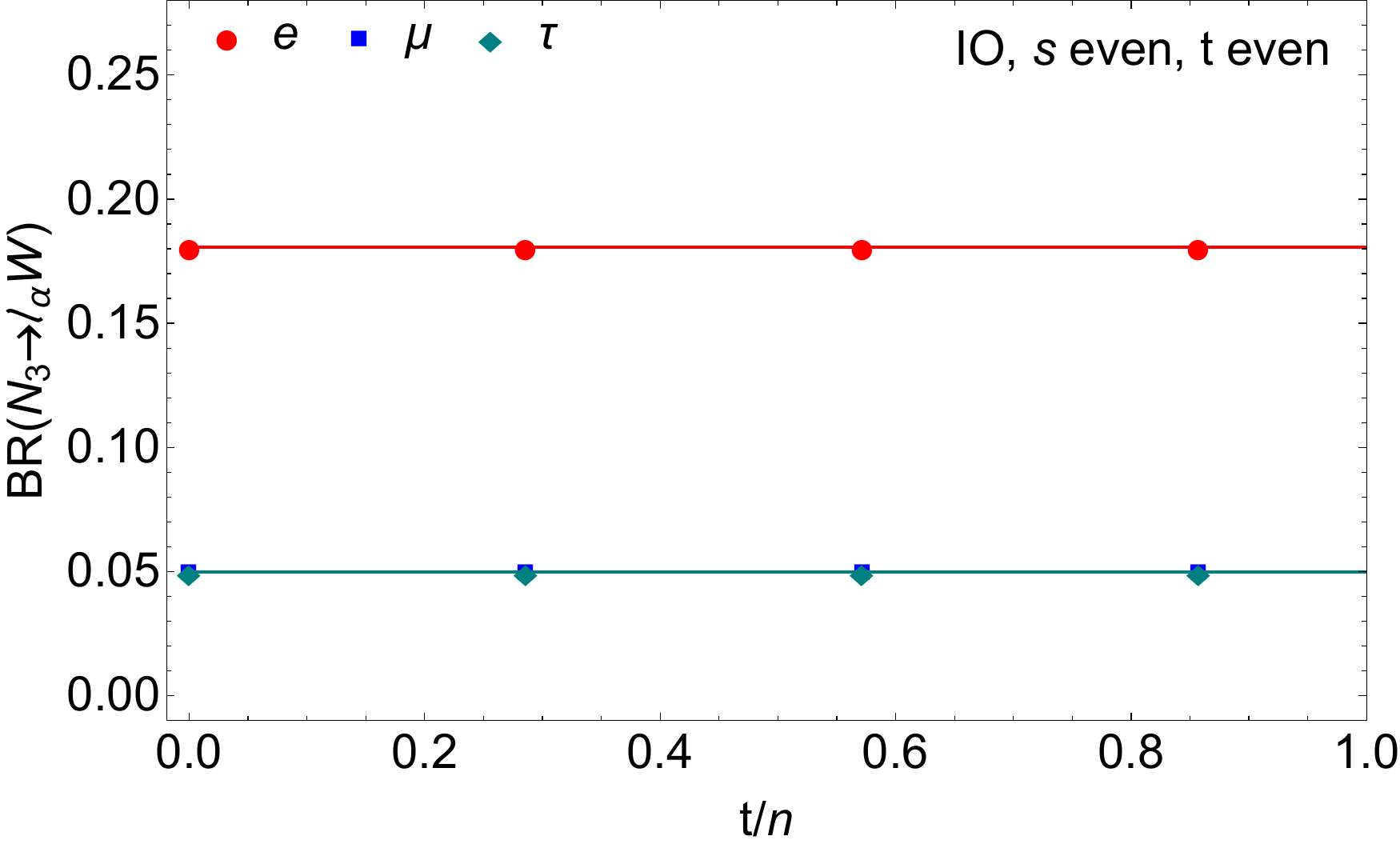}
\\
\includegraphics[width=0.325\textwidth]{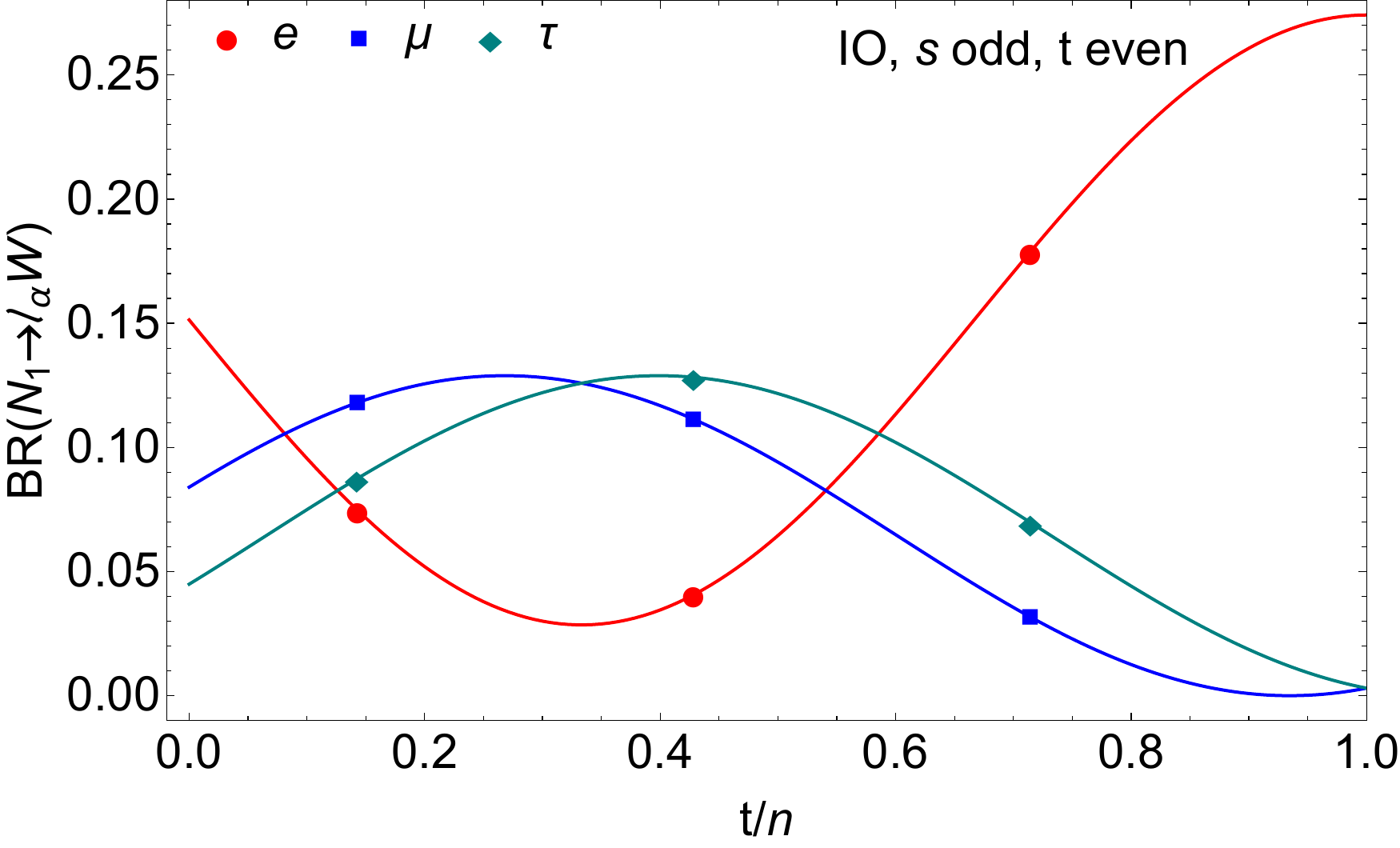}
\includegraphics[width=0.325\textwidth]{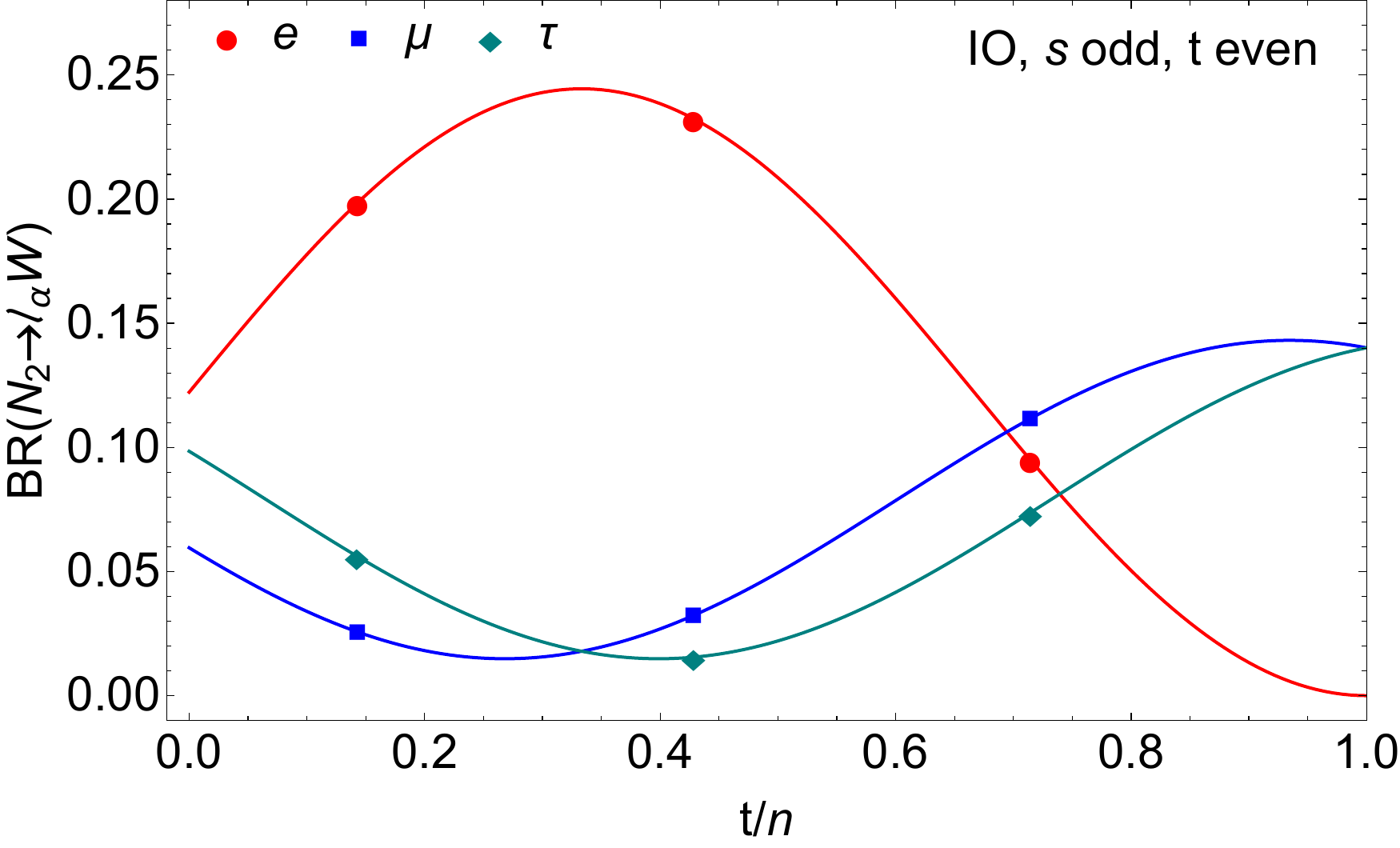}
\includegraphics[width=0.325\textwidth]{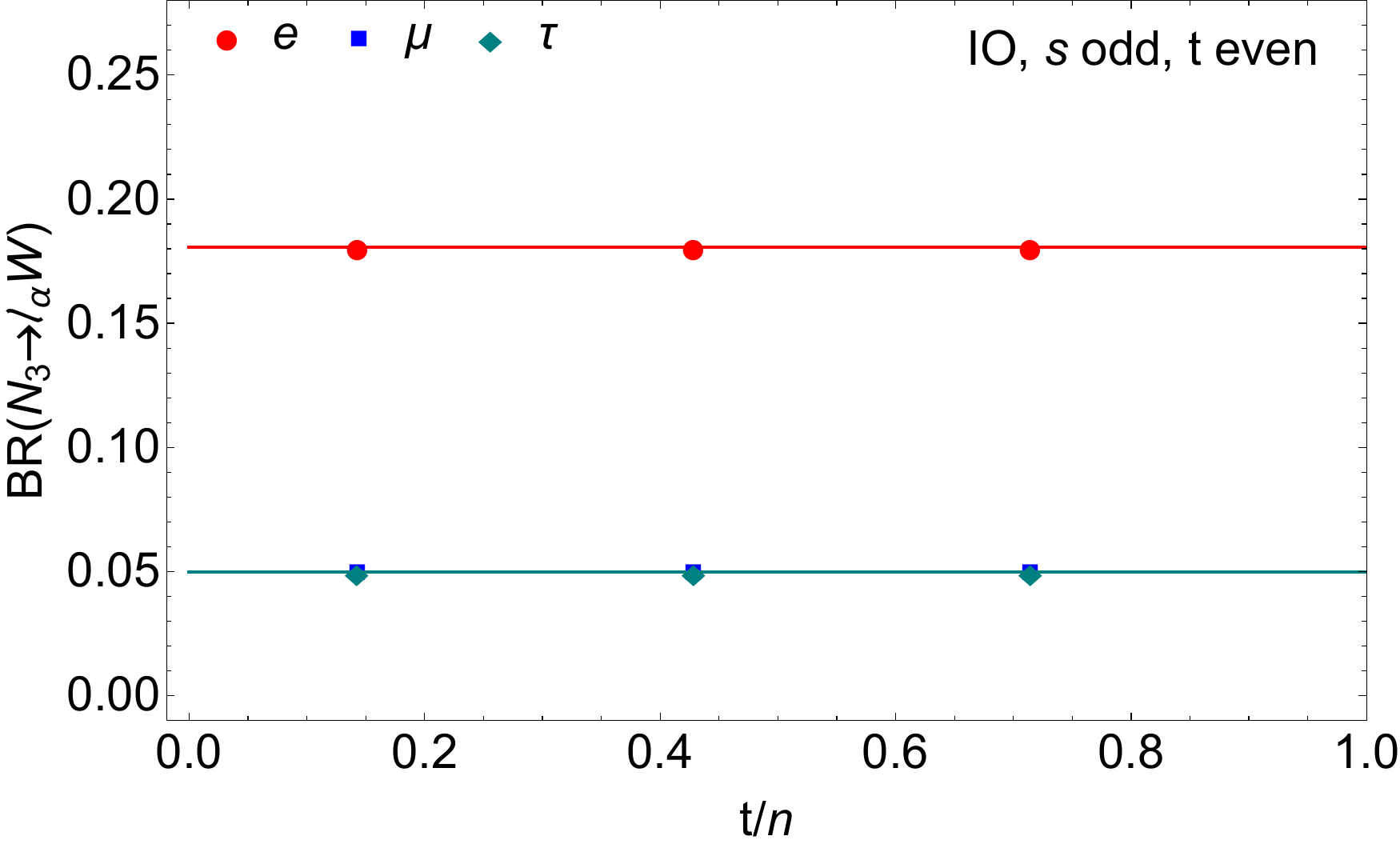}
\caption{For {\it Case 2}. ${\rm BR}(N_i\to \ell_\alpha W)$ for $\alpha=e,\mu,\tau$ as a function of $t/n$ with $t$ even. The top (bottom) two rows are for strong NO (IO) with $s$ even and $s$ odd, respectively. These results do not depend on the specific choice of $n$, but for illustration, we show the results for $u=2s-t=0,n=14$ by the discrete points (4(3) in total, corresponding to $s$ $(0\leq s\leq n-1)$ being even  (odd) and $t$ even). We have fixed  $M_N = 250$ GeV and $\theta_R$ at its ERS value in each case.}
\label{fig:brc2}
\end{figure*}
\begin{figure*}[t!]
\centering
\includegraphics[width=0.325\textwidth]{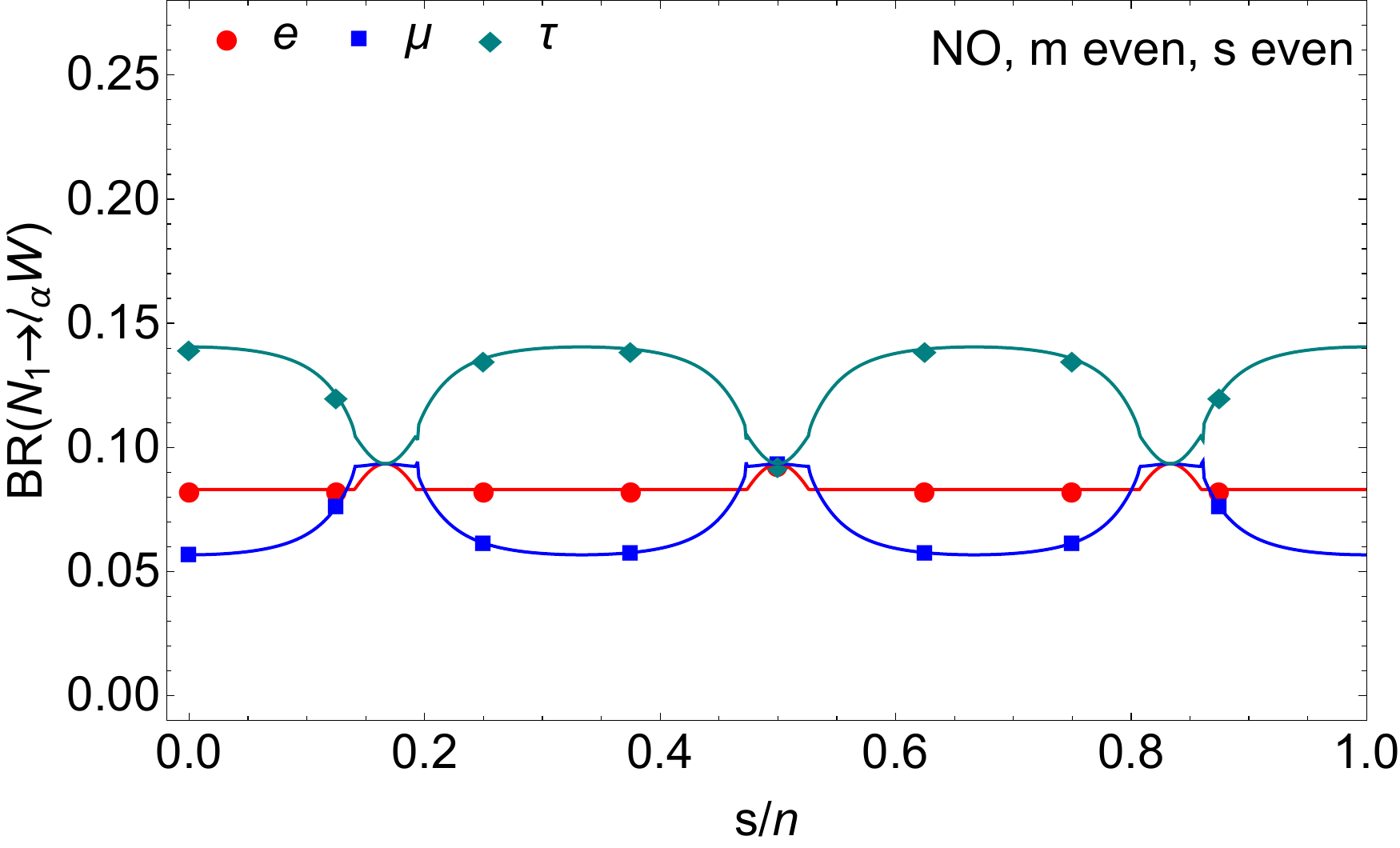}
\includegraphics[width=0.325\textwidth]{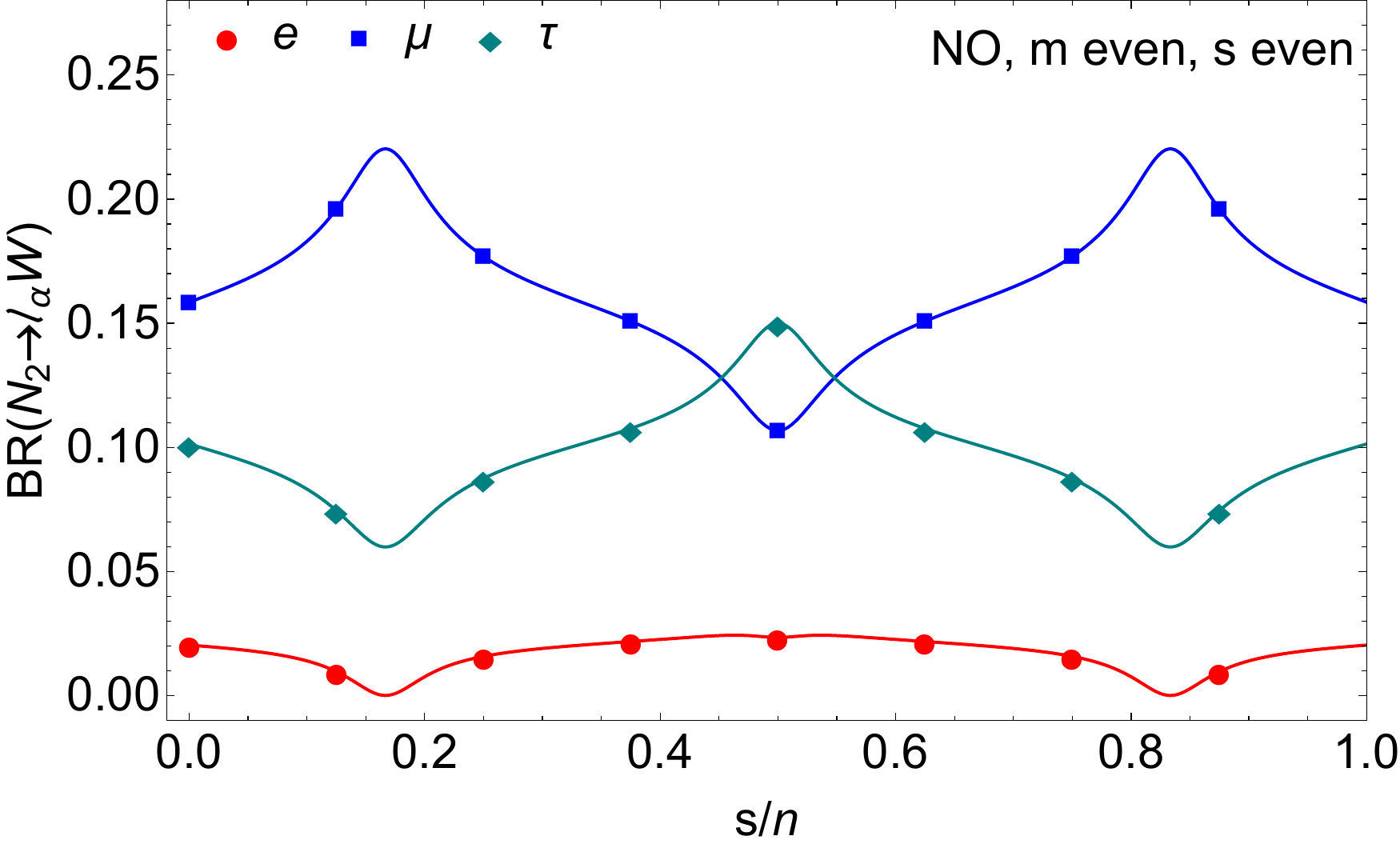}
\includegraphics[width=0.325\textwidth]{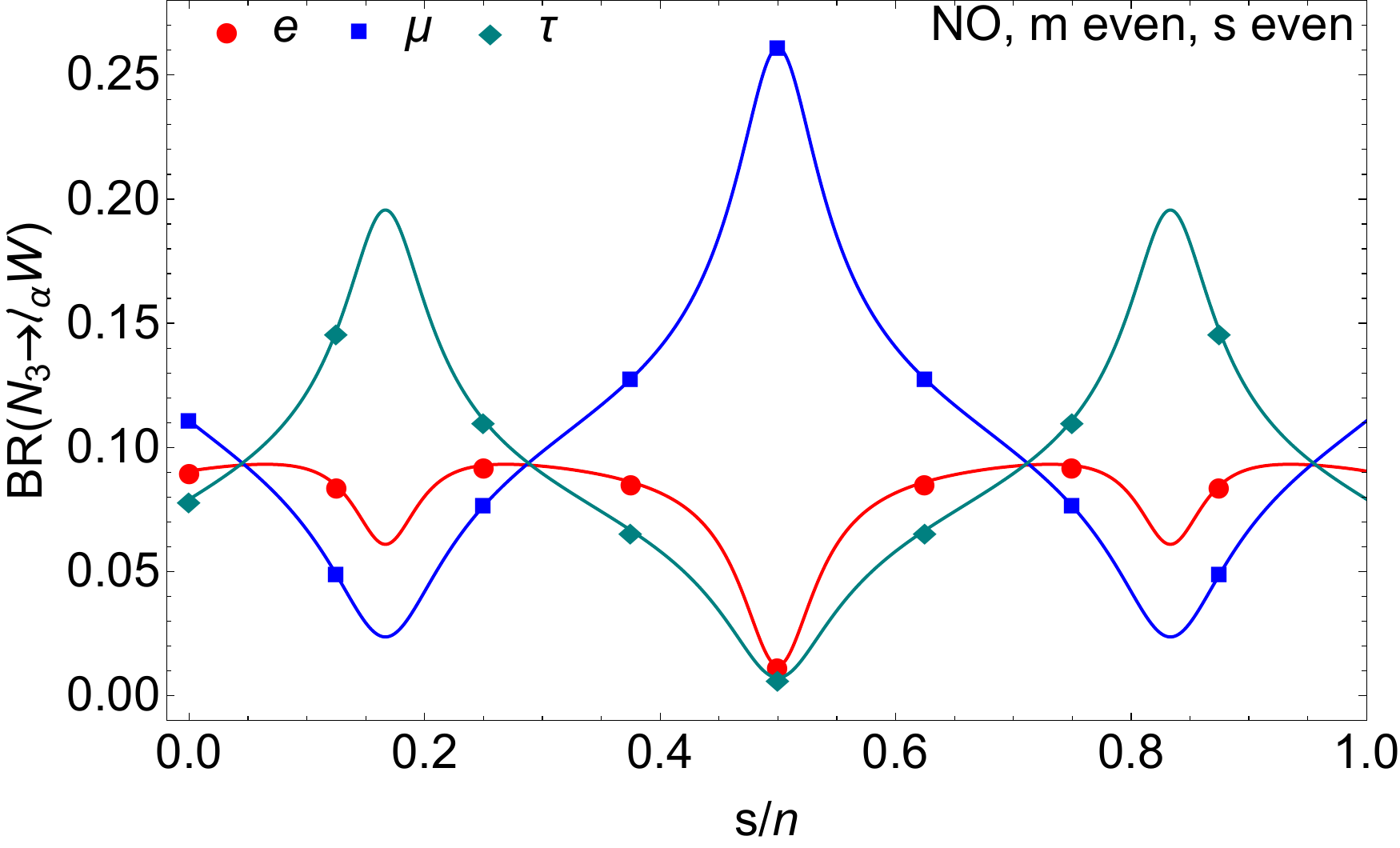} \\
\includegraphics[width=0.325\textwidth]{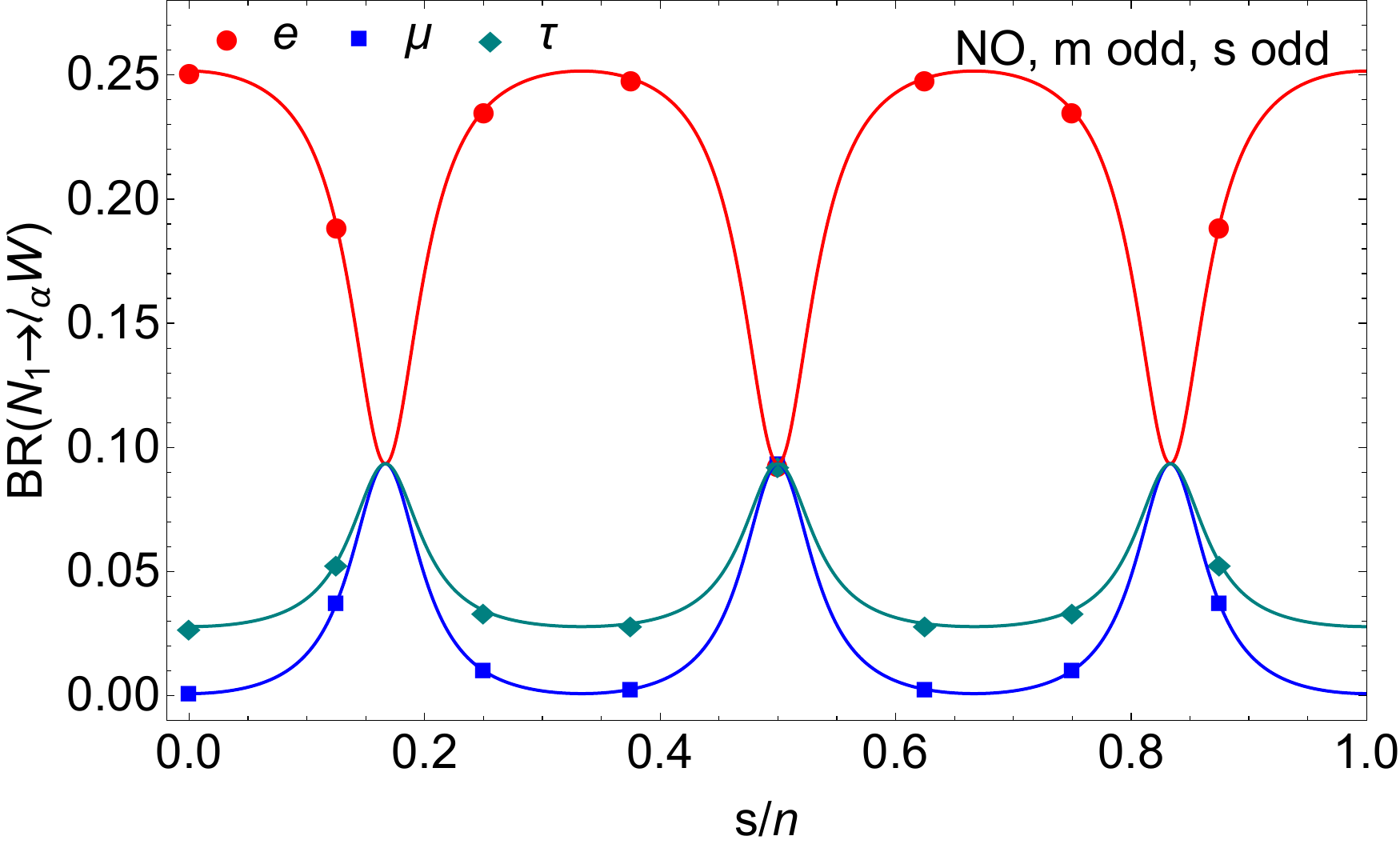}
\includegraphics[width=0.325\textwidth]{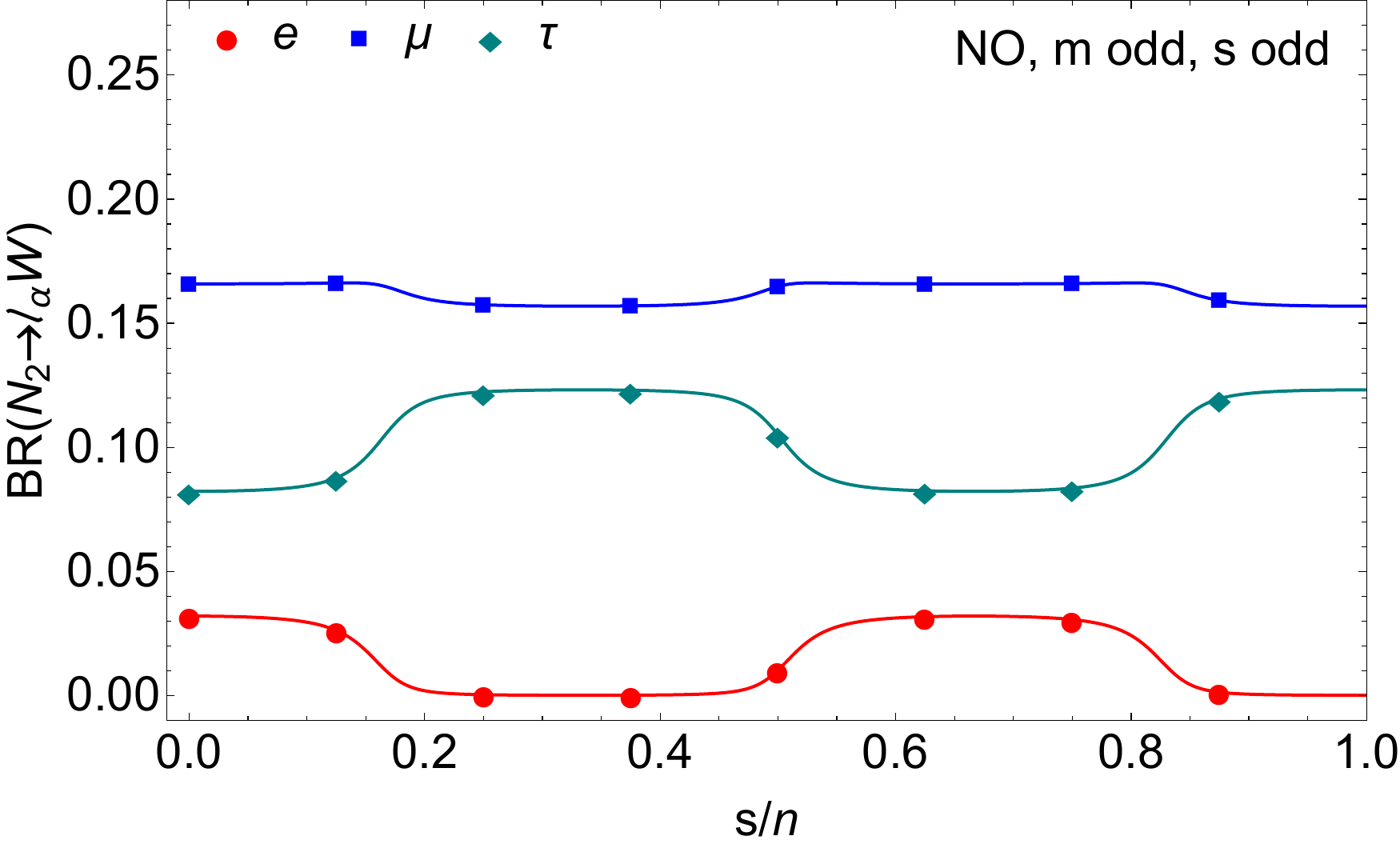}
\includegraphics[width=0.325\textwidth]{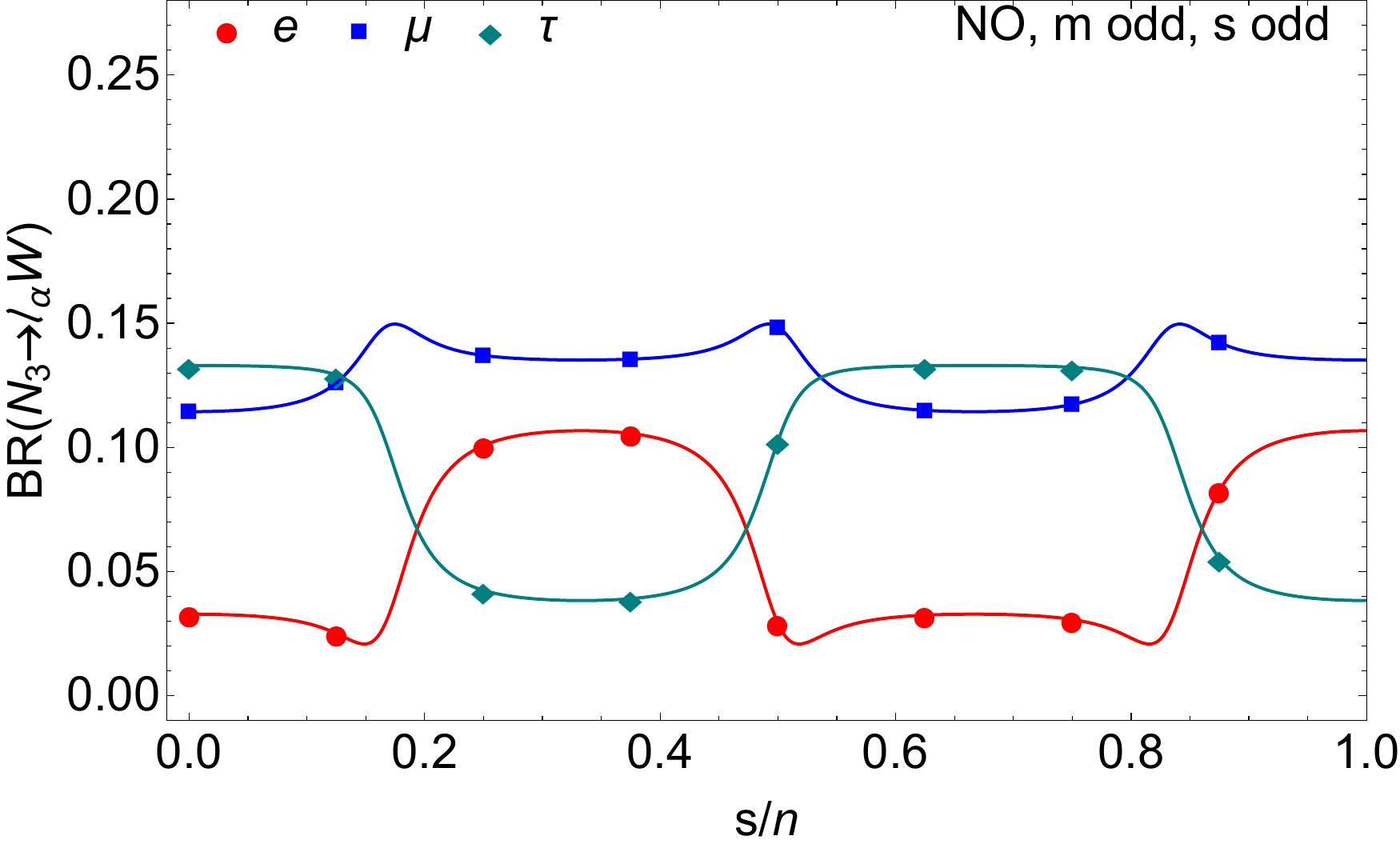}\\
\includegraphics[width=0.325\textwidth]{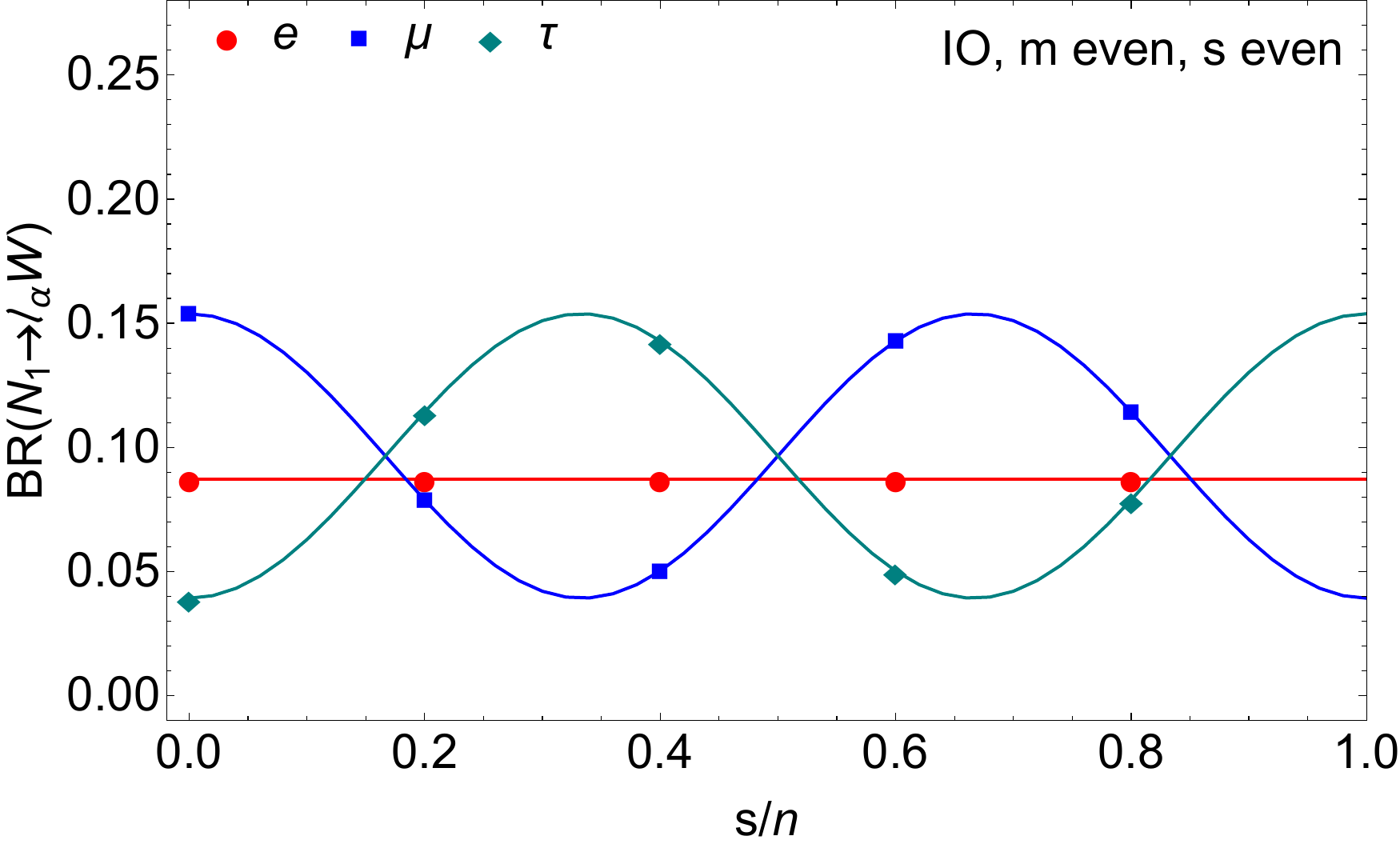}
\includegraphics[width=0.325\textwidth]{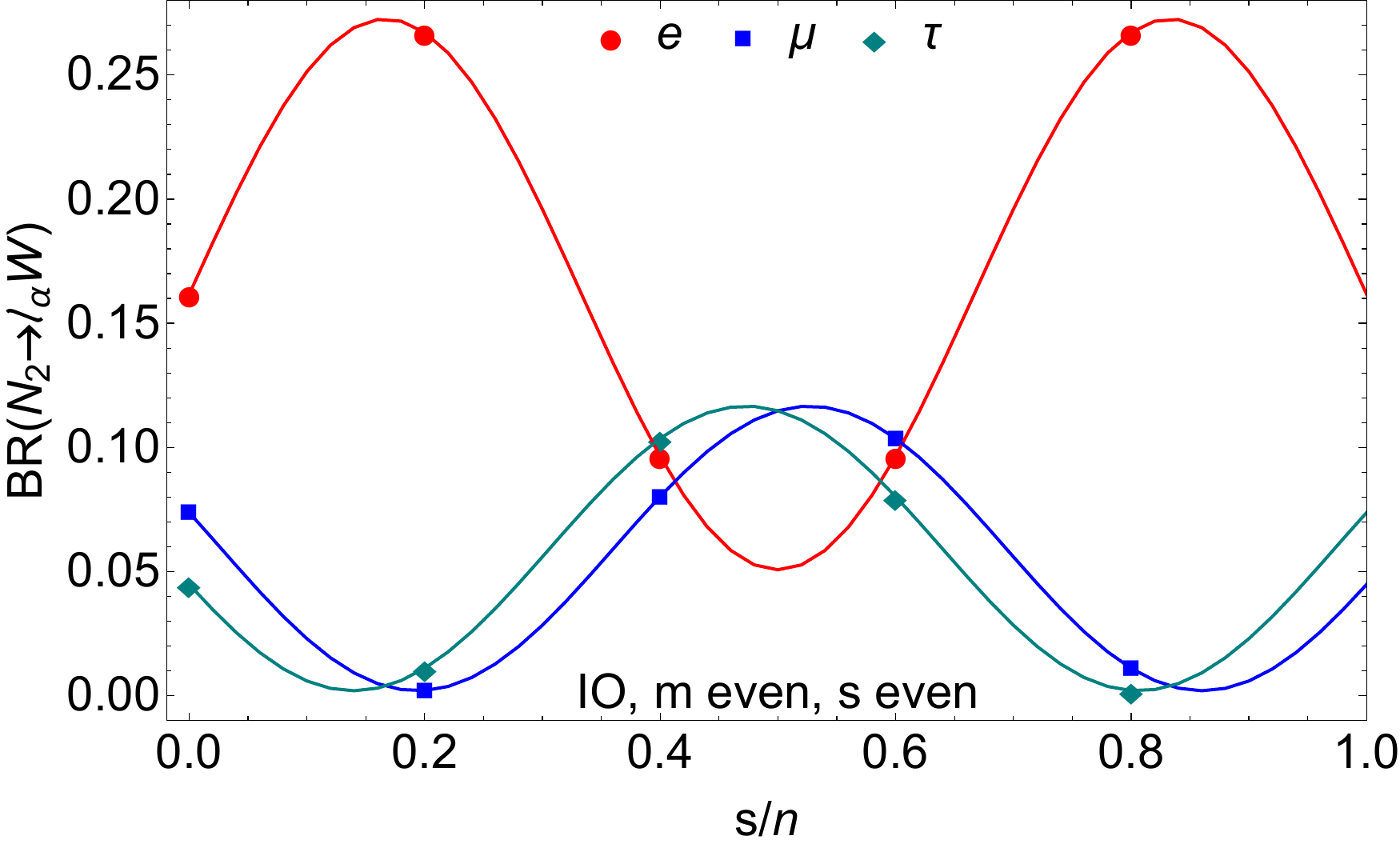}
\includegraphics[width=0.325\textwidth]{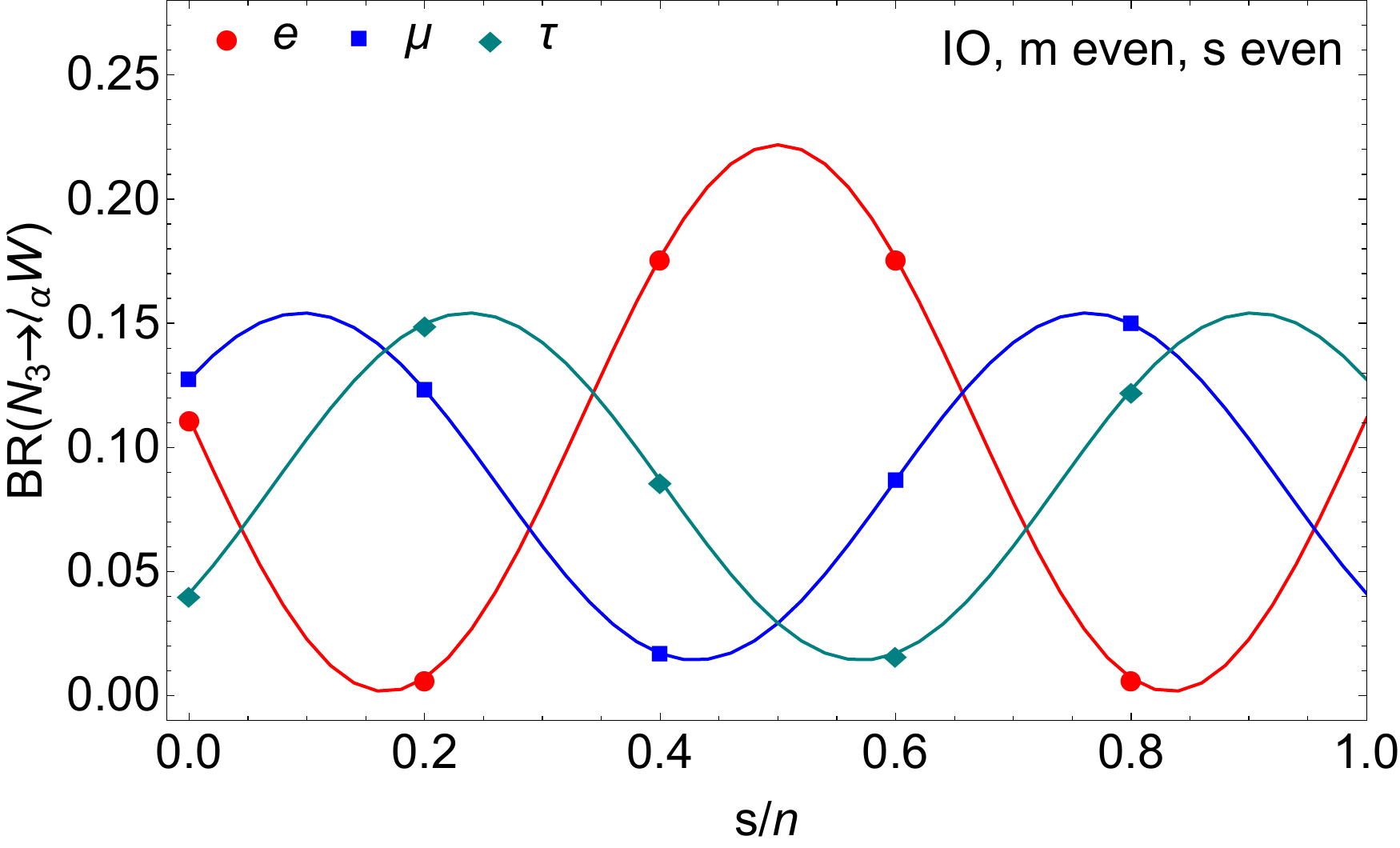}\\
\includegraphics[width=0.325\textwidth]{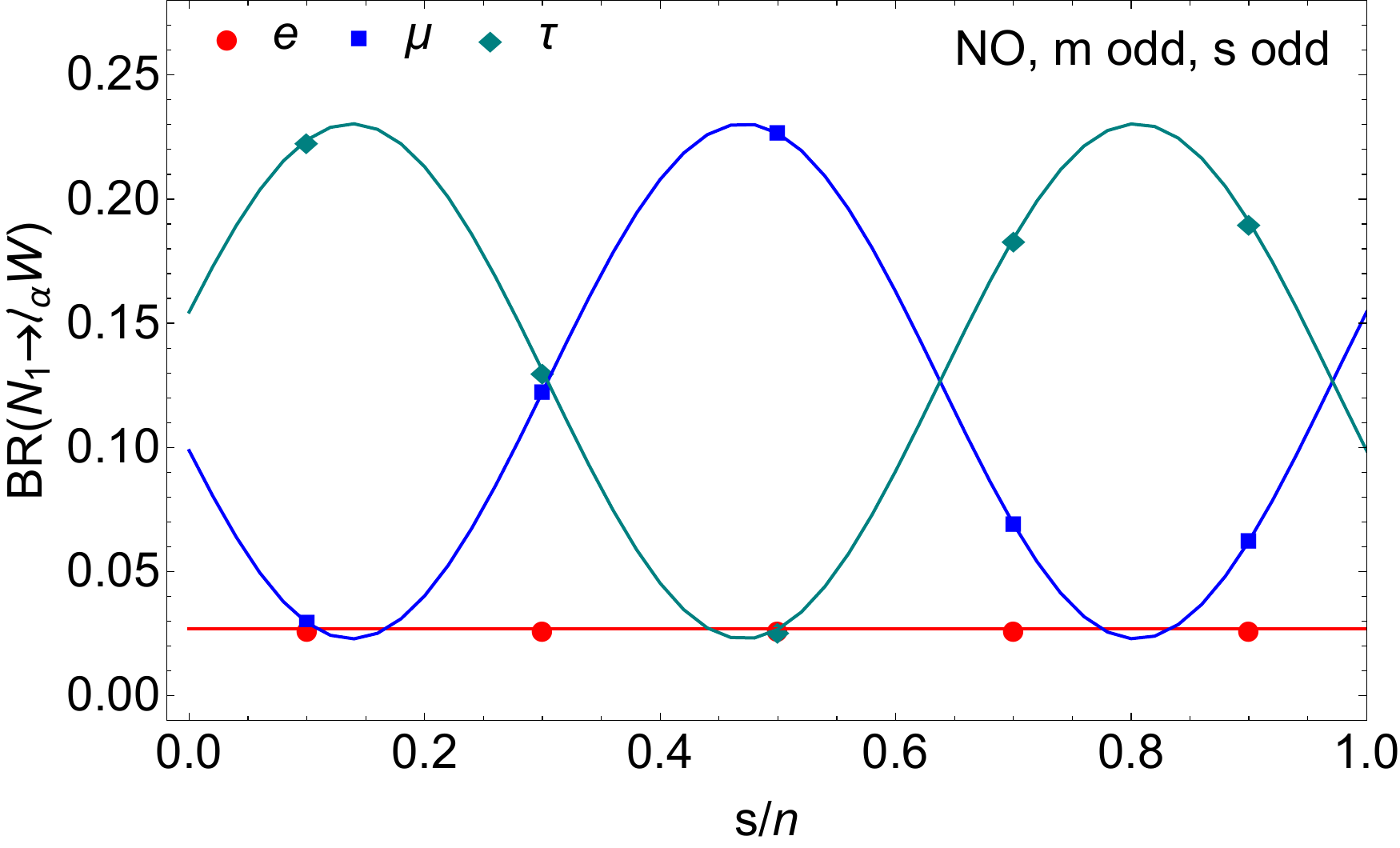}
\includegraphics[width=0.325\textwidth]{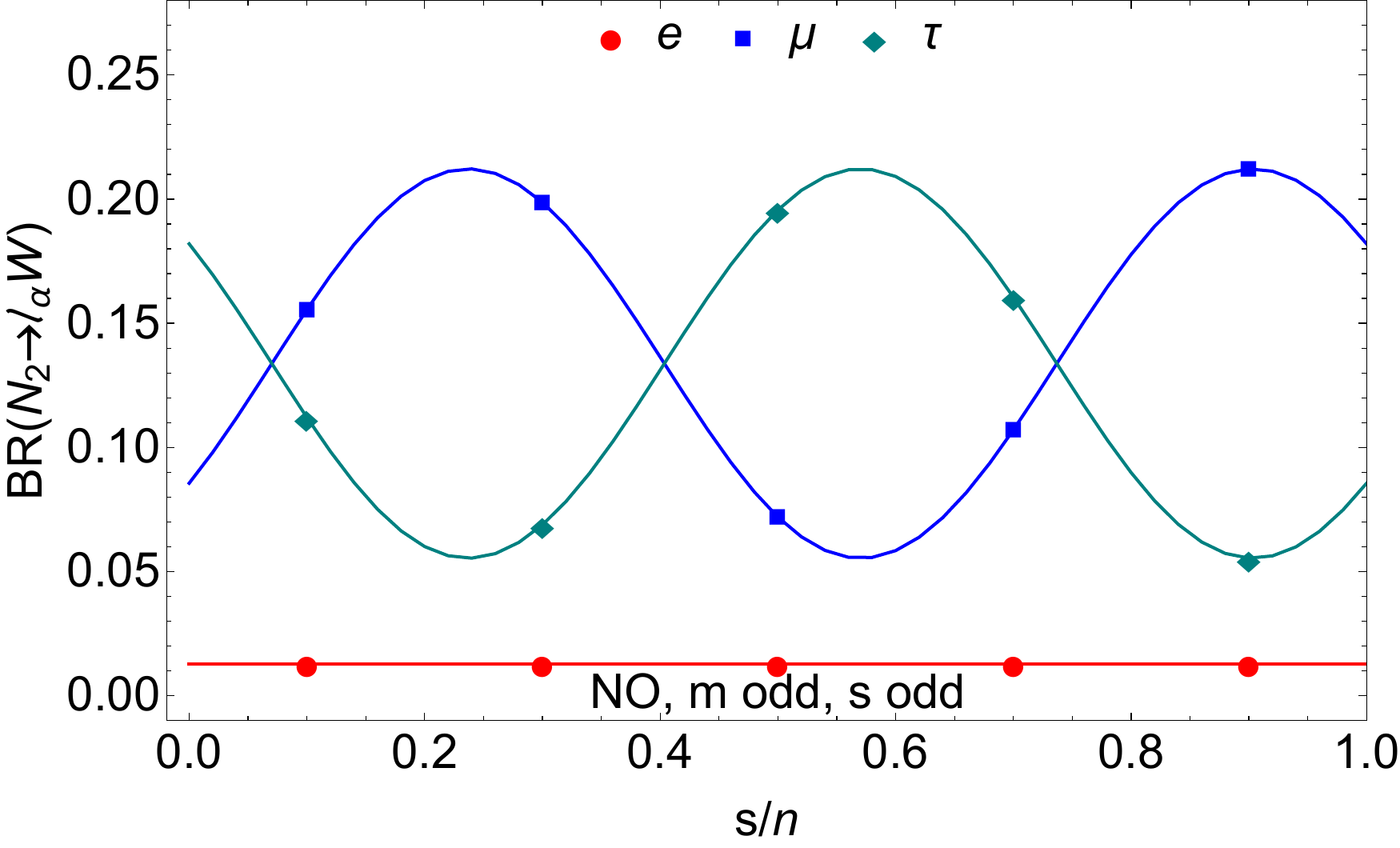}
\includegraphics[width=0.325\textwidth]{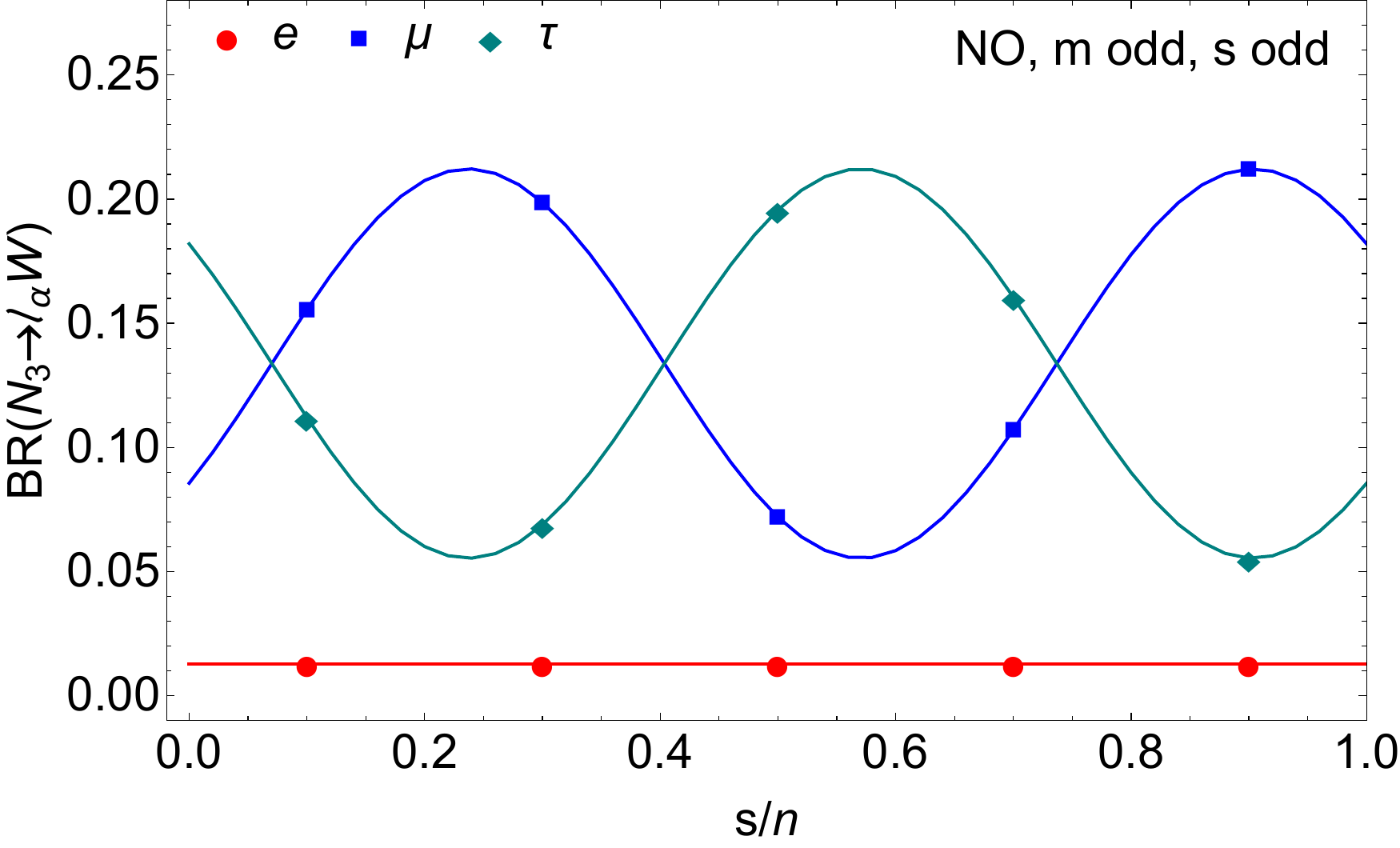}
\caption{For {\it Case 3a} (top two rows, NO only) and {\it Case 3b.1} (bottom two rows, IO and NO). ${\rm BR}(N_{i}\rightarrow \ell_\alpha\, W)$ for $\alpha=e,\mu,\tau$ as a function of $s/n$. The top two rows are for strong NO with ($m$ even and $s$ even) and ($m$ odd and $s$ odd), respectively. The bottom two rows are for strong IO and strong NO with ($m$ even and $s$ even) and ($m$ odd and $s$ odd), respectively. These results do not depend on the specific choice of $n$, but for illustration, we show the results for $n=16(10)$ for Case 3a(3b.1) by the discrete points (corresponding to $0\leq s\leq n-1$, either even or odd). We have fixed $M_N = 250$ GeV and $\theta_R$ at its ERS value for first and third row, while fixed to $\theta_R=0$ for rest, since ERS is absent in these cases.}
\label{fig:brc3}
\end{figure*}
Assuming $M_N>\{m_H,m_W,m_Z\}$, the heavy RHN $N_i$ can decay into $\ell_\alpha\, W,\,\nu_\alpha\, Z$ and $\nu_\alpha\, H$, through its mixing with the SM neutrinos, given by $V_{\alpha i}\simeq (M_DM_R^{-1})_{\alpha i}$. The corresponding partial decay widths for these channels are~\cite{Atre:2009rg}
\begin{widetext}
\begin{subequations}
\begin{alignat}{2}
\Gamma (N_i \rightarrow \ell_\alpha\, W)  \, = \, & \frac{g^2\, v^2}{64\,\pi} \frac{(M_i^2-m_W^2)^2(M_i^2+2m_W^2)}{M_i^5 \, m_W^2} \left|\left(Y_D\right)_{\alpha i}\right|^2,  \\
\Gamma (N_i \rightarrow \nu_\alpha\, Z)  \, = \, & \frac{g^2\, v^2}{128\,\pi\, \cos^2\theta_w} \frac{(M_i^2-m_Z^2)^2(M_i^2+2m_Z^2)}{M_i^5 \, m_Z^2} \left|\left(Y_D\right)_{\alpha i}\right|^2,   \\
\Gamma (N_i \rightarrow \nu_\alpha\, H)  \, = \, & \frac{g^2\, v^2}{128\,\pi} \frac{(M_i^2-m_H^2)^2}{M_i^3 \, m_W^2} \left|\left(Y_D\right)_{\alpha i}\right|^2, 
\end{alignat}
\end{subequations}
where $\theta_w$ is the weak mixing angle. 
Note that for Majorana RHNs, the charge-conjugated final states have the same decay rate; e.g. $\Gamma(N_i\to \ell_\alpha^+ W^-)= \Gamma(N_i\to \ell_\alpha^- W^+)$.   
The BR into the charged-lepton final states $N_i \rightarrow \ell_\alpha^\pm W^\mp$ is given by 
\begin{equation}
   \text{BR}(N_i \rightarrow \ell_\alpha W) \, = \, \frac{\Gamma (N_i \rightarrow \ell_\alpha W)}{2\,[\Gamma (N_i \rightarrow \ell_\alpha\, W)+\Gamma (N_i \rightarrow \nu_\alpha\, Z)+\Gamma (N_i \rightarrow \nu_\alpha\, H)]} \, .
\end{equation}
\end{widetext}
Thus after the RHNs are produced on-shell, their decay BRs are predicted in terms of the underlying Yukawa structure, as shown in Figures~\ref{fig:brc1}, \ref{fig:brc2} and \ref{fig:brc3} for Case 1, Case 2 and Case 3, respectively. We only show those cases having ERS points (see section~\ref{subsec:RHnuwidths}) and tune the $\theta_R$ value to one of the ERS points in each case. We also set $m_0=0$ so that we have either strong NO or strong IO. The RHN mass scale is fixed at a representative value of $M_N=250$ GeV. The behavior of the BRs is independent of the choice of the index $n$; however, for illustration, we choose $n=26$ for which $s/n$ can only take a finite number of values, which are shown by the points for any given flavor.  

\begin{widetext}
Considering the $N_3$ decay at LLP detector, for Case 1 we find
\begin{equation}
\label{eq:br3}
{\rm BR}(N_3\to e^\pm W^\mp): {\rm BR}(N_3\to \mu^\pm W^\mp): {\rm BR}(N_3\to \tau^\pm W^\mp) \, = \, \left\{\begin{array}{ll}
1: 27.7:18.1 & ({\rm NO}) \\
8.5:1:3.7 & ({\rm IO})
\end{array}\right. ,
\end{equation}
\end{widetext}
independent of $\theta_R$ and $s$, and almost independent of $M_N$, if $M_N \gg m_W$. This can be seen from the right column of Figure~\ref{fig:brc1}, where BR$(N_3 \rightarrow \ell_\alpha\, W)$ remains constant for all values of $s/n$. Thus, measuring these BRs at an LLP detector like MATHUSLA for at least two charged lepton flavors $\alpha$ allows a test of the neutrino mass hierarchy at the high-energy frontier. 

Case 1 can also be tested with prompt or displaced vertex signals at the LHC from the
 decays of $N_{1,2}$. However, their BRs
 depend on the chosen CP symmetry $X (s)$ as well as on $\theta_R$. For instance, for $M_N=250$ GeV, $s=2$, $n=26$ and $\theta_R$ at an ERS point, we get 
\begin{widetext}
\begin{eqnarray}
&&\!\!\!\!\!\!\!\!\!\!\!\!\!\!\!\!\!\!\!\!{\rm BR}(N_1\to e^\pm W^\mp): {\rm BR}(N_1\to \mu^\pm W^\mp): {\rm BR}(N_1\to \tau^\pm W^\mp) \, = \, \left\{\begin{array}{ll}
1: 4.9:6.6 & ({\rm NO}) \\
17.3:1:1.6 & ({\rm IO})
\end{array}\right. ,
\\
\label{eq:br12}
&&\!\!\!\!\!\!\!\!\!\!\!\!\!\!\!\!\!\!\!\!{\rm BR}(N_2\to e^\pm W^\mp): {\rm BR}(N_2\to \mu^\pm W^\mp): {\rm BR}(N_2\to \tau^\pm W^\mp) \, = \, \left\{\begin{array}{ll}
1: 17.6:3.0 & ({\rm NO})\\
1:3.3:4.8 & ({\rm IO})
\end{array}\right. .
\end{eqnarray}
These ratios of BRs are independent of $M_N$ for $M_N\gg m_W$. 
\end{widetext}

For Case 2, similar to Case 1, the BR for $N_3$ is independent of $\theta_R$ and $s$, and almost independent of $M_N$, if $M_N \gg M_W$. Moreover, the BRs for $(N_3 \rightarrow \mu\, W)$ and $(N_3 \rightarrow \tau\, W)$ are equal for both NO and IO, as can be seen in the right column of Figure~\ref{fig:brc2}. The BRs of $N_1$ and $N_2$ however depend on the choice of $s$ and $t$, apart from the mass ordering, as shown by the first two columns in Figure~\ref{fig:brc2}.

For Case 3a and 3b.1, the BRs are shown in Figure.~\ref{fig:brc3} top two and bottom two panels respectively. Unlike the previous two cases, none of the BRs is constant and all of them depend on the  the chosen CP symmetry $X (s,m)$ as well as on $\theta_R$, apart from the mass ordering. Furthermore, we notice that the BRs in Case 3a do not have a simple sinusoidal dependence on $s/n$, unlike all other cases. This is because of the strong dependence of the Yukawa parameters on $\theta_{\rm bf}$ which changes with $s$. 

\subsection{Same-sign dilepton signals}
\label{subsec:SSleptonsLHCRHnu}
\begin{figure*}
\centering
\includegraphics[width=0.49\textwidth]{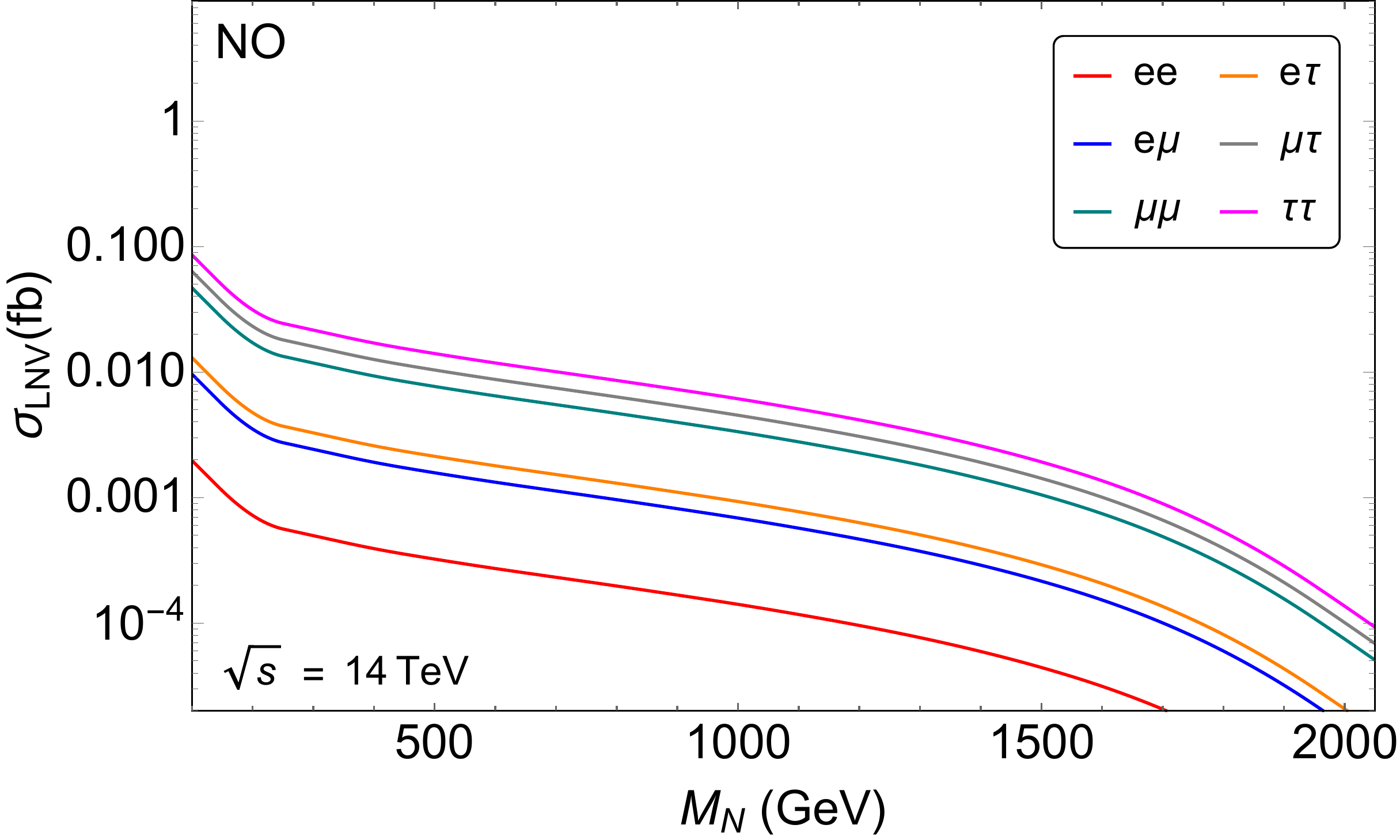}
\includegraphics[width=0.49\textwidth]{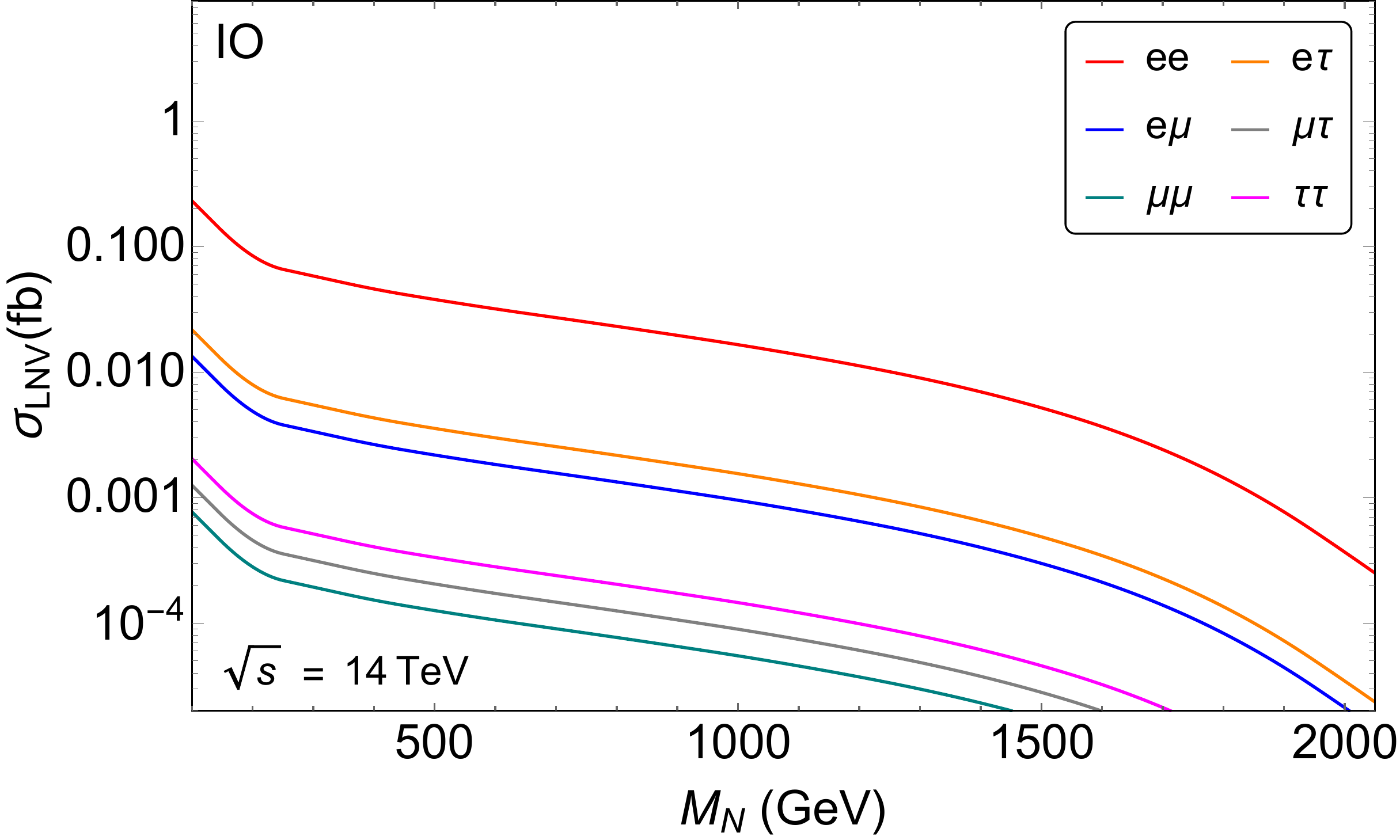}
\caption{LNV signal~\eqref{eq:sigmaLNV} normalized to the $g_{B-L}=1$ case as a function of the RHN mass scale $M_N$ at $\sqrt s=14$ TeV LHC for all possible lepton flavor combinations in the strong NO (left) and strong IO (right) limit. Here $M_{Z'}= 4$ TeV.}
\label{fig:collider1}
\end{figure*}

As shown in Figure~\ref{fig:feyn}, once the Majorana $N_i$'s are produced in pairs via the $Z'$ mediation process in $pp$ collision, their decay into charged leptons leads to the striking  LNV signal
\begin{align}
    pp \to Z'\to N_iN_i\to \ell_\alpha^\pm \ell_\beta^\pm +2W^\mp \to \ell_\alpha^\pm \ell_\beta^\pm +4j \, .
    \label{eq:signal}
\end{align}
Note that for $\alpha\neq \beta$ this process also violates lepton flavor~\cite{Deppisch:2013cya}. 
The process~\eqref{eq:signal} has a much smaller SM background than its lepton number conserving counterpart, namely, $N_iN_i\to \ell_\alpha^\pm \ell_\beta^\mp +W^+W^- \to \ell_\alpha^\pm \ell_\beta^\mp +4j$.
In the narrow-width approximation, the cross section for the LNV process can be written as
\begin{align}
\sigma_\text{LNV}^{\alpha\beta} \, = \, &  c_{\alpha\beta} \sum_i \sigma_{\rm prod}(pp\rightarrow Z'\rightarrow N_i N_i)\text{BR}(N_i \rightarrow \ell_\alpha^\pm W^\mp)  \nonumber \\
& \quad \times \ \text{BR}(N_i \rightarrow \ell_\beta^\pm W^\mp) \ [\text{BR}(W^\mp \rightarrow jj)]^2, 
\label{eq:sigmaLNV}
\end{align}
where $c_{\alpha\beta}=1\, (2)$ for $\alpha=\beta$ ($\alpha\neq \beta$). 
The production cross sections [cf.~Eq.~\eqref{eq:prod} and Figure~\ref{fig:collider}] only depend on the $Z'$ and RHN masses, and are independent of the Yukawa couplings. On the other hand, the BRs for $N_i\rightarrow \ell_\alpha W$ encode the Yukawa structure and will be different for the different cases studied above, as discussed in section~\ref{subsec:RHnuBRs}. Finally, the BR for $W^\mp \rightarrow jj$ is known to be 67.4\% in the SM~\cite{ParticleDataGroup:2020ssz}, which does not change in our setup. For illustration, the results for $\sigma_\text{LNV}$ for $N_1$, normalized to the coupling strength $g_{B-L}=1$, with the Yukawa structure from Case 1 as a function of the RHN mass scale $M_N$ for fixed $M_{Z'} =4$ TeV, $s=2$, $n=26$ is shown in Figure~\ref{fig:collider1} for all possible final-state lepton flavor combinations. The results for other $g_{B-L}$ values can be obtained by simply scaling the cross-sections as $g_{B-L}^2$. We find that for NO, the $\tau \tau$ ($ee$)-channel has the highest (lowest) signal cross section, whereas  for IO, the $ee$ ($\mu\mu$)-channel has the highest (lowest) signal cross section. These results can be understood by analyzing the $N_1$ BRs appearing in Eq.~\eqref{eq:sigmaLNV} from Figure~\ref{fig:brc1}.  For points corresponding to $s=2$, $n=26$ in Figure~\ref{fig:brc1} top left panel (NO, $s$ even), the BR for $N_1 \rightarrow \ell_\alpha W$  is highest (lowest) for the  $\tau$ ($e$)-channel. Similarly, in Figure~\ref{fig:brc1} third row left panel (IO, $s$ even), the BR for $N_1 \rightarrow \ell_\alpha W$  is highest (lowest) for the  $e$ ($\mu$)-channel. From Figure~\ref{fig:collider1}, we see that comparing the LNV final states with different charged-lepton flavor combinations can provide an independent, complementary test of the neutrino mass ordering at the high-energy frontier in our setup. Similar conclusions can be drawn in Cases 2 and 3.

It is also possible to probe the high-energy CP  phases in the Yukawa coupling matrix at colliders using simple observables constructed out of the same-sign dilepton charge asymmetry. In particular, the difference $\sigma_{\rm LNV}^{\alpha,-}$ between and the sum  $\sigma_{\rm LNV}^{\alpha,+}$ of the same-sign charged-lepton final states of a given flavor $\alpha$ can be defined as 
\begin{align}
    \sigma_{\rm LNV}^{\alpha,\pm} \, &= \,  \sum_i \sigma_{\rm prod}(pp\to N_iN_i)\left[{\rm BR}(W\to jj)\right]^2 \nonumber \\
     \quad  \times &\left(\left[{\rm BR}(N_i\to \ell_\alpha^-W^+)\right]^2\pm \left[{\rm BR}(N_i\to \ell_\alpha^+W^-)\right]^2 \right)  \, .
\end{align}
Then the ratio $\sigma_{\rm LNV}^{\alpha,-}/\sigma_{\rm LNV}^{\alpha,+}$ is related to the flavored CP asymmetries $\varepsilon_{i\alpha}$, as pointed out in Refs.~\cite{Bray:2007ru, Blanchet:2009bu, Dev:2019ljp}. Therefore, a measurement of $\sigma_{\rm LNV}^{\alpha,-}/\sigma_{\rm LNV}^{\alpha,+}$ also measures the CP asymmetry in our case for a given set of RHN BRs fixed by the group theory parameters. 

\subsection{Correlation with leptogenesis}
\label{subsec:colliderleptogenesis}

\begin{figure*}[t!]
\centering
\includegraphics[valign=t,scale=0.45]{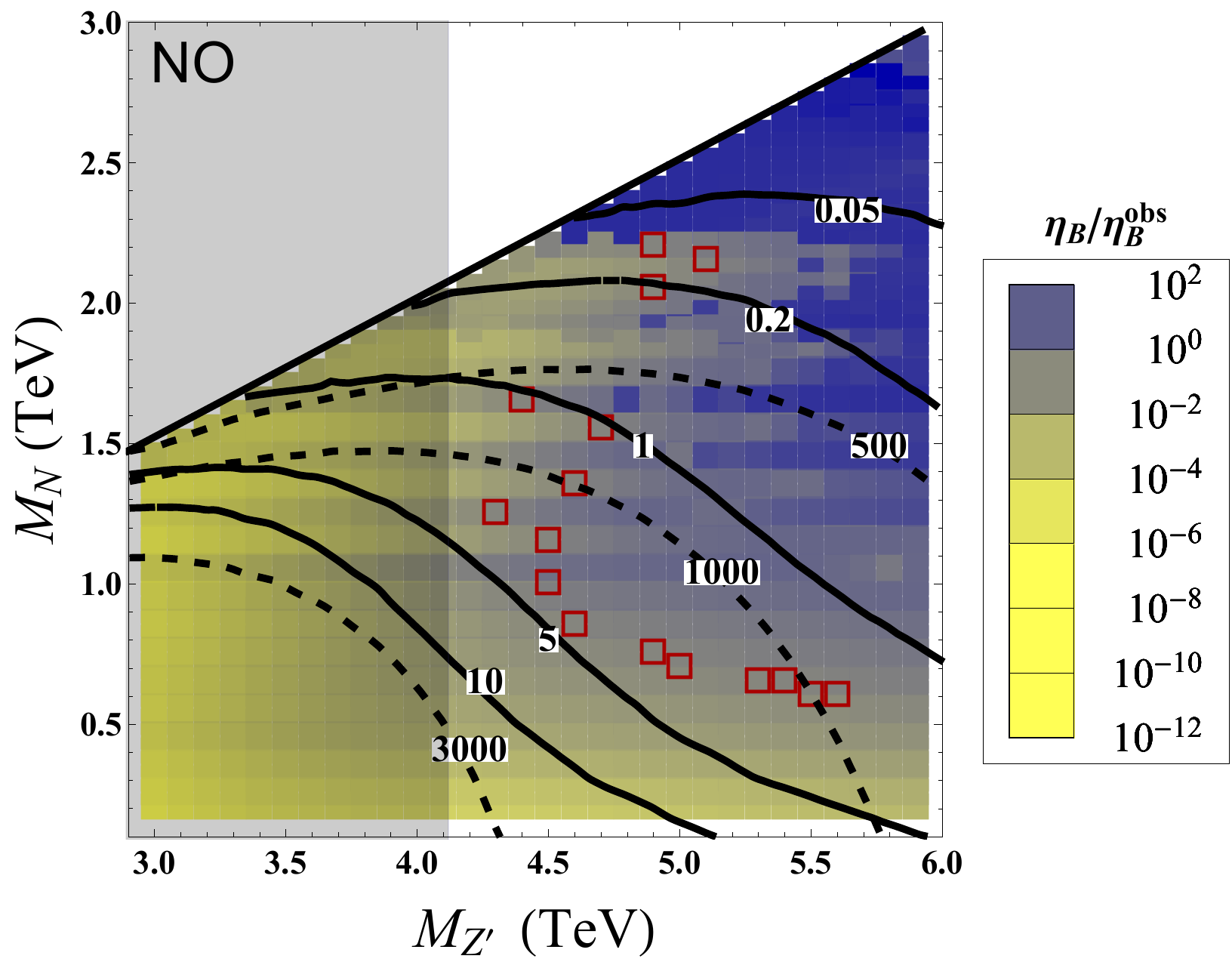}
\includegraphics[valign=t,scale=0.45]{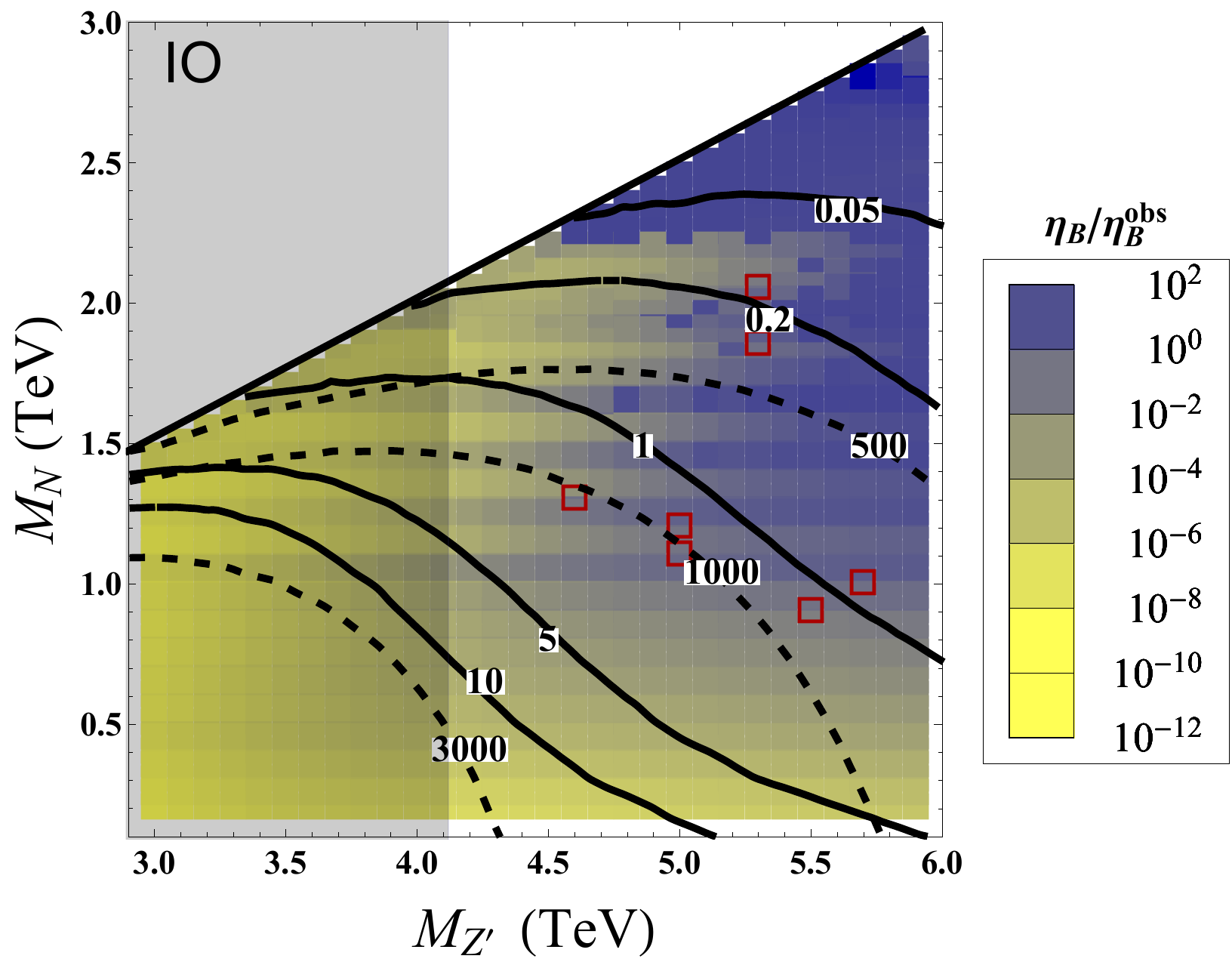}
\caption{For {\it Case 1} prediction of the baryon asymmetry $\eta_B$ relative to the observed value $\eta_B^{\rm obs}$ in the $(M_{Z'},M_N)$ plane for a fixed $g_{B-L}=0.1$. We have fixed $n=26$ and set $s$ to $2 \ (17)$ for strong NO (IO) in the left (right) panel, with the corresponding ERS value of $\theta_R$. The red points correspond to $\eta_B$ within $10\%$ of $\eta_B^{\rm obs}$. The contours show $\sigma_{\mathrm{prod}}$ (in ab) at the $\sqrt s=14$ TeV LHC (solid) and at $\sqrt s=100$ TeV future collider (dashed).}
\label{fig:etaBZp}
\end{figure*}

\begin{figure*}[ht!]
\centering
\includegraphics[valign=t,scale=0.45]{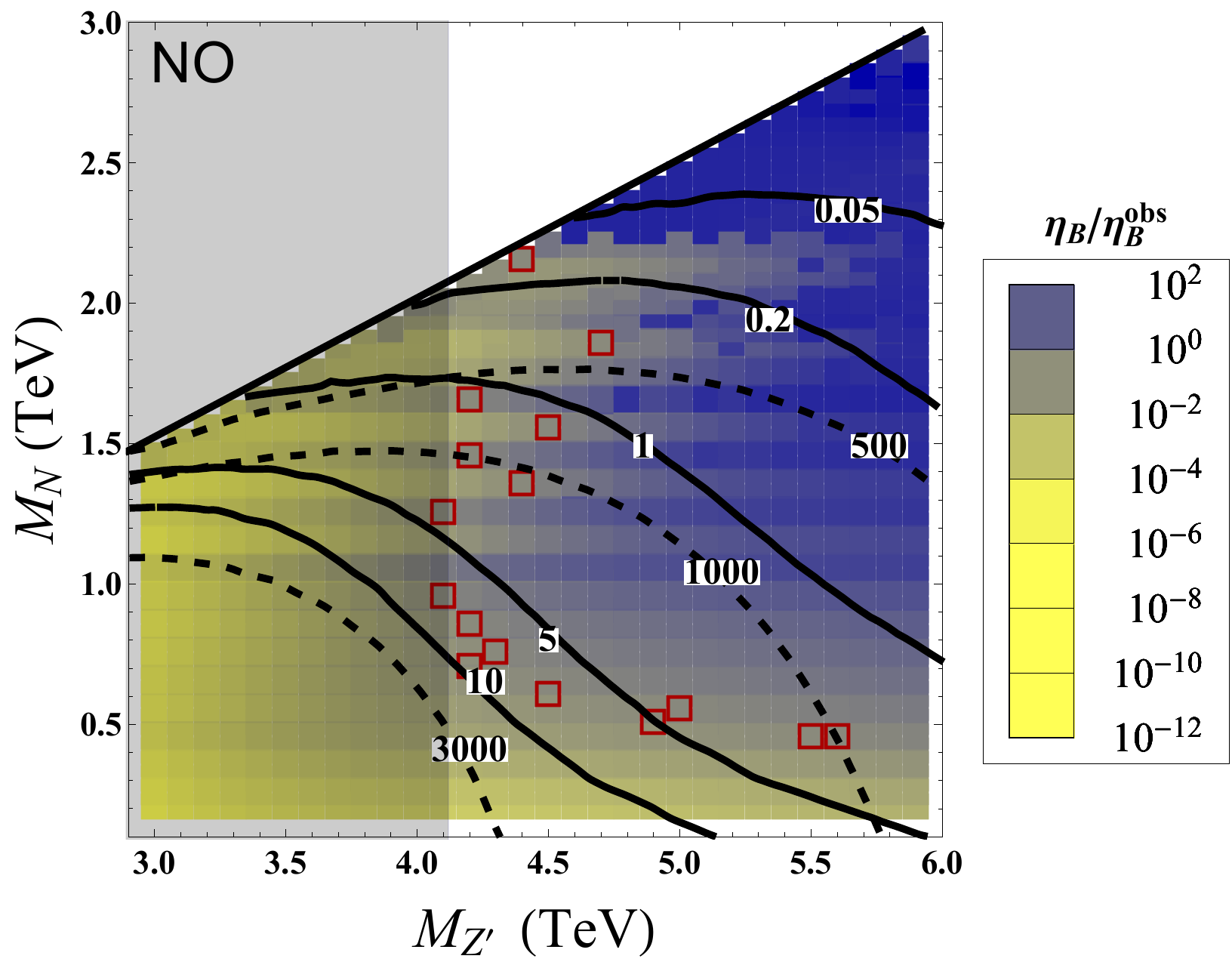}
\includegraphics[valign=t,scale=0.45]{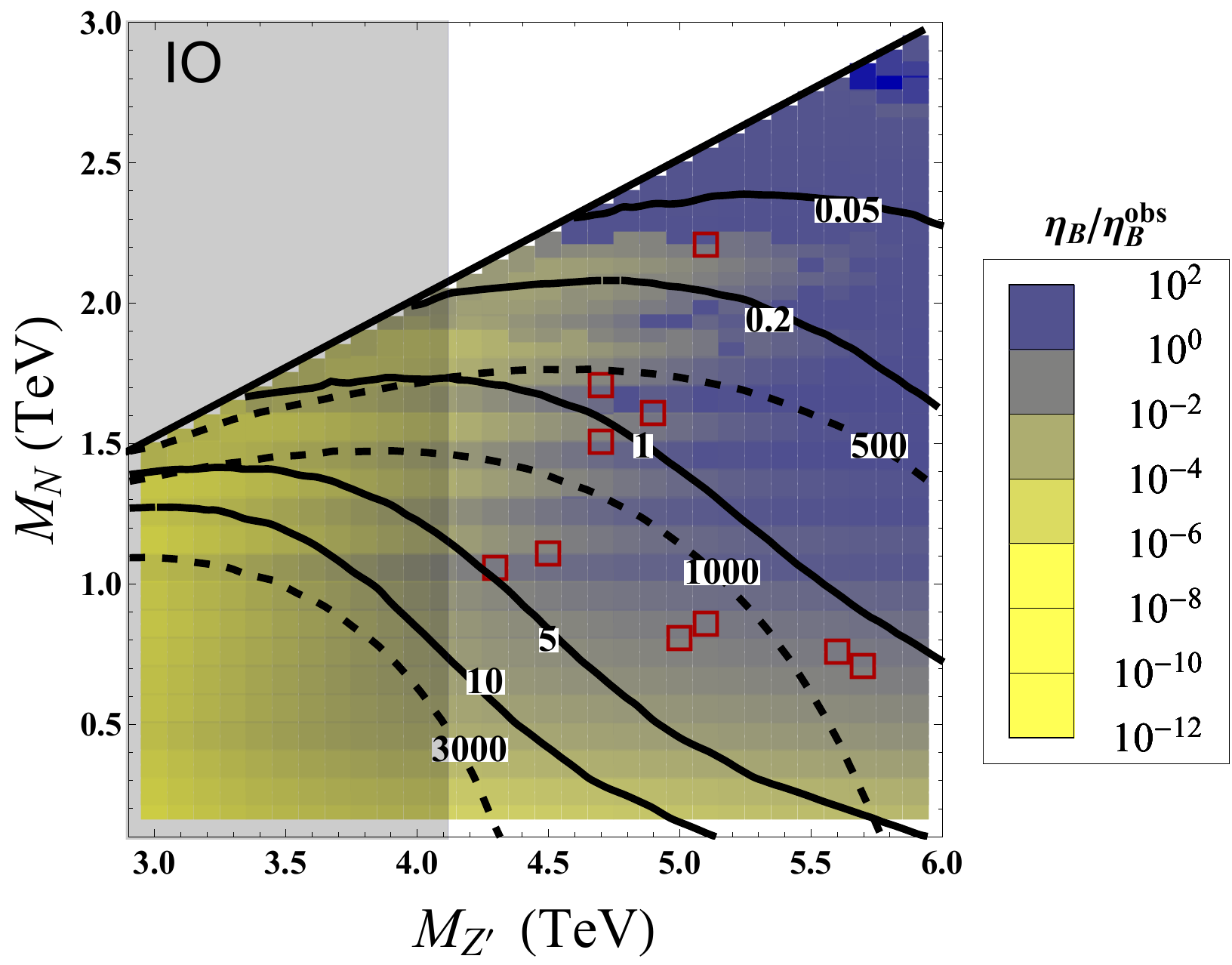}
\caption{For {\it Case 2} prediction of the baryon asymmetry $\eta_B$ relative to the observed value $\eta_B^{\rm obs}$ in the $(M_{Z'},M_N)$ plane for a fixed $g_{B-L}=0.1$. We have fixed $n=14$, $s=1$, $t=2$ (i.e. $u=2s-t=0$) with $\theta_R$ being a point of ERS for strong NO (IO) in the left (right) panel. The red points correspond to $\eta_B$ within $10\%$ of $\eta_B^{\rm obs}$. The contours show $\sigma_{\mathrm{prod}}$ (in ab) at the $\sqrt s=14$ TeV LHC (solid) and at $\sqrt s=100$ TeV future collider (dashed).}
\label{fig:etaBZp2}
\end{figure*}

\begin{figure*}[ht!]
\centering
\includegraphics[valign=t,scale=0.45]{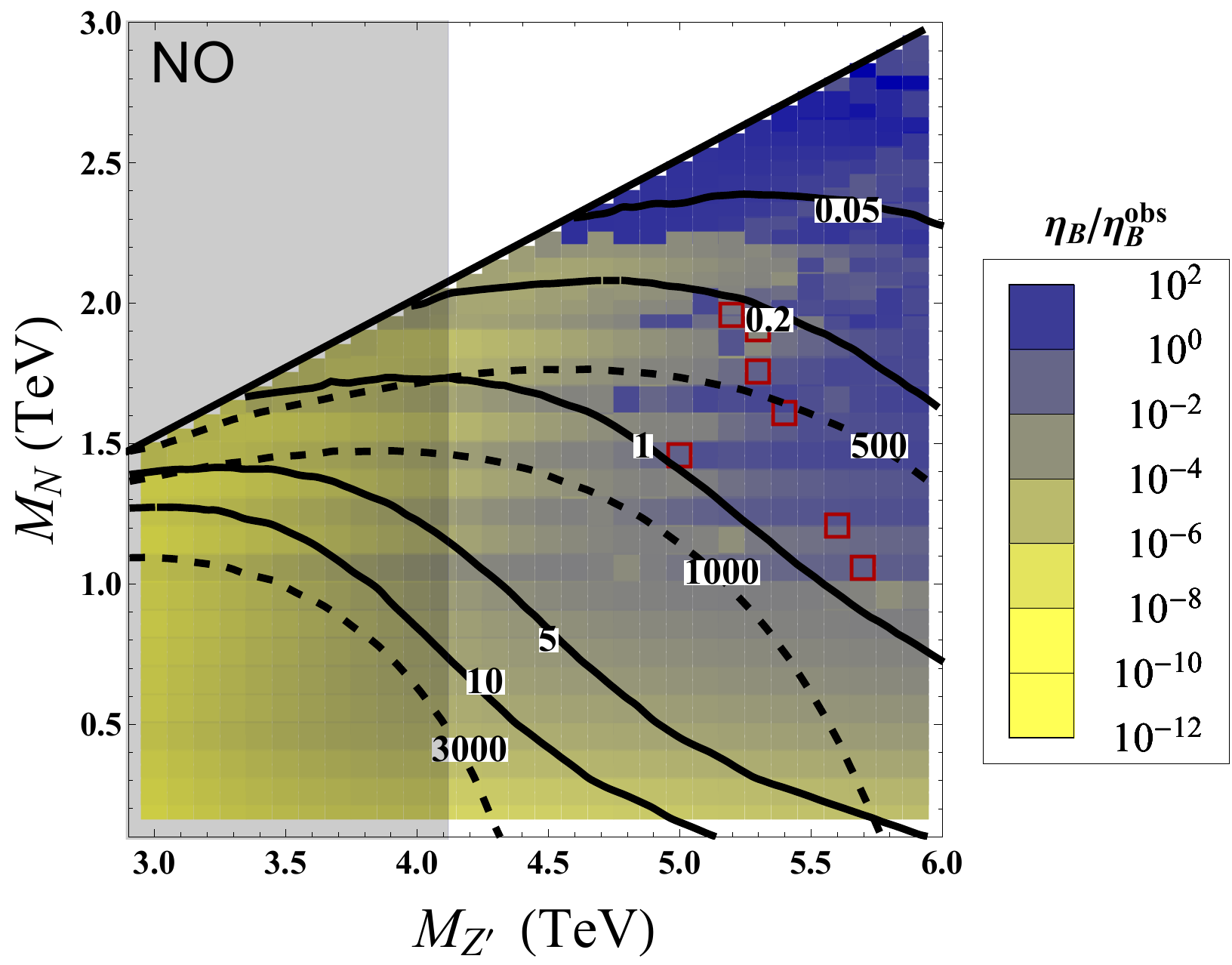}
\includegraphics[valign=t,scale=0.45]{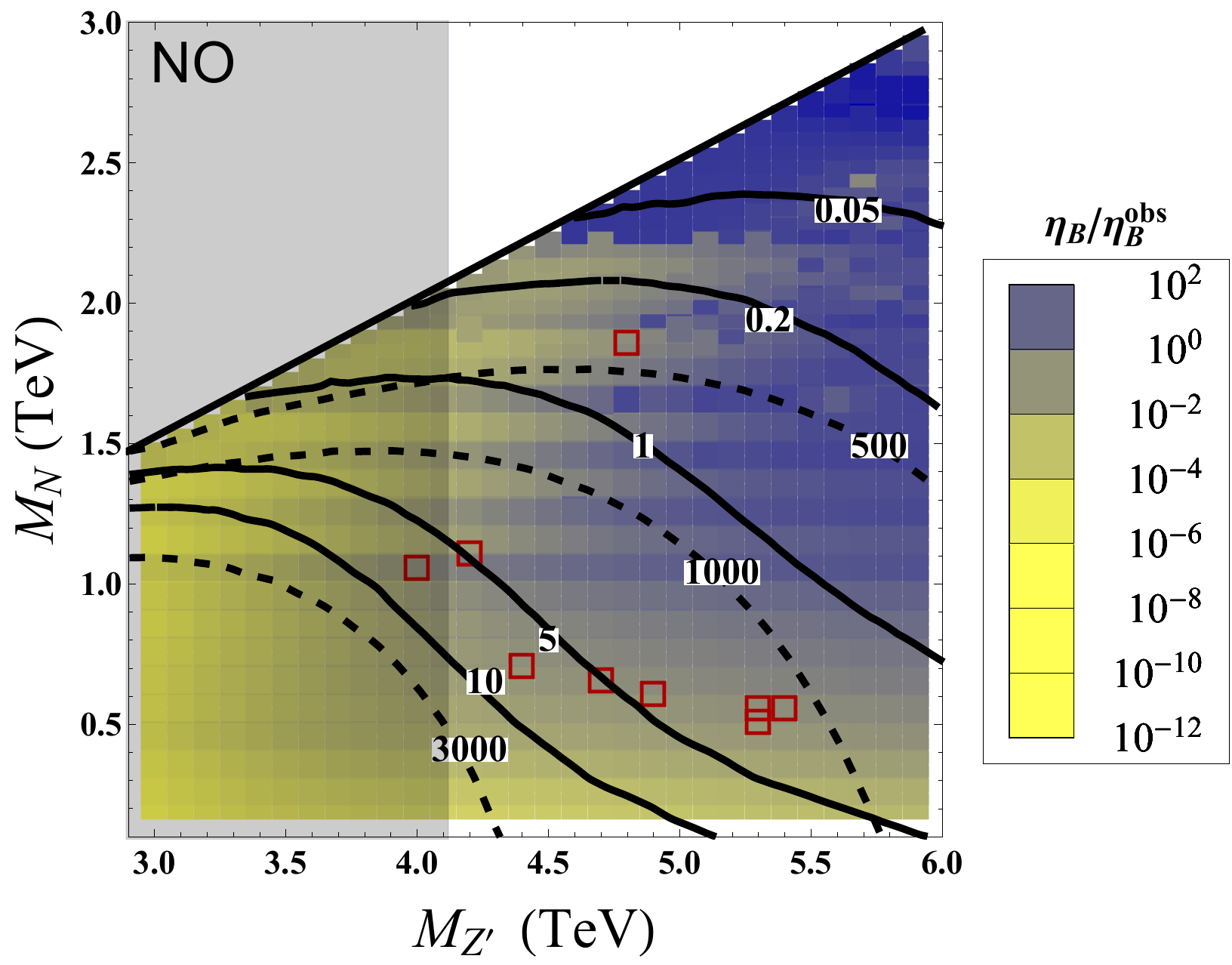}
\caption{{\it Case 3a} (left panel) and {\it Case 3b.1} (right panel) predictions of the baryon asymmetry $\eta_B$ relative to the observed value $\eta_B^{\rm obs}$ in the $(M_{Z'},M_N)$ plane for a fixed $g_{B-L}=0.1$. We have fixed $n=16, \ m=1, \ s=1$ for Case 3a, and $n=10, \ m=5, \ s=5$ for Case 3b.1. In both cases, we take NO with $\theta$ set to zero (not an ERS point). The red points correspond to $\eta_B$ within $10\%$ of $\eta_B^{\rm obs}$. The contours show $\sigma_{\mathrm{prod}}$ (in ab) at the $\sqrt s=14$ TeV LHC (solid) and at $\sqrt s=100$ TeV future collider (dashed).}
\label{fig:etaBZp3}
\end{figure*}

Following the formalism developed in Ref.~\cite{Dev:2014laa}, we compute the baryon asymmetry $\eta_B$ in our scenario [cf.~Eq.~\eqref{eq:etaB}]. For the $U(1)_{B-L}$ case, there are additional washout processes, like $N_i N_i \to Z'\to f\bar{f}$ (where $f$ stands for any SM fermion), mediated by $Z'$~\cite{Blanchet:2009bu, Blanchet:2010kw}. These processes affect both washout and dilution factors in the Boltzmann equations for the RHN and lepton asymmetry number densities, but do not contribute to the CP asymmetry, since the gauge interactions do not violate CP.\footnote{Also, the new $Z'$ gauge boson and the associated scalar sector responsible for $U(1)_{B-L}$ breaking will not alter the electroweak sphaleron transition rate, as long as these new states are heavy enough that they have decoupled from the thermal bath by the time of electroweak phase transition, which will be assumed to be the case here. But if the $U(1)_{B-L}$ phase transition occurs simultaneously with the electroweak phase transition, the SM sphaleron conversion factor of 28/79~\cite{Harvey:1990qw} in Eq.~\eqref{eq:etaB} will be modified to 32/99~\cite{FileviezPerez:2014lnj}.} Therefore, a {\it lower} limit on $M_{Z'}$ follows for a given value of $g_{B-L}$, if successful leptogenesis is demanded.
This is illustrated in Figures~\ref{fig:etaBZp}, \ref{fig:etaBZp2} and \ref{fig:etaBZp3} for Case 1, 2 and 3 respectively. Keeping the collider signals discussed earlier in mind, we particularly focus on the ERS points by suitably choosing $\theta_R$, which leads to one RHN being long-lived. We also fix the gauge coupling $g_{B-L}=0.1$ which corresponds to a dilepton bound of $M_{Z'}\gtrsim 4.1$ TeV~\cite{Das:2021esm} as shown by the vertical shaded region in these plots. Since we are interested in the on-shell production of $N$ via the decay of $Z'$, we do not consider the mass range $M_N>M_{Z'}/2$ as indicated by the white region.

For Case 1, we take $n=26$ and $\theta_R$ being a point of ERS for strong NO (IO) in the left (right) panel of Figure~\ref{fig:etaBZp}. In addition, we choose $s=2$ in order to
maximize the Majorana phase (see Eq.~\eqref{case1sina}) and generate higher $\eta_B$ (as well as $m_{\beta\beta}$, as discussed in section~\ref{sec:leptogenesis0nubb}). The mass splitting $\Delta M_N$, or equivalently, the $\kappa$ value is chosen according to Eq.~\eqref{eq:kappamax} which maximizes the CP asymmetry for each point in the $(M_{Z'},M_N)$ plane, meaning this gives the best-case scenario, since we are interested in the lower bound on $M_{Z'}$. As one can see in Figure~\ref{fig:etaBZp}, successful leptogenesis requires $M_{Z'}\gtrsim 4.3 \, (5)$ TeV for strong NO (IO) in Case 1. 
The red points in the plot give BAU within 10\% of the observed value, which we consider as the allowed range, taking into account the theoretical uncertainties in the calculation. In general, all the points with $\eta_B\gtrsim \eta_B^{\rm obs}$ are allowed in our scenario, because we have maximized the predicted value of $\eta_B$ and it can be easily brought down to match the observed value by moving away from the resonance condition. Another important feature to be noted in Figure~\ref{fig:etaBZp} is that for a given point in the $(M_{Z'},M_N)$ plane, $\eta_B^{\text{NO}} > \eta_B^{\text{IO}}$. This can be qualitatively understood using Eqs.~\eqref{eps1NO} and \eqref{eps1IO}:\footnote{For a comprehensive quantitative comparison of $\eta_B$ in both cases, the efficiency factor for asymmetry production needs to be taken into account, which has been done in our  numerical scans.} 
\begin{align}
    \frac{\eta_B^{\text{NO}}}{\eta_B^{\text{IO}}} \, &\sim \, \frac{\varepsilon_{\text{NO}}}{\varepsilon_{\text{IO}}} \, \approx \, \frac{({y_2 y_3} \, (- y_2^2+ y_3^2 ) \, \sin\theta_{L, \alpha})_\text{NO}}{({y_1 y_2} \, (- y_2^2+y_1^2 ) \, \cos\theta_{L, \alpha})_\text{IO}} \, \nonumber \\ &  \approx \,  \frac{2 \Delta m_{\mathrm{atm}}^2}{\Delta m_{\mathrm{sol}}^2}  \tan\theta_{L, \alpha} \, \gg \, 1 \, .
\end{align}
Furthermore, we compare the results for $\eta_B$ from resonant leptogenesis with the collider sensitivity in the same $(M_{Z'},M_N)$ plane.  Figure~\ref{fig:etaBZp} shows the contours of $\sigma(pp\to Z'\to N_iN_i)$ in ab for both $\sqrt s=14$ TeV LHC (solid lines) and $\sqrt s=100$ TeV collider (dashed lines); see also Figure~\ref{fig:collider}. It turns out that the region of parameter space in Case 1 for strong NO allowing successful leptogenesis yields  $\sigma_{\mathrm{prod}}\lesssim 5$ ab at $\sqrt s=14$ TeV LHC. After taking into account the decay BRs in Eq.~\eqref{eq:sigmaLNV}, and even assuming a fairly low SM background, the number of events is still $<{\cal O}(1)$ with the final target luminosity of 3 ab$^{-1}$. For strong IO, the cross sections are smaller than strong NO case by at least an order of magnitude. 
Therefore, a simultaneous explanation of $\eta_B$ via resonant leptogenesis and an LNV signal at the LHC in our Case 1 is precluded. The same conclusion holds for the Case 2 and Case 3 benchmark points analyzed in Figures~\ref{fig:etaBZp2} and \ref{fig:etaBZp3} respectively.

At a future $100$ TeV collider,  $\sigma_{\mathrm{prod}}$ can reach up to $2000$ ab for the region of successful leptogenesis; taking into account the decay BRs in Eq.~\eqref{eq:sigmaLNV}, and assuming a fairly low SM background, we can expect up to about a thousand  LNV events with 30 ab$^{-1}$ integrated luminosity. The detection prospects at 100 TeV collider might significantly improve by going to higher $Z'$ masses than those shown in Figure~\ref{fig:etaBZp}. This is because the experimental limits on $g_{B-L}$ are relaxed for $M_{Z'}\gtrsim 6$ TeV~\cite{Das:2021esm}. For instance, at $M_{Z'}=7$ TeV, only the LEP-II constraint applies and $g_{B-L}$ is allowed to be as large as one. Since $\sigma_{\mathrm{prod}}$ scales as $g_{B-L}^2$ for $M_N<M_{Z'}/2$, we gain a factor of 100 in the cross-section, at the expense of a mild suppression due to the $Z'$ mass change.  Another way to improve the sensitivity would be by considering other $Z'$ variants, such as the leptophobic case; this will be pursued elsewhere. 

For Case 2, we choose $n=14$, $t=2$, $s=1$ (or $u=2s-t=0$), which gives a good fit to the neutrino oscillation data~\cite{Hagedorn:2014wha}. Here successful leptogenesis at the ERS points requires $M_{Z'}\gtrsim 4.2 \, (4.3)$ TeV for strong NO (IO), with the results for $\sigma_{\mathrm{prod}}$ remaining the same as in Case 1. The comparison between the leptogenesis and collider accessible regions is shown in Figure~\ref{fig:etaBZp2} for both NO (left panel) and IO (right panel). Again, the LHC will not be able to probe the successful leptogenesis region in this case, but a future 100 TeV collider can do.

For Cases 3a and 3b.1, as mentioned in section~\ref{subsec:decay3}, leptogenesis is not viable at the ERS points, because $N_1$ becomes long-lived. Since ERS only occurs for $m$ even and $s$ even case [cf.~Figure~\ref{fig:dlc3}], we choose a different case with $m$ odd and $s$ odd for leptogenesis. In particular, for Case 3a, we fix $n=16$, $m=1$, $s=1$ which gives a good fit to the neutrino data with NO~\cite{Hagedorn:2014wha}, and furthermore, choose $\theta_R=0$ for simplicity. Our results for leptogenesis in the $(M_{Z'},M_{N})$ plane is shown in Figure~\ref{fig:etaBZp3} left panel. We find that successful leptogenesis requires $M_{Z'}\gtrsim 5.0$ TeV.
Similarly, for Case 3b.1, we fix $n=10$, $m=5$, $s=5$ which gives a good fit to the neutrino data with NO~\cite{Hagedorn:2014wha}, and furthermore, choose $\theta_R=0$ for simplicity. Our results for leptogenesis in the $(M_{Z'},M_{N})$ plane is shown in Figure~\ref{fig:etaBZp3} right panel. We find that successful leptogenesis requires $M_{Z'}\gtrsim 4.2$ TeV. The LHC and 100 TeV collider contours for the LNV signal are the same as in Cases 1 and 2. Again, the LHC will not be able to probe the successful leptogenesis region in these cases, but a future 100 TeV collider can do.

\mathversion{bold}
\section{Correlation of BAU with \texorpdfstring{\(0\nu\beta\beta\)}{0nubb}}
\mathversion{normal}
\label{sec:leptogenesis0nubb}

In this section, we discuss the connection between the high- and low-energy CP phases in our flavor model. To this effect, we consider the classic low-energy LNV process of $0\nu\beta\beta$ which can unambiguously  discern the Majorana nature of the neutrinos~\cite{Schechter:1981bd, Dolinski:2019nrj}.  The theory predictions for this yet unobserved process depends explicitly on the low-energy Majorana phases $\alpha_1$ and $\alpha_2$. Therefore, we expect the rate of $0\nu\beta\beta$ process to be correlated with the baryon asymmetry predictions, which depend on the high-energy CP phases that are related to the low-energy CP phases in our model. We will study this correlation for various scenarios of lepton mixing for which leptogenesis has been studied in section~\ref{sec:CPasymmetries}. The following analysis generically applies to Majorana neutrinos and does not depend on the extra $U(1)$ introduced in section~\ref{sec:colliderLep}. 

A nuclear isotope decaying through $0\nu\beta\beta$ process 
\begin{align}
    (A,Z) \, \to \, (A,Z+2)+2e^-
\end{align}
would exhibit an half-life  of 
\begin{equation}
     T_{1/2}^{0\nu} \, = \,  \left[G^{0\nu}\,\left|M^{0\nu}\right|^{2} \left(\frac{m_{\beta\beta}}{m_e}\right)^{2} \right]^{-1}\, ,
\end{equation}
where $G^{0\nu}$ is the phase-space factor~\cite{Kotila:2012zza, Neacsu:2015uja}, $|M^{0\nu}|$ is the nuclear matrix element (NME) for this LNV transition~\cite{Fang:2018tui, Ejiri:2020xmm}, $m_{\beta\beta}$ is the effective Majorana neutrino mass and $m_e$ is the electron mass. The values of $G^{0\nu}$ and $|M^{0\nu}|^2$ cannot be measured independently but can be computed based on the nuclear isotope, whereas $m_{\beta\beta}$ is expressed only in terms of the light neutrino masses and lepton mixing parameters, i.e.\footnote{Here we only consider the canonical light neutrino exchange for the $0\nu\beta\beta$ process~\cite{Racah:1937qq, Furry:1939qr}. The heavy RHNs in the model could also mediate the $0\nu\beta\beta$ process via their mixing $V_{\ell N}$ with the light neutrinos~\cite{Ibarra:2010xw, Mitra:2011qr}; however, for the Yukawa couplings being considered here, these RHN contributions are negligible.} 
\begin{equation}
\label{meedef}
m_{\beta\beta} \, = \, \sum_i \left|U_{ei}^2m_i\right| \, = \, \left| U_{e1}^2 \, m_1 + U_{e2}^2 \, m_2 + U_{e3}^2 \, m_3  \right|\, .
\end{equation}
With the form of $U$ given by Eq.~\eqref{UPMNSdef}, the effective neutrino mass in Eq.~\eqref{meedef} reads
\begin{align}
\label{meedef2}
m_{\beta\beta} \, = \, |& \cos^2 \theta_{12} \, \cos^2 \theta_{13} \, m_1 + \sin^2 \theta_{12} \, \cos^2 \theta_{13} \, e^{i \alpha_1} \, m_2 \nonumber \\ & + \sin^2 \theta_{13} \, e^{i \alpha_2} \, m_3  | \, .
\end{align}
For a strongly hierarchical light neutrino mass spectrum with $m_0\to 0$, the value of $m_{\beta\beta}$ depends on the mass ordering. From Eq.~\eqref{meedef2}, we get the following in the strong NO ($m_1=0$) and strong IO ($m_3=0$) limits respectively:
\begin{widetext}
\begin{subequations}
\begin{alignat}{2}
    m_{\beta\beta}^{\rm NO} \, \approx \, & \left|\sin^2\theta_{12}\cos^2\theta_{13}e^{i\alpha_1}\sqrt{\Delta m^2_{\rm sol}}+\sin^2\theta_{13}e^{i\alpha_2}\sqrt{\Delta m^2_{\rm atm}} \right| \, ,\\
     m_{\beta\beta}^{\rm IO} \, \approx \, & 
     \left|\cos^2\theta_{12}+\sin^2\theta_{12}e^{i\alpha_1}\right|\cos^2\theta_{13}\sqrt{|\Delta m^2_{\rm atm}|} \, .
\end{alignat}
\end{subequations}
\end{widetext}
An upper bound on the effective Majorana neutrino mass has been set by several experiments, using different nuclear isotopes, such as KamLAND-Zen ($^{136}$Xe) \cite{KamLAND-Zen:2022tow}, EXO-200 ($^{136}$Xe) \cite{EXO-200:2019rkq}, GERDA ($^{76}$Ge) \cite{Agostini:2020xta},   CUORE-0 ($^{130}$Te) \cite{Adams:2021rbc}, 
and 
NEMO 3 ($^{100}$Mo among others) \cite{NEMO-3:2019gwo}. The strongest bound on $m_{\beta\beta}$ is currently given by the KamLAND-Zen experiment using $^{136}$Xe isotope~\cite{KamLAND-Zen:2022tow}:
\begin{equation}
\label{meebound}
	m_{\beta\beta}\;<\; \left(36 -156\right)\;\text{meV}\quad\text{at 90\% C.L.}
\end{equation}
with the spread coming from different NME calculations. 
Future tonne-scale experiments like nEXO~\cite{nEXO:2021ujk} and LEGEND~\cite{LEGEND:2021bnm} can extend this down to 
\begin{equation}
\label{meebound2}
	m_{\beta\beta}\;<\; \left\{\begin{array}{ll} \left(4.7 -20.3\right)\;\text{meV} \quad & {\textrm {(nEXO)}} \\
	\left(34-78\right)\;\text{meV} \quad & {\textrm {(LEGEND-200)}} \\
	\left(9-21\right)\;\text{meV} \quad & {\textrm {(LEGEND-1000)}} 
	\end{array}
	\right. \, .
\end{equation}
As we will see below, some of our $m_{\beta\beta}$ predictions for the IO case are already excluded by KamLAND-Zen~\cite{KamLAND-Zen:2022tow} for specific choice of the NMEs. Moreover, the remaining IO parameter space for $m_{\beta\beta}$ is within reach of nEXO and LEGEND sensitivities. However, the corresponding predictions for the NO case are at the level of 1-4 meV, which are out of reach of tonne-scale detectors, but the kilotonne-scale detectors might be able to probe this region~\cite{Avasthi:2021lgy}.

\subsection{Case 1}

In this case the Majorana phase $\alpha_2$ is always trivial, for any choice of $\theta$ and the group parameters $n$ and $s$, while the second Majorana phase $\alpha_1$ can take non-trivial values [cf.~Eq.~\eqref{case1sina}]. Using the form of the PMNS mixing matrix  given in terms of the model parameters in Eq.~\eqref{eq:PMNS1}, we get from Eq.~\eqref{meedef} the effective neutrino mass in strong NO and IO limits respectively as
\begin{subequations}
\begin{alignat}{2}
     m_{\beta\beta}^{\rm NO} & \, \approx \,  \frac{1}{3}\left|\sqrt{\Delta m_{\mathrm{sol}}^2}+2(-1)^{k_1+k_2} \sin^2\theta\,e^{6i\phi_s}\sqrt{\Delta m_{\mathrm{atm}}^2} \right|, \\
     m_{\beta\beta}^{\rm IO} &\approx \frac{1}{3}\left|1+2(-1)^{k_1} \cos^2\theta\,e^{6i\phi_s} \right|\sqrt{\left|\Delta m^2_{\mathrm{atm}}\right|} \, . 
\end{alignat}
\end{subequations}
Thus we see that the value of the effective Majorana neutrino mass $m_{\beta\beta}$, accessible in $0\nu\beta\beta$ experiments, crucially depends on the choice of the CP symmetry and is in this scenario considerably restricted~\cite{Hagedorn:2016lva}. For our choice of $n=26$, $k_1=k_2=0$ and for $\theta=\theta_{\rm bf} \approx 0.18$, using the best fit values for $\Delta m_{\mathrm{sol}}^2$ and $\Delta m_{\mathrm{atm}}^2$~\cite{Esteban:2020cvm, NUFIT}, we get 
\begin{align}
   1.8 \, \mathrm{meV} &\lesssim m_{\beta\beta} \lesssim 4.0 \, \mathrm{meV}~({\textrm {NO}}) \, , \nonumber \\ 16 \, \mathrm{meV} &\lesssim m_{\beta\beta} \lesssim 49 \, \mathrm{meV}~({\textrm {IO}}) \, .
\end{align}
\begin{figure*}[t!]
\centering
\includegraphics[width=0.48\textwidth]{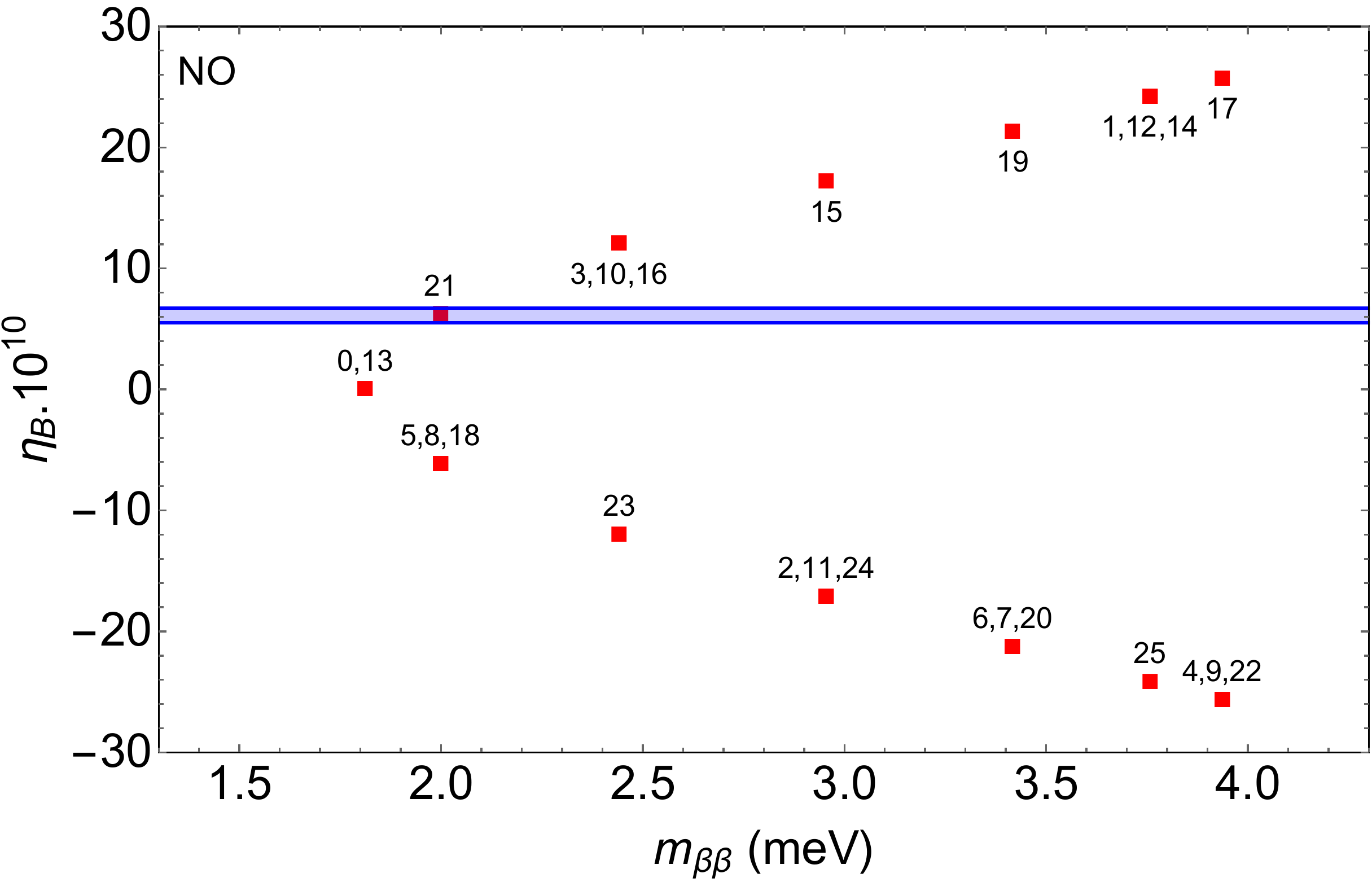}
\hspace{0.1in}
\includegraphics[width=0.48\textwidth]{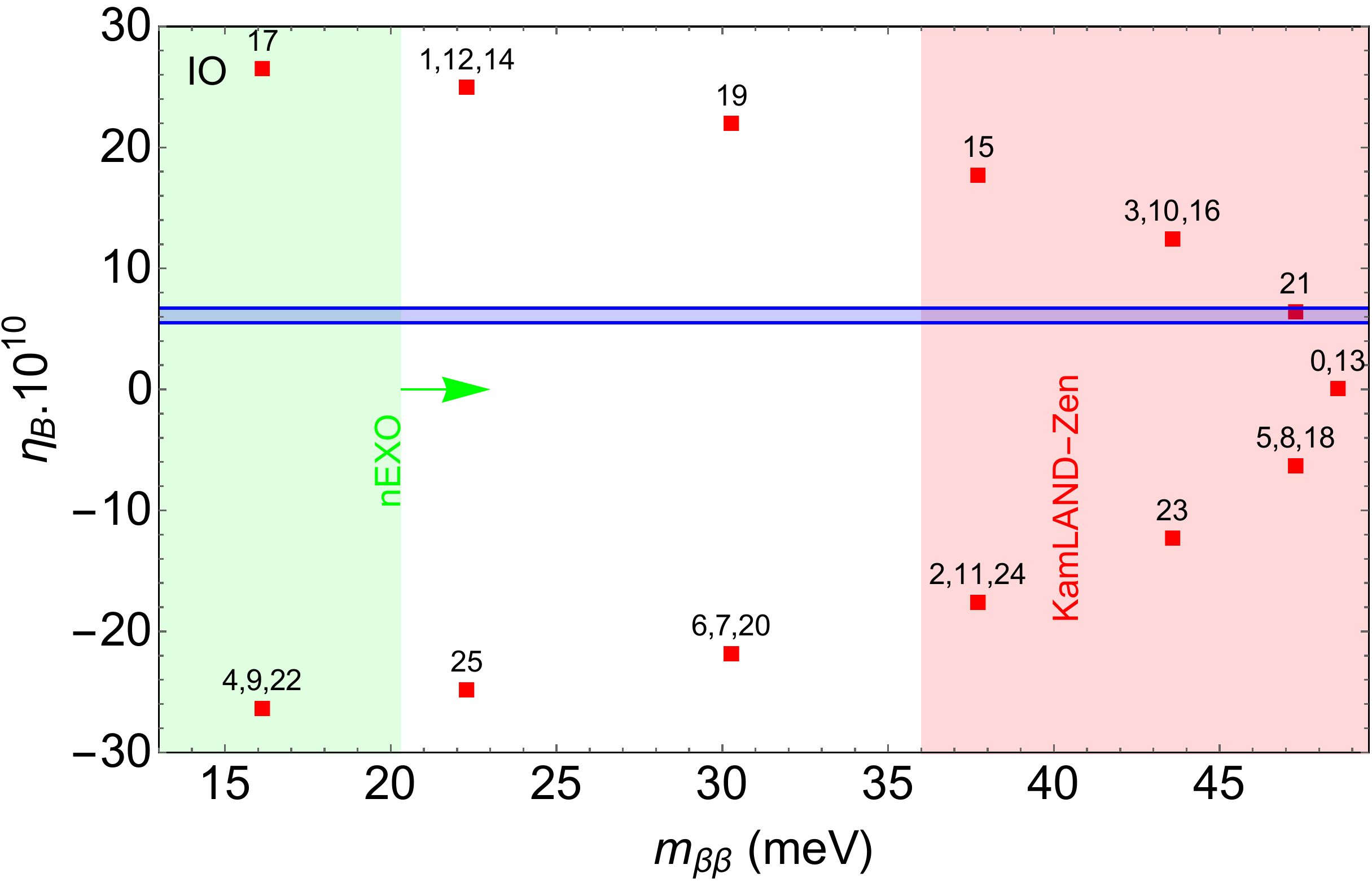}
\caption{Correlation between the predicted BAU $\eta_B$ and the effective neutrino mass $m_{\beta\beta}$ for {\it Case 1} with $n=26$, $k_1=k_2=0$ and $0\leq s \leq n-1$ (as shown by the numbered points). We have chosen the benchmark values for $M_{N}=1 $ TeV and  $\kappa/ \kappa_\text{max} = 8 \times 10^4 \,(6.5 \times 10^2 )$ for strong NO (IO) in the left (right) panel, which according to Figure~\ref{fig:etaBZp} produce the BAU of the correct magnitude. The blue-shaded horizontal bar corresponds to $\eta_B$ within $10\%$ of $\eta_B^{\rm obs}$. The vertical shaded bands for the IO case indicate the smallest $m_{\beta\beta}$ value (including the NME uncertainties) either ruled out by the current KamLAND-Zen bound (red) or accessible to future nEXO (green). }
\label{fig:0vbb1}
\end{figure*}
This is shown in Figure~\ref{fig:0vbb1} by the red points, which correspond to different values of $s$ ranging from 0 to $n-1$. The left (right) panel is for NO (IO). 
For strong IO, some of the admitted values of $m_{\beta\beta}$ are already excluded by the current KamLAND-Zen bound~\cite{KamLAND-Zen:2022tow} [cf.~Eq.~\eqref{meebound}] for the most aggressive NME (as shown by the red shaded region). Most of the remaining values of $m_{\beta\beta}$ for IO can be tested with the proposed experiment LEGEND~\cite{LEGEND:2021bnm} (not shown in Figure~\ref{fig:0vbb1}) 
and all of them can be explored with nEXO~\cite{nEXO:2021ujk} (as shown by the green shaded region); cf.~Eq.~\eqref{meebound2}. 

To illustrate the correlation between the high-and low-energy CP phases, we also plot in Figure~\ref{fig:0vbb1} the predictions for the BAU corresponding to the $s$ values shown here. Here we have fixed $M_N=1$ TeV  for both strong NO and IO. We choose a value for the mass splitting between the RHNs $N_1$ and $N_{2,3}$, i.e.~$\Delta\, M_N=3\kappa M_N$ [cf.~Eq.~\eqref{eq:massesNi}], which generate $\eta_B$ values in the vicinity of $\eta_B^{\rm obs}$. For Figure~\ref{fig:0vbb1}, we have chosen $\kappa = 8 \times 10^4 \kappa_\text{max}$ and $6 \times 10^5 \kappa_\text{max}$ respectively for NO and IO.\footnote{Although Figure~\ref{fig:etaBZp} is for a fixed $s$ value, we do not expect too much variation in $\eta_B$ with respect to $s$, as confirmed in Figure~\ref{fig:0vbb1}.}  The horizontal blue-band corresponds to $\eta_B$ values within 10\% of the observed value. As can be seen in Figure~\ref{fig:0vbb1}, values of $s$ like $s= 21$ can successfully reproduce the observed BAU for both orderings. Note that while some values of $s$ like $s=5,8,18$ can produce the correct magnitude of $\eta_B$ for both orderings but have the wrong sign and should be discarded. It should also be pointed out that since the inclusion of $Z'$ only contributes to the washout of the generated BAU and does not affect $0\nu\beta\beta$ predictions, here we have integrated out the $Z'$ from the low-energy effective theory which leads to higher values of $\eta_B$ than in section~\ref{subsec:colliderleptogenesis}.

\begin{figure*}[t!]
\centering
\includegraphics[width=0.48\textwidth]{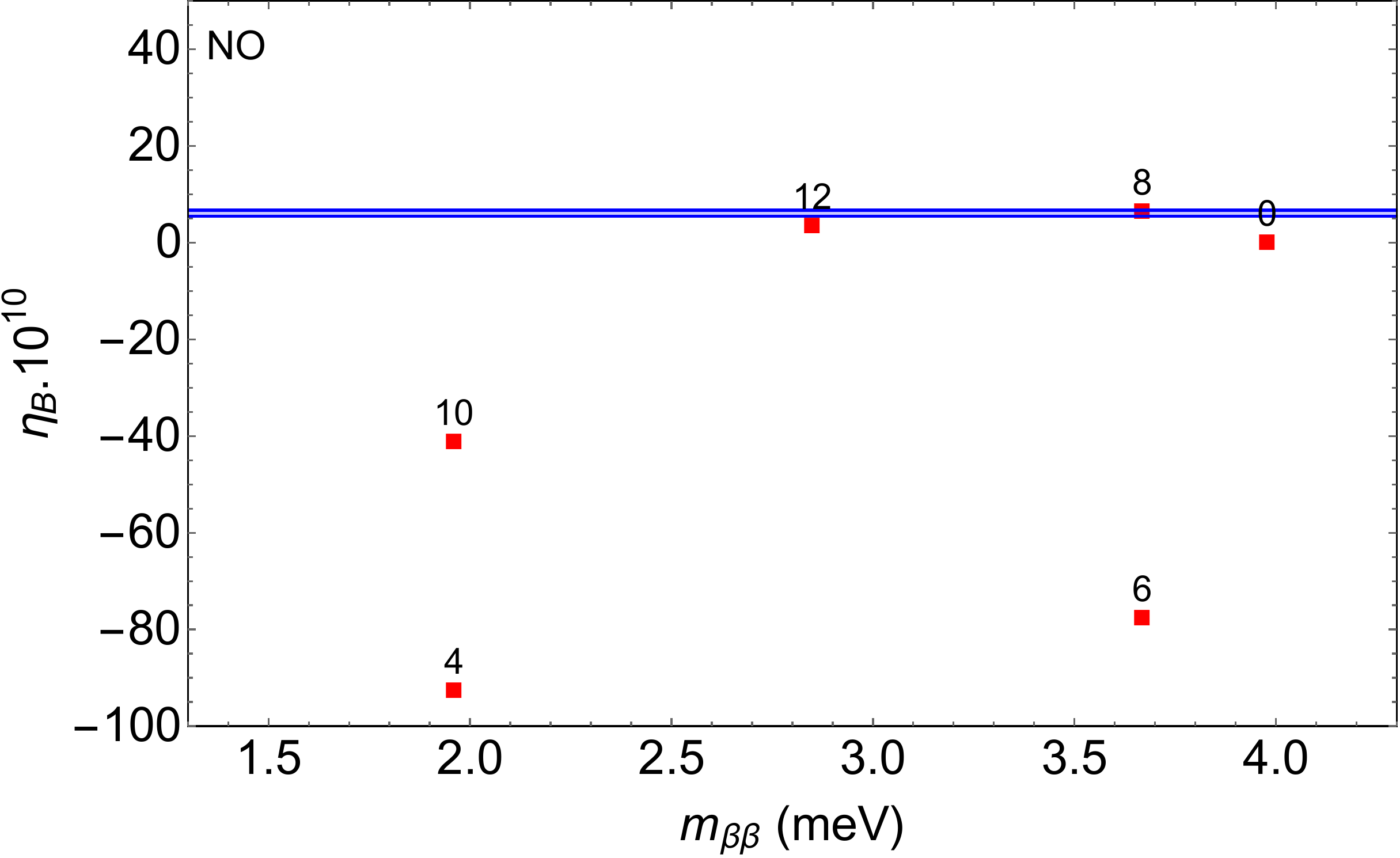}
\hspace{0.1in}
\includegraphics[width=0.48\textwidth]{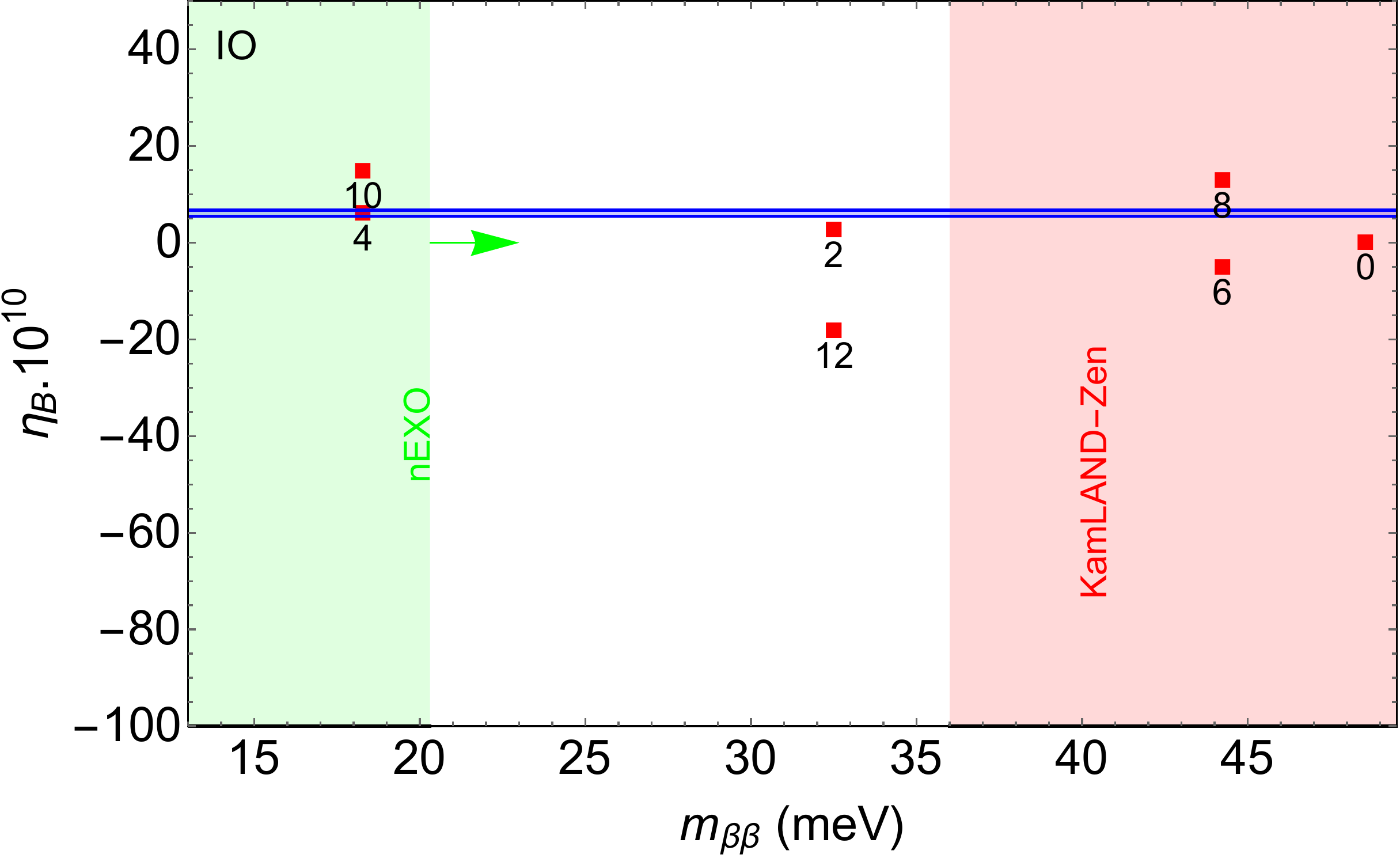}
\caption{Correlation between the predicted BAU $\eta_B$ and the effective neutrino mass $m_{\beta\beta}$ for {it Case 2} with $n=14$, $k_1=k_2=0$ and $0\leq t\leq n-1$ (as shown by the numbered points).  Only even $t$ are allowed because of the condition $u=2s-t=0$. We have chosen the benchmark values for $M_{N}=1 $ TeV and  $\kappa/ \kappa_\text{max} = 7 \times 10^3 \,(1.2 \times 10^3 )$ for strong NO (IO) in the left (right) panel using the results from Figure~\ref{fig:etaBZp2}. The blue-shaded horizontal bar corresponds to $\eta_B$ within $10\%$ of $\eta_B^{\rm obs}$. The vertical shaded bands for the IO case indicate the smallest $m_{\beta\beta}$ value (including the NME uncertainties) either ruled out by the current KamLAND-Zen bound (red) or accessible to future nEXO (green).}
\label{fig:0vbb2}
\end{figure*}

\subsection{Case 2}
In this case both the Majorana phases $\alpha_1$ and $\alpha_2$ can in general have non-trivial values [cf.~Eq.~\eqref{eq:CPcase2}]. Using the PMNS mixing matrix from Eq.~\eqref{eq:PMNScase2}, we get the following for strong NO and IO respectively: 
\begin{widetext}
\begin{subequations}
\begin{alignat}{2}
     m_{\beta\beta}^{\rm NO} & \, \approx \,  \frac{1}{3}\left|\sqrt{\Delta m_{\mathrm{sol}}^2}-2(-1)^{k_1+k_2} e^{i\phi_v}\left(\cos{\theta}\sin{\frac{\phi_u}{2}}-i\sin{\theta}\cos{\frac{\phi_u}{2}}\right)^2\sqrt{\Delta m_{\mathrm{atm}}^2} \right|, \\
     m_{\beta\beta}^{\rm IO} & \, \approx \,  \frac{1}{3}\left|1+(-1)^{k_1} e^{i\phi_v}(\cos{\phi_u}+\cos{2\theta}-i\sin{2\theta}\sin{\phi_u}) \right|\sqrt{\left|\Delta m^2_{\mathrm{atm}}\right|}\, . 
\end{alignat}
\end{subequations}
\end{widetext}
For our choice of $n=14$ and $u=2s-t=0$, only even values of $t$ between 0 and $n-1$ are allowed. Using the best-fit $\theta_L \approx 2.96$, $k_1=k_2=0$, and the best-fit values for the oscillation parameters, we get 
\begin{align}
    2.0 \, \mathrm{meV} &\lesssim m_{\beta\beta} \lesssim 4.0 \, \mathrm{meV}~({\textrm {NO}}) \, , \nonumber \\ 18 \, \mathrm{meV} &\lesssim m_{\beta\beta} \lesssim 49 \, \mathrm{meV}~({\textrm {IO}}) \, .
\end{align}
This is shown in Figure~\ref{fig:0vbb2} by the red points, which correspond to different even values of $t$ ranging from 0 to $n-1$. The left (right) panel is for NO (IO). As in Case 1, for strong IO, some of the admitted values of $m_{\beta\beta}$ are already excluded by the current KamLAND-Zen bound~\cite{KamLAND-Zen:2022tow} (red shaded region) and the remaining values can be tested with the future nEXO~\cite{nEXO:2021ujk} (green shaded region). 

The corresponding values of $\eta_B$ have been generated for $M_{N}=1 $ TeV and  $\kappa/ \kappa_\text{max} = 7 \times 10^3 \,(1.2 \times 10^3 )$ in the NO (IO) case. As can be seen in Figure~\ref{fig:0vbb2}, $s= 8(4)$ can successfully reproduce the observed BAU for strong NO (IO).

\subsection{Case 3}
\begin{figure*}[t!]
\centering
\includegraphics[width=0.48\textwidth]{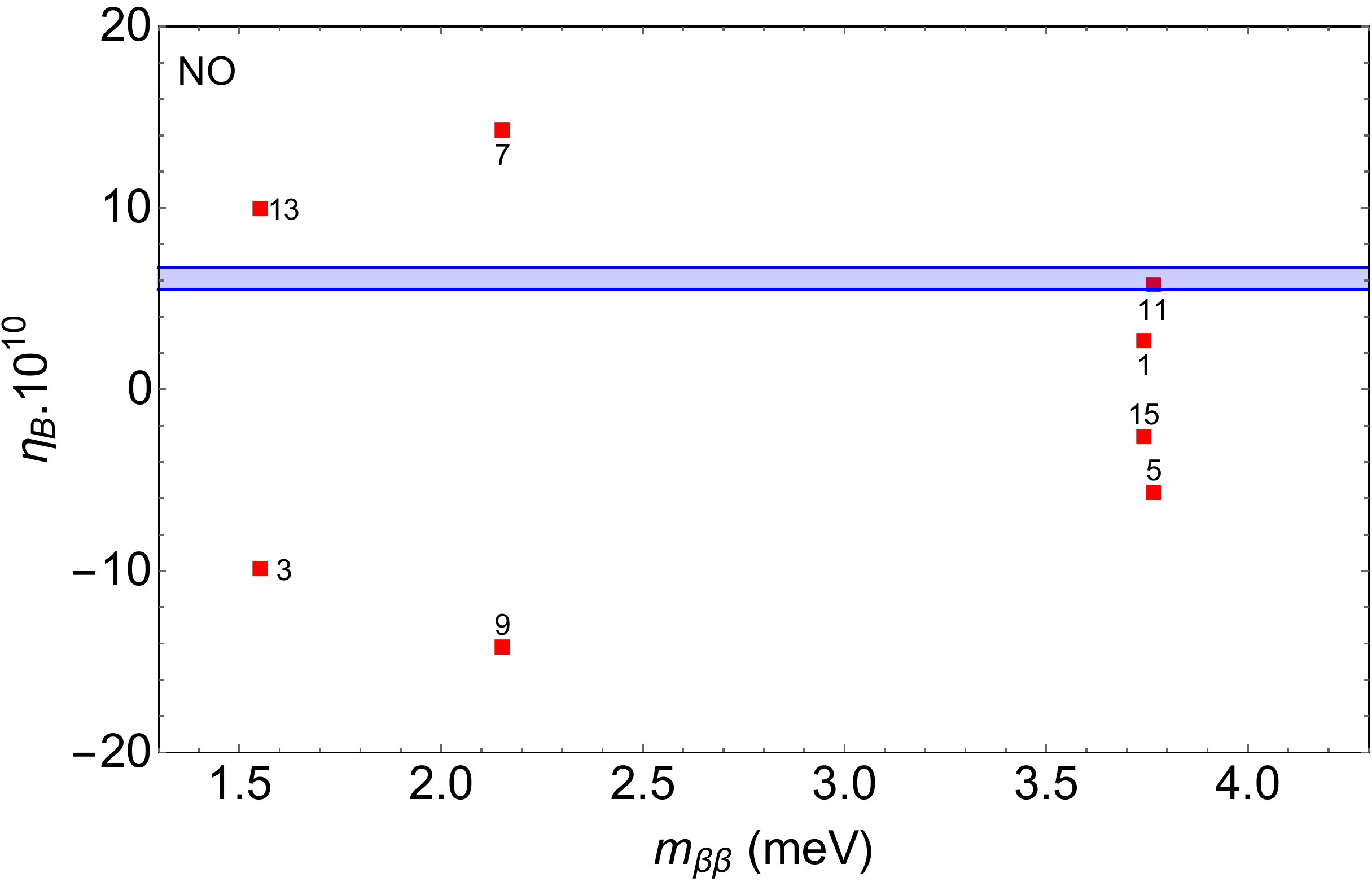}
\hspace{0.1in}
\includegraphics[width=0.48\textwidth]{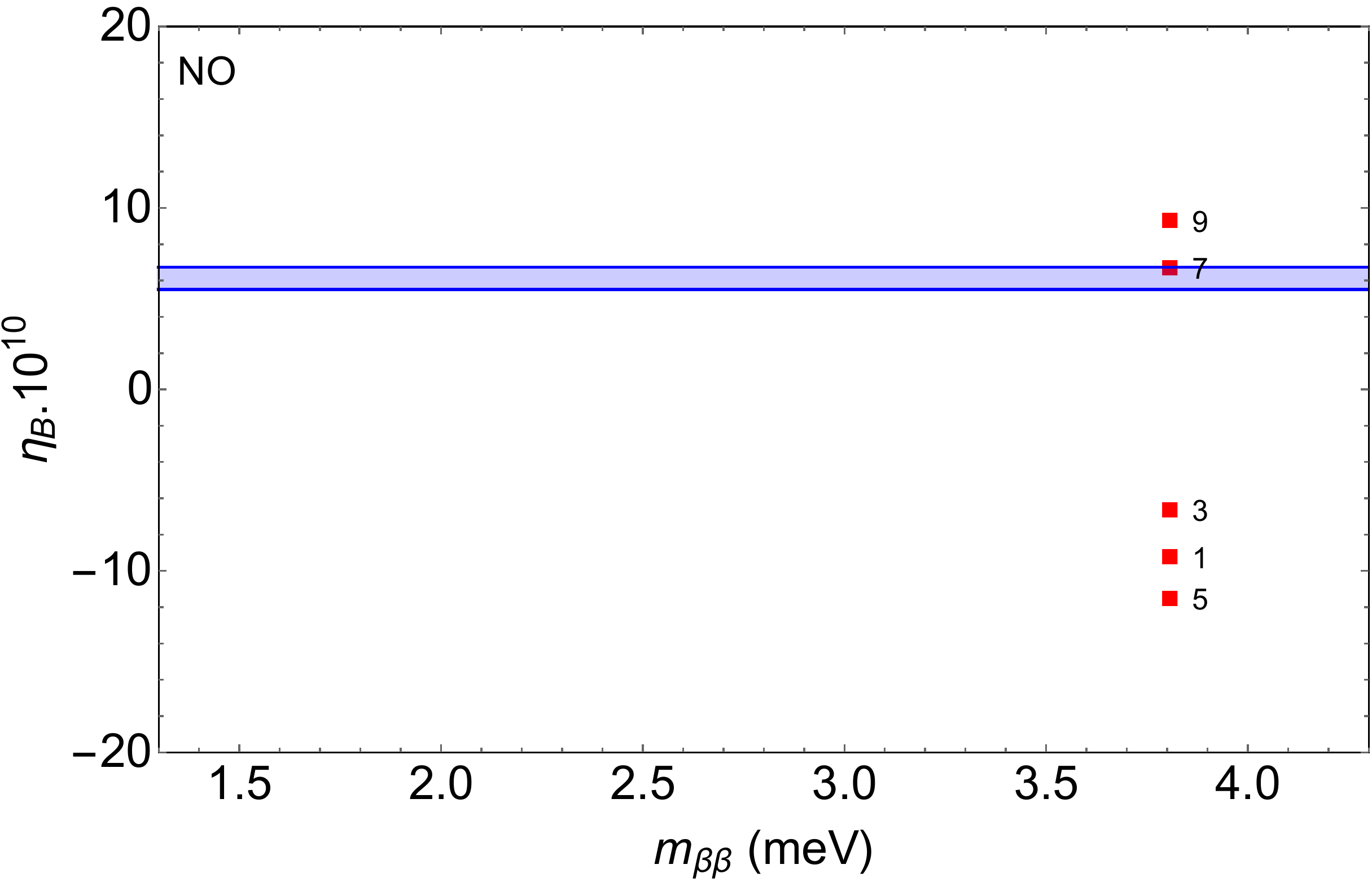}
\caption{Correlation between the predicted BAU $\eta_B$ and the effective neutrino mass $m_{\beta\beta}$ for {\it Case 3a} (left) and {\it Case 3b.1} (right) both with NO and $m$ odd, $s$ odd. We have chosen $n=16 \, (10)$, $k_1=k_2=0$, $m=1\, (5)$, $M_{N}=1 $ TeV and  $\kappa/ \kappa_\text{max} = 1 \times 10^3 \,(1.3 \times 10^5 )$ in the left (right) panel. In Case 3b.1, the blue-shaded horizontal bar corresponds to $\eta_B$ within $10\%$ of $\eta_B^{\rm obs}$. 
}
\label{fig:0vbb3}
\end{figure*}
Following the leptogenesis analysis in Figure~\ref{fig:etaBZp3}, we only consider the NO case here. The effective neutrino mass in terms of the high-energy CP phases is given by 
\begin{widetext}
\begin{subequations}
\begin{alignat}{2}
    (m_{\beta\beta}^{\rm NO})_{\rm 3a} & \, \approx \,  \frac{1}{3}\left| \left(\sqrt{2}\sin{\theta}\cos{\phi_m}+e^{-3i\phi_s}\cos{\theta}\right)^2\sqrt{\Delta m_{\mathrm{sol}}^2}+2(-1)^{k_1+k_2}\sin^2{\phi_m}\sqrt{\Delta m_{\mathrm{atm}}^2} \right|, \\
    (m_{\beta\beta}^{\rm NO})_{\rm 3b.1} & \, \approx \, \frac{1}{3}\left|\sin^2\theta\sqrt{\Delta m_{\mathrm{sol}}^2}+(-1)^{k_1} \cos^2\theta\,\sqrt{\Delta m_{\mathrm{atm}}^2} \right|. 
     \label{eq:mbb3b1}
\end{alignat}
\end{subequations}
\end{widetext}
In case 3a for a given $n$, only certain values of $m$ close to 0 or $n$ reproduce the observed neutrino mixing angles with allowed ordering restricted to NO. Thus, we set $n=16$, $m=1$ with $\theta_L$ computed separately for each value of $s/n$. In case 3b.1 for a given $n$, values of $m\sim n/2 $  reproduce the observed neutrino mixing angles with allowed ordering restricted to NO. Thus, we set $n=10$, $m=5$ with $\theta_L=1.31$. For Case 3a with NO and $k_1=k_2=0$, we get 
\begin{align}
    1.5 \, \mathrm{meV} \lesssim m_{\beta\beta} \lesssim 3.8 \, \mathrm{meV} \, ,
\end{align}
whereas for Case 3b.1 with NO, we get a constant value of $m_{\beta\beta}=3.8$ meV, independent of $s$ and $k_1=0$. This can be understood from Eq.~\eqref{eq:mbb3b1} which indeed is independent of the phase $\phi_s=s/n$ (and thus of the chosen CP transformation). This is also shown in Figure~\ref{fig:0vbb3}, where the red points correspond to different odd $s$ values between 0 and $n-1$ in both cases.  

The corresponding values of $\eta_B$ have been generated for $M_{N}=1 $ TeV and  $\kappa/ \kappa_\text{max} = 1 \times 10^3 \,(1.3 \times 10^5 )$ in Case 3a (3b1). As can be seen in Figure~\ref{fig:0vbb3}, $s= 11(7)$ can successfully reproduce the observed BAU for strong NO in Case 3a (3b1).


\section{Conclusion}
\label{sec:concl}

We have presented a low-scale type-I seesaw scenario with a flavor $G_f$ and CP symmetry that are broken into non-trivial residual symmetries in the charged-lepton and neutrino sectors. We show that this class of models with nearly degenerate heavy right-handed neutrinos can naturally explain the observed baryon asymmetry of the Universe through resonant leptogenesis mechanism. An important distinguishing feature of these models is the possibility of having a relatively long lifetime for one of the RHNs, which is attributed to enhanced residual symmetries. This allows the long-lived RHN to be accessible at the dedicated long-lived particle search facilities, while at the same time, the other two RHNs in the model can be probed via either prompt or displaced vertex signals at the LHC. 

We have studied the correlation of the BAU predictions with the collider signals in a simple $U(1)_{B-L}$ extension of the type-I seesaw. We find that while the LHC might not be able to probe the region allowed by successful leptogenesis, a future 100 TeV collider can easily access these regions, thereby providing a direct test of the leptogenesis mechanism in this framework. The collider prospects turn out to be better in case of normal mass ordering of light neutrinos. 

We have also studied the correlation between the high- and low-energy CP phases in the model. This is exemplified by considering the classic low-energy lepton-number-violating process of neutrinoless double beta decay which crucially depends on the low-energy phases, and connecting it to BAU which depends on the high-energy CP phases. We find that the region allowing successful leptogenesis can be completely tested in future tonne-scale $0\nu\beta\beta$ experiments provided the light neutrino mass ordering is inverted. Therefore, the collider and $0\nu\beta\beta$ experiments will provide complementary probes of these flavor and CP symmetries and their roles in the generation of baryon asymmetry. 

\section*{Acknowledgments}
 We thank Marco Drewes for alerting us to their related work~\cite{Drewes:2022kap} which appeared a few months after ours on arXiv. BD thanks Claudia Hagedorn and Emiliano Molinaro for many useful discussions and for collaboration during the early stages of this work, part of which was presented in section 7.6 of Ref.~\cite{Curtin:2018mvb}. Part of the work presented here also appeared as Chapter 6 in Ref.~\cite{Chauhan:2021get}. BD also acknowledges the local hospitality at MIAPP, UMass Amherst and CP3 Odense, where parts of this work were done during 2016-2018.  The work of BD is supported in part by the US Department of Energy under Grant No.~DE-SC0017987 and by a Fermilab Intensity Frontier Fellowship. This work was partly performed at the Aspen Center for Physics, which is supported by National Science Foundation grant PHY-1607611.

\appendix

\mathversion{bold}
\section{Group Theory of \texorpdfstring{\( \Delta(6\, n^2)\)}{dd} and Representation Matrices}
\mathversion{normal}
\label{appA}

As discussed in Ref.~\cite{Luhn:2007uq}, the discrete groups $\Delta (3 \, n^2)$, $n \geq 2$ integer, can be
described in terms of three generators $a$, $c$ and $d$ fulfilling the relations
\begin{align}
a^3 \, = \, e \; , \;\; c^n \, = \, e \; , \;\; d^n \, = \, e \; , \;\; \nonumber \\ c \, d \, = \, d \, c\; , \;\; a \, c \, a^{-1} \, = \, c^{-1} d^{-1} \; , \;\; a \, d \, a^{-1} \, = \, c \, ,
\end{align}
with $e$ being the identity element of the group. The discrete groups $\Delta (6 \, n^2)$, $n \geq 2$ integer~\cite{Escobar:2008vc}, are obtained
by adding a fourth generator $b$ to the set of $a$, $c$ and $d$. The relations involving $b$ are
\begin{equation}
b^2 \, = \, e \; , \;\; (a \, b)^2\, = \, e \; , \;\; b \, c \, b^{-1} \, = \, d^{-1} \; , \;\; b \, d \, b^{-1} \, = \, c^{-1} \, .
\end{equation}  
In the trivial representation ${\bf 1}$ all elements of the group are represented by the character $1$.
The explicit representation matrices $g ({\bf 3})$ for $a$, $b$, $c$ and $d$ can be chosen in the irreducible, faithful, complex three-dimensional representation ${\bf 3}$ as
\begin{widetext}
\begin{align}
&\label{abcd3}
a ({\bf 3}) \, = \, \left( \begin{array}{ccc}
1 & 0 & 0\\
0 & \omega & 0\\
0 & 0 & \omega^2
\end{array}
\right)
\; , \quad 
b ({\bf 3}) \, = \,  \left( \begin{array}{ccc}
1 & 0 & 0\\
0 & 0 & \omega^2\\
0 & \omega & 0
\end{array}
\right)
\; , \nonumber \\
&
c({\bf 3}) \, = \, \frac 13 \, \left( \begin{array}{ccc}
 1 + 2\cos\phi_n & 1 -\cos\phi_n - \sqrt{3} \sin \phi_n & 1-\cos\phi_n + \sqrt{3} \sin \phi_n \\
  1-\cos\phi_n + \sqrt{3} \sin \phi_n &  1 + 2\cos\phi_n & 1 -\cos\phi_n - \sqrt{3} \sin \phi_n\\
  1 -\cos\phi_n - \sqrt{3} \sin \phi_n &  1-\cos\phi_n + \sqrt{3} \sin \phi_n & 1 + 2\cos\phi_n
\end{array}
\right) \, ,
\end{align}
\end{widetext}
with $\omega=e^{2 \pi i/3}$ and $\phi_n = 2 \pi/n$. The representation for $d$ can be computed via $d ({\bf 3})=a({\bf 3})^2 c ({\bf 3}) a ({\bf 3})$. 

The existence of an irreducible, in general unfaithful, real three-dimensional representation ${\bf 3^\prime}$ requires that all its characters are real.
This cannot be fulfilled in all groups $\Delta (6 \, n^2)$, but only, if the index $n$ is even. In this case the form of the representation matrices $g ({\bf 3^\prime})$ is
\begin{align}
\label{abcd3prime}
a ({\bf 3^\prime}) \, = \, a ({\bf 3}) \;\; , \;\; b ({\bf 3^\prime})\, = \, b ({\bf 3}) \;\; , \nonumber \\
c ({\bf 3^\prime}) \, = \, \frac 13 \, \left(
\begin{array}{ccc}
 -1 & 2 & 2\\
 2 & -1 & 2\\
 2 & 2 &-1
\end{array}
\right) \, 
\end{align}
and $d ({\bf 3^\prime}) = a({\bf 3^\prime})^2 c ({\bf 3^\prime}) a ({\bf 3^\prime})$. Note that the representation matrices $g({\bf 3^\prime})$ do not depend on the index $n$ of the group and 
thus lead to the same representation for all groups $\Delta (6 \, n^2)$ with even $n$. Indeed, we can observe that the group generated by the representation matrices $g({\bf 3^\prime})$ has 24 elements
 and thus corresponds to the group $\Delta (6 \cdot 2^2)= \Delta (24)$. This group is isomorphic to the permutation group $S_4$. This representation together with the one generated by 
 the representation matrices $a ({\bf 3^\prime})$, $c ({\bf 3^\prime})$, $d ({\bf 3^\prime})$ and $-b ({\bf 3^\prime})$ (i.e. the representation matrix $b ({\bf 3^\prime})$ acquires an overall sign~\cite{Escobar:2008vc}) are the only real three-dimensional representations in a generic group $\Delta (6 n^2)$ with even $n$ and $3 \nmid n$.
  To see this we inspect the characters of the three-dimensional representations. Following Ref.~\cite{Escobar:2008vc} we see that the characters 
$\chi ({\bf 3_{\rm gen}})$ of a generic irreducible three-dimensional representation ${\bf 3_{\rm gen}}$ for a certain type of classes is given by $\eta ^{- \rho \, l}$ with $\eta=e^{2 \pi i/n}$,
$\rho=0, .., n-1$ (labeling this type of class of the group $\Delta (6 n^2)$) and $l=1, .. , n-1$ ($l$ labels the different pairs of three-dimensional representations).
  We have to require that all $\eta ^{- \rho \, l}$ for a certain representation labeled by $l$ are real. This is ensured, if $\eta^{-l}$ is real for all powers $\rho$ with $\rho=0,..,n-1$,
 meaning $\eta^{-l}$ should be real itself. Hence, $2 \, l/n$ must be an integer. With the constraint on $l$, $1 \leq l \leq n-1$, we know that there is a 
single solution to $2 \, l/n$ being an integer, namely $l=n/2$, i.~e.~there is a single pair of irreducible three-dimensional representations that are real. 
 In this case their characters are real for all classes, as can be explicitly checked with the help of the character table, shown in Ref.~\cite{Escobar:2008vc}.

\mathversion{bold}
\section{CP symmetries and form of CP transformations}
\mathversion{normal}
\label{appC}

The CP symmetries correspond to automorphisms of the flavor group $\Delta (6 \, n^2)$, see discussion in Ref.~\cite{Hagedorn:2014wha}.
In the present analysis we employ the ones, as used in Ref.~\cite{Hagedorn:2016lva}. These can be obtained as follows: consider the automorphism
\begin{equation}
\label{XP23auto}
a \;\; \rightarrow \;\; a \;\; , \;\;  c \;\; \rightarrow \;\; c^{-1} \;\; , \;\;  d \;\; \rightarrow \;\; d^{-1} \;\; \mbox{and} \;\; b \;\; \rightarrow \;\; b  \, .
\end{equation}
The automorphism in Eq.~(\ref{XP23auto}) can be represented by $X_0 ({\bf 1})=1$ in the trivial representation ${\bf 1}$
and by the matrix
\begin{equation}
X_0 ({\bf 3}) \, = \, X_0 ({\bf 3^\prime}) \, = \, \left(
\begin{array}{ccc}
1 & 0 & 0\\
0 & 0 & 1\\
0 & 1 & 0
\end{array}
\right) 
\end{equation}
in both three-dimensional representations ${\bf 3}$ and ${\bf 3^\prime}$.
In Case 1, the CP transformation $X (s) ({\bf 3})$ has the explicit form
$a({\bf 3})b({\bf 3})c({\bf 3})d({\bf 3})^{2s}\:X_0$.
The form of the CP transformation $X (s) ({\bf 3^\prime})$ in the representation ${\bf 3^\prime}$ depends on whether $s$ is even or odd, i.~e.
\begin{subequations}
\begin{alignat}{2}
X(s~{\rm even}) ({\bf 3^\prime}) \, = \, & \left( \begin{array}{ccc}
 1 & 0 & 0\\
  0 & 1 & 0\\
 0 & 0 & 1
\end{array}
\right) , \label{eq:case1Xin3prime_seven} \\
X(s~{\rm odd}) ({\bf 3^\prime}) \, = \, & \frac 13 \left( \begin{array}{ccc}
 -1 & 2 & 2\\
 2 & -1 & 2\\
 2 & 2 &-1
\end{array}
\right) \, .
\label{eq:case1Xin3prime_sodd}
\end{alignat}
\end{subequations}
In Case 2, the form of the CP transformation $X ({\bf 3}) (s,t)$ in the representation ${\bf 3}$ can be chosen as $c({\bf 3})^sd({\bf 3})^{t}\:X_0$, 
and is more conveniently written in terms of the variables $u=2s-t$ and $v=3t$ with $\phi_u=\pi \, u/n$ and $\phi_v=\pi \, v/n$.
The form of the CP transformation $X (s,t) ({\bf 3^\prime})$ depends on whether $s$ and $t$ are even or odd.
The explicit form of $X (s,t) ({\bf 3^\prime})$, however, does neither contain $s$ nor $t$ as parameters. 
\begin{subequations}
\begin{alignat}{2}
X (s \, \mbox{even},t \, \mbox{even}) ({\bf 3^\prime}) \,  = \, & \left( \begin{array}{ccc}
1 & 0 & 0\\
0 & 0 & 1\\
0 & 1 & 0
\end{array}
\right) \; , \\
X (s \, \mbox{even},t \, \mbox{odd}) ({\bf 3^\prime}) \, = \, & \frac 13 \, \left( \begin{array}{ccc}
 -1 & 2 \, \omega^2 & 2 \, \omega\\
 2 \, \omega^2 & 2 \, \omega & -1\\
 2 \, \omega & -1 & 2 \, \omega^2 
\end{array}
\right) \; , \\
X (s \, \mbox{odd},t \, \mbox{even}) ({\bf 3^\prime}) \, = \, & \frac 13 \, \left( \begin{array}{ccc}
 -1 & 2 & 2\\
 2 & 2 & -1\\
 2 & -1 & 2
\end{array}
\right) \; , \\
X (s \, \mbox{odd},t \, \mbox{odd}) ({\bf 3^\prime})  ({\bf 3^\prime}) \, = \, & \frac 13 \, \left( \begin{array}{ccc}
 -1 & 2 \, \omega & 2 \, \omega^2\\
 2\, \omega & 2\, \omega^2 & -1\\
 2 \, \omega^2  & -1 & 2 \, \omega 
\end{array}
\right) \; .
\end{alignat}
 \end{subequations}
For Case 3a and Case 3b.1, the form of the CP transformation $X (s, m) ({\bf 3})$ is given as~\cite{Hagedorn:2014wha}
\begin{widetext}
\begin{equation}
\label{eq:Xs3Case3}
\small X (s) ({\bf 3}) \, = \, \frac 13 \, e^{- i \, \delta_s} \, \left(
\begin{array}{ccc}
3 \, \cos 3 \, \delta_s + i \, \sin 3 \, \delta_s & -2 \, i \, \omega \,  \sin 3 \, \delta_s & -2 \, i \, \omega^2 \, \sin 3 \, \delta_s\\
 -2 \, i \, \omega  \, \sin 3 \, \delta_s &    \omega^2 \, \left( 3 \, \cos 3 \, \delta_s + i \, \sin 3 \, \delta_s \right) & -2 \, i \, \sin 3 \, \delta_s\\
-2 \, i \, \omega^2  \, \sin 3 \, \delta_s & -2 \, i  \, \sin 3 \, \delta_s & \omega \, \left( 3 \, \cos 3 \, \delta_s + i \, \sin 3 \, \delta_s \right)
\end{array}
\right) 
\end{equation}
with $\delta_s=\pi s/n$.
\end{widetext}
The form of the CP transformation $X (s) ({\bf 3^\prime})$ only depends on whether $s$ is even or odd. In particular, 
\begin{subequations}
\begin{alignat}{2}
X (s \, \mbox{even}) ({\bf 3^\prime}) \, = \, & \left(
\begin{array}{ccc}
1 & 0 & 0\\
0 & \omega^2 & 0\\
0 & 0 & \omega
\end{array}
\right) \; , \\
X (s \, \mbox{odd}) ({\bf 3^\prime}) \, = \, & \frac 13 \, \left(
\begin{array}{ccc}
 -1 & 2 \, \omega & 2 \, \omega^2\\
 2 \, \omega & - \omega^2 & 2 \\
 2 \, \omega^2 & 2 & -\omega
\end{array}
\right) \; .
\end{alignat}
\end{subequations}


\mathversion{bold}
\section{Form of the Representation Matrices for Residual Symmetries}
\mathversion{normal}
\label{appB}

In the following, we list the form of the representation matrices in the representations ${\bf 3}$ and ${\bf 3^\prime}$ for the different residual symmetries,
used in sections~\ref{sec:frame} and \ref{sec:cases}.

In all these cases, the residual flavor symmetry in the charged-lepton sector is generated by $a$ which corresponds to the representation matrix $a ({\bf 3})$ given in Eq.~\eqref{abcd3}.  
The residual flavor symmetry in the neutrino sector is generated by $Z$. In Case 1 and Case 2, $Z$ is chosen as $c^{n/2}$ which is in the representation ${\bf 3}$ of the form
\begin{equation}
\label{Zcn2in3}
Z ({\bf 3}) \, = \, \frac 13 \, \left(
\begin{array}{ccc}
 -1 & 2 & 2\\
 2 & -1 & 2\\
 2 & 2 &-1
\end{array}
\right) \, 
\end{equation}
independent of the index $n$, while the form of $Z=c^{n/2}$ in ${\bf 3^\prime}$ reads either
\begin{equation}
\label{Zcn2_n2even}
Z ({\bf 3^\prime}) \, = \, \left( 
\begin{array}{ccc}
1 & 0 & 0\\
0 & 1 & 0\\
0 & 0 & 1
\end{array}
\right)
 \;\;\; \mbox{for} \;\;\; n/2 \;\; \mbox{even} \, ,
\end{equation}
or
\begin{equation}
\label{Zcn2_n2odd}
Z ({\bf 3^\prime}) \, = \, \frac 13 \, \left(
\begin{array}{ccc}
 -1 & 2 & 2\\
 2 & -1 & 2\\
 2 & 2 &-1
\end{array}
\right)  = Z ({\bf 3}) 
 \;\;\; \mbox{for} \;\;\; n/2 \;\; \mbox{odd.}
\end{equation}

In Case 3a and Case 3b.1, $Z$ is chosen as $b \, c^m d^m$ with $m=0, ...., n-1$. 
 In the representation ${\bf 3}$ it is of the form~\cite{Hagedorn:2014wha}
\begin{widetext}
\begin{equation}
Z (m) ({\bf 3}) \, = \, \frac 13 \, \left(
\begin{array}{ccc}
1+2 \, \cos \gamma_m & \omega^2 \, \left(1-\cos\gamma_m + \sqrt{3}\, \sin \gamma_m \right) & \omega \, \left( 1- \cos \gamma_m -\sqrt{3} \, \sin \gamma_m \right)\\
\omega \, \left( 1-\cos \gamma_m +\sqrt{3}\, \sin \gamma_m \right) & 1- \cos \gamma_m -\sqrt{3} \, \sin \gamma_m & \omega^2 \, \left( 1+ 2\, \cos \gamma_m \right)\\
\omega^2 \, \left(  1- \cos \gamma_m -\sqrt{3} \, \sin \gamma_m  \right) & \omega \, \left( 1+ 2 \, \cos \gamma_m \right) & 1- \cos \gamma_m +\sqrt{3} \, \sin \gamma_m
\end{array}
\right)
\end{equation}
with $\gamma_m= 2 \, \pi m/n$.
\end{widetext}
For the special values, $m = 0$, $m = n$ and $m=n/2$, the form of $Z (m) ({\bf 3})$ simplifies and we find
 \begin{subequations}
\begin{alignat}{2}
Z (m=0) ({\bf 3}) \, = \, & Z (m=n) ({\bf 3}) \, = \,  \left( \begin{array}{ccc}
 1 & 0 & 0\\
 0 & 0 & \omega^2\\
 0 & \omega & 0
\end{array}
\right) \; , \\
Z (m=n/2) ({\bf 3}) \, = \, & \frac 13 \, \left( \begin{array}{ccc}
 -1 & 2 \, \omega^2 & 2 \, \omega \\
2 \, \omega & 2 & -\omega^2\\
2 \, \omega^2 & -\omega & 2 
\end{array}
\right) \; .
\end{alignat}
\end{subequations}
Similarly, we can analyze the form of the representation matrix $Z (m) ({\bf 3^\prime})$. The decisive criterion for this form is whether $m$ is even or odd; 
 otherwise there is no further dependence on the parameter $m$ for $Z (m) ({\bf 3^\prime})$.
\begin{align}
\label{Z3pmevenodd}
Z (m \, \mbox{even}) ({\bf 3^\prime}) \, = \, &
\left( \begin{array}{ccc}
 1 & 0 & 0\\
 0 & 0 & \omega^2\\
 0 & \omega & 0
\end{array}
\right)
 \; , \nonumber \\
 Z (m \, \mbox{odd}) ({\bf 3^\prime}) \, = \, & \frac 13 \, \left(
\begin{array}{ccc}
-1 & 2 \, \omega^2 & 2 \, \omega\\
2 \, \omega & 2 & - \omega^2\\
2 \, \omega^2 & -\omega & 2 
\end{array}
\right)
\; .
\end{align}
We note that $Z (m \, \mbox{even}) ({\bf 3^\prime})$ coincides with $Z (m=0) ({\bf 3}) = Z (m=n) ({\bf 3})$, and $Z (m \, \mbox{odd}) ({\bf 3^\prime})$ coincides with $Z (m=n/2) ({\bf 3})$.
\begin{widetext}
\section{Explicit Form of the Yukawa Couplings}
\label{appD}
\subsection{Case 1} \label{appD1}
The Yukawa parameters $y_f$ in this case (irrespective of $s$) are
\begin{subequations}\label{eq:ycase1}
\begin{alignat}{2}
	y_1^2 \, = \, & \frac{M_N}{2\,v^2}\left(m_1\,-\,m_3\,+\,\sqrt{m_1^2\,+\,m_3^2\,+\,2\,m_1\,m_3\,\cos\left(4\,\theta_R\right)}\right)\sec\left(2\,\theta_R\right)\,, \label{eq:y1case1}\\
	y_2^2 \, = \, & \frac{M_N\,m_2}{v^2}\,, \label{eq:y2case1}\\
	y_3^2 \, = \, & \frac{M_N}{2\,v^2}\left(-\,m_1\,+\,m_3\,+\,\sqrt{m_1^2\,+\,m_3^2\,+\,2\,m_1\,m_3\,\cos\left(4\,\theta_R\right)}\right)\sec\left(2\,\theta_R\right)\, . \label{eq:y3case1}
\end{alignat}
\end{subequations}

\subsection{Case 2} \label{appD2}
\subsubsection*{\texorpdfstring{\(t\)}{t} even:}
\begin{subequations}\label{eq:ycase2teven}
\begin{alignat}{2}
	y_1^2 \, = \, & \frac{M_N\,m_1}{v^2}\,, \label{eq:y1case2teven}\\
	y_2^2 \, = \, & \frac{M_N\,m_2}{v^2}\,, \label{eq:y2case2teven}\\
	y_3^2 \, = \, & \frac{M_N\,m_3}{v^2}\, . \label{eq:y3case2teven}
\end{alignat}	
\end{subequations}

\subsubsection*{\texorpdfstring{\(t\)}{t} odd:}
\begin{subequations}\label{eq:ycase2todd}
\begin{alignat}{2}
	y_1^2 \, = \, & \frac{M_N}{2\,v^2}\left(-\,m_1\,+\,m_3\,+\,\sqrt{m_1^2\,+\,m_3^2\,-\,2\,m_1\,m_3\,\cos\left(4\,\theta_R\right)}\right)\csc\left(2\,\theta_R\right)\,, \label{eq:y1case2todd}\\
	y_2^2 \, = \, & \frac{M_N\,m_2}{v^2}\,, \label{eq:y2case2todd}\\
	y_3^2 \, = \, & \frac{M_N}{2\,v^2}\left(m_1\,-\,m_3\,+\,\sqrt{m_1^2\,+\,m_3^2\,-\,2\,m_1\,m_3\,\cos\left(4\,\theta_R\right)}\right)\csc\left(2\,\theta_R\right)\, . \label{eq:y3case2todd}
\end{alignat}	
\end{subequations}

\subsection{Case 3a}\label{appD3}
\subsubsection*{\texorpdfstring{\(m\)}{m} and \texorpdfstring{\(s\)}{s}, both even or odd:}
\begin{subequations}\label{eq:ycase3aeven}
\begin{alignat}{2}
	y_1^2 \, = \, & \frac{M_N\,m_1}{v^2}\,, \label{eq:y1case3aeven}\\
	y_2^2 \, = \, & \frac{M_N\,m_2}{v^2}\,, \label{eq:y2case3aeven}\\
	y_3^2 \, = \, & \frac{M_N\,m_3}{v^2}\, . \label{eq:y3case3aeven}
\end{alignat}	
\end{subequations}

\subsubsection*{\texorpdfstring{\(m\)}{m} even, \texorpdfstring{\(s\)}{s} odd:}
\begin{subequations}\label{eq:ycase3aevenodd}
\begin{alignat}{2}
	y_1^2 \, = \, & \frac{M_N}{2\,v^2}\left(m_1\,-\,m_2\,+\,\sqrt{m_1^2\,+\,m_2^2\,+\,2\,m_1\,m_2\,\cos\left(4\,\theta_R\right)}\right)\sec\left(2\,\theta_R\right)\,, \label{eq:y1case3aevenodd}\\
	y_2^2 \, = \, & \frac{M_N}{2\,v^2}\left(-\,m_1\,+\,m_2\,+\,\sqrt{m_1^2\,+\,m_2^2\,+\,2\,m_1\,m_2\,\cos\left(4\,\theta_R\right)}\right)\sec\left(2\,\theta_R\right) \,, \label{eq:y2case3aevenodd}\\
	y_3^2 \, = \, & \frac{M_N\,m_3}{v^2}\,. \label{eq:y3case3aevenodd}
\end{alignat}	
\end{subequations}

\subsubsection*{\texorpdfstring{\(m\)}{m} odd, \texorpdfstring{\(s\)}{s} even:}
\begin{subequations}\label{eq:ycase3aoddeven}
\begin{alignat}{2}
	y_1^2 \, = \, & \frac{M_N\,m_1}{v^2}\,, \label{eq:y1case3aoddeven}\\ 
	y_2^2 \, = \, & \frac{M_N}{2\,v^2}\left(m_2\,-\,m_3\,+\,\sqrt{m_2^2\,+\,m_3^2\,+\,2\,m_2\,m_3\,\cos\left(4\,\theta_R\right)}\right)\sec\left(2\,\theta_R\right)\,, \label{eq:y2case3aoddeven}\\
	y_3^2 \, =  \, & \frac{M_N}{2\,v^2}\left(-m_2\,+\,m_3\,+\,\sqrt{m_2^2\,+\,m_3^2\,+\,2\,m_2\,m_3\,\cos\left(4\,\theta_R\right)}\right)\sec\left(2\,\theta_R\right) \,. \label{eq:y3case3aoddeven}
\end{alignat}	
\end{subequations}

\subsection{Case 3b.1} \label{appD4}
\subsubsection*{\texorpdfstring{\(m\)}{m} and \texorpdfstring{\(s\)}{s}, both even or odd:}
\begin{subequations}\label{eq:ycase3beven}
\begin{alignat}{2}
	y_1^2 \, = \, & \frac{M_N\,m_2}{v^2}\,, \label{eq:y1case3beven}\\
	y_2^2 \, = \, & \frac{M_N\,m_3}{v^2}\,, \label{eq:y2case3beven}\\
	y_3^2 \, = \, & \frac{M_N\,m_1}{v^2}\, . \label{eq:y3case3beven}
\end{alignat}	
\end{subequations}

\subsubsection*{\texorpdfstring{\(m\)}{m} even, \texorpdfstring{\(s\)}{s} odd:}
\begin{subequations}\label{eq:ycase3bevenodd}
\begin{alignat}{2}
    y_1^2 \, = \, & \frac{M_N}{2\,v^2}\left(-\,m_1\,+\,m_2\,+\,\sqrt{m_1^2\,+\,m_2^2\,+\,2\,m_1\,m_2\,\cos\left(4\,\theta_R\right)}\right)\sec\left(2\,\theta_R\right) \,, \label{eq:y1case3bevenodd}\\
    y_2^2 \, = \, & \frac{M_N\,m_3}{v^2}\,, \label{eq:y2case3bevenodd}\\
	y_3^2 \, = \, & \frac{M_N}{2\,v^2}\left(m_1\,-\,m_2\,+\,\sqrt{m_1^2\,+\,m_2^2\,+\,2\,m_1\,m_2\,\cos\left(4\,\theta_R\right)}\right)\sec\left(2\,\theta_R\right)\,, \label{eq:y3case3bevenodd}
\end{alignat}	
\end{subequations}

\subsubsection*{\texorpdfstring{\(m\)}{m} odd, \texorpdfstring{\(s\)}{s} even:}
\begin{subequations}\label{eq:ycase3boddeven}
\begin{alignat}{2}
    y_1^2 \, = \, & \frac{M_N}{2\,v^2}\left(m_2\,-\,m_3\,+\,\sqrt{m_2^2\,+\,m_3^2\,+\,2\,m_2\,m_3\,\cos\left(4\,\theta_R\right)}\right)\sec\left(2\,\theta_R\right)\,, \label{eq:y1case3boddeven}\\
    y_2^2 \, =  \, & \frac{M_N}{2\,v^2}\left(-m_2\,+\,m_3\,+\,\sqrt{m_2^2\,+\,m_3^2\,+\,2\,m_2\,m_3\,\cos\left(4\,\theta_R\right)}\right)\sec\left(2\,\theta_R\right) \,, \label{eq:y2case3boddeven}\\
	y_3^2 \, = \, & \frac{M_N\,m_1}{v^2}\,. \label{eq:y3case3boddeven}
\end{alignat}	
\end{subequations}
\end{widetext}

\bibliographystyle{JHEP}
\bibliography{ResL}

\providecommand{\href}[2]{#2}\begingroup\raggedright\begin{thebibliography}{100}

\bibitem{Aghanim:2018eyx}
{\scshape Planck} collaboration, \emph{{Planck 2018 results. VI. Cosmological
  parameters}},
  \href{https://doi.org/10.1051/0004-6361/201833910}{\emph{Astron. Astrophys.}
  {\bfseries 641} (2020) A6}
  [\href{https://arxiv.org/abs/1807.06209}{{\ttfamily 1807.06209}}].

\bibitem{Sakharov:1967dj}
A.~D. Sakharov, \emph{{Violation of CP Invariance, C asymmetry, and baryon
  asymmetry of the universe}},
  \href{https://doi.org/10.1070/PU1991v034n05ABEH002497}{\emph{Pisma Zh. Eksp.
  Teor. Fiz.} {\bfseries 5} (1967) 32}.

\bibitem{Bodeker:2020ghk}
D.~Bodeker and W.~Buchmuller, \emph{{Baryogenesis from the weak scale to the
  grand unification scale}},
  \href{https://doi.org/10.1103/RevModPhys.93.035004}{\emph{Rev. Mod. Phys.}
  {\bfseries 93} (2021) 035004}
  [\href{https://arxiv.org/abs/2009.07294}{{\ttfamily 2009.07294}}].

\bibitem{Fukugita:1986hr}
M.~Fukugita and T.~Yanagida, \emph{{Baryogenesis Without Grand Unification}},
  \href{https://doi.org/10.1016/0370-2693(86)91126-3}{\emph{Phys. Lett. B}
  {\bfseries 174} (1986) 45}.

\bibitem{Buchmuller:2005eh}
W.~Buchmuller, R.~D. Peccei and T.~Yanagida, \emph{{Leptogenesis as the origin
  of matter}},
  \href{https://doi.org/10.1146/annurev.nucl.55.090704.151558}{\emph{Ann. Rev.
  Nucl. Part. Sci.} {\bfseries 55} (2005) 311}
  [\href{https://arxiv.org/abs/hep-ph/0502169}{{\ttfamily hep-ph/0502169}}].

\bibitem{Davidson:2008bu}
S.~Davidson, E.~Nardi and Y.~Nir, \emph{{Leptogenesis}},
  \href{https://doi.org/10.1016/j.physrep.2008.06.002}{\emph{Phys. Rept.}
  {\bfseries 466} (2008) 105}
  [\href{https://arxiv.org/abs/0802.2962}{{\ttfamily 0802.2962}}].

\bibitem{Fong:2012buy}
C.~S. Fong, E.~Nardi and A.~Riotto, \emph{{Leptogenesis in the Universe}},
  \href{https://doi.org/10.1155/2012/158303}{\emph{Adv. High Energy Phys.}
  {\bfseries 2012} (2012) 158303}
  [\href{https://arxiv.org/abs/1301.3062}{{\ttfamily 1301.3062}}].

\bibitem{Kuzmin:1985mm}
V.~A. Kuzmin, V.~A. Rubakov and M.~E. Shaposhnikov, \emph{{On the Anomalous
  Electroweak Baryon Number Nonconservation in the Early Universe}},
  \href{https://doi.org/10.1016/0370-2693(85)91028-7}{\emph{Phys. Lett. B}
  {\bfseries 155} (1985) 36}.

\bibitem{Minkowski:1977sc}
P.~Minkowski, \emph{{$\mu \to e\gamma$ at a Rate of One Out of $10^{9}$ Muon
  Decays?}}, \href{https://doi.org/10.1016/0370-2693(77)90435-X}{\emph{Phys.
  Lett. B} {\bfseries 67} (1977) 421}.

\bibitem{Mohapatra:1979ia}
R.~N. Mohapatra and G.~Senjanovic, \emph{{Neutrino Mass and Spontaneous Parity
  Nonconservation}},
  \href{https://doi.org/10.1103/PhysRevLett.44.912}{\emph{Phys. Rev. Lett.}
  {\bfseries 44} (1980) 912}.

\bibitem{Yanagida:1979as}
T.~Yanagida, \emph{{Horizontal gauge symmetry and masses of neutrinos}},
  {\emph{Conf. Proc. C} {\bfseries 7902131} (1979) 95}.

\bibitem{GellMann:1980vs}
M.~Gell-Mann, P.~Ramond and R.~Slansky, \emph{{Complex Spinors and Unified
  Theories}}, {\emph{Conf. Proc. C} {\bfseries 790927} (1979) 315}
  [\href{https://arxiv.org/abs/1306.4669}{{\ttfamily 1306.4669}}].

\bibitem{Glashow:1979nm}
S.~Glashow, \emph{{The Future of Elementary Particle Physics}},
  \href{https://doi.org/10.1007/978-1-4684-7197-7\_15}{\emph{NATO Sci. Ser. B}
  {\bfseries 61} (1980) 687}.

\bibitem{Schechter:1980gr}
J.~Schechter and J.~W.~F. Valle, \emph{{Neutrino Masses in SU(2) x U(1)
  Theories}}, \href{https://doi.org/10.1103/PhysRevD.22.2227}{\emph{Phys. Rev.
  D} {\bfseries 22} (1980) 2227}.

\bibitem{Davidson:2002qv}
S.~Davidson and A.~Ibarra, \emph{{A Lower bound on the right-handed neutrino
  mass from leptogenesis}},
  \href{https://doi.org/10.1016/S0370-2693(02)01735-5}{\emph{Phys. Lett. B}
  {\bfseries 535} (2002) 25}
  [\href{https://arxiv.org/abs/hep-ph/0202239}{{\ttfamily hep-ph/0202239}}].

\bibitem{Moffat:2018wke}
K.~Moffat, S.~Pascoli, S.~T. Petcov, H.~Schulz and J.~Turner,
  \emph{{Three-flavored nonresonant leptogenesis at intermediate scales}},
  \href{https://doi.org/10.1103/PhysRevD.98.015036}{\emph{Phys. Rev. D}
  {\bfseries 98} (2018) 015036}
  [\href{https://arxiv.org/abs/1804.05066}{{\ttfamily 1804.05066}}].

\bibitem{Pilaftsis:2003gt}
A.~Pilaftsis and T.~E.~J. Underwood, \emph{{Resonant leptogenesis}},
  \href{https://doi.org/10.1016/j.nuclphysb.2004.05.029}{\emph{Nucl. Phys. B}
  {\bfseries 692} (2004) 303}
  [\href{https://arxiv.org/abs/hep-ph/0309342}{{\ttfamily hep-ph/0309342}}].

\bibitem{Dev:2017wwc}
P.~S.~B. Dev, M.~Garny, J.~Klaric, P.~Millington and D.~Teresi, \emph{{Resonant
  enhancement in leptogenesis}},
  \href{https://doi.org/10.1142/S0217751X18420034}{\emph{Int. J. Mod. Phys. A}
  {\bfseries 33} (2018) 1842003}
  [\href{https://arxiv.org/abs/1711.02863}{{\ttfamily 1711.02863}}].

\bibitem{Flanz:1994yx}
M.~Flanz, E.~A. Paschos and U.~Sarkar, \emph{{Baryogenesis from a lepton
  asymmetric universe}},
  \href{https://doi.org/10.1016/0370-2693(94)01555-Q}{\emph{Phys. Lett. B}
  {\bfseries 345} (1995) 248}
  [\href{https://arxiv.org/abs/hep-ph/9411366}{{\ttfamily hep-ph/9411366}}].

\bibitem{Covi:1996wh}
L.~Covi, E.~Roulet and F.~Vissani, \emph{{CP violating decays in leptogenesis
  scenarios}}, \href{https://doi.org/10.1016/0370-2693(96)00817-9}{\emph{Phys.
  Lett. B} {\bfseries 384} (1996) 169}
  [\href{https://arxiv.org/abs/hep-ph/9605319}{{\ttfamily hep-ph/9605319}}].

\bibitem{Flanz:1996fb}
M.~Flanz, E.~A. Paschos, U.~Sarkar and J.~Weiss, \emph{{Baryogenesis through
  mixing of heavy Majorana neutrinos}},
  \href{https://doi.org/10.1016/S0370-2693(96)01337-8}{\emph{Phys. Lett. B}
  {\bfseries 389} (1996) 693}
  [\href{https://arxiv.org/abs/hep-ph/9607310}{{\ttfamily hep-ph/9607310}}].

\bibitem{Pilaftsis:1997dr}
A.~Pilaftsis, \emph{{Resonant CP violation induced by particle mixing in
  transition amplitudes}},
  \href{https://doi.org/10.1016/S0550-3213(97)00469-0}{\emph{Nucl. Phys. B}
  {\bfseries 504} (1997) 61}
  [\href{https://arxiv.org/abs/hep-ph/9702393}{{\ttfamily hep-ph/9702393}}].

\bibitem{Pilaftsis:1997jf}
A.~Pilaftsis, \emph{{CP violation and baryogenesis due to heavy Majorana
  neutrinos}}, \href{https://doi.org/10.1103/PhysRevD.56.5431}{\emph{Phys. Rev.
  D} {\bfseries 56} (1997) 5431}
  [\href{https://arxiv.org/abs/hep-ph/9707235}{{\ttfamily hep-ph/9707235}}].

\bibitem{Pilaftsis:2005rv}
A.~Pilaftsis and T.~E.~J. Underwood, \emph{{Electroweak-scale resonant
  leptogenesis}}, \href{https://doi.org/10.1103/PhysRevD.72.113001}{\emph{Phys.
  Rev. D} {\bfseries 72} (2005) 113001}
  [\href{https://arxiv.org/abs/hep-ph/0506107}{{\ttfamily hep-ph/0506107}}].

\bibitem{Deppisch:2010fr}
F.~F. Deppisch and A.~Pilaftsis, \emph{{Lepton Flavour Violation and theta(13)
  in Minimal Resonant Leptogenesis}},
  \href{https://doi.org/10.1103/PhysRevD.83.076007}{\emph{Phys. Rev. D}
  {\bfseries 83} (2011) 076007}
  [\href{https://arxiv.org/abs/1012.1834}{{\ttfamily 1012.1834}}].

\bibitem{Drewes:2017zyw}
M.~Drewes, B.~Garbrecht, P.~Hernandez, M.~Kekic, J.~Lopez-Pavon, J.~Racker
  et~al., \emph{{ARS Leptogenesis}},
  \href{https://doi.org/10.1142/S0217751X18420022}{\emph{Int. J. Mod. Phys. A}
  {\bfseries 33} (2018) 1842002}
  [\href{https://arxiv.org/abs/1711.02862}{{\ttfamily 1711.02862}}].

\bibitem{Frere:2008ct}
J.-M. Frere, T.~Hambye and G.~Vertongen, \emph{{Is leptogenesis falsifiable at
  LHC?}}, \href{https://doi.org/10.1088/1126-6708/2009/01/051}{\emph{JHEP}
  {\bfseries 01} (2009) 051} [\href{https://arxiv.org/abs/0806.0841}{{\ttfamily
  0806.0841}}].

\bibitem{Blanchet:2009bu}
S.~Blanchet, Z.~Chacko, S.~S. Granor and R.~N. Mohapatra, \emph{{Probing
  Resonant Leptogenesis at the LHC}},
  \href{https://doi.org/10.1103/PhysRevD.82.076008}{\emph{Phys. Rev. D}
  {\bfseries 82} (2010) 076008}
  [\href{https://arxiv.org/abs/0904.2174}{{\ttfamily 0904.2174}}].

\bibitem{Blanchet:2010kw}
S.~Blanchet, P.~S.~B. Dev and R.~N. Mohapatra, \emph{{Leptogenesis with TeV
  Scale Inverse Seesaw in SO(10)}},
  \href{https://doi.org/10.1103/PhysRevD.82.115025}{\emph{Phys. Rev. D}
  {\bfseries 82} (2010) 115025}
  [\href{https://arxiv.org/abs/1010.1471}{{\ttfamily 1010.1471}}].

\bibitem{Iso:2010mv}
S.~Iso, N.~Okada and Y.~Orikasa, \emph{{Resonant Leptogenesis in the Minimal
  B-L Extended Standard Model at TeV}},
  \href{https://doi.org/10.1103/PhysRevD.83.093011}{\emph{Phys. Rev. D}
  {\bfseries 83} (2011) 093011}
  [\href{https://arxiv.org/abs/1011.4769}{{\ttfamily 1011.4769}}].

\bibitem{Okada:2012fs}
N.~Okada, Y.~Orikasa and T.~Yamada, \emph{{Minimal Flavor Violation in the
  Minimal $U(1)_{B-L}$ Model and Resonant Leptogenesis}},
  \href{https://doi.org/10.1103/PhysRevD.86.076003}{\emph{Phys. Rev. D}
  {\bfseries 86} (2012) 076003}
  [\href{https://arxiv.org/abs/1207.1510}{{\ttfamily 1207.1510}}].

\bibitem{Fong:2013gaa}
C.~S. Fong, M.~C. Gonzalez-Garcia, E.~Nardi and E.~Peinado, \emph{{New ways to
  TeV scale leptogenesis}},
  \href{https://doi.org/10.1007/JHEP08(2013)104}{\emph{JHEP} {\bfseries 08}
  (2013) 104} [\href{https://arxiv.org/abs/1305.6312}{{\ttfamily 1305.6312}}].

\bibitem{Dev:2014hro}
P.~S.~B. Dev, C.-H. Lee and R.~N. Mohapatra, \emph{{Leptogenesis Constraints on
  the Mass of Right-handed Gauge Bosons}},
  \href{https://doi.org/10.1103/PhysRevD.90.095012}{\emph{Phys. Rev. D}
  {\bfseries 90} (2014) 095012}
  [\href{https://arxiv.org/abs/1408.2820}{{\ttfamily 1408.2820}}].

\bibitem{Dev:2015khe}
P.~S.~B. Dev, C.-H. Lee and R.~N. Mohapatra, \emph{{TeV Scale Lepton Number
  Violation and Baryogenesis}},
  \href{https://doi.org/10.1088/1742-6596/631/1/012007}{\emph{J. Phys. Conf.
  Ser.} {\bfseries 631} (2015) 012007}
  [\href{https://arxiv.org/abs/1503.04970}{{\ttfamily 1503.04970}}].

\bibitem{Dhuria:2015cfa}
M.~Dhuria, C.~Hati, R.~Rangarajan and U.~Sarkar, \emph{{Falsifying leptogenesis
  for a TeV scale $W^{\pm}_{R}$ at the LHC}},
  \href{https://doi.org/10.1103/PhysRevD.92.031701}{\emph{Phys. Rev. D}
  {\bfseries 92} (2015) 031701}
  [\href{https://arxiv.org/abs/1503.07198}{{\ttfamily 1503.07198}}].

\bibitem{Heeck:2016oda}
J.~Heeck and D.~Teresi, \emph{{Leptogenesis and neutral gauge bosons}},
  \href{https://doi.org/10.1103/PhysRevD.94.095024}{\emph{Phys. Rev. D}
  {\bfseries 94} (2016) 095024}
  [\href{https://arxiv.org/abs/1609.03594}{{\ttfamily 1609.03594}}].

\bibitem{Caputo:2017pit}
A.~Caputo, P.~Hernandez, J.~Lopez-Pavon and J.~Salvado, \emph{{The seesaw
  portal in testable models of neutrino masses}},
  \href{https://doi.org/10.1007/JHEP06(2017)112}{\emph{JHEP} {\bfseries 06}
  (2017) 112} [\href{https://arxiv.org/abs/1704.08721}{{\ttfamily
  1704.08721}}].

\bibitem{Geng:2017foe}
C.-Q. Geng and H.~Okada, \emph{{Neutrino masses, dark matter and leptogenesis
  with $U(1)_{B-L}$ gauge symmetry}},
  \href{https://doi.org/10.1016/j.dark.2018.02.005}{\emph{Phys. Dark Univ.}
  {\bfseries 20} (2018) 13} [\href{https://arxiv.org/abs/1710.09536}{{\ttfamily
  1710.09536}}].

\bibitem{Dev:2017xry}
P.~S.~B. Dev, R.~N. Mohapatra and Y.~Zhang, \emph{{Leptogenesis constraints on
  $B - L$ breaking Higgs boson in TeV scale seesaw models}},
  \href{https://doi.org/10.1007/JHEP03(2018)122}{\emph{JHEP} {\bfseries 03}
  (2018) 122} [\href{https://arxiv.org/abs/1711.07634}{{\ttfamily
  1711.07634}}].

\bibitem{Gu:2017gra}
P.-H. Gu and R.~N. Mohapatra, \emph{{Leptogenesis with TeV Scale $W_R$}},
  \href{https://doi.org/10.1103/PhysRevD.97.075014}{\emph{Phys. Rev. D}
  {\bfseries 97} (2018) 075014}
  [\href{https://arxiv.org/abs/1712.00420}{{\ttfamily 1712.00420}}].

\bibitem{Dev:2019ljp}
P.~S.~B. Dev, R.~N. Mohapatra and Y.~Zhang, \emph{{CP Violating Effects in
  Heavy Neutrino Oscillations: Implications for Colliders and Leptogenesis}},
  \href{https://doi.org/10.1007/JHEP11(2019)137}{\emph{JHEP} {\bfseries 11}
  (2019) 137} [\href{https://arxiv.org/abs/1904.04787}{{\ttfamily
  1904.04787}}].

\bibitem{Borah:2021mri}
D.~Borah, A.~Dasgupta and D.~Mahanta, \emph{{TeV scale resonant leptogenesis
  with L\ensuremath{\mu}-L\ensuremath{\tau} gauge symmetry in light of the muon
  g-2}}, \href{https://doi.org/10.1103/PhysRevD.104.075006}{\emph{Phys. Rev. D}
  {\bfseries 104} (2021) 075006}
  [\href{https://arxiv.org/abs/2106.14410}{{\ttfamily 2106.14410}}].

\bibitem{Liu:2021akf}
W.~Liu, K.-P. Xie and Z.~Yi, \emph{{Testing leptogenesis at the LHC and future
  muon colliders: a $Z'$ scenario}},
  \href{https://arxiv.org/abs/2109.15087}{{\ttfamily 2109.15087}}.

\bibitem{Chun:2017spz}
E.~J. Chun et~al., \emph{{Probing Leptogenesis}},
  \href{https://doi.org/10.1142/S0217751X18420058}{\emph{Int. J. Mod. Phys. A}
  {\bfseries 33} (2018) 1842005}
  [\href{https://arxiv.org/abs/1711.02865}{{\ttfamily 1711.02865}}].

\bibitem{Xing:2020ijf}
Z.-z. Xing, \emph{{Flavor structures of charged fermions and massive
  neutrinos}}, \href{https://doi.org/10.1016/j.physrep.2020.02.001}{\emph{Phys.
  Rept.} {\bfseries 854} (2020) 1}
  [\href{https://arxiv.org/abs/1909.09610}{{\ttfamily 1909.09610}}].

\bibitem{Casas:2001sr}
J.~A. Casas and A.~Ibarra, \emph{{Oscillating neutrinos and $\mu \to e,
  \gamma$}}, \href{https://doi.org/10.1016/S0550-3213(01)00475-8}{\emph{Nucl.
  Phys. B} {\bfseries 618} (2001) 171}
  [\href{https://arxiv.org/abs/hep-ph/0103065}{{\ttfamily hep-ph/0103065}}].

\bibitem{Dev:2017trv}
P.~S.~B. Dev, P.~Di~Bari, B.~Garbrecht, S.~Lavignac, P.~Millington and
  D.~Teresi, \emph{{Flavor effects in leptogenesis}},
  \href{https://doi.org/10.1142/S0217751X18420010}{\emph{Int. J. Mod. Phys. A}
  {\bfseries 33} (2018) 1842001}
  [\href{https://arxiv.org/abs/1711.02861}{{\ttfamily 1711.02861}}].

\bibitem{Branco:2001pq}
G.~C. Branco, T.~Morozumi, B.~M. Nobre and M.~N. Rebelo, \emph{{A Bridge
  between CP violation at low-energies and leptogenesis}},
  \href{https://doi.org/10.1016/S0550-3213(01)00425-4}{\emph{Nucl. Phys. B}
  {\bfseries 617} (2001) 475}
  [\href{https://arxiv.org/abs/hep-ph/0107164}{{\ttfamily hep-ph/0107164}}].

\bibitem{Rebelo:2002wj}
M.~N. Rebelo, \emph{{Leptogenesis without CP violation at low-energies}},
  \href{https://doi.org/10.1103/PhysRevD.67.013008}{\emph{Phys. Rev. D}
  {\bfseries 67} (2003) 013008}
  [\href{https://arxiv.org/abs/hep-ph/0207236}{{\ttfamily hep-ph/0207236}}].

\bibitem{Pascoli:2003uh}
S.~Pascoli, S.~T. Petcov and W.~Rodejohann, \emph{{On the connection of
  leptogenesis with low-energy CP violation and LFV charged lepton decays}},
  \href{https://doi.org/10.1103/PhysRevD.68.093007}{\emph{Phys. Rev. D}
  {\bfseries 68} (2003) 093007}
  [\href{https://arxiv.org/abs/hep-ph/0302054}{{\ttfamily hep-ph/0302054}}].

\bibitem{Xing:2009vb}
Z.-z. Xing, \emph{{Casas-Ibarra Parametrization and Unflavored Leptogenesis}},
  \href{https://doi.org/10.1088/1674-1137/34/1/001}{\emph{Chin. Phys. C}
  {\bfseries 34} (2010) 1} [\href{https://arxiv.org/abs/0902.2469}{{\ttfamily
  0902.2469}}].

\bibitem{Rodejohann:2009cq}
W.~Rodejohann, \emph{{Non-Unitary Lepton Mixing Matrix, Leptogenesis and Low
  Energy CP Violation}},
  \href{https://doi.org/10.1209/0295-5075/88/51001}{\emph{EPL} {\bfseries 88}
  (2009) 51001} [\href{https://arxiv.org/abs/0903.4590}{{\ttfamily
  0903.4590}}].

\bibitem{Antusch:2009gn}
S.~Antusch, S.~Blanchet, M.~Blennow and E.~Fernandez-Martinez,
  \emph{{Non-unitary Leptonic Mixing and Leptogenesis}},
  \href{https://doi.org/10.1007/JHEP01(2010)017}{\emph{JHEP} {\bfseries 01}
  (2010) 017} [\href{https://arxiv.org/abs/0910.5957}{{\ttfamily 0910.5957}}].

\bibitem{Feldman:2012jdx}
G.~J. Feldman, J.~Hartnell and T.~Kobayashi, \emph{{Long-baseline neutrino
  oscillation experiments}},
  \href{https://doi.org/10.1155/2013/475749}{\emph{Adv. High Energy Phys.}
  {\bfseries 2013} (2013) 475749}
  [\href{https://arxiv.org/abs/1210.1778}{{\ttfamily 1210.1778}}].

\bibitem{Dolinski:2019nrj}
M.~J. Dolinski, A.~W.~P. Poon and W.~Rodejohann, \emph{{Neutrinoless
  Double-Beta Decay: Status and Prospects}},
  \href{https://doi.org/10.1146/annurev-nucl-101918-023407}{\emph{Ann. Rev.
  Nucl. Part. Sci.} {\bfseries 69} (2019) 219}
  [\href{https://arxiv.org/abs/1902.04097}{{\ttfamily 1902.04097}}].

\bibitem{Pascoli:2006ie}
S.~Pascoli, S.~T. Petcov and A.~Riotto, \emph{{Connecting low energy leptonic
  CP-violation to leptogenesis}},
  \href{https://doi.org/10.1103/PhysRevD.75.083511}{\emph{Phys. Rev. D}
  {\bfseries 75} (2007) 083511}
  [\href{https://arxiv.org/abs/hep-ph/0609125}{{\ttfamily hep-ph/0609125}}].

\bibitem{Branco:2006ce}
G.~C. Branco, R.~Gonzalez~Felipe and F.~R. Joaquim, \emph{{A New bridge between
  leptonic CP violation and leptogenesis}},
  \href{https://doi.org/10.1016/j.physletb.2006.12.060}{\emph{Phys. Lett. B}
  {\bfseries 645} (2007) 432}
  [\href{https://arxiv.org/abs/hep-ph/0609297}{{\ttfamily hep-ph/0609297}}].

\bibitem{Pascoli:2006ci}
S.~Pascoli, S.~T. Petcov and A.~Riotto, \emph{{Leptogenesis and Low Energy CP
  Violation in Neutrino Physics}},
  \href{https://doi.org/10.1016/j.nuclphysb.2007.02.019}{\emph{Nucl. Phys. B}
  {\bfseries 774} (2007) 1}
  [\href{https://arxiv.org/abs/hep-ph/0611338}{{\ttfamily hep-ph/0611338}}].

\bibitem{Anisimov:2007mw}
A.~Anisimov, S.~Blanchet and P.~Di~Bari, \emph{{Viability of Dirac phase
  leptogenesis}},
  \href{https://doi.org/10.1088/1475-7516/2008/04/033}{\emph{JCAP} {\bfseries
  04} (2008) 033} [\href{https://arxiv.org/abs/0707.3024}{{\ttfamily
  0707.3024}}].

\bibitem{Molinaro:2009lud}
E.~Molinaro and S.~T. Petcov, \emph{{The Interplay Between the 'Low' and 'High'
  Energy CP-Violation in Leptogenesis}},
  \href{https://doi.org/10.1140/epjc/s10052-009-0985-3}{\emph{Eur. Phys. J. C}
  {\bfseries 61} (2009) 93} [\href{https://arxiv.org/abs/0803.4120}{{\ttfamily
  0803.4120}}].

\bibitem{Molinaro:2008cw}
E.~Molinaro and S.~T. Petcov, \emph{{A Case of Subdominant/Suppressed 'High
  Energy' Contribution to the Baryon Asymmetry of the Universe in Flavoured
  Leptogenesis}},
  \href{https://doi.org/10.1016/j.physletb.2008.11.047}{\emph{Phys. Lett. B}
  {\bfseries 671} (2009) 60} [\href{https://arxiv.org/abs/0808.3534}{{\ttfamily
  0808.3534}}].

\bibitem{Moffat:2018smo}
K.~Moffat, S.~Pascoli, S.~T. Petcov and J.~Turner, \emph{{Leptogenesis from Low
  Energy $CP$ Violation}},
  \href{https://doi.org/10.1007/JHEP03(2019)034}{\emph{JHEP} {\bfseries 03}
  (2019) 034} [\href{https://arxiv.org/abs/1809.08251}{{\ttfamily
  1809.08251}}].

\bibitem{Hagedorn:2017wjy}
C.~Hagedorn, R.~N. Mohapatra, E.~Molinaro, C.~C. Nishi and S.~T. Petcov,
  \emph{{CP Violation in the Lepton Sector and Implications for Leptogenesis}},
  \href{https://doi.org/10.1142/S0217751X1842006X}{\emph{Int. J. Mod. Phys. A}
  {\bfseries 33} (2018) 1842006}
  [\href{https://arxiv.org/abs/1711.02866}{{\ttfamily 1711.02866}}].

\bibitem{Chen:2016ptr}
P.~Chen, G.-J. Ding and S.~F. King, \emph{{Leptogenesis and residual CP
  symmetry}}, \href{https://doi.org/10.1007/JHEP03(2016)206}{\emph{JHEP}
  {\bfseries 03} (2016) 206}
  [\href{https://arxiv.org/abs/1602.03873}{{\ttfamily 1602.03873}}].

\bibitem{Hagedorn:2016lva}
C.~Hagedorn and E.~Molinaro, \emph{{Flavor and CP symmetries for leptogenesis
  and 0 \ensuremath{\nu}\ensuremath{\beta}\ensuremath{\beta} decay}},
  \href{https://doi.org/10.1016/j.nuclphysb.2017.03.015}{\emph{Nucl. Phys. B}
  {\bfseries 919} (2017) 404}
  [\href{https://arxiv.org/abs/1602.04206}{{\ttfamily 1602.04206}}].

\bibitem{Li:2017zmk}
C.-C. Li and G.-J. Ding, \emph{{Implications of residual CP symmetry for
  leptogenesis in a model with two right-handed neutrinos}},
  \href{https://doi.org/10.1103/PhysRevD.96.075005}{\emph{Phys. Rev. D}
  {\bfseries 96} (2017) 075005}
  [\href{https://arxiv.org/abs/1701.08508}{{\ttfamily 1701.08508}}].

\bibitem{Samanta:2018efa}
R.~Samanta, R.~Sinha and A.~Ghosal, \emph{{Importance of generalized $\mu\tau$
  symmetry and its CP extension on neutrino mixing and leptogenesis}},
  \href{https://doi.org/10.1007/JHEP10(2019)057}{\emph{JHEP} {\bfseries 10}
  (2019) 057} [\href{https://arxiv.org/abs/1805.10031}{{\ttfamily
  1805.10031}}].

\bibitem{Fong:2021tqj}
C.~S. Fong, M.~H. Rahat and S.~Saad, \emph{{Low-scale resonant leptogenesis in
  SU(5) GUT with T13 family symmetry}},
  \href{https://doi.org/10.1103/PhysRevD.104.095028}{\emph{Phys. Rev. D}
  {\bfseries 104} (2021) 095028}
  [\href{https://arxiv.org/abs/2103.14691}{{\ttfamily 2103.14691}}].

\bibitem{Feruglio:2012cw}
F.~Feruglio, C.~Hagedorn and R.~Ziegler, \emph{{Lepton Mixing Parameters from
  Discrete and CP Symmetries}},
  \href{https://doi.org/10.1007/JHEP07(2013)027}{\emph{JHEP} {\bfseries 07}
  (2013) 027} [\href{https://arxiv.org/abs/1211.5560}{{\ttfamily 1211.5560}}].

\bibitem{Holthausen:2012dk}
M.~Holthausen, M.~Lindner and M.~A. Schmidt, \emph{{CP and Discrete Flavour
  Symmetries}}, \href{https://doi.org/10.1007/JHEP04(2013)122}{\emph{JHEP}
  {\bfseries 04} (2013) 122} [\href{https://arxiv.org/abs/1211.6953}{{\ttfamily
  1211.6953}}].

\bibitem{Chen:2014tpa}
M.-C. Chen, M.~Fallbacher, K.~T. Mahanthappa, M.~Ratz and A.~Trautner,
  \emph{{CP Violation from Finite Groups}},
  \href{https://doi.org/10.1016/j.nuclphysb.2014.03.023}{\emph{Nucl. Phys. B}
  {\bfseries 883} (2014) 267}
  [\href{https://arxiv.org/abs/1402.0507}{{\ttfamily 1402.0507}}].

\bibitem{Luhn:2007uq}
C.~Luhn, S.~Nasri and P.~Ramond, \emph{{The Flavor group Delta(3n**2)}},
  \href{https://doi.org/10.1063/1.2734865}{\emph{J. Math. Phys.} {\bfseries 48}
  (2007) 073501} [\href{https://arxiv.org/abs/hep-th/0701188}{{\ttfamily
  hep-th/0701188}}].

\bibitem{Escobar:2008vc}
J.~A. Escobar and C.~Luhn, \emph{{The Flavor Group Delta(6n**2)}},
  \href{https://doi.org/10.1063/1.3046563}{\emph{J. Math. Phys.} {\bfseries 50}
  (2009) 013524} [\href{https://arxiv.org/abs/0809.0639}{{\ttfamily
  0809.0639}}].

\bibitem{King:2014rwa}
S.~F. King and T.~Neder, \emph{{Lepton mixing predictions including Majorana
  phases from \ensuremath{\Delta}(6n$^2$) flavour symmetry and generalised
  CP}}, \href{https://doi.org/10.1016/j.physletb.2014.07.043}{\emph{Phys. Lett.
  B} {\bfseries 736} (2014) 308}
  [\href{https://arxiv.org/abs/1403.1758}{{\ttfamily 1403.1758}}].

\bibitem{Hagedorn:2014wha}
C.~Hagedorn, A.~Meroni and E.~Molinaro, \emph{{Lepton mixing from
  \ensuremath{\Delta}(3$n^2$) and \ensuremath{\Delta}(6$n^2$) and CP}},
  \href{https://doi.org/10.1016/j.nuclphysb.2014.12.013}{\emph{Nucl. Phys. B}
  {\bfseries 891} (2015) 499}
  [\href{https://arxiv.org/abs/1408.7118}{{\ttfamily 1408.7118}}].

\bibitem{Ding:2014ora}
G.-J. Ding, S.~F. King and T.~Neder, \emph{{Generalised CP and $\Delta(6n^2)$
  family symmetry in semi-direct models of leptons}},
  \href{https://doi.org/10.1007/JHEP12(2014)007}{\emph{JHEP} {\bfseries 12}
  (2014) 007} [\href{https://arxiv.org/abs/1409.8005}{{\ttfamily 1409.8005}}].

\bibitem{Ding:2015rwa}
G.-J. Ding and S.~F. King, \emph{{Generalized CP and $\Delta (3n^2)$ Family
  Symmetry for Semi-Direct Predictions of the PMNS Matrix}},
  \href{https://doi.org/10.1103/PhysRevD.93.025013}{\emph{Phys. Rev. D}
  {\bfseries 93} (2016) 025013}
  [\href{https://arxiv.org/abs/1510.03188}{{\ttfamily 1510.03188}}].

\bibitem{Alimena:2019zri}
J.~Alimena et~al., \emph{{Searching for long-lived particles beyond the
  Standard Model at the Large Hadron Collider}},
  \href{https://doi.org/10.1088/1361-6471/ab4574}{\emph{J. Phys. G} {\bfseries
  47} (2020) 090501} [\href{https://arxiv.org/abs/1903.04497}{{\ttfamily
  1903.04497}}].

\bibitem{Alimena:2021mdu}
D.~Acosta et~al., \emph{{Review of opportunities for new long-lived particle
  triggers in Run 3 of the Large Hadron Collider}},
  \href{https://arxiv.org/abs/2110.14675}{{\ttfamily 2110.14675}}.

\bibitem{Deppisch:2015qwa}
F.~F. Deppisch, P.~S.~B. Dev and A.~Pilaftsis, \emph{{Neutrinos and Collider
  Physics}}, \href{https://doi.org/10.1088/1367-2630/17/7/075019}{\emph{New J.
  Phys.} {\bfseries 17} (2015) 075019}
  [\href{https://arxiv.org/abs/1502.06541}{{\ttfamily 1502.06541}}].

\bibitem{Cai:2017mow}
Y.~Cai, T.~Han, T.~Li and R.~Ruiz, \emph{{Lepton Number Violation: Seesaw
  Models and Their Collider Tests}},
  \href{https://doi.org/10.3389/fphy.2018.00040}{\emph{Front. in Phys.}
  {\bfseries 6} (2018) 40} [\href{https://arxiv.org/abs/1711.02180}{{\ttfamily
  1711.02180}}].

\bibitem{ZurbanoFernandez:2020cco}
I.~Zurbano~Fernandez et~al., \emph{{High-Luminosity Large Hadron Collider
  (HL-LHC): Technical design report}},
  \href{https://doi.org/10.23731/CYRM-2020-0010}{\emph{CERN-2020-010} (2020) }.

\bibitem{FCC:2018vvp}
{\scshape FCC} collaboration, \emph{{FCC-hh: The Hadron Collider}: {Future
  Circular Collider Conceptual Design Report Volume 3}},
  \href{https://doi.org/10.1140/epjst/e2019-900087-0}{\emph{Eur. Phys. J. ST}
  {\bfseries 228} (2019) 755}.

\bibitem{Feruglio:2019ybq}
F.~Feruglio and A.~Romanino, \emph{{Lepton flavor symmetries}},
  \href{https://doi.org/10.1103/RevModPhys.93.015007}{\emph{Rev. Mod. Phys.}
  {\bfseries 93} (2021) 015007}
  [\href{https://arxiv.org/abs/1912.06028}{{\ttfamily 1912.06028}}].

\bibitem{Ishimori:2010au}
H.~Ishimori, T.~Kobayashi, H.~Ohki, Y.~Shimizu, H.~Okada and M.~Tanimoto,
  \emph{{Non-Abelian Discrete Symmetries in Particle Physics}},
  \href{https://doi.org/10.1143/PTPS.183.1}{\emph{Prog. Theor. Phys. Suppl.}
  {\bfseries 183} (2010) 1} [\href{https://arxiv.org/abs/1003.3552}{{\ttfamily
  1003.3552}}].

\bibitem{King:2013eh}
S.~F. King and C.~Luhn, \emph{{Neutrino Mass and Mixing with Discrete
  Symmetry}}, \href{https://doi.org/10.1088/0034-4885/76/5/056201}{\emph{Rept.
  Prog. Phys.} {\bfseries 76} (2013) 056201}
  [\href{https://arxiv.org/abs/1301.1340}{{\ttfamily 1301.1340}}].

\bibitem{GonzalezFelipe:2003fi}
R.~Gonzalez~Felipe, F.~R. Joaquim and B.~M. Nobre, \emph{{Radiatively induced
  leptogenesis in a minimal seesaw model}},
  \href{https://doi.org/10.1103/PhysRevD.70.085009}{\emph{Phys. Rev. D}
  {\bfseries 70} (2004) 085009}
  [\href{https://arxiv.org/abs/hep-ph/0311029}{{\ttfamily hep-ph/0311029}}].

\bibitem{Turzynski:2004xy}
K.~Turzynski, \emph{{Degenerate minimal seesaw and leptogenesis}},
  \href{https://doi.org/10.1016/j.physletb.2004.03.071}{\emph{Phys. Lett. B}
  {\bfseries 589} (2004) 135}
  [\href{https://arxiv.org/abs/hep-ph/0401219}{{\ttfamily hep-ph/0401219}}].

\bibitem{Branco:2005ye}
G.~C. Branco, R.~Gonzalez~Felipe, F.~R. Joaquim and B.~M. Nobre,
  \emph{{Enlarging the window for radiative leptogenesis}},
  \href{https://doi.org/10.1016/j.physletb.2005.11.070}{\emph{Phys. Lett. B}
  {\bfseries 633} (2006) 336}
  [\href{https://arxiv.org/abs/hep-ph/0507092}{{\ttfamily hep-ph/0507092}}].

\bibitem{Ahn:2006rn}
Y.~H. Ahn, C.~S. Kim, S.~K. Kang and J.~Lee, \emph{{mu- tau Symmetry and
  Radiatively Generated Leptogenesis}},
  \href{https://doi.org/10.1103/PhysRevD.75.013012}{\emph{Phys. Rev. D}
  {\bfseries 75} (2007) 013012}
  [\href{https://arxiv.org/abs/hep-ph/0610007}{{\ttfamily hep-ph/0610007}}].

\bibitem{Babu:2008kp}
K.~S. Babu, Y.~Meng and Z.~Tavartkiladze, \emph{{Common Origin for CP Violation
  in Cosmology and in Neutrino Oscillations}},
  \href{https://arxiv.org/abs/0812.4419}{{\ttfamily 0812.4419}}.

\bibitem{Dev:2015wpa}
P.~S.~B. Dev, P.~Millington, A.~Pilaftsis and D.~Teresi, \emph{{Corrigendum to
  ''Flavour Covariant Transport Equations: an Application to Resonant
  Leptogenesis''}},
  \href{https://doi.org/10.1016/j.nuclphysb.2015.06.015}{\emph{Nucl. Phys. B}
  {\bfseries 897} (2015) 749}
  [\href{https://arxiv.org/abs/1504.07640}{{\ttfamily 1504.07640}}].

\bibitem{Zhao:2020bzx}
Z.-h. Zhao, \emph{{Renormalization group evolution induced leptogenesis in the
  minimal seesaw model with the trimaximal mixing and mu-tau reflection
  symmetry}}, \href{https://doi.org/10.1007/JHEP11(2021)170}{\emph{JHEP}
  {\bfseries 11} (2021) 170}
  [\href{https://arxiv.org/abs/2003.00654}{{\ttfamily 2003.00654}}].

\bibitem{ParticleDataGroup:2020ssz}
{\scshape Particle Data Group} collaboration, \emph{{Review of Particle
  Physics}}, \href{https://doi.org/10.1093/ptep/ptaa104}{\emph{PTEP} {\bfseries
  2020} (2020) 083C01}.

\bibitem{Esteban:2020cvm}
I.~Esteban, M.~C. Gonzalez-Garcia, M.~Maltoni, T.~Schwetz and A.~Zhou,
  \emph{{The fate of hints: updated global analysis of three-flavor neutrino
  oscillations}}, \href{https://doi.org/10.1007/JHEP09(2020)178}{\emph{JHEP}
  {\bfseries 09} (2020) 178}
  [\href{https://arxiv.org/abs/2007.14792}{{\ttfamily 2007.14792}}].

\bibitem{NUFIT}
{\scshape NuFIT} collaboration. \url{http://www.nu-fit.org}.

\bibitem{Harrison:2002er}
P.~F. Harrison, D.~H. Perkins and W.~G. Scott, \emph{{Tri-bimaximal mixing and
  the neutrino oscillation data}},
  \href{https://doi.org/10.1016/S0370-2693(02)01336-9}{\emph{Phys. Lett. B}
  {\bfseries 530} (2002) 167}
  [\href{https://arxiv.org/abs/hep-ph/0202074}{{\ttfamily hep-ph/0202074}}].

\bibitem{Drewes:2022kap}
M.~Drewes, Y.~Georis, C.~Hagedorn and J.~Klari\'c, \emph{{Low-scale
  leptogenesis with flavour and CP symmetries}},
  \href{https://arxiv.org/abs/2203.08538}{{\ttfamily 2203.08538}}.

\bibitem{Feruglio:2013hia}
F.~Feruglio, C.~Hagedorn and R.~Ziegler, \emph{{A realistic pattern of lepton
  mixing and masses from $S_4$ and CP}},
  \href{https://doi.org/10.1140/epjc/s10052-014-2753-2}{\emph{Eur. Phys. J. C}
  {\bfseries 74} (2014) 2753}
  [\href{https://arxiv.org/abs/1303.7178}{{\ttfamily 1303.7178}}].

\bibitem{Lin:2009bw}
Y.~Lin, \emph{{Tri-bimaximal Neutrino Mixing from A(4) and $\theta_{13}$
  \textasciitilde{} theta(C)}},
  \href{https://doi.org/10.1016/j.nuclphysb.2009.08.018}{\emph{Nucl. Phys. B}
  {\bfseries 824} (2010) 95} [\href{https://arxiv.org/abs/0905.3534}{{\ttfamily
  0905.3534}}].

\bibitem{Buchmuller:2004nz}
W.~Buchmuller, P.~Di~Bari and M.~Plumacher, \emph{{Leptogenesis for
  pedestrians}}, \href{https://doi.org/10.1016/j.aop.2004.02.003}{\emph{Annals
  Phys.} {\bfseries 315} (2005) 305}
  [\href{https://arxiv.org/abs/hep-ph/0401240}{{\ttfamily hep-ph/0401240}}].

\bibitem{Dev:2014laa}
P.~S.~B. Dev, P.~Millington, A.~Pilaftsis and D.~Teresi, \emph{{Flavour
  Covariant Transport Equations: an Application to Resonant Leptogenesis}},
  \href{https://doi.org/10.1016/j.nuclphysb.2014.06.020}{\emph{Nucl. Phys. B}
  {\bfseries 886} (2014) 569}
  [\href{https://arxiv.org/abs/1404.1003}{{\ttfamily 1404.1003}}].

\bibitem{Harvey:1990qw}
J.~A. Harvey and M.~S. Turner, \emph{{Cosmological baryon and lepton number in
  the presence of electroweak fermion number violation}},
  \href{https://doi.org/10.1103/PhysRevD.42.3344}{\emph{Phys. Rev. D}
  {\bfseries 42} (1990) 3344}.

\bibitem{Dev:2014oar}
P.~S.~B. Dev, P.~Millington, A.~Pilaftsis and D.~Teresi,
  \emph{{Kadanoff\textendash{}Baym approach to flavour mixing and oscillations
  in resonant leptogenesis}},
  \href{https://doi.org/10.1016/j.nuclphysb.2014.12.003}{\emph{Nucl. Phys. B}
  {\bfseries 891} (2015) 128}
  [\href{https://arxiv.org/abs/1410.6434}{{\ttfamily 1410.6434}}].

\bibitem{Kartavtsev:2015vto}
A.~Kartavtsev, P.~Millington and H.~Vogel, \emph{{Lepton asymmetry from mixing
  and oscillations}},
  \href{https://doi.org/10.1007/JHEP06(2016)066}{\emph{JHEP} {\bfseries 06}
  (2016) 066} [\href{https://arxiv.org/abs/1601.03086}{{\ttfamily
  1601.03086}}].

\bibitem{Klaric:2021cpi}
J.~Klari\'c, M.~Shaposhnikov and I.~Timiryasov, \emph{{Reconciling resonant
  leptogenesis and baryogenesis via neutrino oscillations}},
  \href{https://doi.org/10.1103/PhysRevD.104.055010}{\emph{Phys. Rev. D}
  {\bfseries 104} (2021) 055010}
  [\href{https://arxiv.org/abs/2103.16545}{{\ttfamily 2103.16545}}].

\bibitem{Branco:1986gr}
G.~C. Branco, L.~Lavoura and M.~N. Rebelo, \emph{{Majorana Neutrinos and {CP}
  Violation in the Leptonic Sector}},
  \href{https://doi.org/10.1016/0370-2693(86)90307-2}{\emph{Phys. Lett. B}
  {\bfseries 180} (1986) 264}.

\bibitem{Branco:2005jr}
G.~C. Branco, M.~N. Rebelo and J.~I. Silva-Marcos, \emph{{Leptogenesis, Yukawa
  textures and weak basis invariants}},
  \href{https://doi.org/10.1016/j.physletb.2005.11.067}{\emph{Phys. Lett. B}
  {\bfseries 633} (2006) 345}
  [\href{https://arxiv.org/abs/hep-ph/0510412}{{\ttfamily hep-ph/0510412}}].

\bibitem{Jenkins:2009dy}
E.~E. Jenkins and A.~V. Manohar, \emph{{Algebraic Structure of Lepton and Quark
  Flavor Invariants and CP Violation}},
  \href{https://doi.org/10.1088/1126-6708/2009/10/094}{\emph{JHEP} {\bfseries
  10} (2009) 094} [\href{https://arxiv.org/abs/0907.4763}{{\ttfamily
  0907.4763}}].

\bibitem{Deppisch:2013jxa}
F.~F. Deppisch, J.~Harz and M.~Hirsch, \emph{{Falsifying High-Scale
  Leptogenesis at the LHC}},
  \href{https://doi.org/10.1103/PhysRevLett.112.221601}{\emph{Phys. Rev. Lett.}
  {\bfseries 112} (2014) 221601}
  [\href{https://arxiv.org/abs/1312.4447}{{\ttfamily 1312.4447}}].

\bibitem{Deppisch:2015yqa}
F.~F. Deppisch, J.~Harz, M.~Hirsch, W.-C. Huang and H.~P\"as, \emph{{Falsifying
  High-Scale Baryogenesis with Neutrinoless Double Beta Decay and Lepton Flavor
  Violation}}, \href{https://doi.org/10.1103/PhysRevD.92.036005}{\emph{Phys.
  Rev. D} {\bfseries 92} (2015) 036005}
  [\href{https://arxiv.org/abs/1503.04825}{{\ttfamily 1503.04825}}].

\bibitem{Deppisch:2017ecm}
F.~F. Deppisch, L.~Graf, J.~Harz and W.-C. Huang, \emph{{Neutrinoless Double
  Beta Decay and the Baryon Asymmetry of the Universe}},
  \href{https://doi.org/10.1103/PhysRevD.98.055029}{\emph{Phys. Rev. D}
  {\bfseries 98} (2018) 055029}
  [\href{https://arxiv.org/abs/1711.10432}{{\ttfamily 1711.10432}}].

\bibitem{Pilaftsis:1991ug}
A.~Pilaftsis, \emph{{Radiatively induced neutrino masses and large Higgs
  neutrino couplings in the standard model with Majorana fields}},
  \href{https://doi.org/10.1007/BF01482590}{\emph{Z. Phys. C} {\bfseries 55}
  (1992) 275} [\href{https://arxiv.org/abs/hep-ph/9901206}{{\ttfamily
  hep-ph/9901206}}].

\bibitem{Tommasini:1995ii}
D.~Tommasini, G.~Barenboim, J.~Bernabeu and C.~Jarlskog, \emph{{Nondecoupling
  of heavy neutrinos and lepton flavor violation}},
  \href{https://doi.org/10.1016/0550-3213(95)00201-3}{\emph{Nucl. Phys. B}
  {\bfseries 444} (1995) 451}
  [\href{https://arxiv.org/abs/hep-ph/9503228}{{\ttfamily hep-ph/9503228}}].

\bibitem{Gluza:2002vs}
J.~Gluza, \emph{{On teraelectronvolt Majorana neutrinos}}, {\emph{Acta Phys.
  Polon. B} {\bfseries 33} (2002) 1735}
  [\href{https://arxiv.org/abs/hep-ph/0201002}{{\ttfamily hep-ph/0201002}}].

\bibitem{Kersten:2007vk}
J.~Kersten and A.~Y. Smirnov, \emph{{Right-Handed Neutrinos at CERN LHC and the
  Mechanism of Neutrino Mass Generation}},
  \href{https://doi.org/10.1103/PhysRevD.76.073005}{\emph{Phys. Rev. D}
  {\bfseries 76} (2007) 073005}
  [\href{https://arxiv.org/abs/0705.3221}{{\ttfamily 0705.3221}}].

\bibitem{Xing:2009in}
Z.-z. Xing, \emph{{Naturalness and Testability of TeV Seesaw Mechanisms}},
  \href{https://doi.org/10.1143/PTPS.180.112}{\emph{Prog. Theor. Phys. Suppl.}
  {\bfseries 180} (2009) 112}
  [\href{https://arxiv.org/abs/0905.3903}{{\ttfamily 0905.3903}}].

\bibitem{Gavela:2009cd}
M.~B. Gavela, T.~Hambye, D.~Hernandez and P.~Hernandez, \emph{{Minimal Flavour
  Seesaw Models}},
  \href{https://doi.org/10.1088/1126-6708/2009/09/038}{\emph{JHEP} {\bfseries
  09} (2009) 038} [\href{https://arxiv.org/abs/0906.1461}{{\ttfamily
  0906.1461}}].

\bibitem{He:2009ua}
X.-G. He, S.~Oh, J.~Tandean and C.-C. Wen, \emph{{Large Mixing of Light and
  Heavy Neutrinos in Seesaw Models and the LHC}},
  \href{https://doi.org/10.1103/PhysRevD.80.073012}{\emph{Phys. Rev. D}
  {\bfseries 80} (2009) 073012}
  [\href{https://arxiv.org/abs/0907.1607}{{\ttfamily 0907.1607}}].

\bibitem{Adhikari:2010yt}
R.~Adhikari and A.~Raychaudhuri, \emph{{Light neutrinos from massless texture
  and below TeV seesaw scale}},
  \href{https://doi.org/10.1103/PhysRevD.84.033002}{\emph{Phys. Rev. D}
  {\bfseries 84} (2011) 033002}
  [\href{https://arxiv.org/abs/1004.5111}{{\ttfamily 1004.5111}}].

\bibitem{Ibarra:2010xw}
A.~Ibarra, E.~Molinaro and S.~T. Petcov, \emph{{TeV Scale See-Saw Mechanisms of
  Neutrino Mass Generation, the Majorana Nature of the Heavy Singlet Neutrinos
  and $(\beta\beta)_{0\nu}$-Decay}},
  \href{https://doi.org/10.1007/JHEP09(2010)108}{\emph{JHEP} {\bfseries 09}
  (2010) 108} [\href{https://arxiv.org/abs/1007.2378}{{\ttfamily 1007.2378}}].

\bibitem{Ibarra:2011xn}
A.~Ibarra, E.~Molinaro and S.~T. Petcov, \emph{{Low Energy Signatures of the
  TeV Scale See-Saw Mechanism}},
  \href{https://doi.org/10.1103/PhysRevD.84.013005}{\emph{Phys. Rev. D}
  {\bfseries 84} (2011) 013005}
  [\href{https://arxiv.org/abs/1103.6217}{{\ttfamily 1103.6217}}].

\bibitem{Mitra:2011qr}
M.~Mitra, G.~Senjanovic and F.~Vissani, \emph{{Neutrinoless Double Beta Decay
  and Heavy Sterile Neutrinos}},
  \href{https://doi.org/10.1016/j.nuclphysb.2011.10.035}{\emph{Nucl. Phys. B}
  {\bfseries 856} (2012) 26} [\href{https://arxiv.org/abs/1108.0004}{{\ttfamily
  1108.0004}}].

\bibitem{Lee:2013htl}
C.-H. Lee, P.~S.~B. Dev and R.~N. Mohapatra, \emph{{Natural TeV-scale
  left-right seesaw mechanism for neutrinos and experimental tests}},
  \href{https://doi.org/10.1103/PhysRevD.88.093010}{\emph{Phys. Rev. D}
  {\bfseries 88} (2013) 093010}
  [\href{https://arxiv.org/abs/1309.0774}{{\ttfamily 1309.0774}}].

\bibitem{CarcamoHernandez:2019kjy}
A.~E. C\'arcamo~Hern\'andez, M.~Gonz\'alez and N.~A. Neill, \emph{{Low scale
  type I seesaw model for lepton masses and mixings}},
  \href{https://doi.org/10.1103/PhysRevD.101.035005}{\emph{Phys. Rev. D}
  {\bfseries 101} (2020) 035005}
  [\href{https://arxiv.org/abs/1906.00978}{{\ttfamily 1906.00978}}].

\bibitem{Lopez-Pavon:2015cga}
J.~Lopez-Pavon, E.~Molinaro and S.~T. Petcov, \emph{{Radiative Corrections to
  Light Neutrino Masses in Low Scale Type I Seesaw Scenarios and Neutrinoless
  Double Beta Decay}},
  \href{https://doi.org/10.1007/JHEP11(2015)030}{\emph{JHEP} {\bfseries 11}
  (2015) 030} [\href{https://arxiv.org/abs/1506.05296}{{\ttfamily
  1506.05296}}].

\bibitem{Fernandez-Martinez:2015hxa}
E.~Fernandez-Martinez, J.~Hernandez-Garcia, J.~Lopez-Pavon and M.~Lucente,
  \emph{{Loop level constraints on Seesaw neutrino mixing}},
  \href{https://doi.org/10.1007/JHEP10(2015)130}{\emph{JHEP} {\bfseries 10}
  (2015) 130} [\href{https://arxiv.org/abs/1508.03051}{{\ttfamily
  1508.03051}}].

\bibitem{Fernandez-Martinez:2016lgt}
E.~Fernandez-Martinez, J.~Hernandez-Garcia and J.~Lopez-Pavon, \emph{{Global
  constraints on heavy neutrino mixing}},
  \href{https://doi.org/10.1007/JHEP08(2016)033}{\emph{JHEP} {\bfseries 08}
  (2016) 033} [\href{https://arxiv.org/abs/1605.08774}{{\ttfamily
  1605.08774}}].

\bibitem{Das:2017nvm}
A.~Das and N.~Okada, \emph{{Bounds on heavy Majorana neutrinos in type-I seesaw
  and implications for collider searches}},
  \href{https://doi.org/10.1016/j.physletb.2017.09.042}{\emph{Phys. Lett. B}
  {\bfseries 774} (2017) 32}
  [\href{https://arxiv.org/abs/1702.04668}{{\ttfamily 1702.04668}}].

\bibitem{Bolton:2019pcu}
P.~D. Bolton, F.~F. Deppisch and P.~S.~B. Dev, \emph{{Neutrinoless double beta
  decay versus other probes of heavy sterile neutrinos}},
  \href{https://doi.org/10.1007/JHEP03(2020)170}{\emph{JHEP} {\bfseries 03}
  (2020) 170} [\href{https://arxiv.org/abs/1912.03058}{{\ttfamily
  1912.03058}}].

\bibitem{Keung:1983uu}
W.-Y. Keung and G.~Senjanovic, \emph{{Majorana Neutrinos and the Production of
  the Right-handed Charged Gauge Boson}},
  \href{https://doi.org/10.1103/PhysRevLett.50.1427}{\emph{Phys. Rev. Lett.}
  {\bfseries 50} (1983) 1427}.

\bibitem{Datta:1993nm}
A.~Datta, M.~Guchait and A.~Pilaftsis, \emph{{Probing lepton number violation
  via majorana neutrinos at hadron supercolliders}},
  \href{https://doi.org/10.1103/PhysRevD.50.3195}{\emph{Phys. Rev. D}
  {\bfseries 50} (1994) 3195}
  [\href{https://arxiv.org/abs/hep-ph/9311257}{{\ttfamily hep-ph/9311257}}].

\bibitem{Han:2006ip}
T.~Han and B.~Zhang, \emph{{Signatures for Majorana neutrinos at hadron
  colliders}}, \href{https://doi.org/10.1103/PhysRevLett.97.171804}{\emph{Phys.
  Rev. Lett.} {\bfseries 97} (2006) 171804}
  [\href{https://arxiv.org/abs/hep-ph/0604064}{{\ttfamily hep-ph/0604064}}].

\bibitem{delAguila:2007qnc}
F.~del Aguila, J.~A. Aguilar-Saavedra and R.~Pittau, \emph{{Heavy neutrino
  signals at large hadron colliders}},
  \href{https://doi.org/10.1088/1126-6708/2007/10/047}{\emph{JHEP} {\bfseries
  10} (2007) 047} [\href{https://arxiv.org/abs/hep-ph/0703261}{{\ttfamily
  hep-ph/0703261}}].

\bibitem{Atre:2009rg}
A.~Atre, T.~Han, S.~Pascoli and B.~Zhang, \emph{{The Search for Heavy Majorana
  Neutrinos}}, \href{https://doi.org/10.1088/1126-6708/2009/05/030}{\emph{JHEP}
  {\bfseries 05} (2009) 030} [\href{https://arxiv.org/abs/0901.3589}{{\ttfamily
  0901.3589}}].

\bibitem{Dev:2013wba}
P.~S.~B. Dev, A.~Pilaftsis and U.-k. Yang, \emph{{New Production Mechanism for
  Heavy Neutrinos at the LHC}},
  \href{https://doi.org/10.1103/PhysRevLett.112.081801}{\emph{Phys. Rev. Lett.}
  {\bfseries 112} (2014) 081801}
  [\href{https://arxiv.org/abs/1308.2209}{{\ttfamily 1308.2209}}].

\bibitem{Alva:2014gxa}
D.~Alva, T.~Han and R.~Ruiz, \emph{{Heavy Majorana neutrinos from $W\gamma$
  fusion at hadron colliders}},
  \href{https://doi.org/10.1007/JHEP02(2015)072}{\emph{JHEP} {\bfseries 02}
  (2015) 072} [\href{https://arxiv.org/abs/1411.7305}{{\ttfamily 1411.7305}}].

\bibitem{Das:2015toa}
A.~Das and N.~Okada, \emph{{Improved bounds on the heavy neutrino productions
  at the LHC}}, \href{https://doi.org/10.1103/PhysRevD.93.033003}{\emph{Phys.
  Rev. D} {\bfseries 93} (2016) 033003}
  [\href{https://arxiv.org/abs/1510.04790}{{\ttfamily 1510.04790}}].

\bibitem{Das:2016hof}
A.~Das, P.~Konar and S.~Majhi, \emph{{Production of Heavy neutrino in
  next-to-leading order QCD at the LHC and beyond}},
  \href{https://doi.org/10.1007/JHEP06(2016)019}{\emph{JHEP} {\bfseries 06}
  (2016) 019} [\href{https://arxiv.org/abs/1604.00608}{{\ttfamily
  1604.00608}}].

\bibitem{Das:2017gke}
A.~Das, P.~Konar and A.~Thalapillil, \emph{{Jet substructure shedding light on
  heavy Majorana neutrinos at the LHC}},
  \href{https://doi.org/10.1007/JHEP02(2018)083}{\emph{JHEP} {\bfseries 02}
  (2018) 083} [\href{https://arxiv.org/abs/1709.09712}{{\ttfamily
  1709.09712}}].

\bibitem{CMS:2018jxx}
{\scshape CMS} collaboration, \emph{{Search for heavy Majorana neutrinos in
  same-sign dilepton channels in proton-proton collisions at $ \sqrt{s}=13 $
  TeV}}, \href{https://doi.org/10.1007/JHEP01(2019)122}{\emph{JHEP} {\bfseries
  01} (2019) 122} [\href{https://arxiv.org/abs/1806.10905}{{\ttfamily
  1806.10905}}].

\bibitem{ATLAS:2019kpx}
{\scshape ATLAS} collaboration, \emph{{Search for heavy neutral leptons in
  decays of $W$ bosons produced in 13 TeV $pp$ collisions using prompt and
  displaced signatures with the ATLAS detector}},
  \href{https://doi.org/10.1007/JHEP10(2019)265}{\emph{JHEP} {\bfseries 10}
  (2019) 265} [\href{https://arxiv.org/abs/1905.09787}{{\ttfamily
  1905.09787}}].

\bibitem{Alonso:2012ji}
R.~Alonso, M.~Dhen, M.~B. Gavela and T.~Hambye, \emph{{Muon conversion to
  electron in nuclei in type-I seesaw models}},
  \href{https://doi.org/10.1007/JHEP01(2013)118}{\emph{JHEP} {\bfseries 01}
  (2013) 118} [\href{https://arxiv.org/abs/1209.2679}{{\ttfamily 1209.2679}}].

\bibitem{Deppisch:2013cya}
F.~F. Deppisch, N.~Desai and J.~W.~F. Valle, \emph{{Is charged lepton flavor
  violation a high energy phenomenon?}},
  \href{https://doi.org/10.1103/PhysRevD.89.051302}{\emph{Phys. Rev. D}
  {\bfseries 89} (2014) 051302}
  [\href{https://arxiv.org/abs/1308.6789}{{\ttfamily 1308.6789}}].

\bibitem{Abada:2021zcm}
A.~Abada, J.~Kriewald and A.~M. Teixeira, \emph{{On the role of leptonic CPV
  phases in cLFV observables}},
  \href{https://doi.org/10.1140/epjc/s10052-021-09754-w}{\emph{Eur. Phys. J. C}
  {\bfseries 81} (2021) 1016}
  [\href{https://arxiv.org/abs/2107.06313}{{\ttfamily 2107.06313}}].

\bibitem{Davidson:1978pm}
A.~Davidson, \emph{{$B-L$ as the fourth color within an $\mathrm{SU}(2)_L
  \times \mathrm{U}(1)_R \times \mathrm{U}(1)$ model}},
  \href{https://doi.org/10.1103/PhysRevD.20.776}{\emph{Phys. Rev. D} {\bfseries
  20} (1979) 776}.

\bibitem{Marshak:1979fm}
R.~E. Marshak and R.~N. Mohapatra, \emph{{Quark - Lepton Symmetry and B-L as
  the U(1) Generator of the Electroweak Symmetry Group}},
  \href{https://doi.org/10.1016/0370-2693(80)90436-0}{\emph{Phys. Lett. B}
  {\bfseries 91} (1980) 222}.

\bibitem{Buchmuller:1991ce}
W.~Buchmuller, C.~Greub and P.~Minkowski, \emph{{Neutrino masses, neutral
  vector bosons and the scale of B-L breaking}},
  \href{https://doi.org/10.1016/0370-2693(91)90952-M}{\emph{Phys. Lett. B}
  {\bfseries 267} (1991) 395}.

\bibitem{Basso:2008iv}
L.~Basso, A.~Belyaev, S.~Moretti and C.~H. Shepherd-Themistocleous,
  \emph{{Phenomenology of the minimal B-L extension of the Standard model: Z'
  and neutrinos}},
  \href{https://doi.org/10.1103/PhysRevD.80.055030}{\emph{Phys. Rev. D}
  {\bfseries 80} (2009) 055030}
  [\href{https://arxiv.org/abs/0812.4313}{{\ttfamily 0812.4313}}].

\bibitem{FileviezPerez:2009hdc}
P.~Fileviez~Perez, T.~Han and T.~Li, \emph{{Testability of Type I Seesaw at the
  CERN LHC: Revealing the Existence of the B-L Symmetry}},
  \href{https://doi.org/10.1103/PhysRevD.80.073015}{\emph{Phys. Rev. D}
  {\bfseries 80} (2009) 073015}
  [\href{https://arxiv.org/abs/0907.4186}{{\ttfamily 0907.4186}}].

\bibitem{Kang:2015uoc}
Z.~Kang, P.~Ko and J.~Li, \emph{{New Avenues to Heavy Right-handed Neutrinos
  with Pair Production at Hadronic Colliders}},
  \href{https://doi.org/10.1103/PhysRevD.93.075037}{\emph{Phys. Rev. D}
  {\bfseries 93} (2016) 075037}
  [\href{https://arxiv.org/abs/1512.08373}{{\ttfamily 1512.08373}}].

\bibitem{Cox:2017eme}
P.~Cox, C.~Han and T.~T. Yanagida, \emph{{LHC Search for Right-handed Neutrinos
  in $Z^\prime$ Models}},
  \href{https://doi.org/10.1007/JHEP01(2018)037}{\emph{JHEP} {\bfseries 01}
  (2018) 037} [\href{https://arxiv.org/abs/1707.04532}{{\ttfamily
  1707.04532}}].

\bibitem{Han:2021pun}
C.~Han, T.~Li and C.-Y. Yao, \emph{{Searching for heavy neutrino in terms of
  tau lepton at future hadron collider}},
  \href{https://doi.org/10.1103/PhysRevD.104.015036}{\emph{Phys. Rev. D}
  {\bfseries 104} (2021) 015036}
  [\href{https://arxiv.org/abs/2103.03548}{{\ttfamily 2103.03548}}].

\bibitem{ALEPH:2013dgf}
{\scshape ALEPH, DELPHI, L3, OPAL, LEP Electroweak} collaboration,
  \emph{{Electroweak Measurements in Electron-Positron Collisions at
  W-Boson-Pair Energies at LEP}},
  \href{https://doi.org/10.1016/j.physrep.2013.07.004}{\emph{Phys. Rept.}
  {\bfseries 532} (2013) 119}
  [\href{https://arxiv.org/abs/1302.3415}{{\ttfamily 1302.3415}}].

\bibitem{ATLAS:2019erb}
{\scshape ATLAS} collaboration, \emph{{Search for high-mass dilepton resonances
  using 139 fb$^{-1}$ of $pp$ collision data collected at $\sqrt{s}=$13 TeV
  with the ATLAS detector}},
  \href{https://doi.org/10.1016/j.physletb.2019.07.016}{\emph{Phys. Lett. B}
  {\bfseries 796} (2019) 68}
  [\href{https://arxiv.org/abs/1903.06248}{{\ttfamily 1903.06248}}].

\bibitem{CMS:2019tbu}
{\scshape CMS} collaboration, \emph{{Search for a narrow resonance in high-mass
  dilepton final states in proton-proton collisions using
  140$~\mathrm{fb}^{-1}$ of data at $\sqrt{s}=13~\mathrm{TeV}$}},
  {\emph{CMS-PAS-EXO-19-019} (2019) }.

\bibitem{Das:2021esm}
A.~Das, P.~S.~B. Dev, Y.~Hosotani and S.~Mandal, \emph{{Probing the minimal
  $U(1)_X$ model at future electron-positron colliders via the fermion
  pair-production channel}},
  \href{https://arxiv.org/abs/2104.10902}{{\ttfamily 2104.10902}}.

\bibitem{Appelquist:2002mw}
T.~Appelquist, B.~A. Dobrescu and A.~R. Hopper, \emph{{Nonexotic Neutral Gauge
  Bosons}}, \href{https://doi.org/10.1103/PhysRevD.68.035012}{\emph{Phys. Rev.
  D} {\bfseries 68} (2003) 035012}
  [\href{https://arxiv.org/abs/hep-ph/0212073}{{\ttfamily hep-ph/0212073}}].

\bibitem{Das:2017flq}
A.~Das, N.~Okada and D.~Raut, \emph{{Enhanced pair production of heavy Majorana
  neutrinos at the LHC}},
  \href{https://doi.org/10.1103/PhysRevD.97.115023}{\emph{Phys. Rev. D}
  {\bfseries 97} (2018) 115023}
  [\href{https://arxiv.org/abs/1710.03377}{{\ttfamily 1710.03377}}].

\bibitem{Faraggi:1996kk}
A.~E. Faraggi and M.~Masip, \emph{{Leptophobic Z-prime from superstring derived
  models}}, \href{https://doi.org/10.1016/S0370-2693(96)01187-2}{\emph{Phys.
  Lett. B} {\bfseries 388} (1996) 524}
  [\href{https://arxiv.org/abs/hep-ph/9604302}{{\ttfamily hep-ph/9604302}}].

\bibitem{GomezDumm:1997br}
D.~Gomez~Dumm, \emph{{Leptophobic character of the Z-prime in an SU(3)(C) x
  SU(3)(L) x U(1)(X) model}},
  \href{https://doi.org/10.1016/S0370-2693(97)00947-7}{\emph{Phys. Lett. B}
  {\bfseries 411} (1997) 313}
  [\href{https://arxiv.org/abs/hep-ph/9709245}{{\ttfamily hep-ph/9709245}}].

\bibitem{Malinsky:2005bi}
M.~Malinsky, J.~C. Romao and J.~W.~F. Valle, \emph{{Novel supersymmetric SO(10)
  seesaw mechanism}},
  \href{https://doi.org/10.1103/PhysRevLett.95.161801}{\emph{Phys. Rev. Lett.}
  {\bfseries 95} (2005) 161801}
  [\href{https://arxiv.org/abs/hep-ph/0506296}{{\ttfamily hep-ph/0506296}}].

\bibitem{Buckley:2011mm}
M.~R. Buckley, D.~Hooper and J.~L. Rosner, \emph{{A Leptophobic Z' And Dark
  Matter From Grand Unification}},
  \href{https://doi.org/10.1016/j.physletb.2011.08.014}{\emph{Phys. Lett. B}
  {\bfseries 703} (2011) 343}
  [\href{https://arxiv.org/abs/1106.3583}{{\ttfamily 1106.3583}}].

\bibitem{Sirunyan:2018xlo}
{\scshape CMS} collaboration, \emph{{Search for narrow and broad dijet
  resonances in proton-proton collisions at $ \sqrt{s}=13 $ TeV and constraints
  on dark matter mediators and other new particles}},
  \href{https://doi.org/10.1007/JHEP08(2018)130}{\emph{JHEP} {\bfseries 08}
  (2018) 130} [\href{https://arxiv.org/abs/1806.00843}{{\ttfamily
  1806.00843}}].

\bibitem{ATLAS:2019bov}
{\scshape ATLAS} collaboration, \emph{{Search for New Phenomena in Dijet Events
  using 139 fb$^{-1}$ of $pp$ collisions at $\sqrt{s}$ = 13TeV collected with
  the ATLAS Detector}}, {\emph{ATLAS-CONF-2019-007} (2019) }.

\bibitem{Alloul:2013bka}
A.~Alloul, N.~D. Christensen, C.~Degrande, C.~Duhr and B.~Fuks,
  \emph{{FeynRules 2.0 - A complete toolbox for tree-level phenomenology}},
  \href{https://doi.org/10.1016/j.cpc.2014.04.012}{\emph{Comput. Phys. Commun.}
  {\bfseries 185} (2014) 2250}
  [\href{https://arxiv.org/abs/1310.1921}{{\ttfamily 1310.1921}}].

\bibitem{Amrith:2018yfb}
S.~Amrith, J.~M. Butterworth, F.~F. Deppisch, W.~Liu, A.~Varma and D.~Yallup,
  \emph{{LHC Constraints on a $B-L$ Gauge Model using Contur}},
  \href{https://doi.org/10.1007/JHEP05(2019)154}{\emph{JHEP} {\bfseries 05}
  (2019) 154} [\href{https://arxiv.org/abs/1811.11452}{{\ttfamily
  1811.11452}}].

\bibitem{Alwall:2014hca}
J.~Alwall, R.~Frederix, S.~Frixione, V.~Hirschi, F.~Maltoni, O.~Mattelaer
  et~al., \emph{{The automated computation of tree-level and next-to-leading
  order differential cross sections, and their matching to parton shower
  simulations}}, \href{https://doi.org/10.1007/JHEP07(2014)079}{\emph{JHEP}
  {\bfseries 07} (2014) 079} [\href{https://arxiv.org/abs/1405.0301}{{\ttfamily
  1405.0301}}].

\bibitem{FASER:2018eoc}
{\scshape FASER} collaboration, \emph{{FASER\textquoteright{}s physics reach
  for long-lived particles}},
  \href{https://doi.org/10.1103/PhysRevD.99.095011}{\emph{Phys. Rev. D}
  {\bfseries 99} (2019) 095011}
  [\href{https://arxiv.org/abs/1811.12522}{{\ttfamily 1811.12522}}].

\bibitem{Curtin:2018mvb}
D.~Curtin et~al., \emph{{Long-Lived Particles at the Energy Frontier: The
  MATHUSLA Physics Case}},
  \href{https://doi.org/10.1088/1361-6633/ab28d6}{\emph{Rept. Prog. Phys.}
  {\bfseries 82} (2019) 116201}
  [\href{https://arxiv.org/abs/1806.07396}{{\ttfamily 1806.07396}}].

\bibitem{Das:2019fee}
A.~Das, P.~S.~B. Dev and N.~Okada, \emph{{Long-lived TeV-scale right-handed
  neutrino production at the LHC in gauged $U(1)_X$ model}},
  \href{https://doi.org/10.1016/j.physletb.2019.135052}{\emph{Phys. Lett. B}
  {\bfseries 799} (2019) 135052}
  [\href{https://arxiv.org/abs/1906.04132}{{\ttfamily 1906.04132}}].

\bibitem{Chiang:2019ajm}
C.-W. Chiang, G.~Cottin, A.~Das and S.~Mandal, \emph{{Displaced heavy neutrinos
  from $Z'$ decays at the LHC}},
  \href{https://doi.org/10.1007/JHEP12(2019)070}{\emph{JHEP} {\bfseries 12}
  (2019) 070} [\href{https://arxiv.org/abs/1908.09838}{{\ttfamily
  1908.09838}}].

\bibitem{Bray:2007ru}
S.~Bray, J.~S. Lee and A.~Pilaftsis, \emph{{Resonant CP violation due to heavy
  neutrinos at the LHC}},
  \href{https://doi.org/10.1016/j.nuclphysb.2007.07.002}{\emph{Nucl. Phys. B}
  {\bfseries 786} (2007) 95}
  [\href{https://arxiv.org/abs/hep-ph/0702294}{{\ttfamily hep-ph/0702294}}].

\bibitem{FileviezPerez:2014lnj}
P.~Fileviez~Perez, S.~Ohmer and H.~H. Patel, \emph{{Minimal Theory for
  Lepto-Baryons}},
  \href{https://doi.org/10.1016/j.physletb.2014.06.057}{\emph{Phys. Lett. B}
  {\bfseries 735} (2014) 283}
  [\href{https://arxiv.org/abs/1403.8029}{{\ttfamily 1403.8029}}].

\bibitem{Schechter:1981bd}
J.~Schechter and J.~W.~F. Valle, \emph{{Neutrinoless Double beta Decay in SU(2)
  x U(1) Theories}},
  \href{https://doi.org/10.1103/PhysRevD.25.2951}{\emph{Phys. Rev. D}
  {\bfseries 25} (1982) 2951}.

\bibitem{Kotila:2012zza}
J.~Kotila and F.~Iachello, \emph{{Phase space factors for double-$\beta$
  decay}}, \href{https://doi.org/10.1103/PhysRevC.85.034316}{\emph{Phys. Rev.
  C} {\bfseries 85} (2012) 034316}
  [\href{https://arxiv.org/abs/1209.5722}{{\ttfamily 1209.5722}}].

\bibitem{Neacsu:2015uja}
A.~Neacsu and M.~Horoi, \emph{{An effective method to accurately calculate the
  phase space factors for $\beta^- \beta^-$ decay}},
  \href{https://doi.org/10.1155/2016/7486712}{\emph{Adv. High Energy Phys.}
  {\bfseries 2016} (2016) 7486712}
  [\href{https://arxiv.org/abs/1510.00882}{{\ttfamily 1510.00882}}].

\bibitem{Fang:2018tui}
D.-L. Fang, A.~Faessler and F.~Simkovic, \emph{{0\ensuremath{\nu}$\beta\beta$
  -decay nuclear matrix element for light and heavy neutrino mass mechanisms
  from deformed quasiparticle random-phase approximation calculations for
  $^{76}$Ge, $^{82}$Se, $^{130}$Te, $^{136}$Xe , and $^{150}$Nd with isospin
  restoration}}, \href{https://doi.org/10.1103/PhysRevC.97.045503}{\emph{Phys.
  Rev. C} {\bfseries 97} (2018) 045503}
  [\href{https://arxiv.org/abs/1803.09195}{{\ttfamily 1803.09195}}].

\bibitem{Ejiri:2020xmm}
H.~Ejiri, \emph{{Neutrino-Mass Sensitivity and Nuclear Matrix Element for
  Neutrinoless Double Beta Decay}},
  \href{https://doi.org/10.3390/universe6120225}{\emph{Universe} {\bfseries 6}
  (2020) 225}.

\bibitem{Racah:1937qq}
G.~Racah, \emph{{On the symmetry of particle and antiparticle}},
  \href{https://doi.org/10.1007/BF02961321}{\emph{Nuovo Cim.} {\bfseries 14}
  (1937) 322}.

\bibitem{Furry:1939qr}
W.~H. Furry, \emph{{On transition probabilities in double
  beta-disintegration}},
  \href{https://doi.org/10.1103/PhysRev.56.1184}{\emph{Phys. Rev.} {\bfseries
  56} (1939) 1184}.

\bibitem{KamLAND-Zen:2022tow}
{\scshape KamLAND-Zen} collaboration, \emph{{First Search for the Majorana
  Nature of Neutrinos in the Inverted Mass Ordering Region with KamLAND-Zen}},
  \href{https://arxiv.org/abs/2203.02139}{{\ttfamily 2203.02139}}.

\bibitem{EXO-200:2019rkq}
{\scshape EXO-200} collaboration, \emph{{Search for Neutrinoless Double-$\beta$
  Decay with the Complete EXO-200 Dataset}},
  \href{https://doi.org/10.1103/PhysRevLett.123.161802}{\emph{Phys. Rev. Lett.}
  {\bfseries 123} (2019) 161802}
  [\href{https://arxiv.org/abs/1906.02723}{{\ttfamily 1906.02723}}].

\bibitem{Agostini:2020xta}
{\scshape GERDA} collaboration, \emph{{Final Results of GERDA on the Search for
  Neutrinoless Double-$\beta$ Decay}},
  \href{https://doi.org/10.1103/PhysRevLett.125.252502}{\emph{Phys. Rev. Lett.}
  {\bfseries 125} (2020) 252502}
  [\href{https://arxiv.org/abs/2009.06079}{{\ttfamily 2009.06079}}].

\bibitem{Adams:2021rbc}
{\scshape CUORE} collaboration, \emph{{High sensitivity neutrinoless
  double-beta decay search with one tonne-year of CUORE data}},
  \href{https://arxiv.org/abs/2104.06906}{{\ttfamily 2104.06906}}.

\bibitem{NEMO-3:2019gwo}
{\scshape NEMO-3} collaboration, \emph{{Detailed studies of $^{100}$Mo
  two-neutrino double beta decay in NEMO-3}},
  \href{https://doi.org/10.1140/epjc/s10052-019-6948-4}{\emph{Eur. Phys. J. C}
  {\bfseries 79} (2019) 440}
  [\href{https://arxiv.org/abs/1903.08084}{{\ttfamily 1903.08084}}].

\bibitem{nEXO:2021ujk}
{\scshape nEXO} collaboration, \emph{{nEXO: neutrinoless double beta decay
  search beyond 10$^{28}$ year half-life sensitivity}},
  \href{https://doi.org/10.1088/1361-6471/ac3631}{\emph{J. Phys. G} {\bfseries
  49} (2022) 015104} [\href{https://arxiv.org/abs/2106.16243}{{\ttfamily
  2106.16243}}].

\bibitem{LEGEND:2021bnm}
{\scshape LEGEND} collaboration, \emph{{The Large Enriched Germanium Experiment
  for Neutrinoless $\beta\beta$ Decay}: {LEGEND-1000 Preconceptual Design
  Report}},  \href{https://arxiv.org/abs/2107.11462}{{\ttfamily 2107.11462}}.

\bibitem{Avasthi:2021lgy}
A.~Avasthi et~al., \emph{{Kilotonne-scale xenon detectors for neutrinoless
  double beta decay and other new physics searches}},
  \href{https://arxiv.org/abs/2110.01537}{{\ttfamily 2110.01537}}.

\bibitem{Chauhan:2021get}
G.~Chauhan, \emph{{Probing New Physics Beyond the Standard Model via New
  Neutrino Interactions}}, Ph.D. thesis, Washington U., St. Louis, 2021.

\end{thebibliography}\endgroup

\end{document}